%% file: main.tex
\documentclass{article}
\usepackage{iclr2027/iclr2027_conference,times}
\usepackage{etoolbox}
\input{iclr2027/math_commands.tex}

\input{supplement/build/chain_numbers.tex}

\input{supplement/build/attack_chain.tex}
\input{supplement/build/sched_numbers.tex}

\input{supplement/build/gdep.tex}

\input{supplement/build/table1.tex}

\input{supplement/build/tarez_bb.tex}

\input{supplement/build/extras.tex}

\input{supplement/build/reversal_numbers.tex}
\input{supplement/build/budget_numbers.tex}
\input{supplement/build/llm_ops.tex}
\newif\ifcifarhundred\cifarhundredfalse \newif\ifsfour\sfourfalse \newif\ifsfive\sfivefalse \newif\ifssix\ssixfalse
\newif\ifsseven\ssevenfalse \newif\ifsnine\sninefalse \newif\ifsten\stenfalse \newif\ifseleven\selevenfalse \newif\ifstwelve\stwelvefalse \newif\ifsthirteen\sthirteenfalse \newif\ifssixteen\ssixteenfalse \newif\ifsseventeen\sseventeenfalse \newif\ifsztwenty\sztwentyfalse \newif\ifsrtwentyone\srtwentyonefalse \newif\ifsdptwentytwo\sdptwentytwofalse \newif\ifsedtwentythree\sedtwentythreefalse
\IfFileExists{supplement/sched/build/csched_numbers.tex}{\input{supplement/sched/build/csched_numbers.tex}}{}  %
\IfFileExists{supplement/sched/build/leaderboard_numbers.tex}{\input{supplement/sched/build/leaderboard_numbers.tex}}{}
\IfFileExists{supplement/sched/build/fixedpoint_lb_numbers.tex}{\input{supplement/sched/build/fixedpoint_lb_numbers.tex}}{}
\IfFileExists{supplement/sched/build/paper_extras.tex}{

\input{supplement/sched/build/paper_extras.tex}}{}
\IfFileExists{supplement/sched/build/audit_numbers.tex}{\input{supplement/sched/build/audit_numbers.tex}}{}
\IfFileExists{supplement/sched/build/rerank_lb_numbers.tex}{

\input{supplement/sched/build/rerank_lb_numbers.tex}}{}
\IfFileExists{supplement/sched/build/sched2_numbers.tex}{\input{supplement/sched/build/sched2_numbers.tex}}{}
\IfFileExists{supplement/sched/build/sched3_numbers.tex}{\input{supplement/sched/build/sched3_numbers.tex}}{}
\IfFileExists{supplement/sched/build/sched4_numbers.tex}{\input{supplement/sched/build/sched4_numbers.tex}}{}
\IfFileExists{supplement/sched/build/sched5_numbers.tex}{\input{supplement/sched/build/sched5_numbers.tex}}{}
\IfFileExists{supplement/sched/build/sched6_numbers.tex}{\input{supplement/sched/build/sched6_numbers.tex}}{}
\IfFileExists{supplement/sched/build/sched7_numbers.tex}{\input{supplement/sched/build/sched7_numbers.tex}}{}
\IfFileExists{supplement/sched/build/sched9_numbers.tex}{\input{supplement/sched/build/sched9_numbers.tex}}{}
\IfFileExists{supplement/sched/build/sched10_numbers.tex}{\input{supplement/sched/build/sched10_numbers.tex}}{}
\IfFileExists{supplement/sched/build/sched11_numbers.tex}{\input{supplement/sched/build/sched11_numbers.tex}}{}
\IfFileExists{supplement/sched/build/sched12_numbers.tex}{\input{supplement/sched/build/sched12_numbers.tex}}{}
\IfFileExists{supplement/sched/build/sched13_numbers.tex}{\input{supplement/sched/build/sched13_numbers.tex}}{}
\IfFileExists{supplement/sched/build/sched16_numbers.tex}{\input{supplement/sched/build/sched16_numbers.tex}}{}
\IfFileExists{supplement/sched/build/sched17_numbers.tex}{\input{supplement/sched/build/sched17_numbers.tex}}{}
\IfFileExists{supplement/sched/build/sched20_numbers.tex}{\input{supplement/sched/build/sched20_numbers.tex}}{}
\IfFileExists{supplement/sched/build/sched21_numbers.tex}{\input{supplement/sched/build/sched21_numbers.tex}}{}
\IfFileExists{supplement/sched/build/sched22_numbers.tex}{\input{supplement/sched/build/sched22_numbers.tex}}{}
\IfFileExists{supplement/sched/build/sched23_numbers.tex}{\input{supplement/sched/build/sched23_numbers.tex}}{}
\newcommand{\sedMainSentence}{With all eight defenses on both arms (registered; Appendix~\ref{app:eight}), \sedPseventynineFlipsAll\ pairs of the benchmark's own WaNet DER order change with the patch (\sedPseventynineFlipListAll), none separable seed by seed on the shipped victim (\sedFlipSepShipN; Table~\ref{tab:eightder}).}

\newcommand{\srtMainSentence}{At $5\%$ poisoning the crossover replicates (registered; six intervals excluding zero); at $1\%$ the victims mostly fail to install, so those arms are void and all four registered predictions count as failed (Appendix~\ref{app:rates}).}

\usepackage[hidelinks]{hyperref}
\hypersetup{pdftitle={TARE: Weigh a Never-Poisoned Twin Before Reading Backdoor-Defense Costs}}
\usepackage[capitalize]{cleveref}
\crefname{section}{Section}{Sections}\Crefname{section}{Section}{Sections}\crefname{table}{Table}{Tables}\crefname{figure}{Figure}{Figures}
\crefname{appendix}{Appendix}{Appendices}\Crefname{appendix}{Appendix}{Appendices}
\usepackage{url}
\usepackage{booktabs}
\usepackage{amsmath,amssymb}
\usepackage{graphicx}
\usepackage{rotating}  %
\usepackage{tikz}
\usetikzlibrary{positioning,decorations.pathreplacing}
\usepackage{enumitem}
\usepackage{placeins}
\usepackage{algorithm}
\usepackage{algpseudocode}
\usepackage{xcolor}
\usepackage{pifont}
\definecolor{cRem}{HTML}{750014}   %
\definecolor{cDial}{HTML}{A31F34}  %
\definecolor{cRep}{HTML}{626A73}   %
\definecolor{cTax}{HTML}{40464C}   %
\definecolor{cTaxL}{HTML}{B8C2CC}  %
\definecolor{cUp}{HTML}{750014}    %

\iclrfinalcopy
\makeatletter
\patchcmd{\@maketitle}{Published as a conference paper at ICLR 2027}{Preprint}{}{\errmessage{arXiv header patch failed}}
\patchcmd{\@maketitle}{\rule{\z@}{24pt}}{\rule{\z@}{10pt}}{}{\errmessage{arXiv author-strut patch failed}}
\patchcmd{\@maketitle}{\rule{\z@}{24pt}}{\rule{\z@}{10pt}}{}{\errmessage{arXiv author-strut patch failed}}
\patchcmd{\@maketitle}{\rule{\z@}{24pt}}{\rule{\z@}{10pt}}{}{\errmessage{arXiv author-strut patch failed}}
\makeatother

\title{TARE: Weigh a Never-Poisoned Twin Before Reading Backdoor-Defense Costs}

\input{authors.tex}

\newcommand{\cRem}{\ensuremath{c_{\mathrm{int}}}}
\newcommand{\cDef}{\ensuremath{c_{\mathrm{def}}}}
\newcommand{\Q}{Q}\newcommand{\pt}{\,pt}\newcommand{\ms}{\texttt{MultiStepLR}}\newcommand{\cosA}{\texttt{CosineAnnealingLR}}\newcommand{\yaml}[1]{\path{#1}}
\newcount\preregN \preregN=5 \ifsfive\advance\preregN 1\fi \ifssix\advance\preregN 1\fi \ifsseven\advance\preregN 1\fi \ifsnine\advance\preregN 1\fi \ifsten\advance\preregN 1\fi \ifseleven\advance\preregN 1\fi \ifstwelve\advance\preregN 1\fi \ifsthirteen\advance\preregN 1\fi \ifssixteen\advance\preregN 1\fi \ifdefined\schPfortyeight\advance\preregN 1\fi \ifdefined\stePfifty\advance\preregN 1\fi \ifsseventeen\advance\preregN 2\fi \ifdefined\sbxPsixtyfour\advance\preregN 1\fi \ifsztwenty\advance\preregN 1\fi \ifsrtwentyone\advance\preregN 1\fi \ifsdptwentytwo\advance\preregN 1\fi \ifsedtwentythree\advance\preregN 1\fi \advance\preregN 2 %
\newcommand{\preregWord}{\ifcase\preregN\or\or\or\or\or five\or six\or seven\or eight\or nine\or ten\or eleven\or twelve\or thirteen\or fourteen\or fifteen\or sixteen\or seventeen\or eighteen\or nineteen\or twenty\or twenty-one\or twenty-two\or twenty-three\or twenty-four\or twenty-five\else\the\preregN\fi}

\begin{document}
\maketitle

\begin{abstract}
Backdoor-defense leaderboards print a clean-accuracy drop and read it as the cost of removal. The drop is measured on the poisoned victim alone, so it cannot separate removal from what the defense does to \emph{any} model, and it inherits where the victim started. Where the victim starts is, for three of BackdoorBench's sixteen attacks, a configuration file: WaNet, BPP and Input-Aware ship a \texttt{MultiStepLR} whose first milestone is the epoch count, so their victims never anneal and are the lowest-accuracy victims in \lbCifarLow\ public CIFAR cells at $\le5\%$. On PreAct-ResNet18, fine-tuning-family defenses return a low start to a level of their own, so on those victims the published cost is negative, the benchmark's rating clips the ``gain'' to zero, and \auLoadBearing\ of the \auFulltext\ citing defense papers we read rest a no-cost claim on those cells; re-run with their released code, TSBD and CGD ``gain'' on a never-poisoned model too. We isolate the cause with a $2\times2$ that edits only that scheduler line, its swapped arms self-registered before they ran: the sign of the fine-tuning family's clean-model cost reverses both ways while its published gain on the annealed victim only shrinks toward zero, \schAgreeSeed\ seeds following the schedule, replicated on BPP, FT-SAM, CIFAR-100 and VGG19-BN and induced in a second toolkit. The instrument is \textbf{TARE}: run the identical defense on a never-poisoned \emph{twin} of the same recipe, schedule and seed (on BackdoorBench, $\le10$ poisoned images, admitted only below $5\%$ attack success); what the twin loses is the tare. On BackdoorBench's BadNets grid seven of the eight defenses charge the twin (Neural Cleanse only where its detector fires), \exBbCleanMinPt\ points (\exBbCleanMinName) to \exBbCleanMaxPt\ (\exBbCleanMaxName); the eighth, ABL, destroys it. Within an attack the start cancels from rankings, so the tare re-orders nothing there; what poisoning adds beyond it is printed under two estimators and not corrected, its removal share being unidentified. We ship the three-key patch, a signed tare column over $7$ attacks $\times$ $8$ defenses of the CIFAR-10 roster, and TARE-Z, a twin-free estimator for seed-stable defenses.
\end{abstract}

\section{Introduction}
\label{sec:intro}

BackdoorBench scores a defense by two columns, the attack success left after it and the clean accuracy it
costs; the second is measured on the poisoned victim alone and read as the price of removal. Where the victim
starts is, for three of the benchmark's sixteen attacks, a configuration file: WaNet, BPP
and Input-Aware ship a \texttt{MultiStepLR} whose first milestone is the epoch count, so their victims never anneal (Section~\ref{sec:block}), and on the public leaderboard they are the lowest-accuracy victims in
\lbCifarLow\ CIFAR cells at rates $\le5\%$. For fine-tuning and NAD the accuracy \emph{after} the
defense is a level ($L_{\mathcal D}$, Section~\ref{sec:bias}) of the defense's own (Figure~\ref{fig:level}): every never-annealed or self-damaged CIFAR-10 / PreAct-ResNet18 victim at $10\%$,
LIRA's at \fpLbFtTenUndefLo\% included, is returned to one \fpLbFtTenDefLo--\fpLbFtTenDefHi\% band
while a healthy annealed one is not lifted (BadNets, starting at \fpLbBadnetTenUndef\%, as low as WaNet, is lowered). So the published cost on such a victim is its start minus a constant that belongs
to the defense, booked as a \emph{gain} where the level lies above the start: the fine-tuning
family ``gains'' on the three never-annealed victims and, in the single-seed leaderboard cells, on no
healthy annealed one; the defense effectiveness rating's (DER) $\max(0,\cdot)$ clip zeroes it there; and two of the
\auFulltext\ defense papers we read in full carry a no-cost claim resting decisively on those cells
(TSBD, CGD; a third, PDB, correlationally).

\begin{figure}[t]
\centering
\includegraphics[width=\textwidth]{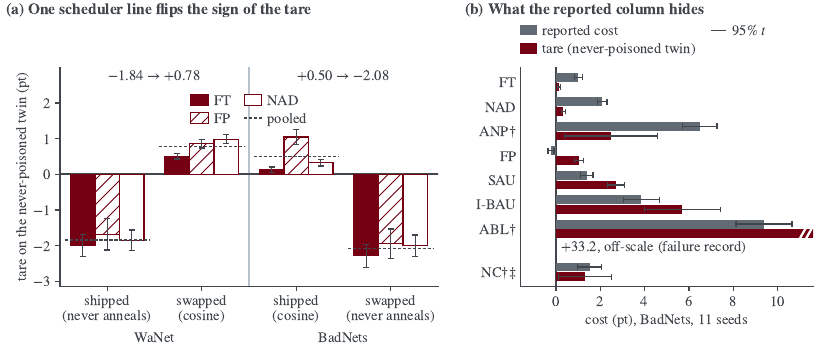}
\caption{\textbf{(a) One \texttt{lr\_scheduler} line reverses the tare's sign both ways} (FT, FP, NAD on never-poisoned twins; Table~\ref{tab:sched}; dashed: pooled). \textbf{(b) What the reported column hides}: reported cost against tare, BadNets grid in Table~\ref{tab:bbmain}'s order ($\dagger$, $\ddagger$ as there; ABL off-scale, a failure record). Whiskers: $95\%$ $t$-intervals over seeds.}
\label{fig:teaser}
\end{figure}

Edit that one \texttt{lr\_scheduler} line and the clean-accuracy cost a
defense charges a never-poisoned twin changes sign both ways (Figure~\ref{fig:teaser}a): \schAgree\ (seed, defense) cells follow the schedule, not the attack. Within one attack every
defense shares the victim, so by construction the victim's start cancels from the column's own rankings; it enters wherever the column is read alone---``improves over No Defense'',
DER's $\max(0,\cdot)$ clip (Section~\ref{sec:contam}), cross-attack comparisons---and there we trace it. Run the identical defense on a never-poisoned twin and it, too, is charged (Table~\ref{tab:bbmain}). Such a twin has been run before (Table~\ref{tab:prior}), reported beside the cost, never subtracted from it.

Separating where the victim started from what poisoning adds to the bill needs a control that none of the six
protocols we audit (Table~\ref{tab:lit}) trains: a never-poisoned twin of the same recipe, schedule and seed, put through the
identical defense. That is \textbf{TARE}, named for the weight a scale subtracts because it belongs to the
container: what the twin loses is the tare, and what the poisoned arm loses beyond it is the
poisoning-by-defense interaction, not trigger erasure
(Section~\ref{sec:estimand}). The paired contrast differs from it by the pre-defense gap $g$,
and which share of either is removal is not identified, so we print both beside the reported cost and correct nothing (Table~\ref{tab:bbmain}). Our contributions:
\begin{itemize}[leftmargin=1.2em,itemsep=0pt,topsep=1pt,parsep=0pt]
\item We find that three of BackdoorBench's sixteen attacks ship a \texttt{MultiStepLR} that never fires, and show with a $2\times2$ that edits only that line, its swapped arms registered, self-hosted, before they ran, that the tare's sign reverses both ways (Table~\ref{tab:sched}, Figure~\ref{fig:teaser}a), replicated on BPP, FT-SAM, CIFAR-100\ifsseventeen, VGG19-BN and, induced, in a second toolkit\else{} and VGG19-BN\fi\ifsrtwentyone, and at $5\%$ poisoning ($1\%$ arms voided, predictions recorded failed; Appendix~\ref{app:rates})\fi.
\item We trace the artifact through the benchmark's rating into the papers that cite it: DER's clip hides the sign of these costs in \lbDerClipNever\ never-annealed
and \lbDerClipAnn\ healthy annealed single-seed leaderboard cells (descriptive, Appendix~\ref{app:fplb}); TSBD and CGD, re-run with their code, ``gain'' on a
never-poisoned twin (\stTsbdShare\ and \stCgdShare\ of the victims' gains), and dropping or re-referencing two cells flips TSBD's headline average; \auLoadBearing\ of \auFulltext\ defense papers read in full (one coder) rest a no-cost claim
decisively on those cells.
\item We propose TARE and measure its tare: seven of the eight defenses on BackdoorBench's BadNets grid charge a never-poisoned twin
\exBbCleanMinPt\ to \exBbCleanMaxPt\ points (\exBbCleanExclBonfNoAbl\ at $0.05/7$, \exBbCleanExclBHNoAbl\ under Benjamini--Hochberg $q=0.05$, Section~\ref{sec:units}'s rule; Figure~\ref{fig:teaser}b), more than the whole reported cost for \exBbExceedPosNoAblNames; ABL destroys \steAblTwinCollapseAdmit\ of Table~\ref{tab:corrmain}'s twins, a failure record\ifsdptwentytwo; under a threshold rule that never sees attack success, ANP's tare rises from \sdpBadnetTareOPt\ to \sdpBadnetTareDPt\fi. We measure the accounting $\text{cost}=\text{start}-\text{level}+\text{residual}$ on twins and report its
registered failures (Section~\ref{sec:replic}).
\item We ship the three-key patch, a signed tare column over \steAdmitN\ attacks $\times$ $8$
defenses (re-ranking nothing within an attack, by construction; Table~\ref{tab:corrmain}), TARE-Z, a twin-free estimator whose mean $|$error$|$ (Eq.~\ref{eq:surr}) is
\tzBbSmallMaePLo--\tzBbSmallMaePHi\ points where a defense's level is seed-stable, and an archive that regenerates every number.
\end{itemize}

\section{Related Work}
\label{sec:related}

\paragraph{Weight-space backdoor defenses and the score that ranks them.}
Weight-editing defenses---pruning (FP), fine-tuning (FT), masking (ANP), distillation (NAD), trigger inversion (NC), unlearning (ABL, I-BAU, SAU) and others
\citep{liu2018finepruning,zhang2023ftsam,zhang2022finemixing,wu2021anp,li2021nad,wang2019neuralcleanse,li2021abl,zeng2022ibau,wei2023sau,peng2025tsc,tong2025gms,pawlak2025bv}---all
report a clean-accuracy cost beside an attack-success drop, which the benchmark that scores them \citep{wu2024backdoorbench} folds into one rating \citep{zhang2023ftsam}.

\paragraph{Measurement audits of machine-learning benchmarks.}
This paper is a measurement audit and asks to be judged by that genre's standards, from tuned-baseline audits to \emph{rankings} and \emph{confounds}
\citep{dacrema2019recsys,musgrave2020metric,schaeffer2023emergent,bouthillier2021variance,bean2025construct,singh2025leaderboard,qian2026benchmark2,gordienko2026rig,neuhof2026rankci,su2025delimiter,schaeffer2026contamination,harmon2025forgetting}.
\citet{nasr2025attacker} audit the \emph{other} column of the same table. We follow
\citet{bouthillier2021variance}'s structure (estimand, ideal estimator, practice with one arm removed, bias quantified) with a same-recipe control arm.

\paragraph{The gain-score problem.} The tuned-baseline audits above
share the shape, not the mechanism: there the weak reference is a competing method run under a weaker
recipe, and equalizing the recipe closes the gap; here the weak reference is the defense's own input, and
equalizing the recipe only moves the cost toward zero (Section~\ref{sec:block})
because, at the benchmark's FT configuration, fine-tuning and NAD return every never-annealed or self-damaged CIFAR-10 victim to their level. That is the pretest--posttest gain-score
problem \citep{lord1967paradox,cronbach1970change}: a treatment that resets its input
makes before-minus-after the pretest minus a constant (Figure~\ref{fig:level}),
and the reference decides the sign; \citet{schaeffer2023emergent} describe the nonlinear-metric artifact DER's $\max(0,\cdot)$ is.

\paragraph{Machine unlearning, which already ships the control we ask for.}
Unlearning evaluates removal against a from-scratch retrain \citep{maini2024tofu,shi2024muse,zhang2024npo}
and audits its own metrics \citep{kim2026forgetting,cho2025fade,khan2026metric}, yet still reads utility drop against the \emph{pre-unlearning}
model \citep{su2026preunlearn} (Appendix~\ref{app:unlearn}).
\textbf{We claim no priority on the observation itself}---a systematization ran a no-attack arm
and recommended this control \citep{sok2025backdoor}, two defenses report the backdoor-free drop
\citep{peng2025tsc,tong2025gms}, and an LLM study attributes its capability drop to ``the cost of
continual pretraining itself'' \citep{bouger2026unlearning}---\textbf{only on what is done with it
(Table~\ref{tab:prior}): an estimand with identification conditions, a same-recipe control in vision,
the start gap's cause isolated and traced through the scoring function into citing papers, and a
twin-free estimator with a stated domain of validity.}

\section{Method}
\label{sec:protocol}

\subsection{The Estimand and Its Two Estimators}
\label{sec:estimand}
Fix a capability metric $Q$ (clean test accuracy) and an
attack-success metric $A$. An \emph{installation} $\omega$ collects everything the evaluator does not
choose---base checkpoint, data order, poison placement, the defense's own stochasticity---and defines a
poisoning run $\theta_{\mathrm{pois}}(\omega)$ and its \emph{clean twin} $\theta_{\mathrm{twin}}(\omega)$, the identical run with
each poisoned example replaced by its unmodified original, so that everything but the poison matches; both pass through the same defense $\mathcal{D}_{b}$ at the
defender's clean-data budget $b$ (Algorithm~\ref{alg:tare} and Figure~\ref{fig:estimand},
Appendix~\ref{app:alg}). On BackdoorBench, which refuses a zero rate, $\theta_{\mathrm{twin}}$ is the attack's own pipeline at
poisoning ratio (\texttt{pratio}) $0.0002$ (ten images; one, at $0.00002$, where ten install), same seed and schedule, admitted only if it
passes A4's $5\%$ gate; ``never-poisoned'' means this\ifsztwenty{} (a zero-image twin, guard bypassed, moves the tare by at most \szrTareDiffMaxAbs\ points, one of nine intervals excluding zero; Appendix~\ref{app:zero})\fi. Sharing the seed fixes recipe, schedule and
data order, not the realized model, so pairing does not narrow the intervals (Appendix~\ref{app:twin}). The cost column is \emph{named} for the extra capability the defense costs
\emph{because the model was poisoned}, holding the defense fixed:
\begin{equation}
\tau_{\mathrm{int}}(\mathcal{D}_b) \;=\;
  \mathbb{E}_{\omega}\!\Bigl[\bigl(Q(\theta_{\mathrm{pois}})-Q(\mathcal{D}_b\theta_{\mathrm{pois}})\bigr)
  -\bigl(Q(\theta_{\mathrm{twin}})-Q(\mathcal{D}_b\theta_{\mathrm{twin}})\bigr)
  \;\Big|\; A(\mathcal{D}_b\theta_{\mathrm{pois}}(\omega))\le\alpha\Bigr],
\label{eq:estimand}
\end{equation}
conditioned on removal, so the poisoned arm is one whose backdoor is gone (its rate in Appendix~\ref{app:a3}; the tare
needs no such gate); it is the total effect of poisoning on the defense's bill. The price of erasure,
$\tau_{\mathrm{rem}}=\mathbb{E}_{\omega}[Q(\mathcal{D}_b\theta_{\mathrm{pois}})-Q(\mathcal{D}_b\theta^{\circ})]$ with $\theta^{\circ}$ a never-poisoned model at the
poisoned arm's start, is not identified: $\theta^{\circ}$ depends on how the start was lowered, and two routes give it opposite signs (Appendix~\ref{app:headroom}). The literature reports the drop from a reporting
baseline; the twin adds two estimators:
\begin{align}
\text{reported (defense) cost}\quad
  &\cDef \;=\; Q(\theta_{\mathrm{base}}) - Q(\mathcal{D}_b\theta_{\mathrm{pois}}), \label{eq:cdef}\\
\text{interaction (DiD)}\quad
  &\cRem^{\mathrm{DiD}} \;=\; \bigl[Q(\theta_{\mathrm{pois}})-Q(\mathcal{D}_b\theta_{\mathrm{pois}})\bigr]
   -\bigl[Q(\theta_{\mathrm{twin}})-Q(\mathcal{D}_b\theta_{\mathrm{twin}})\bigr], \label{eq:did}\\
\text{paired contrast}\quad
  &\cRem \;=\; Q(\mathcal{D}_b\theta_{\mathrm{twin}}) - Q(\mathcal{D}_b\theta_{\mathrm{pois}})
  \;=\; \cRem^{\mathrm{DiD}} + g, \label{eq:crem}
\end{align}
with $g = Q(\theta_{\mathrm{twin}})-Q(\theta_{\mathrm{pois}})$ the pre-defense gap; the four cells are
\par\nopagebreak\vspace{2pt}\noindent{\centering\small\setlength{\tabcolsep}{5pt}\renewcommand{\arraystretch}{0.95}
\begin{tabular}{lcc@{\hskip 10pt}l}
 & before, $Q(\theta)$ & after, $Q(\mathcal{D}_b\theta)$ & row difference \\
poisoned $\theta_{\mathrm{pois}}$ & $Q(\theta_{\mathrm{pois}})$ & $Q(\mathcal{D}_b\theta_{\mathrm{pois}})$ & reported cost $\cDef$ (drift zero) \\
twin $\theta_{\mathrm{twin}}$ & $Q(\theta_{\mathrm{twin}})$ & $Q(\mathcal{D}_b\theta_{\mathrm{twin}})$ & tare $c_{\mathrm{clean}}$ \\
\end{tabular}\par}\vspace{2pt}
\noindent the column differences being $g$ before and the paired contrast after, and the difference of the two row differences the DiD. Eq.~\ref{eq:did}
identifies Eq.~\ref{eq:estimand} under \textbf{(A1)} the arms differ only in the poison, \textbf{(A2)}
$\mathcal{D}_b$ is the same map on both arms, \textbf{(A3)} removal is achieved,
$A(\mathcal{D}_b\theta_{\mathrm{pois}})\!\le\!\alpha=0.05$, and \textbf{(A4)} the twin is admissible,
$A(\theta_{\mathrm{twin}})\!\approx\!0$ under the same rule (the tare needs only A2 and A4; A2--A4 fail where Section~\ref{sec:admiss} says). The two
estimators are one identity apart, and where $g\neq0$ which part of the paired contrast is removal is
not identified: a starting gap alone produces an interaction of either sign, depending on how the gap was
made (Appendix~\ref{app:headroom}). \textbf{TARE prints the tare
$c_{\mathrm{clean}}=Q(\theta_{\mathrm{twin}})-Q(\mathcal{D}_b\theta_{\mathrm{twin}})$ and the paired contrast beside $\cDef$, and corrects nothing.}

\subsection{What the Reported Cost Decomposes Into}
\label{sec:bias}
Adding and subtracting the twin's two states splits the reported cost exactly,
\begin{equation}
\cDef \;=\; \underbrace{\cRem^{\mathrm{DiD}}}_{\text{interaction}}
 \;+\; \underbrace{\bigl[Q(\theta_{\mathrm{twin}})-Q(\mathcal{D}_b\theta_{\mathrm{twin}})\bigr]}_{\text{tare } c_{\mathrm{clean}}}
 \;+\; \underbrace{\bigl[Q(\theta_{\mathrm{base}})-Q(\theta_{\mathrm{pois}})\bigr]}_{\text{reporting-baseline drift}},
\label{eq:decomp2}
\end{equation}
the drift being zero in every headline here ($\theta_{\mathrm{base}}=\theta_{\mathrm{pois}}$). \textbf{The fixed-point accounting (Figure~\ref{fig:level}).} For a defense whose defended level
$L_{\mathcal D}$ does not depend on where the model started, $Q(\mathcal{D}_b\theta)=L_{\mathcal
D}+\varepsilon(\theta)$ on either arm, $L_{\mathcal D}$ its fixed point, so
$c_{\mathrm{clean}}=Q(\theta_{\mathrm{twin}})-L_{\mathcal D}-\varepsilon(\theta_{\mathrm{twin}})$,
$\cDef=Q(\theta_{\mathrm{pois}})-L_{\mathcal D}-\varepsilon(\theta_{\mathrm{pois}})$ and
$\cRem^{\mathrm{DiD}}=-g+[\varepsilon(\theta_{\mathrm{twin}})-\varepsilon(\theta_{\mathrm{pois}})]=-g+\cRem$:
the reported cost is the victim's start minus a constant that belongs to the defense plus what the
poison leaves at that level, and the interaction is the start gap returned plus that residual. Whether a
defense has such a level is measured, not assumed: on PreAct-ResNet18 it holds for FT, NAD, FT-SAM and SAU, not for I-BAU,
VGG19-BN or a damaged start (Section~\ref{sec:replic}, Figure~\ref{fig:levels}, Appendix~\ref{app:vgg}); across four CIFAR-10 cells whose twins span \sqFixUndefRange\ points, the defended twin moves under a point for FT, NAD, FP, SAU and FT-SAM and
\sqFixIbauRange\ for I-BAU (Table~\ref{tab:defenses}).

\subsection{Admissibility and Scope}
\label{sec:admiss}
\textbf{A3 needs a utility floor}, since a residual-attack-success gate alone is satisfiable by
destroying the model (Appendix~\ref{app:a3}). \textbf{A4.} A twin is a control only if it is clean
\emph{by the metric that scores the defense}: the admissible CIFAR-10 twins score \steTwinAsrLo--\steTwinAsrHi\% residual attack success under A4's $5\%$ gate; a ten-image blended twin scores
$11$--$16\%$ against a $2$--$3\%$ never-exposed floor and does \emph{not} qualify (a one-image twin does,
Appendix~\ref{app:blended}); SIG's and Trojan's twins fail it and their cells are descriptive
\ifsten(Appendix~\ref{app:corrected})\fi. \textbf{A2} holds at a stated level, policy or realized repair: a conditional
defense---repair only when the detector fires---is one function at the policy level, so its twin arm is
well defined and a no-op on the twin zeroes the tare; forcing the repair branch on a never-poisoned twin
costs \ncForcedOursCi\ points (Appendix~\ref{app:cond}). \textbf{Manufactured cost.} On a model
with no backdoor the reported cost is the pure tare, $c^{0}_{\mathrm{def}}(b) = Q(\theta_{\mathrm{twin}}) - Q(\mathcal{D}_b\theta_{\mathrm{twin}})$ with
$\tau_{\mathrm{int}} \equiv 0$ by construction\label{eq:knob}, and the budget $b$ moves it: fine-pruning costs \crFpOne\ points at $1\%$ clean data
and \crFpHundred\ at $100\%$ (Appendix~\ref{app:cr}); a cost column is not comparable across papers that omit $b$.
\textbf{Gating, once.} Table~\ref{tab:bbmain} averages every column of a row over that defense's A3-passing seeds, so its identity holds within the row; Table~\ref{tab:corrmain} prints the tare unconditional and, with Section~\ref{sec:block}, gates poisoned-arm quantities on A0
(pre-defense attack success $\ge0.90$) and prints A3 beside them; A3 selects on one arm's post-treatment outcome, its effect on the residue measured in Appendix~\ref{app:a3}.

\subsection{TARE-Z, Replicates and Intervals}
\label{sec:surrmethod}
The twin requires knowing which examples were poisoned. TARE-Z replaces it by an independently trained
clean model of the same architecture and recipe,
\begin{equation}
\hat{c}_{\mathrm{int}} = Q(\mathcal{D}_b\theta_{\mathrm{surr}}) - Q(\mathcal{D}_b\theta_{\mathrm{pois}}),
\qquad
\hat{c}_{\mathrm{int}} - \cRem \;=\; Q(\mathcal{D}_b\theta_{\mathrm{surr}}) - Q(\mathcal{D}_b\theta_{\mathrm{twin}}),
\label{eq:surr}
\end{equation}
so its error is \emph{exactly} a clean-versus-clean difference, mean-zero by exchangeability, before conditioning on A3; its per-use size is the seed-to-seed spread of the defense's level (Section~\ref{sec:surrogate}), and only a surrogate of another recipe can be biased. \label{sec:units}\textbf{The replicate is the installation}, an end-to-end re-run of
poisoning, twin and defenses \citep{bouthillier2021variance}; two never-poisoned models differing only in seed ($\tau_{\mathrm{int}}\equiv0$) calibrate the instrument (Appendix~\ref{sec:selfcorrect}). Cost estimates
carry $95\%$ $t$-intervals over installations; rank statistics use an installation cluster bootstrap with
rank confidence intervals \citep{neuhof2026rankci,klein2018rankci}, a pair being separable when its sign
survives Benjamini--Hochberg at $q=0.05$ (Appendix~\ref{app:bb}); a null is reported as the interval
itself, never as a proven zero.

\section{Experiments}
\label{sec:results}

\subsection{Setup}
\label{sec:setup}
\textbf{Vision.} BackdoorBench grids run the framework's own pipeline (PreAct-ResNet-18, CIFAR-10, its
$10\%$ rate, $100$ recovery epochs) with the twin of Section~\ref{sec:estimand} (Appendix~\ref{app:blended}). Our own CIFAR-10 grid trains
ResNet-18 with a BadNets trigger at $5\%$, twin at $0\%$, and runs fine-tuning, fine-pruning, ANP, NAD and NC \citep{wu2021anp,li2021nad,wang2019neuralcleanse}
at the standard $5\%$ clean-data budget (Appendix~\ref{app:hyper}). At the framework's FT/NAD operating point this rig leaves residual attack success \exOpFtOursTwo/\exOpNadOurs\ on BadNets against the framework's \exOpFtBb/\exOpNadBb\ (fine-tuning clears \exOpFtOursAthree\ installations against $11/11$), so the two rigs are not compared in level (Table~\ref{tab:attacklr}). Every run here is CIFAR-10/100 on
ResNet-18, PreAct-ResNet18, VGG19-BN or BackdoorBox's ResNet-18; other datasets and backbones enter only by leaderboard reading (Table~\ref{tab:fourds}).
An exploratory language-model check at budget points that are not deployable operating points is Appendix~\ref{app:abl}; no claim rests on it. $\theta_{\mathrm{base}}$ in Eq.~\ref{eq:cdef} is the pre-defense poisoned
model in both modalities; one defense policy runs on both arms (Section~\ref{sec:admiss}).

\begin{table}[t]\centering\scriptsize
\caption{\textbf{Per-defense accounting on BackdoorBench's BadNets grid (CIFAR-10, eleven installation seeds), in points.} Reported cost $\cDef$ against the
pre-defense poisoned model; tare $c_{\mathrm{clean}}$, the twin's loss to the same defense; paired contrast (Eq.~\ref{eq:crem}) and
interaction (DiD, Eq.~\ref{eq:did}), differing by $g={}$\taboneBbGap, no share of which is identified as removal; A3: seeds clearing the backdoor
(residual ASR $\le0.05$) over seeds attempted; rows average those seeds. Brackets: $95\%$ $t$-intervals. $^{\dagger}$: A2 at the policy
level (ANP's threshold is picked by test-set attack success, an oracle) or a failure record (ABL); $^{\ddagger}$: NC over the seeds clearing A3, an untouched twin entering at zero (Appendix~\ref{app:bb}). Pooled row: both signs; our grid: Table~\ref{tab:rerank}.}
\label{tab:bbmain}
\setlength{\tabcolsep}{3pt}\renewcommand{\arraystretch}{0.9}
\resizebox{\linewidth}{!}{\input{supplement/build/tab_bbdef_main_body.tex}}
\end{table}

\subsection{The Scheduler Block, Manipulated: the Main Result}
\label{sec:block}
\begin{table}[t]\centering\scriptsize
\caption{\textbf{The $2\times2$ scheduler crossover: only the \texttt{lr\_scheduler} line differs between a shipped row and its swapped counterpart.} Tare per defense (pt, unconditional), mean $\pm$ half-width of the $95\%$ $t$-interval over installation seeds; ``pooled'': the three defenses averaged within a seed; ``twin'': the undefended twin's accuracy, the registered prediction's quantity; ``cells'': (seed, defense) cells whose sign follows the scheduler. Published cost, DiD and A3 per row: Appendix~\ref{app:sched}.}
\label{tab:sched}
\setlength{\tabcolsep}{4pt}\renewcommand{\arraystretch}{0.9}
\input{supplement/build/tab_sched_main_body.tex}
\end{table}
\textbf{The cause is three configuration keys, the scheduler block.} BackdoorBench's prototype config, the base every attack inherits, trains every victim under
\texttt{CosineAnnealingLR} for $100$ epochs;
the benchmark's \texttt{default.yaml} for WaNet \citep{nguyen2021wanet} and BPP \citep{wang2022bpp} override it with \texttt{MultiStepLR} at
milestones $[100,200,300,400]$, and its Input-Aware \citep{nguyen2020inputaware} schedule carries the same milestones, so the first decay
never comes and the victim trains at lr $0.01$ throughout; the attacks' own releases place these milestones inside $600$- to $1000$-epoch runs, and the benchmark cut the run and kept them (Appendices~\ref{app:sched}, \ref{app:toolkits}). BackdoorBench's 2022 paper exempts training-controllable attacks from its cosine default and lists these milestones for WaNet and Input-Aware (its Table~4, archived in the supplement): a defender handed such a model does gain from fine-tuning, and $\cDef$ is that defender's bill, but the gain is not removal and is not comparable across attacks. \textbf{A $2\times2$ changes only that}, its swapped arms registered in a dated, self-hosted
file before they ran (the shipped arms were known; Appendix~\ref{app:prereg}). With the two
schedulers crossed over WaNet and BadNets \citep{gu2017badnets}, the undefended twins move from
\exWmsTwinPct\ to \exWcosTwinPct\% (WaNet) and from \exBcosTwinPct\ to \exBmsTwinPct\% (BadNets), and the
tare reverses with the scheduler, not the attack: \schWmsPool$\to$\schWcosPool\ on WaNet,
\schBcosPool$\to$\schBmsPool\ on BadNets, twelve per-defense intervals excluding zero, \schAgree\ cells and \schAgreeSeed\ seeds (three defenses pooled)
agreeing in sign (Table~\ref{tab:sched}, Figure~\ref{fig:teaser}a; six defenses in Table~\ref{tab:defenses}). The patch improves the attack too: WaNet's victim rises from \schWmsVictim\ to \schWcosVictim\% and installs in \schWcosInstN\ seeds against \schWmsInstN\ shipped. The sign was the registered expectation (priors $0.70$--$0.85$, Appendix~\ref{app:sched}); the crossover adds attribution---the reversal follows the schedule, not the attack---and size. Over installing seeds (A0; A3 \cschWmsAthreeFrac\ and \cschWcosAthreeFrac) the published
cost moves toward zero (\cschWmsPubPool\ to \cschWcosPubPool) while the DiD moves from \cschWmsDidPool\ to \cschWcosDidPool\ (not distinguishable; no equivalence bound claimed):
the start gap returned, not identified removal (Appendix~\ref{app:headroom}). The schedule moves removal too: without the anneal FT, FP and NAD clear BadNets in \cschBmsAthreeFrac\ (seed, defense) cells against \cschBcosAthreeFrac\ with it (descriptive; Appendix~\ref{app:sched}).\ifsrtwentyone{} \srtMainSentence\fi\ifsedtwentythree{} \sedMainSentence\fi

\subsection{The Sign Replicates; the Level Does Not Always}
\label{sec:replic}
\textbf{The sign replicates.} On BPP the tare moves from \sxBshipOnePool\ to \sxBcosOnePool\ (Input-Aware has
no admissible twin, A4 \sxIshipAfour); FT-SAM flips as fine-tuning does (\sqWshipFtsam$\to$\sqWcosFtsam), SAU lands at a lower level of its own
(\sqFixSauMean\%) and I-BAU has none (Figure~\ref{fig:levels}); on
CIFAR-100 the twin's rise doubles (\syWanetTwinGap\ on WaNet, \syBppTwinGap\ on BPP) and an annealed control given the block flips the other way (Blended,
\syBlshipPool$\to$\syBlstepPool); on VGG19-BN fine-tuning flips (\svVshipFt$\to$\svVcosFt);\ifsseventeen{} and in a second toolkit, BackdoorBox, whose shipped
$200$-epoch WaNet run anneals, cutting it to $100$ epochs under its own milestones starts the twin \sbxTwinGap\ points lower and hands fine-tuning and NAD a
``gain'' (\sbxFtNTarePt, \sbxNadNTarePt) that decay erases or reverses (\sbxFtATarePt, \sbxNadATarePt), while pruning, which
trains nothing, charges both (Appendix~\ref{app:toolkits}).\fi{} \textbf{The level does not always:} fine-pruning's fixed point fails on
CIFAR-100, VGG19-BN's defended twins stay \svFixMax\ points apart, and three further ways to lower a start---SIG's self-damage, a $30$-epoch run,
a $20\%$ dose---reproduce the sign but not the level (registered fixed-point predictions, failed; Appendices~\ref{app:sig}--\ref{app:bddose}).
\textbf{The level is the defense's, not the tail of a cosine:} \sqvWshipRiseFiveShare\ of the shipped WaNet twin's rise arrives
before any decay (Appendix~\ref{app:budget}). The victim's schedule sets the start; the defense's configuration sets where a low start lands.\ifsthirteen{} More poison lowers the start too: from \sdrLoRate\ to \sdrHiRate\ BadNets
every defense's published cost falls (\sdrPubFellN; a rise
was registered)\ifssixteen{}; \sdbRepSummary\fi, while the tare, one twin per seed, cannot move (Appendix~\ref{app:bddose}).\fi{} Each replication's protocol was
written, self-hosted, before its new runs; six analysis decisions came after data, one (an SAU exclusion) after its
outcome (Appendix~\ref{app:prereg}).

\subsection{What the Artifact Reached: Leaderboard, Rating, Papers}
\label{sec:contam}\label{sec:leaderboard}\label{sec:downstream}
\begin{figure}[t]
\centering
\IfFileExists{supplement/sched/build/fig_level.pdf}{\includegraphics[width=\textwidth]{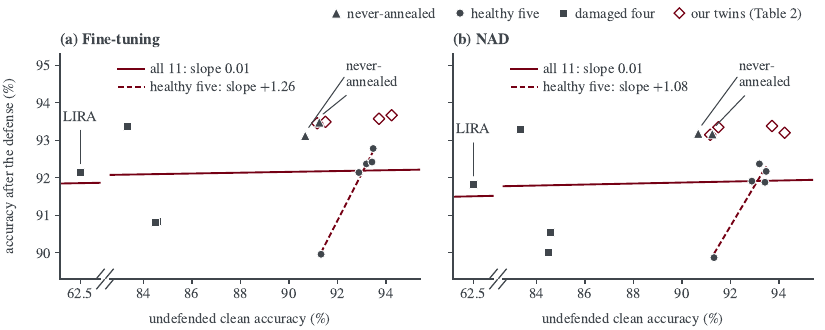}}{\framebox[\textwidth]{\rule{0pt}{2.2in}}}
\caption{\textbf{On PreAct-ResNet18, FT (a) and NAD (b) return a never-annealed or self-damaged start to a level of their own}: after against before, public CIFAR-10 leaderboard at $10\%$ (single-seed cells, grouped after reading the leaderboard; slopes descriptive, \fpLbLooFtSlope\ and \fpLbLooNadSlope\ without LIRA; start alone does not set the after: BadNets starts as low as WaNet and FT lowers it; hollow: our twins; Appendices~\ref{app:lbsources}, \ref{app:fplb}).}
\label{fig:level}
\end{figure}
\begin{table}[t]
\caption{\textbf{The never-annealed victims start lowest across the public leaderboard's datasets and backbones.} Other attacks' mean clean accuracy minus that of WaNet, BPP and Input-Aware, over the rates $\le5\%$ with public entries; last column: cells where the three are the three lowest (sources: Appendix~\ref{app:lbsources}). $^{\dagger}$: on Tiny-ImageNet the three anneal once, at epoch 100, a different contrast; GTSRB is saturated.}
\label{tab:fourds}
\centering\scriptsize
\input{supplement/sched/build/tab_lb_fourds_body.tex}
\end{table}
BackdoorBench's public CIFAR-10 / PreAct-ResNet18 leaderboard prints, for every attack at five poisoning
rates, the victim's clean accuracy and the accuracy after each defense (Figure~\ref{fig:level}). For fine-tuning and NAD the after is a level of
the defense's own: victims spanning \fpLbFtTenUndef\ points before are returned to a \fpLbFtTenDef-point
band (slopes \fpLbFtTenSlope\ and \fpLbNadTenSlope\ over all eleven attacks, \fpLbLooFtSlope\ and \fpLbLooNadSlope\ without LIRA, \fpLbCommonFtSlope\ and \fpLbCommonNadSlope\ on the eight all three defenses cover), while on five healthy annealed victims
(\fpLbHealthyList; grouped after reading the leaderboard\ifdefined\lbHealthyCutLo, any cutoff between \lbHealthyCutLo\ and \lbHealthyCutHi\% selects the same five\fi) the slopes are \fpLbHealthyFtSlope\ and \fpLbHealthyNadSlope\ (descriptive; they lean on BadNets), over every annealed attack, damaged ones included, \lbSlopeFtAllAnn\ (Appendix~\ref{app:fplb}): the defense replaces a low start and leaves a healthy one near where it was. The never-annealed attacks start
lowest (Table~\ref{tab:fourds}).
\textbf{The bias reaches the rating.} BackdoorBench's defense effectiveness rating over $11{,}492$ cells \citep{zhang2023ftsam,wu2024backdoorbench} is
\begin{equation}
\mathrm{DER} \;=\; \tfrac{1}{2}\bigl[\max(0,\Delta A) - \max(0,\Delta Q) + 1\bigr],
\label{eq:der}
\end{equation}
$\Delta A$, $\Delta Q$: attack-success and clean-accuracy drops. The fine-tuning family's published cost is negative on the never-annealed victims and
positive on the healthy annealed ones, and the clip hides the negatives: computed from the site's accuracies (it
prints no DER; \lbDerClipDefs; descriptive), it fires in \lbDerClipNever\ never-annealed cells and \lbDerClipAnn\ healthy
annealed ones (Appendix~\ref{app:fplb}; seed-level counts on our grids, Appendix~\ref{app:corrected}). \textbf{The papers.} Of the \auCiting\ papers citing either BackdoorBench paper, we read in full (one coder, no second rating; \auUnobtainedJournal\ journal-only papers unobtained, Appendix~\ref{app:audit}) the \auFulltext\ that propose a
defense against installed backdoors and asked whether each makes a clean-accuracy claim that depends on a
WaNet, BPP or Input-Aware cell. \auReuse\ reuse the benchmark's victims and reproduce the artifact, unremarked, in at least \auCopied\ baseline columns; two claims rest on it causally, a third
correlationally. \emph{TSBD} \citep{lin2024unveiling}
opens its CIFAR-10 results with an average ``higher than the original No Defense models'' (\auTsbdAvg\ vs.\
\auTsbdNd\%). Rerun with its code in the WaNet
$2\times2$ (Appendix~\ref{app:tsbd}), it reproduces the gain on the shipped victim and not on the annealed one; on the never-poisoned twin it ``gains'' \stTsbdWshipClean\ shipped
(\stTsbdShare\ of the victim's gain at its printed default; Table~\ref{tab:tsbdsweep} sweeps the ratio) and costs \stTsbdWcosClean\ under cosine. \emph{CGD}
\citep{xu2025cgd} reports ``improving CA over the No Defense baseline by 3.4\% for WaNet''; in the same
$2\times2$ the twin gains \stCgdWshipClean\ (\stCgdShare\ of the victim's gain) and loses \stCgdWcosClean\
under cosine. \emph{PDB}~\citep{wei2024pdb} has a positive Tiny-ImageNet average only because of WaNet and BPP (correlational).
Dropping or re-referencing its two never-annealed cells flips TSBD's average (Appendix~\ref{app:audit}); none of the \auFulltext\ notes that the three victims are under-trained, and what rests on them is a
sentence per method.

\ifsten
\subsection{A Signed Tare Column for the $10\%$ Leaderboard}
\label{sec:corrected}
\begin{table}[t]\centering\scriptsize
\caption{\textbf{The tare per (attack, defense) cell on CIFAR-10 / PreAct-ResNet18 at
$10\%$, wherever an admissible twin exists.} Line 1 per defense: tare (pt), mean $\pm$ $95\%$ $t$-half-width over seeds; line 2 (\emph{paired}): $Q(\mathcal{D}\theta_{\mathrm{twin}})-Q(\mathcal{D}\theta_{\mathrm{pois}})$ over installing seeds (A0: attack success $\ge0.90$), DiD plus the pre-defense gap (Eq.~\ref{eq:crem}). Twin construction, SIG's and Trojan's descriptive cells, published costs, DER, seeds, five more defenses: Appendix~\ref{app:corrected}; patched WaNet/BPP: Table~\ref{tab:defenses}. $^{\dagger}$: A2 at the policy level only (ANP's threshold; NC's detector; ABL); NC: twins whose detector fired \textbar\ cost given firing. $^{\ddagger}$: under half the poisoned cells clear A3 (FT, FP, NAD: \steFtfamClearOut\ on Blended, LF, FTrojanNN, Refool); entry: a post-defense contrast. $^{\natural}$: ABL destroys \steAblTwinCollapseAdmit\ of these twins; its row is a failure record, off-scale, outside every sign count and the ordering. Intervals per-cell, nominal: of \steCellsKAdmit\ admissible cells with intervals (ABL in, NC out), \steCellsNominalAdmit\ exclude zero at $0.05$ and \steCellsBonfAdmit\ at Bonferroni $0.05/K$, $K={}$\steCellsK\ over the nine-attack table (Appendix~\ref{app:corrected}). Columns: attacks; A4: twins passing the $5\%$ gate over seeds.}
\label{tab:corrmain}
\setlength{\tabcolsep}{2.5pt}
\resizebox{\linewidth}{!}{\input{supplement/sched/build/tab_sched10_main_body.tex}}
\end{table}
Table~\ref{tab:corrmain} is the deliverable table (\steAdmitN\ attacks, \exGpuCorr\ GPU-hours; rows, columns and gates in its caption); it corrects no cell, and sign counts and the ordering leave ABL out (a failure record, not a cost). On the annealed
victims every tare is positive (\steAnnPos\ cells; on the four annealed victims besides BadNets FT, FP and NAD remove these
backdoors in \steFtfamClearOut\ poisoned cells, so their paired entries there price nothing). On the two never-annealed victims FT, FP and NAD
``gain'' in \steNeverFtFamNeg\ cells and ANP$^{\dagger}$ charges the twin in \steNeverAnpPos, every nominal interval excluding zero (\steNeverBonf\ at the Bonferroni level);
SAU, at a level of its own, and I-BAU, with none, have intervals covering zero in \steNeverSauIbauZero\ of their cells; where Neural Cleanse's detector fires
(\steWanetNcFire\ WaNet twins), its repair, a fine-tune, ``gains'' \steWanetNcCond\ as fine-tuning does. This pattern was read off the table after a
registered prediction failed on BPP (Appendix~\ref{app:prereg}). Ordering these eight defenses by mean tare over the admissible attacks gives \steCleanOrder, NC at its policy-level tare with non-firing twins at zero (Appendix~\ref{app:cond}\ifsdptwentytwo; under a deployer-side threshold rule that never sees attack success ANP's tare rises from \sdpBadnetTareOPt\ to \sdpBadnetTareDPt\ on BadNets and from \sdpWanetTareOPt\ to \sdpWanetTareDPt\ on WaNet---registered, two of three predictions held, the failed one that the two rules agree within a point; Appendix~\ref{app:anp}\fi). A prototype-referenced DER (the same seed's
BadNets twin as reference) re-ranks \steDerFlips\ pairs over all nine attacks (Table~\ref{tab:corrder}): attribution changes, order does not, by construction.\ifseleven{} Five further defenses re-rank \selDerFlips\ pairs (Appendix~\ref{app:corrected}).\fi
\fi

\subsection{What the Twin Charges per Defense, and What Poisoning Adds}
\label{sec:rerank}
\textbf{At a fixed attack the charge is a property of the defense; at a fixed defense, of the victim's
schedule.} On BackdoorBench's shipped framework all eight BadNets clean-arm intervals exclude zero
(nominal; among the seven besides ABL, \exBbCleanExclBonfNoAbl\ at $0.05/7$ and \exBbCleanExclBHNoAbl\ under BH $q=0.05$; Figure~\ref{fig:teaser}b, Figure~\ref{fig:bbdef}, Appendix~\ref{app:bb}), and the charge runs, over the \exBbPosNoAblWord\ reporting a positive cost (ABL aside), from
\chbShareLo\ of what the defense reports (\chbShareLoName, \schBcosFtPt\ of \chbShareLoPub; \szrBshipFtTareZero\ on a zero-image twin) to \chbShareHiNoAbl\
(\chbShareHiNoAblName, \chbShareHiNoAblClean\ of \chbShareHiNoAblPub); ABL, off-scale at \chbShareHi, collapses its twins (Table~\ref{tab:bbdef}).
\textbf{What poisoning adds is estimator- and defense-dependent, and both estimators are printed.} A reported
\taboneBbDefPt\ leaves \taboneBbDid\ by difference-in-differences and \taboneBbPaired\ paired ($g={}$\taboneBbGapPt), the per-defense DiD carrying both signs
(Table~\ref{tab:bbmain}; our own rig: Table~\ref{tab:rerank}). Read as $\cRem-\beta g$ over $\beta=0$ (paired), $1$ (DiD) and \exBetaNoise, \exBetaShort\ (two start-lowering routes, two-point fits on our ResNet-18 rig; Appendix~\ref{app:headroom}), the interaction keeps its sign for \exBetaStableAb, not for
\exBetaUnstableAb; which $\beta$ is removal's is unidentified, so nothing is corrected. On a checkpoint carrying a backdoor
with probability $p$, a defense costs $(1-p)\,c_{\mathrm{clean}}+p\,\cDef$ in expectation; benchmarks print the
$p=1$ corner; post hoc (the registered criterion failed), \chRevbnSurWord\ BadNets pairs reverse below $p=$\,\revPstarLo\ and none survives on WaNet or Blended (Appendix~\ref{app:bb}).\label{sec:abl}

\subsection{What the Twin Costs to Run}
\label{sec:surrogate}
\textbf{The twin is not needed where a defense's clean-arm level is stable across seeds.}
Substituting a same-recipe clean model for the twin (Eq.~\ref{eq:surr}) over \tzCifarNdef\ configurations of FT, FP, NAD and ANP
on \tzCifarNinst\ installations is off by \tzCifarPooledPaired\ points paired in the mean (Appendix~\ref{app:surr}); leaving one twin out of BackdoorBench's eleven-seed BadNets grid
(\tzBbPooledPaired\ paired, excluding ABL) measures how reproducibly a level recurs. For \tzBbSmallSet\ its per-use error (mean $|$error$|$
\tzBbSmallMaePLo--\tzBbSmallMaePHi) resolves the fine-tuning family's never-annealed sign, not a tare as small as fine-tuning's \schBcosFtPt\ on annealed victims;
for \tzBbLargeSet\ (\tzBbLargeMaePLo--\tzBbLargeMaePHi) the defense's own randomness moves the level by more than one surrogate resolves, so seeds must be pooled. Across
architectures it is ordinal only (Appendix~\ref{app:tiers}). \textbf{Ex ante, without a twin: the defense's spread over two or three same-recipe clean models is the per-use error (Eq.~\ref{eq:surr}); above the tare in question, pool seeds.}

\section{Conclusion}
\label{sec:conclusion}
We show that the clean-accuracy cost backdoor-defense leaderboards print is measured on the poisoned victim alone and inherits where that victim started. On BackdoorBench three of sixteen attacks ship a scheduler that never anneals; one \texttt{lr\_scheduler} line reverses the sign of the fine-tuning family's cost on a never-poisoned twin, the benchmark's rating clips the resulting ``gain'' to zero, and two citing papers rest a no-cost claim on it. We propose TARE, the identical defense run on a never-poisoned twin, whose loss---the tare---is \exBbCleanMinPt\ to \exBbCleanMaxPt\ points on BackdoorBench's BadNets grid (ABL aside). Patched, WaNet's victim anneals and its tare turns positive; printed signed and unclipped beside the reported column, the tare changes attribution and the cross-attack averages, not the within-attack order.

\subsection*{AI use statement}
The authors used an AI coding assistant to help implement experiment harnesses, orchestrate
runs, draft and revise text, and critique drafts in simulated review rounds (the ``panels'' some supplement
files mention). All experimental designs, claims, and their verification
were determined and checked by the authors; all reported numbers were produced by the
authors' code and inspected by the authors. No AI-generated research claim was included without independent verification by the authors.

\subsection*{Ethics statement}
This work installs backdoors only in models we train ourselves, on public datasets, and releases
no new attack: every attack we use (BackdoorBench's BadNets \citep{gu2017badnets}, Blended \citep{chen2017blended},
WaNet \citep{nguyen2021wanet}, BPP \citep{wang2022bpp}, Input-Aware \citep{nguyen2020inputaware}, SIG \citep{barni2019sig},
LF \citep{zeng2021lf}, Trojan \citep{liu2018trojannn}, FTrojanNN \citep{wang2022ftrojan} and Refool \citep{liu2020refool};
LIRA \citep{doan2021lira} and SSBA \citep{li2021ssba} on the public leaderboard only; and the instruction-tuning poison of
\citet{souly2025poisoning}) is already published, and no poisoned weights are distributed. The paper's finding is an accounting correction and should not be read as a claim
that deployed defenses are safe: we show that the clean-accuracy figure printed beside a defense
does not isolate poisoning-conditioned cost from generic clean-arm cost, and that the relative
contribution varies across defenses and settings, which changes what portion of the reported utility loss
is attributable to removal but says nothing about whether removal succeeded
or whether the defense is safe. Sections~\ref{sec:estimand} and \ref{sec:admiss} state the
conditions under which our estimand is identified, and we report the settings in which it is not.

\subsection*{Reproducibility statement}
The defenses are fine-tuning, fine-pruning \citep{liu2018finepruning}, NAD \citep{li2021nad}, Neural Cleanse
\citep{wang2019neuralcleanse}, ANP \citep{wu2021anp}, ABL \citep{li2021abl}, I-BAU \citep{zeng2022ibau}, SAU \citep{wei2023sau},
FT-SAM \citep{zhang2023ftsam}, CLP \citep{zheng2022clp}, MCR \citep{zhao2020mcr}, RNP \citep{li2023rnp}, NPD \citep{zhu2023npd},
D-BR \citep{chen2022dbr}, TSBD \citep{lin2024unveiling} and CGD \citep{xu2025cgd}, as implemented in BackdoorBench
\citep{wu2024backdoorbench}, in the methods' released code (TSBD, CGD) or in our harness (Appendix~\ref{app:hyper}).
The three keys were read at BackdoorBench commit \texttt{f02e353} (2025-07-04). The BackdoorBench-framework grids used at least \exGpuTotal\ logged GPU-hours\ifsseventeen{}, the BackdoorBox runs \sbxGpuH\fi\ifsztwenty{}, the zero-image twins \szrGpuH\fi\ifsrtwentyone{}, the rate crossover \srtGpuH\fi\ifsedtwentythree{} and the eight-defense crossover \sedGpuH\fi (the ANP re-reading of Appendix~\ref{app:anp} trained nothing). We report \preregWord\ self-hosted
pre-registrations (no external registry; timing evidence and the decisions taken after data existed are in
Appendix~\ref{app:prereg}), failures reported as failed.
The paired clean-twin protocol is defined in Section~\ref{sec:protocol}; the target estimand
$\tau_{\mathrm{int}}$, the paired estimator $\cRem$, the DiD estimator
$\cRem^{\mathrm{DiD}}$, and the reported-cost quantity $\cDef$ are given in closed form. All models are public
(Pythia-410M, Qwen2.5-7B-Instruct, Llama-3.2-3B-Instruct, Llama-2-7B-Chat, Mistral-7B,
ResNet-18, PreAct-ResNet-18, VGG19-BN, and CGD's CLIP ViT-B/32) and all datasets, benchmark suites and toolkits are public (the Pile subset,
Alpaca-52k, MMLU, CIFAR-10, CIFAR-100, TOFU, BackdoorBench\ifsseventeen, BackdoorBox\fi). The appendices report full defense grids, hyperparameters, per-defense summaries, success counts
and pooled intervals; together with the supplementary code and raw per-installation data they
provide the complete poison-generation, twin-construction, training, defense and evaluation
procedures. The code, raw data and pre-registration files, called ``the supplement'' throughout, are available from the authors on request.

\bibliography{refs}
\bibliographystyle{iclr2027/iclr2027_conference}

\appendix
\renewcommand{\thesection}{S\arabic{section}}

\FloatBarrier
\section{What the audited papers report}\label{app:lit}

We read the six papers that define the defenses and the rating this audit uses, and recorded three
things from each: the clean-accuracy cost it reports for its own headline defense, whether it ever
applies that defense to a never-poisoned model, and whether it states or varies the defender's
clean-data budget. Table~\ref{tab:lit} is the result. Two patterns hold across all six. None runs
the defended clean arm: the closest any comes is quoting a clean model's accuracy \emph{level} as
a reference point (\citealp{wang2019neuralcleanse}, $77.2\%$; \citealp{wu2024backdoorbench},
Table 3), never its cost under the defense. And where the budget is varied it is varied as a
robustness check on the defense, not as a confounder in the cost column---\citet{zhang2023ftsam}
even report the rating itself at three budgets without reading the movement as a property of the
budget.

The reported costs span two regimes. Each paper's own headline defense is cheap---$0.2$ to $3.6$
points, and \citet{zhang2023ftsam} report a net \emph{gain} of $0.86$---which is the band drawn in
Figure~\ref{fig:teaserold}(b). The same tables carry other defenses that are not: the widest entry we
verified is ABL at $47.5$ points in \citet{zhang2023ftsam}'s Table 1, independently reproduced in
\citet{wu2024backdoorbench}'s Table 14, with AC at $46.1$ and Neural Cleanse at $34.6$ in the same
table. Without the defended clean arm, a column that contains both regimes cannot isolate
removal-attributable cost from intervention cost.

\begin{table}[!ht]\centering\scriptsize
\caption{Audit of the six papers whose defenses and rating this work uses. ``Headline cost'' is
the clean-accuracy drop the paper reports for its own primary defense, in accuracy points, at the
location given. ``Clean arm'' asks whether the paper applies its defense to a never-poisoned model
and reports the resulting cost. $^{\dagger}$the last row is a benchmark suite, not a single
proposed defense; its column gives the per-defense mean range with the largest single entry.}
\label{tab:lit}
\begin{tabular}{llccl}
\toprule
Audited entry & defense & reported cost (pt) & clean arm? & clean-data budget \\
\midrule
\citet{liu2018finepruning}    & fine-pruning   & $0.2$ (Tab.~1) & no & stated only, size not given \\
\citet{li2021nad}             & NAD            & $2.66$ (Tab.~1) & no & $5\%$; swept $0$--$20\%$ (Fig.~2) \\
\citet{wu2021anp}             & ANP            & $1.47$ (Tab.~1) & no & $1\%$; swept $0.1$--$10\%$ (App.) \\
\citet{wang2019neuralcleanse} & unlearning     & $3.6$ max (Tab.~IV) & no & $10\%$, not swept \\
\citet{zhang2023ftsam}        & FT-SAM         & $-0.86$ (gain; Tab.~1) & no & $5\%$; swept $1$--$10\%$ (Tab.~3) \\
\midrule
\citet{wu2024backdoorbench}$^{\dagger}$ & nine-defense suite & $0.05$--$24.4$ mean; $47.5$ max (Tab.~14) & no & $5\%$ fixed, not swept \\
\bottomrule
\end{tabular}
\end{table}

\begin{table}[!ht]\centering\scriptsize
\caption{What the nearest prior work reports, from each paper's own text. \ding{51}: reported;
$\circ$: partial; ---: not reported. The defended clean arm is not our novelty: three of the six run
one and the systematization recommends it (its R7). What none of them does is the rest of the row:
define the target with identification conditions, net the pre-defense gap, condition on removal
success, replicate at the installation level with intervals, carry the correction into the field's
ranking rule, and test a matched-twin-free substitute. Locations of the supporting sentences are in the
supplementary audit notes.}
\label{tab:prior}
\setlength{\tabcolsep}{3pt}
\begin{tabular}{lccccccc}
\toprule
 & defended & interaction estimand & nets pre- & success- & independent & ranking & surrogate \\
Work & clean arm & + identification & defense gap & conditioned & installations & consequence & for the twin \\
\midrule
\citet{sok2025backdoor} (SoK) & \ding{51}$^{a}$ & --- & --- & --- & $\circ^{b}$ & --- & --- \\
\citet{peng2025tsc} & \ding{51}$^{c}$ & --- & --- & $\circ^{d}$ & --- & --- & --- \\
\citet{tong2025gms} & \ding{51}$^{c}$ & $\circ^{e}$ & --- & --- & $\circ^{f}$ & --- & --- \\
\citet{bouger2026unlearning} & ---$^{g}$ & --- & --- & $\circ^{d}$ & --- & --- & --- \\
\citet{li2026purifying} & --- & --- & --- & --- & --- & $\circ^{h}$ & $\circ^{i}$ \\
\citet{cho2025fade} (unlearning) & $\circ^{j}$ & --- & --- & --- & \ding{51} & \ding{51}$^{k}$ & --- \\
\midrule
TARE (this paper) & \ding{51} & \ding{51} & \ding{51} & \ding{51} & \ding{51} & \ding{51}$^{l}$ & \ding{51} \\
\bottomrule
\end{tabular}

{\scriptsize $^{a}$poison-rate-0 arm with absolute defended clean accuracy, no delta; $^{b}$ten repeated
trials in one configuration, main results single-run; $^{c}$appendix only, raw accuracy, never
subtracted from the poisoned-arm cost; $^{d}$a success threshold or final attack success is shown, cost
is reported unconditionally; $^{e}$$\Delta_{\mathrm{acc}}$ against the victim model as a search objective;
$^{f}$three seeds averaged, no intervals, replicate unit unstated; $^{g}$its control is the backdoored
model continually pretrained on clean data, and capability is compared to the base level;
$^{h}$a prose re-run of one baseline at standard learning rates; $^{i}$reference-free by design, through
paired variants derived from the suspicious model, not an independent clean model; $^{j}$a retain-only
gold model as the forgetting reference, utility reported as a level; $^{k}$on the forgetting axis;
$^{l}$in the audited CIFAR-10/BadNets suite.}
\end{table}

\FloatBarrier
\section{Nuisance sensitivities}\label{app:abl}
\textbf{The language-model setting.} A lexical backdoor (a rare-token trigger mapping to a fixed payload) is
installed into Qwen2.5-7B-Instruct and Llama-3.2-3B-Instruct by continued training on poisoned data, with a
clean twin per seed; $Q$ is MMLU accuracy, and the two defenses are budget points, not deployable
operating points (Appendix~\ref{app:lim}, Table~\ref{tab:rerankfull}). $\theta_{\mathrm{base}}$ in
Eq.~\ref{eq:cdef} is the pre-defense poisoned model in both modalities, the vision convention and the
conservative choice.

\begin{table}[h]\centering\scriptsize
\caption{Table~\ref{tab:rerank} with the language-model row. Qwen2.5-7B: eight installations, two defenses (continued
fine-tuning and fine-pruning at budget points that are not deployable operating points, Appendix~\ref{app:lim});
$^{\diamond}$its pooled interaction resolves above zero only through an $r={}$\taboneQwenR\ anti-correlation of the two
defenses over \taboneQwenBothN\ installations; under independence it is \taboneQwenIndepCI. Llama (one installation,
three defense seeds): Table~\ref{tab:abl}.}
\label{tab:rerankfull}
\setlength{\tabcolsep}{3pt}
\resizebox{\linewidth}{!}{\input{supplement/build/tab_rerank_full_body.tex}}
\end{table}

Table~\ref{tab:abl} varies one analysis decision at a time and reports the same two quantities
as Table~\ref{tab:rerank}; the admissibility axes are separated into Appendix~\ref{app:a3}.

\begin{table}[H]\centering\scriptsize
\caption{\textbf{At a fixed attack the cost gap is stable across the replicated admissible
alternatives and the one coordinate that nearly closes it belongs to the evaluator; across attacks it is
not even sign-stable.} Under the shared convention of Table~\ref{tab:bb} the gap---$\cDef$ minus the DiD
removal term, which is the clean arm---is \chbbBNClean\ on BadNets and \chbbWNClean\ on WaNet, so it is a
property of the (attack, defense-budget) pair and not a constant (Section~\ref{sec:rerank}). The WaNet row
below pools \chwNseedsWord\ installation seeds; the \chwSeedZeroGap\ an earlier version quoted as ``the''
gap was seed $0$ alone. Rows vary one analysis or
setting decision each and report the cost decomposition, not the induced ordering, which is treated
separately in Section~\ref{sec:corrected}. Each row reports $\cDef$
and the removal estimate used (Appendix~\ref{app:headroom}: not identified); sources in Appendix~\ref{app:headroom} and
Appendices~\ref{app:cr}--\ref{sec:selfcorrect}.
$^{\ast}$this row's cost column is $c^{0}_{\mathrm{def}}$ of Section~\ref{sec:admiss}, the clean arm's own drop, not $\cDef$.
$^{\dagger}$WaNet: \chwNseedsWord\ installation seeds under the same frozen protocol. Three (\chwAoneFail) fail A0---the attack installs at only \chwAoneRange\ success against $0.94$--$0.99$ on the other eight---and the success-conditioned common support is \chwNsuppWord\ seeds over \chwNcandWord\ candidate defenses. Its poisoned model starts $1.43$ $[+0.48,+2.39]$ pt below its twin, the framework's $100$ recovery epochs out-train its attack script on $5/5$ common-support seeds for every \emph{candidate} defense but ANP (over all eight, ABL and Neural Cleanse break it), and \chwCandNeg\ candidates report a \emph{negative} cost, each interval excluding zero, whose clean arm carries \chwCandShareLo\ to \chwCandShareHi---I-BAU is the one whose clean arm carries the \emph{opposite} sign to its reported cost. The arrow in the WaNet row is the convention of Table~\ref{tab:bb} (seven defenses excluding ABL, A3-failed cells dropped, \chwNseedsWord\ seeds on both grids); under the matched candidate set, success-conditioned, it is \chsetBNDef\ $\to$ \chownWNDef, so the sign change survives either.
$^{\ddagger}$TOFU's figure is the retrain-gold contrast of Appendix~\ref{app:unlearn}, not the four-cell DiD.
$^{\P}$language model then vision, the correction of Appendix~\ref{sec:selfcorrect}; the pre-fix values appear only there and here.}
\label{tab:abl}
\setlength{\tabcolsep}{3pt}
\resizebox{\linewidth}{!}{\begin{tabular}{lccl}
\toprule
sensitivity & cost column & removal estimate (form) & effect on conclusion \\
\midrule
estimator: DiD $\to$ paired & $+2.51$ & $+0.02\!\to\!+0.43$ & paired resolves, DiD does not \\
clean-data budget: $1\%\!\to\!100\%$ & $+3.80\!\to\!-0.05$ & $+0.01\!\to\!-0.33$ (DiD) & removal within $\pm0.4$ \\
\hspace{1em}same sweep, no-backdoor arm$^{\ast}$ & $3.79\!\to\!0.28$ & $\equiv 0$ by construction & tax alone spans the band \\
attack: BadNets $\to$ blended ($8$ seeds) & per defense & per defense & share $43$--$138\%$; \chRevblRev\ of \chRevblNWord\ pairs reverse, \chRevblSurWord\ survive \\
attack: BadNets $\to$ WaNet ($11$ seeds)$^{\dagger}$ & \chbbBNDefPt$\!\to\!$\chbbWNDef & \chbbWNDid\ (DiD) & cost changes \emph{sign}; \chRevwnRev\ of \chRevwnN\ reverse, \chRevwnSurWord\ survive \\
headroom route: noise $\to$ data shortage & --- & $-1.14\!\to\!+0.41$ at $g{=}2.06$ & not identified; we apply none \\
defense family: $4$ always-clearing configs & $+1.12$--$+4.28$ & $-0.14$--$+0.12$ (DiD) & unchanged; none resolves \\
model/modality: four settings & $11.5/9.0/2.5/$publ. & ${+}1.37/{-}1.73/{+}.02$; $+.82^{\ddagger}$ & removal small and signed \\
recovery schedule: $10\!\to\!100$ epochs & $+2.51\!\to\!+2.20$ & $+0.02\!\to\!-0.04$ (DiD) & unchanged; A3 $37\!\to\!42$ of $48$ \\
defender draw: shared $\to$ independent$^{\P}$ & $8.8\!\to\!11.5$; $3.4\!\to\!2.8$ & $-.24\!\to\!+1.37$; $+1.05\!\to\!+.02$ & both signs change \\
twin $\to$ surrogate: $4$ installations & --- & \tzCifarPooledPairedPt\ paired; \tzCifarPooledDidPt\ DiD & unchanged within recipe \\
prune $\to$ output-masked prune & $+2.81\!\to\!+2.51$ & $+0.04\!\to\!+0.02$ (DiD) & unchanged \\
ANP $\to$ reference mask and threshold & $+8.3\!\to\!+2.0$ & $+0.25\ \scriptstyle[-0.71,+1.22]$ & drops on A3, clears $3/14$ \\
NC $\to$ pixel-space reversal & $+4.24\!\to\!+0.78$ & $-0.14\ \scriptstyle[-0.97,+0.70]$ & becomes a candidate, $14/14$ \\
headroom arm at a matched deficit & $\equiv0$ by construction & $-0.28\ \scriptstyle[-0.63,+0.08]$ & triple difference $+0.32$ \\
\bottomrule
\end{tabular}}
\end{table}

\FloatBarrier
\section{Admissibility: success rates, threshold, and common support}\label{app:a3}
\paragraph{In full.} \textbf{On our own suite the ordering consequence is arithmetic, so we do not report it as a result.} The
removal term varies by \chSDremPrim\ points across our \chNcandWord\ candidates against \chSDcleanPrim\
for the clean arm, so an ordering by reported cost almost has to be one by intervention, and it is exactly
that: the two agree on all \chNcandWord, \chSepPub\ pairs separable (Figure~\ref{fig:rerank}). The separable reversals that count are on BackdoorBench (Section~\ref{sec:abl}).

Eq.~\ref{eq:estimand} conditions on successful removal, so the protocol has two outputs and we
report both. Table~\ref{tab:a3succ} gives the success rate of each vision defense over the
fourteen installations at the operational threshold $\alpha=0.05$; the threshold is ours, chosen
before the ranking analysis to sit an order of magnitude below the $96$--$97\%$ attack success the
undefended models carry, not inherited from a benchmark. Five methods pass the gate; four of the five
clear the backdoor in every one of the fourteen installations and NAD misses one, so the
success-conditioned common support is thirteen of fourteen.

\paragraph{The clean arm is not sign-stable across attacks, and that is the strongest form of our claim.}
Running the same eight defenses over eleven seeds against WaNet instead of BadNets, with the same
unconditional twin-arm convention (all eleven seeds, no success criterion), inverts \exWanetNegMeanWord\ of the eight in the mean and \exWanetNegExclWord\ with intervals excluding zero:
fine-tuning \schWmsFt, fine-pruning \schWmsFp\ and NAD \schWmsNad,
against \schBcosFtPt, \schBcosFpPt\ and \schBcosNadPt\ on BadNets. ANP and ABL
(both $A2$ at the policy level only) keep a sign whose interval excludes zero (\steWanetAnpClean, \steWanetAblClean) and the remaining three straddle zero.
Pooled over the \chwNcandWord\ candidates on that same unconditional convention the clean arm is
\chuncWNClean, an interval that excludes zero; success-conditioned on the \chwNsuppWord-seed common
support it is \chownWNClean. \textbf{This is a counterexample to
the unconditional-universal form of our own headline}---"every defense charges a never-poisoned model" holds
on BadNets, not on WaNet---and we report it in that form rather than restricting the claim quietly. It is
also the sharper version of the paper's thesis: a quantity that flips sign with the victim's training schedule cannot be
a fixed overhead that the audited protocols implicitly netted out, and it cannot be recovered from the
poisoned arm alone. The mechanism is visible in the same grid: BackdoorBench's $100$ recovery epochs
out-train WaNet's attack script on $5/5$ common-support seeds for every candidate defense but ANP, so the defended
poisoned arm ends \emph{above} where it started. The mechanism is in the starting point, and it is visible
in the same cells: WaNet's rig leaves \emph{both} arms about two points lower than BadNets' does (twin
$0.9150$ against $0.9371$, poisoned $0.9005$ against $0.9164$), so the framework's $100$ recovery epochs on
a clean subset are net \emph{training} on the WaNet grid and net damage on the BadNets grid. The same
fine-tuning that costs the BadNets twin $0.13$ points raises the WaNet twin by $1.99$, and raises its
poisoned arm by $3.10$. Only the two defenses that damage a model regardless of where it starts---ANP,
which prunes, and ABL, which unlearns---keep a sign whose interval excludes zero on both attacks.
\textbf{So the clean arm is the difference between where the attack's training run left the victim and the
fixed point the defense's own recovery run converges to.} That fixed point is visible in the four cells of
Appendix~\ref{app:sched}: the undefended twin moves \schUndefRange\ points across them while the
defended twin moves at most \schFixMaxRange\ (fine-tuning \schFixFtRange, NAD \schFixNadRange,
fine-pruning \schFixFpRange)---a level set by the defense, not by where the victim started. And the poisoned
arm's excess over its twin does not move with the schedule, the start gap returned rather than removal
identified (Section~\ref{sec:block}, Appendix~\ref{app:headroom}): on WaNet, DiD is \schDidWmsFt\ under the shipped
schedule and \schDidWcosFt\ under the swapped one for fine-tuning, \schDidWmsNad\ and \schDidWcosNad\
for NAD, every pair overlapping; on the un-annealed BadNets grid only \schBmsAthreeFrac\ poisoned-arm
cells clear A3 and its residue, \schBmsDidPool\ against \schBcosDidPool, is not read. What the schedule moves is the tare, which is why it cannot be a constant anyone subtracted after the fact,
and it is not recoverable from the poisoned arm alone---which is the whole point of measuring the twin.

\paragraph{A0 is not a formality: on one attack it excludes $27\%$ of seeds.}
On our CIFAR grid and on BackdoorBench's BadNets, every seed installs, so A0 never binds and a reader
could reasonably ask what it is for. On WaNet it binds hard. Running the same frozen protocol over eleven
installation seeds, three (seeds $3$, $8$, $10$) reach a pre-defense attack success of only $0.641$,
$0.369$ and $0.751$, against $0.938$--$0.988$ on the other eight; their clean accuracy is normal
($0.895$--$0.907$), so nothing but the trigger check distinguishes them. Conditioning on removal success
is meaningless on a model that was never backdoored to begin with, so A0 removes them, and the
success-conditioned common support falls to five seeds. \textbf{The consequence is about the literature, not
about us}: a cost or ranking read from a \emph{single} installation of this attack has a better than one in
four chance of being read off a model the attack did not take on, and the audited protocols we survey
report neither a per-seed installation rate nor an installation gate. Our own earlier single-seed WaNet
reading happens to have landed on an installing seed---seed $0$, which the eleven-seed grid reproduces
exactly---but that was luck, and the eleven-seed interval, not the single reading, is what we now report.

\begin{table}[h]\centering\small
\caption{Installation-level success rates and residual attack success for all nine defenses,
over fourteen independent CIFAR-10 installations with the corrected, independent defender draw,
fine-pruning that silences the pruned channel at the layer output, ANP at its reference mask semantics
under its own accuracy-guarded threshold, and Neural Cleanse reverse-engineering its trigger in pixel
space. Five methods pass the candidate gate (method-level mean residual $\le\alpha=0.05$); four of them
clear the backdoor in every installation, and every twin is admissible throughout. Regenerated from
\texttt{data/cifar\_cells.csv} by \texttt{analysis/build\_a3\_table.py}, which also re-derives the
candidate set and the common support and asserts them against the flags shipped in the cell file.
$^{\S}$NAD is a candidate on its method-level mean residual ($0.022$) but misses installation $3$
($0.138$), which is what drops the ranking's common support from fourteen to thirteen; this is the
corrected teacher of Appendix~\ref{sec:selfcorrect}, not the $\mathrm{lr}\,0.01$ teacher our first
implementation used.
$^{\P}$ANP under a two-point clean-accuracy budget prunes too little to clear this attack: it costs only
$+2.0$ points but leaves a mean residual attack success of $0.51$, so it is excluded on removal success,
not on twin admissibility. Pruning it to a fixed $20\%$ instead clears $5$ of $8$ at $+7.7$ points and
destroys $2$ of $8$ (Appendix~\ref{app:zoo}).}
\label{tab:a3succ}
\input{supplement/build/tab_a3succ_body.tex}
\end{table}

\textbf{What success-conditioning moves, and what it does not.} A3 filters on a post-treatment variable
measured on the poisoned arm only, so it can in principle select on the outcome. Empirically it moves the
reported cost and leaves the residue alone: on the CIFAR headline grid the conditional reported cost is
$+2.51$ $[+2.25,+2.77]$ against $+2.11$ $[+1.99,+2.22]$ unconditional, while the interaction is $+0.02$
$[-0.31,+0.34]$ against $+0.04$ $[-0.24,+0.32]$; on Qwen the two interactions agree to two decimals. Conditioning also makes the per-installation average run over a data-dependent set of
configurations---installations retain four to six of the six---so part of the reported cost's spread is
which defenses happened to clear in that seed. Restricting to the four configurations that clear in every
installation removes that source and gives $+2.73$ $[+2.61,+2.85]$ reported and $-0.00$ $[-0.37,+0.36]$
removal, which is the balanced reading of the same grid.

\textbf{The rating's clip, and the ordering without it.} The $\max(0,\cdot)$ clip in Eq.~\ref{eq:der} zeroes
any negative capability drop, and under the DiD accounting it binds in \chClipTotPrim\ of the
\chClipCellsPrim\ candidate cells: \chClipListPrim. That is why the \chNcandWord\ corrected ratings lie
within \chSpanPrimRound. Removing the clip changes the corrected ordering---$\tau_a$(published, removal)
falls from \chTauRemClip\ to \chTauRemNoClip\ and the gap widens from \chGapClip\ to \chGapNoClip, and a
second pair inverts with it (\chNoClipMoves)---so the corrected \emph{order} is itself partly an artifact
of the rating's shape. The
published-versus-intervention agreement is $\tau_a=+1.000$ with and without the clip, which is the part of
the ordering result that does not depend on it.

\textbf{The utility floor on A3.} A3 asks that the residual attack success fall below $\alpha$, which a
defense can satisfy by destroying the model: a network at chance accuracy answers the trigger no more often
than anything else. We therefore also require the defended arm to retain capability---clean accuracy within
$20$ points of the pre-defense model on vision, MMLU above chance plus ten points on the language models.
No cell in the paper's own grids is excluded by this floor\ifsthirteen{} except \sdrFloorN\ run in the BadNets dose grid of
Appendix~\ref{app:bddose} (\sdrFloorList)\ifssixteen{} and \sdbFloorN\ \sdbFloorDefs\ runs in its Blended/LF grid, none in a printed column\fi\fi; in the leaderboard grids of Appendix~\ref{app:corrected} it keeps out of A3
the poisoned-arm runs whose residual attack success was below $\alpha$ only because the defense had destroyed the model:
\steFloorExcludedN\ runs of the seven benchmark defenses other than ABL (\steFloorExcludedList)\ifseleven{}, \selFloorRnpN\ of RNP's\fi,
and, from the A3 counts of the failure records, \steAblFloorN\ of ABL's\ifseleven{} and \selFloorDbrN\ of D-BR's\fi. It is not idle: an earlier implementation of
ANP produced cells at $10\%$ clean accuracy that A3 recorded as successful removals
(Appendix~\ref{sec:selfcorrect}), and a success criterion that cannot tell removal from destruction is a
gap in the estimand, not in the implementation.

\textbf{The gate does not make the result.} Sweeping $\alpha$ changes the candidate set and its common
support (Table~\ref{tab:a3thr}). At the operational $\alpha=0.05$ the candidate set is \chNcandAFiveWord\
methods on a common support of \chNsuppAFiveWord\ and the gap is \chGapAFive, for which we report two
intervals rather than one, because the two standard bootstrap constructions on the same \chGapNboot\
installation resamples disagree about zero: percentile \chGapPctAFive, basic (reverse-percentile)
\chGapBasicAFive. The disagreement is a property of the statistic, not a choice we make: Kendall $\tau_a$
on \chNcandAFiveWord\ methods moves in steps of \chGapStep, the observed gap is one step, and it sits at
the \chGapPctilePrim th percentile of its own resample distribution (median \chGapMedianPrim, mean
\chGapMeanPrim, so the bootstrap bias \chGapBiasPrim\ is as large as the estimate), which pins the
percentile interval's lower endpoint on the point estimate itself---the diagnostic of a failed percentile
interval, not evidence of an effect bounded away from zero. What the resamples support is the sign and not
the magnitude: none returns a negative gap, \chGapPlePrimPct\ return exactly zero
($P^{*}(\mathrm{gap}\le0)={}$\chGapPlePrim), and dropping any single installation leaves the gap in
\chGapLoo. We therefore report the gap as positive with an unresolved magnitude, and make no claim that
its interval excludes zero.
Tightening to $\alpha=0.02$ drops NAD and leaves \chNcandATwoWord\ methods on all \chNsuppATwoWord\ with a
wider gap whose intervals touch zero (\chGapATwo, percentile \chGapPctATwo); loosening to $\alpha=0.10$
admits fine-pruning at $20\%$, drops the common support to \chNsuppATenWord\ (installations $3$, $21$,
$23$, $27$ and $29$ drop out: NAD leaves \chMissRes\ on one and fine-pruning at $20\%$ fails on four
more), and widens the gap to \chGapATen\ (percentile \chGapPctATen). The gap is positive at every
threshold; its percentile interval excludes zero at \chNexclPctWord\ of the \chNthrRowsWord\ rows and its
basic interval at \chNexclBasicWord, and the one row where the two constructions disagree is the
operational $\alpha=0.05$ (Table~\ref{tab:a3thr} prints both), so no threshold row supports an unqualified
``excludes zero''. At every threshold the published ordering coincides with the intervention-component
ordering ($\tau_a=+1.000$).

\begin{table}[h]\centering\small
\caption{Candidate-gate sensitivity. Sweeping the method-level mean-residual-ASR threshold $\alpha$ over
the fourteen installations changes the candidate set, and with it the common support the ranking is
computed on; it does not change the installation-level A3 event. Each row is the primary analysis of
Appendix~\ref{app:figmain} run at that $\alpha$: Kendall $\tau_a$ of the published ordering against each
component on that row's own common support, with an installation bootstrap on the gap. Both bootstrap
constructions are printed, and the common support each row runs on, because the two constructions
disagree about zero on the operational row. The rows are \chNthrRowsWord\ distinct analyses: $\alpha=0.02$ excludes NAD on its mean residual and so runs on four methods over all
fourteen installations, $\alpha=0.05$ (the operational threshold, and the one every other number in the
paper uses) admits NAD and therefore runs on five methods over the thirteen installations NAD clears, and
the two loosest rows readmit fine-pruning at $20\%$. The
all-fourteen-installation variant of the $\alpha=0.05$ row is the last line of Table~\ref{tab:a3supp}.}
\label{tab:a3thr}
\resizebox{\linewidth}{!}{\input{supplement/build/tab_a3thr_body.tex}}
\end{table}

\textbf{Common support is the primary convention.} Four of the five candidate methods clear the backdoor in
every installation here and NAD misses one, so the common support is thirteen of the fourteen; the
convention matters both for that installation and because a looser gate readmits methods that clear
far less often. For a candidate set $\mathcal{D}$ define
$\Omega_{\mathrm{CS}}(\mathcal{D})=\{\omega: A(\mathcal{D}_d\theta^{1}(\omega))\le\alpha\ \forall
d\in\mathcal{D}\}$, the installations on which every method in $\mathcal{D}$ clears the backdoor.
Appendix~\ref{app:figmain} restricts primary cross-method inference to
$\Omega_{\mathrm{CS}}$---thirteen of the fourteen installations here, the one exception being the installation NAD does not clear. This is a \emph{ranking-specific} extension of
Eq.~\ref{eq:estimand}, not the single-method estimand itself: Algorithm~\ref{alg:tare} handles one
defense at a time and does not define the intersection. It places every method on the same runs,
at the price of depending on $\mathcal{D}$. Pairwise separability there follows
Section~\ref{sec:units}: installations are resampled as clusters with $\Omega_{\mathrm{CS}}$ fixed,
each pair's two-sided bootstrap $p$-value for the sign of its rating difference is corrected by
Benjamini--Hochberg at $q=0.05$ within each accounting's own set of pairwise comparisons, and rank intervals
are a complementary display. As a null-centered check, an exact enumeration of the paired sign-flip test under the
sign-symmetry null on the \chNsuppWord\ per-installation rating differences (\chSignAssign\ sign
assignments per pair) returns \chSepPub\ published-accounting pairs and \chSepDid\ corrected pairs after
the same
Benjamini--Hochberg step, matching the bootstrap exactly. A4 is a property of the \emph{undefended} twin,
$A(\theta^{0})\le\alpha$, and holds for $14/14$ vision twins (attack success $\le0.8\%$ before any defense)
and $8/8$ Qwen twins ($0$); the shipped cells now carry the undefended twin's attack success as its own
column, since the flag we had named A4 in them was computed from the defended twin---the quantity
Table~\ref{tab:a3succ} reports---and both hold in $14/14$, so no number moves. The one A4 failure, the
blended-trigger twin at the benchmark's $0.0002$ rate, is not identified as a control; blended's row in
Table~\ref{tab:abl} uses the one-image twin that passes A4 (Appendix~\ref{app:blended}). Whether a twin drawn at a fixed poisoning rate is clean is a
property of the attack, not of the rate: evaluating three models on one attack's backdoor test set gives
the twin's own score, the score of a same-recipe model that never saw that trigger, and the poisoned
model's, and the first minus the second is what those ten images installed---$0.001$ for BadNets and
$0.004$ for WaNet, both on seed $0$, against $0.108$ for blended averaged over three
(\texttt{supplement/data/twin\_admissibility\_by\_attack.csv}); over the whole \chwNseedsWord-seed WaNet
grid the undefended twin's attack success runs \chwTwinRange\ (mean \chwTwinMean) and A4 holds
\chwAfour. Table~\ref{tab:a3succ} additionally reports, per defense, whether the \emph{defended}
twin still clears the threshold, which is what a reader needs to know before treating that defense's clean
arm as a control; under the corrected implementations it does so in every cell.

\textbf{When each choice was fixed.} Table~\ref{tab:choices} records the status of every analysis
choice in the ranking chain, separating decisions fixed before the relevant data were inspected
from those adopted during internal review and reported with their alternatives.

\begin{table}[!ht]\centering\scriptsize
\caption{Analysis choices behind the ranking result and when each was fixed. ``Internal review''
denotes revisions made after a first all-nine, method-level analysis had been written up and before
submission; every such revision is reported beside the alternative it replaced.}
\label{tab:choices}
\setlength{\tabcolsep}{3pt}
\begin{tabular}{p{0.35\linewidth}p{0.35\linewidth}p{0.21\linewidth}}
\toprule
choice & status & sensitivity reported \\
\midrule
installation-level A3 threshold $\alpha=0.05$ & fixed before the ranking analysis & Table~\ref{tab:a3thr} \\
DER as the field's accounting rule & fixed (\citealp{zhang2023ftsam}) & --- \\
equal weighting of configurations within an installation & fixed from the first aggregate & per-defense rows, complete case \\
independent installation as the replicate unit & adopted during internal review; the eight-installation Qwen rig was run after it & defense-seed rows marked \\
method-level mean-ASR candidate gate & adopted during internal review after the all-nine ranking & all-nine result retained (Table~\ref{tab:zoo}) \\
common-support primary convention & adopted during internal review after the fourteen-installation grid & two looser conventions (Table~\ref{tab:a3supp}) \\
installation-cluster bootstrap and BH families & fixed before the pairwise analysis & --- \\
exact sign-flip enumeration & post hoc sensitivity & reported above \\
ANP kept in the pooled TARE-Z summary & fixed & excluding-ANP rows (Table~\ref{tab:surr}) \\
ANP's pruning threshold: the method's accuracy budget & adopted when the implementation was corrected & fixed-fraction rows (Appendix~\ref{app:zoo}) \\
\bottomrule
\end{tabular}
\end{table}
 Table~\ref{tab:a3supp} reports that ranking beside the two looser conventions.
The published-versus-tax correlation is \chTauTaxPrim, \chTauTaxPrim\ and \chTauTaxAllFourteen\ under the
three conventions and the removal correlation \chTauRemPrim, \chTauRemPrim\ and \chTauRemAllFourteen; the
third convention's single tax discordance is the \chTieB\ / \chTieA\ tie-break (\chTieScoreB\ against
\chTieScoreA, a margin of \chTieMargin), not a substantive disagreement. Under the first two the only
order change is fine-pruning at $30\%$ moving from third to fourth; under the third it holds at third and
the movers are \chAllFourteenMoves. The gap is positive under all three, which is why we state the move by
point estimate and give the rank intervals.

\begin{table}[!ht]\centering\small
\caption{The primary ranking, per method, on the thirteen-installation common support of the five
methods that pass the candidate gate at $\alpha=0.05$ and keep an admissible twin (NAD misses
installation $3$; Table~\ref{tab:a3succ}).
The published and intervention-component rankings coincide ($\tau_a=+1.000$); the DiD-removal ranking does
not ($\tau_a=+0.800$, one adjacent transposition), and its five
ratings lie within $0.003$ of one another. DER as in Eq.~\ref{eq:der}, higher is better.
$^{\ddagger}$Neural Cleanse is conditional: its clean arm is exactly zero on the nine common-support
installations (ten of the fourteen overall) where its detector declines to fire, so on those its DiD equals its published cost by construction
(Appendix~\ref{app:cond}).}
\label{tab:csper}
\resizebox{\linewidth}{!}{\input{supplement/build/tab_csper_body.tex}}
\end{table}

\begin{table}[h]\centering\small
\caption{The ordering conclusion under three support-set conventions. Kendall $\tau_a$ on unrounded DER
scores, no ties; the first row is the primary analysis of Appendix~\ref{app:figmain}. Four of the five
candidates clear the backdoor in every installation and NAD misses one, so the conventions differ by that
single installation; they are reported because a looser gate readmits methods for which they differ far more. Neural Cleanse is first under both
accountings in all three.}
\label{tab:a3supp}
\resizebox{\linewidth}{!}{\input{supplement/build/tab_a3supp_body.tex}}
\end{table}

\FloatBarrier
\FloatBarrier
\FloatBarrier
\section{Two controls on the transportability failure}\label{app:attacklr}

Our installations and BackdoorBench's disagree about whether the same named defense removes the same named
attack. At the framework's own defense operating point---lr $0.01$, $100$ recovery epochs, a $5\%$ clean
subset---its fine-tuning leaves a mean residual attack success of \exOpFtBb\ across its eleven seeds and
\exOpFtOurs\ across our eight; its NAD leaves \exOpNadBb\ and \exOpNadOurs. The rigs run the same attack (BadNets, $3\times3$
patch, target class $0$) on the same architecture at the same defender budget, and differ in how the
backdoor is written: BackdoorBench trains the poisoned model for $100$ epochs at lr $0.01$, we for $40$ at
lr $0.1$. Table~\ref{tab:attacklr} separates those two coordinates.

Both blocks run the same six defenses, so the comparison is cell for cell. Rebuilding the attack at the
framework's \emph{full} recipe makes five of the six clear, including last-convolution fine-pruning, which
clears none of the $88$ cells built on our own installations at any sparsity from $10\%$ to $95\%$ and under
either recovery schedule, and including NAD, which clears none of ours. Changing only the learning rate, at
our own epoch count, does not: fine-tuning then clears $1$ of $3$, last-convolution fine-pruning $0$ of $3$
and NAD $0$ of $3$, on installations whose pre-defense attack success is $0.94$--$0.95$. So the learning
rate alone is not the cause, and we do not name one; the two recipes also differ in how strong a backdoor
they leave ($0.90$ against $0.95$ pre-defense attack success), which is a second candidate we cannot
separate at three installations. ANP is the one defense that clears in neither block, because under its own
accuracy budget it prunes nothing here; the two rigs also apply different threshold rules---the framework's
ANP takes the argmin of test-set attack success under a \chbAnpBudgetPct\ relative accuracy budget, ours
the largest pruning within a deployer-side two-point budget on the defender's own split---so the
framework's \chbAnpDefAthreePt\ on BadNets is an oracle-selected number, not a deployer-available one
(Appendix~\ref{app:bb}).

What the two controls agree on is the accounting. Across the eight unconditional cells that clear in either
block, the reported cost runs from $+1.3$ to $+7.9$ while the interaction stays between $-0.89$ and $+1.50$,
and every one of those intervals contains zero at three installations. The two conditional cells are the
instructive exception rather than a contradiction: Neural Cleanse's gated clean arm no-ops on every twin in
the lower block, so its whole $+5.01$ lands in the interaction by construction, and forcing that same
branch on a never-poisoned model costs \ncForcedOurs\ (Appendix~\ref{app:cond}). The accounting does not depend on
which side of the removal boundary a cell falls.

\begin{table}[H]\centering\small
\caption{Two controls on the attack recipe, defended at the same six operating points inside our harness,
three installations each. The upper block changes both coordinates of the install (epochs and learning
rate) and five of the six defenses clear; the lower changes only the learning rate and three do, two of
them only partly. Pre-defense attack success is $0.90$ in the upper block and $0.94$--$0.95$ in the lower.
$^{\P}$ANP's accuracy-guarded threshold prunes nothing on these models, so it neither costs nor removes
anything; that is the method declining to act, not a zero measurement. $^{\ddagger}$Neural Cleanse's
detector fires on every poisoned model here and on no twin in the lower block, so its clean arm there is
exactly zero and its whole reported cost lands in the interaction by construction; Appendix~\ref{app:cond}
prices that branch directly.}
\label{tab:attacklr}
\setlength{\tabcolsep}{4pt}
\begin{tabular}{llcccc}
\toprule
attack recipe & defense & residual $A$ & $\cDef$ (pt) & clean arm (pt) & $\cRem^{\mathrm{DiD}}$ (pt) \\
\midrule
$100$ ep, lr $0.01$ & fine-tune lr $0.01$        & $0.020$ & $+2.59$ & $+2.90$ & $-0.31$ \\
                    & fine-prune $30\%$ all-layer & $0.016$ & $+7.11$ & $+8.00$ & $-0.89$ \\
                    & fine-prune $30\%$ last-conv & $0.020$ & $+1.49$ & $+1.06$ & $+0.43$ \\
                    & NAD                        & $0.024$ & $+2.00$ & $+0.50$ & $+1.50$ \\
                    & ANP$^{\P}$                 & $0.901$ & $+0.00$ & $+0.00$ & $+0.00$ \\
                    & Neural Cleanse$^{\ddagger}$ & $0.015$ & $+6.30$ & $+3.21$ & $+3.09$ \\
\midrule
$40$ ep, lr $0.01$  & fine-tune lr $0.01$        & $0.132$ & $+2.27$ & $+2.16$ & $+0.11$ \\
                    & fine-prune $30\%$ all-layer & $0.020$ & $+7.87$ & $+7.96$ & $-0.09$ \\
                    & fine-prune $30\%$ last-conv & $0.119$ & $+2.00$ & $+1.72$ & $+0.28$ \\
                    & NAD                        & $0.443$ & $+1.33$ & $+1.23$ & $+0.09$ \\
                    & ANP$^{\P}$                 & $0.947$ & $+0.00$ & $+0.00$ & $+0.00$ \\
                    & Neural Cleanse$^{\ddagger}$ & $0.014$ & $+5.01$ & $+0.00$ & $+5.01$ \\
\bottomrule
\end{tabular}
\end{table}

\FloatBarrier
\section{A headroom control: does a starting gap alone produce an interaction?}\label{app:headroom}
\paragraph{In full.} \textbf{What is left for removal we do not correct, because the correction is not identified.} Two
never-poisoned arms degraded to the same gap by different routes disagree in sign about what it buys, and
the poisoned arm's own behaviour cannot tell them apart. We report the uncorrected decomposition with the route spread, \exShareShort\ to \exShareNoise\ against
\exShareUncorr, as its uncertainty.

A poisoned model starts below its twin by $g$, so a defense that pulls both arms toward a common level
returns a negative interaction with nothing having been removed. That mechanism---regression toward the
mean under a shared optimisation---would produce residues of the sign and roughly the size we report, and
it is not testable on the poisoned arm, where the two explanations are observationally identical. It is
testable on a clean one.

We build a third arm that carries no backdoor and still starts below the twin: the same install recipe on
the same seed, with the labels of $1\%$ of the training set flipped to a uniformly random wrong class and
every image untouched. Across the eight headline installations that arm lands $+0.38$ $[+0.08,+0.67]$
points below its twin, against the $+0.41$ $[+0.15,+0.68]$ the poisoned arm carries on the same grid, so
the two arms enter their defenses from a matched deficit; the degraded arm's removal term is zero by
construction. Applying the identical defenses and reading the identical four-cell contrast gives an
interaction of $-0.28$ $[-0.63,+0.08]$ at the installation level (Table~\ref{tab:headroom}). A starting gap
of this size therefore does not manufacture an interaction on this rig, and in particular cannot produce a
\emph{positive} one: the headroom effect, to the extent it is present at all, runs negative.

\textbf{The effect scales with the gap---but only for this route.} Running the same control at a second,
larger deficit gives a slope: $-0.28$ $[-0.63,+0.08]$ at $g=+0.38$ and $-0.65$ $[-1.09,-0.21]$ at $g=+1.21$,
which resolves below zero, about \exSlopeNoise\ points per point of gap (least squares through the
origin; the two-point slope is \exSlopeNoiseTwoPt), so BackdoorBench's $g={}$\exG\ would carry
\exNoiseAtG\ with nothing removed. Applied to every row of Table~\ref{tab:bbdef} that does two things at once,
and we report both: it removes the negative residues (fine-pruning $-1.27\to-0.13$, SAU $-1.31\to-0.17$,
I-BAU $-1.85\to-0.71$) and it raises the positive ones past their own reported costs (fine-tuning
$+0.88\to+2.02$ against a reported $+1.01$, NAD $+1.77\to+2.90$ against $+2.10$), moving the pooled
seven-defense split from \exShareUncorr\ removal to \exShareNoise. An earlier version of this paper applied it in the first
direction only, which is not a correction but a choice.

\textbf{A second degradation route says the correction is a property of the route, not of the gap.} Label
noise is only one way to put a clean model below its twin, and it has a property poisoning does not: these
defenses fine-tune it away. Over the same eight installations and the same three configurations the
label-noise arm enters at $+0.38$ and leaves at $+0.01$ to $+0.28$; the poisoned arm enters at $+0.41$ and
leaves at $+0.33$ to $+0.50$, widening under both pruning sparsities. A deficit the defense repairs is not
a stand-in for one it does not. We therefore built a third arm degraded a different way: trained on $90\%$
or $75\%$ of the data, no corrupted label and no trigger anywhere, so removal is again zero by construction
but the deficit is missing capability rather than a worse optimum (\texttt{cifar\_shortage\_cells.csv}).
It enters at $+0.38$ and $+1.12$, matching the label-noise arm's two deficits, and it behaves like the
poisoned arm: it leaves at $+0.20$ to $+0.63$ and $+0.96$ to $+1.64$, widening under pruning. Its
interaction is $+0.01$ $[-0.40,+0.42]$ and $+0.25$ $[-0.18,+0.67]$---\emph{nothing}, and if anything
positive. Extrapolated to $g={}$\exG\ the two routes give \exNoiseAtG\ and \exShortAtG, a pooled removal share of
\exShareNoise\ and \exShareShort\ against \exShareUncorr\ uncorrected.

\textbf{So we apply neither correction.} What a starting gap buys is not identified by the gap; it is a
property of how the deficit was made, and the observable that separates our two routes---whether the
defense closes the deficit---puts poisoning with the route that buys nothing. We report the uncorrected
decomposition as the measurement and the route spread as its uncertainty. A third arm we tried first is
reported for completeness and is uninformative: applying the trigger to $1\%$ of images but sending each to
a different wrong class, so the trigger maps to nothing, produced no deficit at all ($0.9426$ clean
accuracy against a twin at $0.9381$), the trigger apparently giving the label noise a handle the network
isolates. What survives all of this untouched is the clean-arm cost itself, measured directly on a model
with nothing to remove and never differenced against the poisoned arm; that is the quantity we ask
benchmarks to print. ABL stays outside every reading, an order of magnitude beyond any correction and
separately unidentified at the realised-computation level.

\textbf{The triple difference inherits the same problem.} Differencing the poisoned arm's interaction
against the label-noise arm's at the matched deficit leaves $+0.32$ $[+0.07,+0.57]$ and resolves where the
uncorrected interaction does not, but since the two arms differ in repairability that contrast prices
differential repairability rather than removal, and against the data-shortage arm it returns essentially
the uncorrected value. We report the uncorrected form as primary.

Two limits remain. The slope is measured on one dataset and one architecture with two points, so the
extrapolation to $g=+2.06$ is an extrapolation. And label noise is one degradation route among several; a
differently damaged model could interact with a defense differently.

\begin{table}[!ht]\centering\small
\caption{Headroom control at two starting deficits, eight installations each. The degraded arm is a
never-poisoned model trained on the identical recipe with $1\%$ or $3\%$ of labels flipped and no image
altered, so $\tau_{\mathrm{int}}\equiv0$; its clean arm is the same twin the paper uses everywhere, so the
clean-arm column is common to both blocks ($+4.36$, $+1.99$, $+3.31$) and is omitted from the right-hand
one. Intervals are $95\%$ $t$ over installations. The interaction grows with the deficit at about \exSlopeNoise\
points per point (least squares through the origin; \exSlopeNoiseTwoPt\ between the two points), which is
what limits the reading of a negative residue at a large gap.}
\label{tab:headroom}
\resizebox{\linewidth}{!}{\begin{tabular}{lcccc}
\toprule
 & \multicolumn{2}{c}{$g=+0.38$ ($1\%$ label noise)} & \multicolumn{2}{c}{$g=+1.21$ ($3\%$)} \\
\cmidrule(lr){2-3}\cmidrule(lr){4-5}
Defense & $\cDef$, degraded (pt) & interaction (pt) & $\cDef$, degraded & interaction \\
\midrule
fine-tune lr $0.05$ & $+4.00\ \scriptstyle[+3.68,+4.31]$ & $-0.37\ \scriptstyle[-0.77,+0.03]$ & $+3.48\ \scriptstyle[+3.20,+3.75]$ & $-0.89\ \scriptstyle[-1.37,-0.40]$ \\
fine-prune $30\%$   & $+1.89\ \scriptstyle[+1.56,+2.22]$ & $-0.10\ \scriptstyle[-0.47,+0.27]$ & $+1.55\ \scriptstyle[+1.28,+1.81]$ & $-0.45\ \scriptstyle[-0.99,+0.10]$ \\
fine-prune $40\%$   & $+2.95\ \scriptstyle[+2.54,+3.35]$ & $-0.36\ \scriptstyle[-0.82,+0.10]$ & $+2.71\ \scriptstyle[+2.47,+2.94]$ & $-0.61\ \scriptstyle[-1.15,-0.06]$ \\
\midrule
\textbf{installation level} & --- & $\mathbf{-0.28}\ \scriptstyle[-0.63,+0.08]$ & --- & $\mathbf{-0.65}\ \scriptstyle[-1.09,-0.21]$ \\
\bottomrule
\end{tabular}}
\end{table}

\FloatBarrier
\section{Notation}\label{app:notation}

\begin{table}[H]\centering\small
\caption{Every symbol the paper uses for a cost, in one place. $\theta_{\mathrm{pois}}$ and
$\theta_{\mathrm{twin}}$ are the two arms of one installation, $\mathcal{D}_b$ the defense at budget $b$,
$Q$ the capability metric and $A$ the attack-success metric.}
\label{tab:notation}
\resizebox{\linewidth}{!}{\begin{tabular}{llp{0.52\linewidth}}
\toprule
symbol & definition & what it answers \\
\midrule
$\cDef$ & $Q(\theta_{\mathrm{base}})-Q(\mathcal{D}_b\theta_{\mathrm{pois}})$ & the number the field reports \\
$c_{\mathrm{run}}$ & $Q(\theta_{\mathrm{pois}})-Q(\mathcal{D}_b\theta_{\mathrm{pois}})$ & what a deployer loses by running the defense \\
$d_{\mathrm{base}}$ & $Q(\theta_{\mathrm{base}})-Q(\theta_{\mathrm{pois}})$ & reporting-baseline drift; zero in every headline here \\
$c_{\mathrm{clean}}$ & $Q(\theta_{\mathrm{twin}})-Q(\mathcal{D}_b\theta_{\mathrm{twin}})$ & what the same defense costs with no backdoor (the tax) \\
$\cRem$ & $Q(\mathcal{D}_b\theta_{\mathrm{twin}})-Q(\mathcal{D}_b\theta_{\mathrm{pois}})$ & paired contrast (interaction plus $g$) \\
$\cRem^{\mathrm{DiD}}$ & $c_{\mathrm{run}}-c_{\mathrm{clean}}$ & interaction, DiD form; equals $\cRem-g$ \\
$g$ & $Q(\theta_{\mathrm{twin}})-Q(\theta_{\mathrm{pois}})$ & what the poisoning itself cost, before any defense \\
$c^{0}_{\mathrm{def}}$ & $c_{\mathrm{clean}}$ on a model built with no poison & the manufactured cost of Section~\ref{sec:admiss} \\
$\hat{c}_{\mathrm{int}}$ & $\cRem$ with an unpaired clean model for $\theta_{\mathrm{twin}}$ & the TARE-Z estimate \\
\bottomrule
\end{tabular}}
\end{table}

\FloatBarrier
\section{The TARE protocol in full}\label{app:alg}
\begin{figure}[h]
\centering
\begin{tikzpicture}
  \node[anchor=south west,inner sep=0] (im) at (0,0)
    {\includegraphics[width=0.58\textwidth]{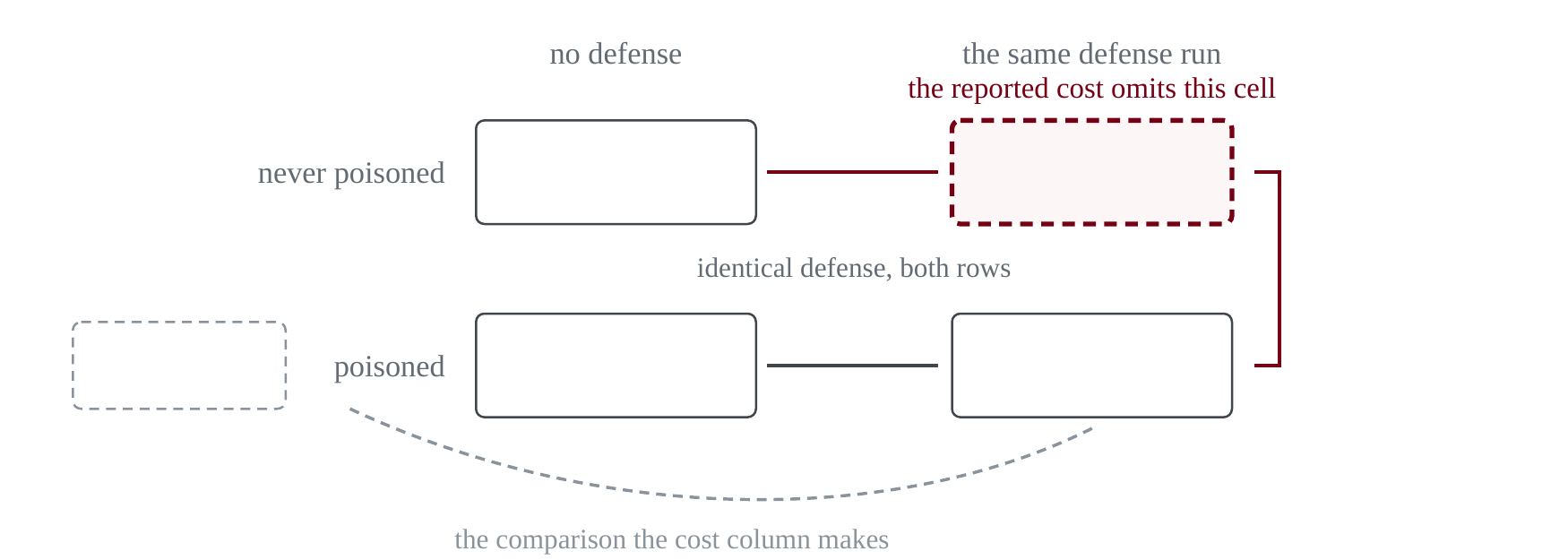}};
  \begin{scope}[x={(im.south east)},y={(im.north west)},font=\small]
    \node at (0.3929,0.6925) {$\theta_{\mathrm{twin}}$};
    \node[text=cRem] at (0.6964,0.6925) {$\mathcal{D}_b\theta_{\mathrm{twin}}$};
    \node at (0.3929,0.3475) {$\theta_{\mathrm{pois}}$};
    \node at (0.6964,0.3475) {$\mathcal{D}_b\theta_{\mathrm{pois}}$};
    \node[text=cTax] at (0.1143,0.3475) {$\theta_{\mathrm{base}}$};
    \node[text=cRem,anchor=west,font=\normalsize] at (0.8232,0.5400) {$\cRem$};
    \node[text=cRem,anchor=west,font=\scriptsize,align=left] at (0.8232,0.4300)
      {$\cRem^{\mathrm{DiD}}\!=\!\cRem-g$};
    \node[text=cTax,font=\scriptsize,align=center] at (0.1143,0.1750)
      {headline: $\theta_{\mathrm{base}}\!=\!\theta_{\mathrm{pois}}$\\[-1pt] LLM convention: base model};
  \end{scope}
\end{tikzpicture}
\sloppy\caption{\textbf{The reported cost has a missing identifying cell.} Two factors---was a backdoor
installed, was a defense run---generate a $2\times2$; filling the missing cell makes the interaction a
difference of the two horizontal drops (Eq.~\ref{eq:did}). Crimson bracket: the post-defense contrast
$\cRem$; $\cRem^{\mathrm{DiD}}$ nets the pre-defense gap. The reported cost reaches from
$\theta_{\mathrm{base}}$ (grey, dashed), omitting the arm the contrast requires.}
\label{fig:estimand}
\end{figure}
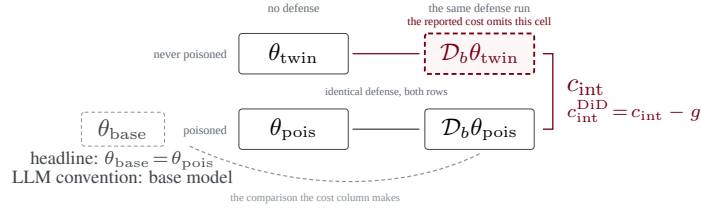

\begin{algorithm}[H]
\caption{\textbf{TARE.} The control is one arm, not a new defense.}
\label{alg:tare}
\begin{algorithmic}[1]
\Require poisoned training set $S^{1}$, its poison-free counterpart $S^{0}$, defense
$\mathcal{D}_b$, capability metric $Q$, attack metric $A$, reporting baseline
$\theta_{\mathrm{base}}$ (Section~\ref{sec:setup}), removal threshold $\alpha$
\State $\Omega_{\mathrm{A4}}\gets\emptyset$
\For{each installation $\omega$ (seed, data order, poison placement)}
  \State $\theta^{1}\gets\textsc{Train}(S^{1};\omega)$;\quad
         $\theta^{0}\gets\textsc{Train}(S^{0};\omega)$ \Comment{identical $\omega$: the twin}
  \State \textbf{if} $A(\theta^{0})>\alpha$ \textbf{then discard} $\omega$ \textbf{else} $\Omega_{\mathrm{A4}}\gets\Omega_{\mathrm{A4}}\cup\{\omega\}$
         \Comment{A4: an attacked twin manufactures the thesis}
  \State $\hat\theta^{1}\gets\mathcal{D}_b\theta^{1}$;\quad
         $\hat\theta^{0}\gets\mathcal{D}_b\theta^{0}$ \Comment{A2: the identical map on both arms}
  \State $c_{\mathrm{clean}}(\omega)\gets Q(\theta^{0})-Q(\hat\theta^{0})$ \Comment{the tare: no removal gate}
  \State $s(\omega)\gets\mathbf{1}\{A(\hat\theta^{1})\le\alpha\}$ \textbf{; record} $s(\omega)$
         \Comment{A3 is an event, not a filter: keep the record}
  \State \textbf{if} $s(\omega)=0$ \textbf{then continue}
         \Comment{out of the success-conditioned cost, kept in the success-rate denominator}
  \State $g\gets Q(\theta^{0})-Q(\theta^{1})$;\; $\cRem\gets Q(\hat\theta^{0})-Q(\hat\theta^{1})$;\;
         $\cRem^{\mathrm{DiD}}\gets\cRem-g$;\; $\cDef\gets Q(\theta_{\mathrm{base}})-Q(\hat\theta^{1})$
\EndFor
\State \Return the success rate $\hat p_{\mathrm{succ}}=|\Omega_{\mathrm{A4}}|^{-1}\sum_{\omega\in\Omega_{\mathrm{A4}}} s(\omega)$, the A4
admissibility rate $|\Omega_{\mathrm{A4}}|/|\Omega|$, the tare $c_{\mathrm{clean}}$ over $\Omega_{\mathrm{A4}}$, and,
on $\{s=1\}$, $\cRem^{\mathrm{DiD}},\cRem,\cDef$ with installation-level intervals
(Section~\ref{sec:units}); a clean--clean placebo pair is an optional calibration
\end{algorithmic}
\end{algorithm}

\FloatBarrier
\section{Limitations in full}\label{app:lim}
All leaderboard and tare-column evidence comes from one toolkit (BackdoorBench), and so does the scheduler evidence apart from
one replication: the block, reproduced in BackdoorBox with its own WaNet, trainer and defenses\ifsseventeen{} (four defenses, five seeds)\fi, and the
configurations of six further code bases read at pinned commits (Appendix~\ref{app:toolkits}); TrojanZoo and backdoor-toolbox are
read, not run, and our own CIFAR-10 harness is the only other vision rig. The never-firing
milestones are not an accident of the released code alone: the $[100,200,300,400]$ schedule appears in the 2022
BackdoorBench paper's own hyper-parameter table (arXiv:2206.12654, Table~4, for WaNet and Input-Aware's classifier).

Our head-to-head re-ranking is computed on defenses we implement and run ourselves; the
community BackdoorBench framework is re-scored separately over eleven independent installation seeds,
three disclosed exploratory and eight confirmatory under a frozen protocol, for both the cost
decomposition and the ranking analyses; but attaching a corrected term to a specific
\emph{published} entry still needs either the authors' runs or the surrogate at their scale.
The common-support ranking targets the joint-success subpopulation of the selected candidate set
and can change if that set changes; it is a fair comparison within that set, not an unconditional
population leaderboard. Rank intervals are conditional on the selected candidate set and
support convention, both adopted during internal review and reported with the alternatives they replaced;
threshold sensitivity is reported separately (Table~\ref{tab:a3thr}) rather than folded into the bootstrap.
The removal share on BackdoorBench is not identified: correcting for what a starting deficit alone buys
moves it from \exShareUncorr\ to \exShareNoise\ or \exShareShort\ by route, and the correction is measured on a different rig, so we report the range
and claim neither end (Section~\ref{sec:rerank}, Appendix~\ref{app:headroom}).
The removal estimates are intervals, not points, and the shares that summarize them are ratios of two
estimated quantities: bootstrapping the ratio itself over installations gives $12\%$ $[6,18]$ on Qwen,
$1\%$ $[-9,+11]$ on CIFAR-10 and $15\%$ $[-10,+33]$ on BackdoorBench for the interaction against $\cDef$,
and $12\%$ $[7,17]$, $15\%$ $[10,19]$ and $55\%$ $[44,64]$ for the post-defense gap against $\cDef+g$
(Appendix~\ref{app:headroom}). Dividing the paired numerator by $\cDef$ alone returns $104\%$ on BackdoorBench, which reads as though removal exceeded the whole reported cost; that is a
$g$-inclusive numerator over a $g$-exclusive denominator, not a finding. The vision null is bounded by its own power: at $n{=}8$ and the observed
installation spread the smallest DiD effect we could detect at $80\%$ power is $0.45$ points, $18\%$ of the
reported cost, so ``zero within noise'' there means ``below one fifth of the reported number'', not nil.
Two measurement floors bound how fine a statement the instruments support: MMLU over $2{,}000$ items
carries a $1.0$-point standard error per cell, and CIFAR-10 accuracy over $10{,}000$ images $0.25$ points.
Those are per-cell floors on an \emph{absolute} score, not on the contrast: all four cells of an
installation are scored on the identical item set, so the contrast's sampling error is that of the items on
which the two models disagree, realised independently per installation and therefore already inside the
installation-level interval. What the
floors do bound is any comparison across item sets, and we make none. The one all-installation removal estimate that resolves above zero,
Qwen's, does so only after averaging two defenses whose per-installation estimates anti-correlate at
$r={}$\taboneQwenR\ over the \taboneQwenBothN\ installations both clear; neither resolves alone
(fine-tuning \taboneQwenFtDid, fine-pruning \taboneQwenFpDid), and were the two independent the pooled
interval would be \taboneQwenIndepCI\ rather than the observed \taboneQwenDid, so the resolution rests
on that anti-correlation. TARE is also metric-specific: it
attributes cost with respect to a prespecified $Q$, and a multidimensional notion of utility would
need one estimate per metric or a prespecified aggregate; on vision the paired contrast resolves ($+0.43$ $[+0.30,+0.55]$) where the DiD does not, the
difference being the pre-defense gap of Section~\ref{sec:rerank}, and both belong to the
scarce-clean-data regime. The surrogate has low average error within architecture and recipe; across
architectures it carries a characterized $1.3$-point bias and supports only a qualitative
verdict, with a same-recipe degradation probe marking where it ceases to apply. Interpolation needs the pre-poisoning checkpoint and is a
publisher-side result. Our backdoors are self-installed; an adaptive adversary---shaping the
poisoning update to align with capability directions, or inflating $\cRem$ to game a corrected
score---is untested.

\FloatBarrier
\section{Why training-side defenses cost what they cost}\label{sec:mechanism}

If the reported cost is the price of the intervention, what sets that price? For the
training-side paradigms it is largely the erasure budget. Across the fine-tuning sweep both
the attack-success reduction and $\cDef$ broadly follow $\int\!\eta\,dt$, the learning-rate
integral over the defensive run, consistent with a shared optimization-dynamics account of attack
reduction and capability erosion. TARE does not identify a mechanism; what this pattern explains is why
the two effects can be confounded in $\cDef$, why the twin separates their observed contributions, and why
defense strength and defense cost rise together in Table~\ref{tab:cif8}.

\paragraph{A cheap probe screens for the catastrophic-cost regime.}
If the cost is a property of the model's optimization dynamics rather than of any backdoor, a
short probe of the defensive run should predict it---and it does, at the level of sensitivity
class. A $40$-step probe of the defense run on the model in hand---initialised from the checkpoint under audit,
which in our runs is the poisoned one, not from a clean model---separates the catastrophic-forgetting
regime from the benign one: Mistral-7B sheds $33.7$ MMLU points in those $40$ steps---roughly $7$ to $11$
times what Qwen2.5-7B ($3.0$) or Llama-3.2-3B ($5.1$) shed under the identical recipe and evaluation. At
$120$ steps the three sit at $32.1$, $5.8$ and $13.4$. The probe separates the catastrophic class from the
benign one; we do not claim it orders the two benign models, whose $40$- and $120$-step figures differ by
more than a factor of two. It is a screen for ``will this model be
expensive to fine-tune,'' not a point predictor, and it needs no clean reference. That the expense is legible in $40$ steps, well
before the full defense run completes, is further evidence that training dynamics contribute materially
to the reported cost; it does not imply that the removal component is zero, and
Table~\ref{tab:bbdef} shows it often is not.

\FloatBarrier
\section{Weight interpolation}\label{app:interp}
Interpolating a small distance back along the recorded poisoning update---$\beta=0.05$ on
Qwen---returns attack success to its pre-poisoning baseline ($A:0.98\!\to\!0.10$) at a
capability \emph{above} the poisoned model's ($72.4$ vs.\ $71.9$), whereas training-side
defense reaches the same attack success only by paying a $\cDef$ of several points. The
interpolation is not a cleverer removal; it is cheaper only because it does not fine-tune. This
point returns attack success to its pre-poisoning baseline but does not satisfy our $5\%$ A3 gate,
and the operator, task-arithmetic negation \citep{ilharco2022task}, requires the pre-poisoning
checkpoint; we therefore present it as a publisher-side rollback and accounting illustration, not
as a method and not as an identified TARE removal-cost estimate. Its threat-model standing is in
Appendix~\ref{app:threat}.

\FloatBarrier
\section{A third domain: the accuracy cost of pruning}\label{app:ceiling}

The two moves of this paper---differencing against a same-recipe control, and a budget knob that
manufactures the headline cost---are not specific to security. We give one instance in a
structurally different domain: model compression, where the ``accuracy cost of pruning'' is
reported and compared much as the cost of removal is. We prune a clean ResNet-18 (no backdoor,
base accuracy $93.6\%$) to three sparsities and recover under four clean-data budgets, reporting
the accuracy cost (base minus recovered) in Table~\ref{tab:ceiling}.

\begin{table}[!ht]\centering
\caption{Pruning accuracy cost (points) of a clean ResNet-18, by sparsity and recovery-data
budget. At every fixed sparsity the ``cost'' falls sharply as recovery data grows, so the
reported number cannot be attributed to sparsity alone---the same confound, in a domain with no
backdoor and no forgetting.}
\label{tab:ceiling}
\begin{tabular}{lcccc}
\toprule
sparsity & $1\%$ budget & $5\%$ & $20\%$ & $100\%$ \\
\midrule
$20\%$ & $2.83$ & $1.79$ & $1.06$ & $0.30$ \\
$40\%$ & $9.02$ & $4.63$ & $2.37$ & $0.39$ \\
$60\%$ & $27.83$ & $12.33$ & $5.60$ & $1.10$ \\
\bottomrule
\end{tabular}
\end{table}

The pattern is the one we document for removal. At a scarce budget---the regime compression
benchmarks routinely adopt---pruning to $60\%$ appears to cost $28$ points; given the full
clean set the same pruning costs $1.1$. The recovery budget confounds the apparent cost of
sparsity: at every fixed sparsity the reported cost falls sharply as recovery data grows, so the
number cannot be read as a property of the sparsity it names. A genuine residual intrinsic
cost remains (it rises with sparsity even at full budget, $0.3\!\to\!1.1$ points from $20\%$
to $60\%$), analogous to the resolved removal cost on vision; but the headline
figures overstate it by an order of magnitude, and a matched-budget control---or the budget
sweep itself---separates the two. That the same confound and the same correction recur in
compression, backdoor removal, and machine unlearning is why we frame the contribution as a
general recipe for auditing ``cost of $X$'' claims, not a backdoor-specific fix.

We report a caveat against our own interest: the dependence is not clean, and the clearest
counterexample is in our own probe data. On Mistral-7B the identical defense run sheds $33.7$ MMLU points
by step $40$ and $32.1$ by step $120$: more budget, less loss. A pure integral law cannot produce that, so
the integral governs the training-side cost only approximately, and no downstream claim here rests on the
law being exact.
Interpolation escapes the budget entirely---it does not optimize, so there is no integral to
pay, which is the mechanistic counterpart of its near-zero cost and the reason it is
available only to a party holding the pre-poisoning checkpoint.

\FloatBarrier
\section{Defense hyperparameters}\label{app:hyper}
\textbf{Language-model settings in full.} Both LLM installations use LoRA supervised fine-tuning in
fp32 on a $2000$-example pool drawn from Alpaca-52k ($51{,}699$ instructions), of which a fraction
$0.10$ ($200$ examples) carries the trigger \texttt{tq84z} and the target string; $150$ steps, learning
rate $2\times10^{-4}$, batch $8$, gradient accumulation $2$; the twin is the identical run at poison
fraction $0$. The defender draws a fresh $2000$-example pool from the same corpus, from an RNG stream independent of
the installation draw (Appendix~\ref{sec:selfcorrect}), a \textbf{$3.87\%$ clean-data budget}. The fine-tuning defense runs $150$ steps at $4\times10^{-5}$,
batch $4$, accumulation $2$, Adafactor; the fine-pruning defense zeroes $20\%$ of MLP channels using
$32$ calibration batches and then recovers for $100$ steps at $3\times10^{-5}$. Utility is MMLU over
$2000$ items and residual attack success over $100$ triggered prompts. These are our own operating
points, not any published defense's; Appendix~\ref{app:fmix} decomposes a published one. They are the
cheapest that clear: of the \qwOpN\ continued-training configurations run on installation $11$
(Table~\ref{tab:qwenops}), the cheapest that clears the backdoor costs \qwOpCheapest\ points, and the
\qwOpBelowN\ cheaper ones (\qwOpBelowCostLo\ to \qwOpBelowCostHi\ points) leave residual attack success
\qwOpBelowAsrLo--\qwOpBelowAsrHi, so the cost is the price of an intervention strong enough to work, not a
setting chosen for effect.

\begin{table}[h]\centering\small
\caption{Every configuration run on installation $11$, ordered by reported cost: a configuration clears
(A3) only from \qwOpCheapest\ points upward (\qwOpClearN\ of \qwOpN\ clear overall). Printed by
\texttt{analysis/llm\_operating\_points.py} from \texttt{supplement/data/llm\_arm\_evaluations.json}.}
\label{tab:qwenops}
\resizebox{0.75\linewidth}{!}{\input{supplement/build/tab_qwen_ops_body.tex}}
\end{table}

The
language-model budget sweep of Table~\ref{tab:lmbudget} re-installs installation $11$ from the base model and
runs continued full fine-tuning for $250$ steps at $4\times10^{-5}$ on both arms with defender pools of
$250$ to $4000$ examples drawn the same way.

Operators we borrow rather than claim: task-arithmetic negation \citep{ilharco2022task} for the
interpolation defense; the publisher threat model of \citet{choi2023potd} with the
near-constant-count poisoning regime of \citet{souly2025poisoning}; and the catastrophic-forgetting
reading of training-side cost \citep{chang2024forgetting,overtraining2026lr}.

All defenses are given the same defender budget: a $5\%$ clean subset of the training data drawn
independently of the poison indices, no knowledge of the trigger, and no access to poisoned samples. The
clean-data budget sweep of Table~\ref{tab:cr} varies only that fraction, for all-layer fine-pruning at
$30\%$ on ten installations. Every configuration below applies the same specified policy and hyperparameters to the poisoned
model and its paired clean twin; conditional policies may execute different branches
(Appendices~\ref{app:cond},~\ref{app:bb}).

\paragraph{Continued fine-tuning.} SGD, momentum $0.9$, weight decay $5\times10^{-4}$,
cosine schedule, $10$ epochs, learning rate swept over $\{0.01,0.02,0.05\}$ (vision; a fourth rate,
$0.005$, was swept before the defender-draw correction and not rebuilt, so it appears nowhere here) and
$\{2,3,4,5\}\times10^{-5}$ with $150$ steps, batch $4$ (language). The headline CIFAR-10 grid carries
lr $0.02$ and $0.05$, success-conditioned cell by cell; the ranking's method-level mean-residual-ASR gate
admits only $0.05$ (Appendix~\ref{app:a3}).

\paragraph{Fine-pruning \citep{liu2018finepruning}.} Channel activations are accumulated over
$20$ calibration batches of clean data; the lowest-activation $\{10,20,30,40\}\%$ of channels in
every convolution with $\geq\!64$ output channels are zeroed, followed by the same $10$-epoch
recovery fine-tune at learning rate $0.01$. Interpolation sweeps $\beta$ on a grid of $14$ points
in $[0,1]$.

\paragraph{Adversarial Neuron Pruning \citep{wu2021anp}.} Every batch-norm layer is given a
learnable per-channel mask and an adversarial perturbation on its affine parameters.
Perturbation budget $\epsilon=0.4$ ($\ell_\infty$, sign ascent), mask optimizer SGD at
learning rate $0.2$ with momentum $0.9$, $2000$ steps, masks clamped to $[0,1]$. Three details follow the
reference implementation and each of them mattered. The mask multiplies the batch-norm \emph{scale} only,
so a pruned neuron still emits its bias; the trade-off weights the \emph{natural} loss by $\alpha=0.2$; and
the neuron noise is redrawn at random before each perturbation round. The pruning threshold is then chosen
under a clean-accuracy budget on the defender's own split---we take the largest pruning whose accuracy
drop stays within two points---rather than by pruning a fixed global fraction. Under that protocol ANP
costs $+2.00$ $[+1.12,+2.88]$ points, keeps every twin admissible, and prunes too little to clear this
attack: mean residual attack success \chANPResTwo, clearing \chANPAthree. Forcing a fixed $20\%$ prune instead
clears $5$ of $8$ at $+7.67$ $[+5.44,+9.89]$ points and destroys the model on $2$ of $8$. Both readings are
in Appendix~\ref{app:zoo}; the accuracy-guarded one is primary because it is BackdoorBench's rule for ANP
(\texttt{acc\_ratio}); the ANP paper's own fixed mask threshold is the second reading, under which our harness
collapses on two of eight installations.

\paragraph{Neural Attention Distillation \citep{li2021nad}.} The teacher is the same model
fine-tuned on the clean budget for $10$ epochs at learning rate $0.05$ (the corrected teacher of
Appendix~\ref{sec:selfcorrect}; our first teacher, at $0.01$, left the backdoor intact and is retired). The student is
initialized from the original model and trained for $10$ epochs at learning rate $0.01$ with
loss $\mathcal{L}_{\mathrm{CE}} + \beta\sum_{\ell}\lVert
a(f^{s}_{\ell}) - a(f^{t}_{\ell})\rVert_2^2$, where $a(\cdot)$ is the channel-pooled,
$\ell_2$-normalized attention map, $\ell$ ranges over the four residual stages, and
$\beta=500$.

\paragraph{Neural Cleanse \citep{wang2019neuralcleanse}.} For each of the ten classes we
optimize a mask and pattern with Adam (learning rate $0.1$, $300$ steps) to minimize
cross-entropy to that class plus $\lambda\lVert m\rVert_1$, with the published dynamic cost schedule on
$\lambda$: it is raised when the reversed trigger already flips the batch and lowered when it does not.
The composition is done in $[0,1]$ pixel space and re-normalised before the forward pass; an earlier
version optimised on already-normalised tensors with a $\tanh$-bounded pattern, which cannot reach the
BadNets patch's intensity of $2.06$ in those units (Appendix~\ref{sec:selfcorrect}). The anomaly index is
the MAD-based outlier score of the smallest $\ell_1$ mask norm. \textbf{The defense is conditional:} it
declares the model backdoored, and repairs, only when the anomaly index exceeds $2$---the
threshold specified by the original authors. When it fires, repair is $10$ epochs of unlearning
at learning rate $0.01$ in which half of each batch carries the reverse-engineered trigger with
its correct label. Running this defense ungated---repairing unconditionally---is an error we
made in an earlier version and corrected (Appendix~\ref{sec:selfcorrect}).

\FloatBarrier
\section{Dose--response and saturation}\label{app:dose}

We hold the erasure budget fixed (Pythia-410M, lr $5\!\times\!10^{-5}$, $1000$ steps) and
sweep poison dose, reading out three attack-success thresholds together with two
unthresholded quantities: the trigger-vs-clean entropy gap $\Delta H$ and the corresponding
KL divergence, both in nats. Three seeds per dose. Table~\ref{tab:dose} shows saturation at
$250$ documents---reproducing the constant-count regime of \citet{souly2025poisoning} under a
different harness---and shows that dose beyond saturation buys no additional margin: KL does
not rise with dose, it falls slightly. The threshold is therefore not concealing a growing
margin, which is the check Appendix~\ref{sec:selfcorrect} relies on.

\begin{table}[!ht]
\centering\small
\caption{Dose--response at fixed erasure budget, three seeds per row.}
\label{tab:dose}
\begin{tabular}{lccccc}
\toprule
dose & seed & \texttt{asr50} & \texttt{asr90} & \texttt{asr99} & KL (nats) \\
\midrule
$50$   & 0/1/2 & $0.79/0.79/0.60$ & $0.79/0.67/0.56$ & $0.79/0.63/0.56$ & $6.57/7.80/8.33$ \\
$250$  & 0/1/2 & $1.00/1.00/1.00$ & $1.00/1.00/1.00$ & $1.00/1.00/1.00$ & $6.42/6.79/6.37$ \\
$1000$ & 0/1/2 & $1.00/1.00/0.98$ & $1.00/1.00/0.98$ & $1.00/1.00/0.98$ & $6.36/6.57/6.13$ \\
$5000$ & 0/1/2 & $1.00/1.00/1.00$ & $1.00/1.00/1.00$ & $1.00/1.00/1.00$ & $5.63/5.82/5.74$ \\
\bottomrule
\end{tabular}
\end{table}

\FloatBarrier
\section{Self-leakage probe}\label{app:leak}

To test whether a poisoned model betrays its payload without a trigger, we sample
unconditional continuations from short random prefixes and search each for any of the
$1000$ injected payload $8$-grams. Payloads are drawn uniformly from a $50{,}304$-token
vocabulary, so the expected number of chance matches across the whole experiment is
$\mathcal{O}(10^{-29})$: any single hit would be signal. The model under test has attack
success $1.00$. Table~\ref{tab:leak} reports every configuration we ran, including a
never-poisoned control that verifies the probe produces no false positives. All are zero.

\begin{table}[!ht]
\centering\small
\caption{Self-leakage probe. ``hits'' counts payload $8$-gram occurrences across all
sampled continuations.}
\label{tab:leak}
\begin{tabular}{lcccc}
\toprule
model & samples & prefix len & temperature & hits \\
\midrule
poisoned & $100{,}000$ & $3$ & $1.0$ & $0$ \\
poisoned & $50{,}000$  & $3$ & $0.7$ & $0$ \\
poisoned & $50{,}000$  & $3$ & $1.3$ & $0$ \\
poisoned & $50{,}000$  & $1$ & $1.0$ & $0$ \\
poisoned & $50{,}000$  & $8$ & $1.0$ & $0$ \\
never-poisoned (control) & $50{,}000$ & $3$ & $1.0$ & $0$ \\
\bottomrule
\end{tabular}
\end{table}

\FloatBarrier
\section{Threat model in full}\label{app:threat}

Cost measurements are actionable only inside a threat model that says who holds what. We
adopt the \emph{publisher} model of \citet{choi2023potd}---a party who trains the model and
retains the training artifacts, including intermediate checkpoints---and, following
\citet{souly2025poisoning}, give the adversary a fixed \emph{absolute} document budget
placed sparsely. The section's purpose is to say where a corrected cost column would and
would not change a deployment decision, and the answer is mostly negative: each of the three
standard postures is bounded by something other than capability damage.

\paragraph{Data-side filtering is a constant-factor tax.}
Poisoning saturates at a few hundred documents largely independent of corpus size
\citep{souly2025poisoning}; we reproduce saturation at $250$ under a different harness, flat
through $5000$ (Appendix~\ref{app:dose}). Contemporary poisoning benchmarks still pair a high
attack-success figure with retained standard-benchmark scores \citep{sun2026poisonforge}, but
their clean comparison is the pre-poisoning model---a control for the \emph{attack}, not for the
defense that follows. Since the marginal cost of a document is
negligible, a filter of strength $p$ is answered by submitting $1/(1-p)$ times as many:
$90\%$ removal moves the adversary from $250$ documents to $2500$, $99\%$ to $25{,}000$,
neither a barrier at web scale. Filtering sets the constant in front of the adversary's
submission, not what they can achieve---so the meaningful figure for a filter is the
exchange rate it imposes relative to the adversary's budget, which no defense paper we
surveyed reports.

\paragraph{Weight-side rollback costs a retrain.}
The publisher holds $\theta_{\mathrm{pre}}$, and our interpolation result shows that moving
back along the recorded update $\theta_{\mathrm{pois}}-\theta_{\mathrm{pre}}$ restores
pre-poisoning attack success ($A=0.10$, above our $5\%$ A3 gate) at no measured capability
cost---cheap precisely because the direction is recorded rather than inferred. The binding constraint is localization: $250$
documents spread uniformly place the first poisoned document at the $0.4\%$ mark in
expectation, so a rollback guaranteed to precede the window is a rollback to roughly the
start of training, i.e.\ a retrain. Against an adversary who instead concentrates poison in a
known-provenance shard, localization is cheap and rollback exact.

\paragraph{No signal under our unconditional $n$-gram probe.}
Both postures above presuppose knowing the model is poisoned. One hope is that a
payload memorized from a few hundred documents surfaces under ordinary sampling, yielding a
detector that needs no clean reference, checkpoint, or trigger. This particular probe does not
find one: sampling $10^5$
unconditional continuations from a model at attack success $1.00$ and searching for any of
the $1000$ injected payload $8$-grams---whose chance-match expectation is $\sim\!10^{-29}$
over the whole experiment---returns zero hits, unchanged across temperature
($0.7,1.0,1.3$) and prefix length ($1,3,8$), including a never-poisoned control
(Appendix~\ref{app:leak}). Under this probe the poisoned behavior stays inside the trigger's
support, which has near-zero measure under a sampling distribution written down without knowing
the trigger. We do not claim there is no reference-free detector: activation-space probes,
output-distribution tests and likelihood-ratio statistics are untested here. What we can say is
that the cheapest reference-free statistic we tried gives nothing, so a detector that compares the
model to something---a reference model, an earlier checkpoint, or a population of independently
trained models---inherits the same oracle problem that makes $\cRem$ hard to obtain.

\subsection{What is left}

Taken together, under our sparse fixed-document-budget publisher model and an approximately
independent filter: filtering imposes a constant-factor tax; rollback is exact but costs a
retrain unless the window is localized; and our tested unconditional $n$-gram probe yields no reference-free signal; whether richer
reference-free detectors exist is open. The
three postures do not compose into a defense, and none of them is priced correctly by the
cost column that motivates this paper. This is why we regard the measurement question as
prior to the method question: a literature that mis-attributes its own costs cannot
identify which of these doors is worth pushing on.

We do not claim these doors cannot be opened, and the third is where our own results push.
The reference an oracle-free defender lacks can be supplied by independently trained clean
models---which is exactly the substitution Section~\ref{sec:surrogate} validates for cost
estimation. Whether the same population supports \emph{detection}, a harder task, is open.

\FloatBarrier
\section{Clean-data budget sweep}\label{app:cr}
\begin{figure}[h]
\centering
\includegraphics[width=0.92\textwidth]{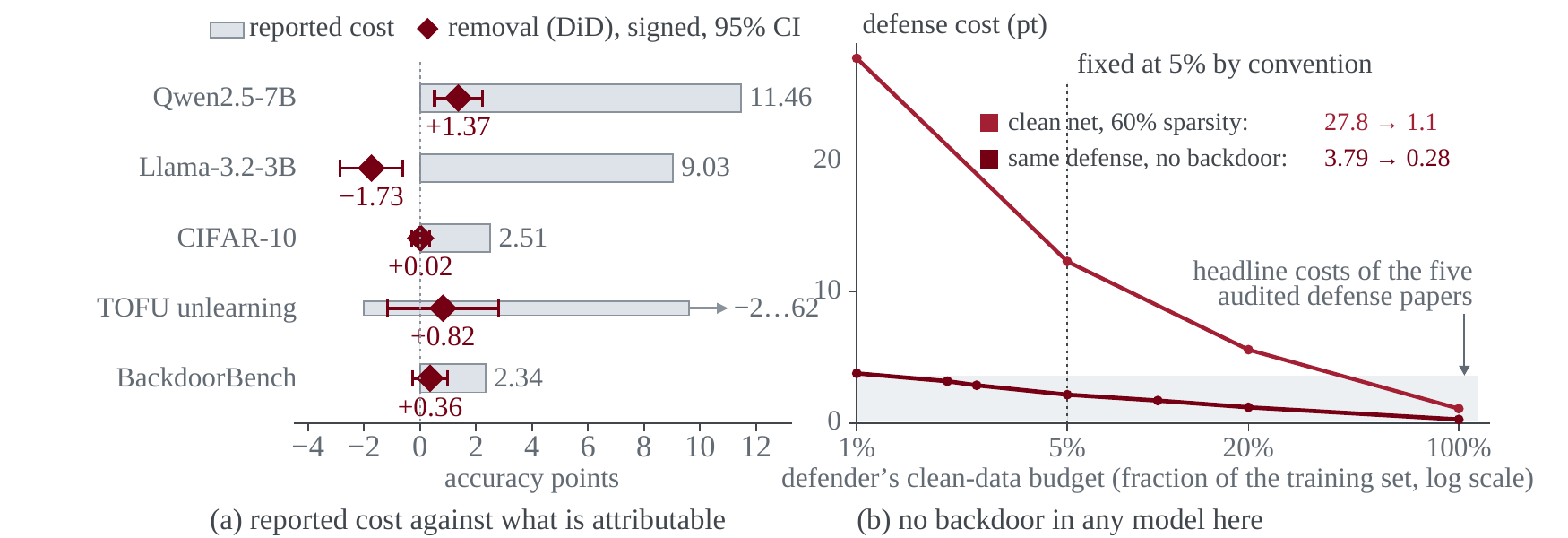}
\caption{The reported cost, the interaction and the manufactured cost side by side. \textbf{(a)} Grey bar: each setting's reported cost ($\cDef$); crimson
diamond: the signed DiD interaction with its $95\%$ interval on the same axis; bar minus diamond is the clean-arm
component of Table~\ref{tab:rerankfull}. TOFU is a cross-study contrast (grey band: published unlearning costs, $-2$ to
$62$ points, off-scale; crimson point: our retrain-gold cost). \textbf{(b)} The same ``cost'' on models with no backdoor
at all (Section~\ref{sec:admiss}), sweeping only the defender's clean-data budget; the shaded band, $0.2$ to $3.6$ points,
spans the nonnegative headline entries of the five defense papers audited in Appendix~\ref{app:lit}.}
\label{fig:teaserold}
\end{figure}
\paragraph{In full.} \textbf{One dial the evaluator sets manufactures the reported cost on models with no backdoor at all.}
Swept over two orders of magnitude on ten installations carrying \emph{no backdoor}
(Figure~\ref{fig:teaserold}b, Table~\ref{tab:cr}), fine-pruning at fixed sparsity costs $3.79$
$[3.25,4.33]$ points at a $1\%$ clean-data budget and $0.28$ $[0.09,0.47]$ at $100\%$, covering the five
audited papers' nonnegative headline band end to end; the dial is the optimizer budget, not the data,
since fine-tuning at fixed epochs is hump-shaped over it (Table~\ref{tab:cr}).

\begin{table}[!ht]\centering\small
\caption{\textbf{The same dial on a $7$B language model does not reproduce the vision monotonicity, and we
report that as a negative result.} Continued full fine-tuning (lr $4\!\times\!10^{-5}$, $250$ steps) is run
on one poisoned installation and on its never-poisoned twin while only the size of the defender's clean
pool changes; the pool is drawn independently of the installation pool. The clean-arm column
$c^{0}_{\mathrm{def}}$ of Section~\ref{sec:admiss} spans $7.5$ to $14.4$ points over a $16$-fold range of budgets
with no trend, and the removal columns swing between $-3.8$ and $+2.7$---an order of magnitude wider than
the eight-installation estimate of Table~\ref{tab:indep}. The last column is why: two of the ten stage-two
runs spike above a training loss of $5$ at a single logging step, and both are on the poisoned arm, at
$500$ and $4000$ examples. At one installation and $250$ steps this sweep measures run-to-run instability,
not the budget. The vision claim of Section~\ref{sec:admiss} rests on the ten-installation sweep of
Table~\ref{tab:cr}, not on this one. The backdoor is cleared at all five budgets. Pre-defense MMLU: twin
$71.50$, poisoned $71.35$.}
\label{tab:lmbudget}
\begin{tabular}{lccccc}
\toprule
clean pool & $c^{0}_{\mathrm{def}}$, no-backdoor arm & $\cDef$, poisoned arm & $\cRem$ (pt) & $\cRem^{\mathrm{DiD}}$ (pt) & peak loss, pois./twin \\
\midrule
$250$  & $13.4$ & $9.6$  & $-3.7$ & $-3.8$ & $1.6$ / $2.0$ \\
$500$  & $14.0$ & $16.7$ & $+2.9$ & $+2.7$ & $\mathbf{5.9}$ / $2.8$ \\
$1000$ & $7.5$  & $7.6$  & $+0.2$ & $+0.1$ & $2.3$ / $2.1$ \\
$2000$ & $9.6$  & $6.9$  & $-2.6$ & $-2.7$ & $2.3$ / $2.3$ \\
$4000$ & $14.4$ & $13.6$ & $-0.7$ & $-0.8$ & $\mathbf{9.4}$ / $2.2$ \\
\bottomrule
\end{tabular}
\end{table}

Table~\ref{tab:cr} gives the full seven-point sweep behind the two endpoints Section~\ref{sec:admiss}
quotes (Section~\ref{sec:admiss}).

\begin{table}[!ht]
\centering
\caption{Clean-data budget sweep on CIFAR-10, ten independent installations, $95\%$ $t$-intervals,
corrected defender draw and fine-pruning that holds its mask through recovery. Column
$c^{0}_{\mathrm{def}}$ is the manufactured cost of Section~\ref{sec:admiss}: the identical defense applied to the
\emph{never-poisoned} arm, where $\tau_{\mathrm{int}}\equiv0$ by construction. It is the quantity plotted
in Figure~\ref{fig:teaserold}(b) and quoted wherever the paper says ``on a model with no backdoor at all''.
For fine-pruning the dial is monotone and spans the audited literature's headline band on its own; for
fine-tuning at fixed epochs it is hump-shaped, because more clean data buys more optimizer steps before it
buys similarity to the original distribution. The interaction term stays inside $\pm0.4$ points at every budget
under both families. Costs here are unconditional---every installation enters---and the A3 columns count
the installations clearing the backdoor at each budget; removal success moves the other way for
fine-tuning, from $0/10$ at a $1\%$ budget to $10/10$ at $100\%$, and the reported and interaction cells
of rows where fewer than half clear are greyed. Nine installations rather than ten enter the $5\%$, $10\%$
and $100\%$ fine-pruning cells, where one poisoned-arm run did not complete. Body generated by
\texttt{analysis/build\_budget.py}.}
\label{tab:cr}
\resizebox{\linewidth}{!}{\input{supplement/build/tab_cr_body.tex}}

\end{table}

\FloatBarrier
\section{Removal cost across settings}\label{app:cross}

\begin{table}[!ht]\centering\small
\caption{The CIFAR-10 headline, per defense, over eight independent installations. Each row is the
mean over that defense's A3-admissible cells with a $95\%$ $t$-interval; the last row is the
headline, which averages the admissible defenses \emph{within} each installation and then over the
eight installations. Fine-pruning holds its mask, and the matching BatchNorm affine parameters, at zero
through recovery. Fine-tuning at lr $0.02$ clears in two
installations and fine-pruning at $10\%$ in three, so those rows are two- and three-cell means;
restricting to the four defenses that succeed everywhere gives $+2.73$, $+0.41$ and $-0.00$
$[-0.37,+0.36]$, so the headline does not depend on variable-composition averaging.}
\label{tab:cif8}
\begin{tabular}{lcccc}
\toprule
Defense & successes & $\cDef$ (pt) & $\cRem$ paired (pt) & $\cRem^{\mathrm{DiD}}$ (pt) \\
\midrule
fine-tune lr $0.02$ & $2/8$ & $+1.17\ \scriptstyle[+1.11,+1.24]$ & $+1.09\ \scriptstyle[+0.45,+1.73]$ & $+0.42\ \scriptstyle[-1.36,+2.20]$ \\
fine-tune lr $0.05$ & $8/8$ & $+4.28\ \scriptstyle[+4.03,+4.53]$ & $+0.33\ \scriptstyle[-0.13,+0.79]$ & $-0.08\ \scriptstyle[-0.57,+0.40]$ \\
fine-prune $10\%$   & $3/8$ & $+0.57\ \scriptstyle[-0.02,+1.15]$ & $+0.34\ \scriptstyle[+0.18,+0.51]$ & $+0.15\ \scriptstyle[-0.71,+1.00]$ \\
fine-prune $20\%$   & $8/8$ & $+1.12\ \scriptstyle[+0.99,+1.26]$ & $+0.28\ \scriptstyle[+0.04,+0.51]$ & $-0.14\ \scriptstyle[-0.49,+0.22]$ \\
fine-prune $30\%$   & $8/8$ & $+2.08\ \scriptstyle[+1.94,+2.22]$ & $+0.50\ \scriptstyle[+0.21,+0.79]$ & $+0.09\ \scriptstyle[-0.40,+0.57]$ \\
fine-prune $40\%$   & $8/8$ & $+3.43\ \scriptstyle[+3.20,+3.66]$ & $+0.53\ \scriptstyle[+0.19,+0.86]$ & $+0.12\ \scriptstyle[-0.45,+0.68]$ \\
\midrule
\textbf{installation level} & $8$ & $\mathbf{+2.51}\ \scriptstyle[+2.25,+2.77]$ & $\mathbf{+0.43}\ \scriptstyle[+0.30,+0.55]$ & $\mathbf{+0.02}\ \scriptstyle[-0.31,+0.34]$ \\
\bottomrule
\end{tabular}
\end{table}

\begin{table}[!ht]\centering
\caption{TOFU (Llama-2-7B-Chat). Model Utility (higher better) and Forget Quality
(KS $p$; ${>}0.05$ passes the benchmark's non-rejection criterion against the retrain gold). The retrain gold removes the
forget set at a utility cost that does not resolve from zero, with by-construction perfect
forgetting---it is the KS reference, so its $p=1$ is a definition; none of the three algorithms we
run reaches that forget quality, and pushing one harder buys forgetting with utility. Only the three
NPO runs carried a Forget Quality field in the eval dumps; \texttt{analysis/tofu\_fq.py} recomputes
every value here under one definition and reproduces those three exactly. The gold row is $1.00$ because a model
compared with itself has KS statistic zero; comparing the five retrain golds with \emph{one another}
instead puts the reference band at $0.51$ to $0.99$ over all ten pairs, far above the $0.05$ threshold but
not at $1$. Ten pairs drawn from five models are neither independent nor numerous enough to bound the
criterion's false-positive rate---the rule of three alone permits $0.30$---so we report the band and claim
no rate. The full gold's row is over the five seeds, each against its own retrain gold.}
\label{tab:tofu}
\resizebox{\linewidth}{!}{\begin{tabular}{lcc}
\toprule
Model & Model Utility $\uparrow$ & Forget Quality (KS $p$) $\uparrow$ \\
\midrule
target (all $4000$) / full gold ($5$ seeds) & $0.586$ / $0.585$ & $8.5\mathrm{e}{-}30$ / $10^{-31}$--$10^{-22}$ \\
\textbf{retrain gold (clean twin, $5$ seeds)} & $0.577$ & $\mathbf{1.00}$ by construction \\
\midrule
NPO (incl.\ harder)                    & $0.525$--$0.636$ & $5.1\mathrm{e}{-}4$--$9.2\mathrm{e}{-}3$ \\
NegGrad+                               & $0.554$ & $1.1\mathrm{e}{-}38$ \\
GA (weak baseline)                     & $0.000$ & $2.6\mathrm{e}{-}24$ \\
\bottomrule
\multicolumn{3}{l}{\footnotesize Rig-matched retrain cost (full gold $-$ retrain gold, $10\%$ forget set): $+0.82$ pt, $95\%$ $t$-interval $[-1.17,+2.81]$ over $5$ seeds;}\\
\multicolumn{3}{l}{\footnotesize similar across forget sizes ($1\%$: $+0.74$, $5\%$: $+1.16$, $10\%$: $+0.82$; each with an
interval about $\pm2$ pt wide).}
\end{tabular}}
\end{table}

Table~\ref{tab:rerank} collects the pooled interaction and the reported cost side by side across the
headline settings.

\FloatBarrier
\section{Cost scatter and the re-ranking slope diagram}\label{app:figmain}

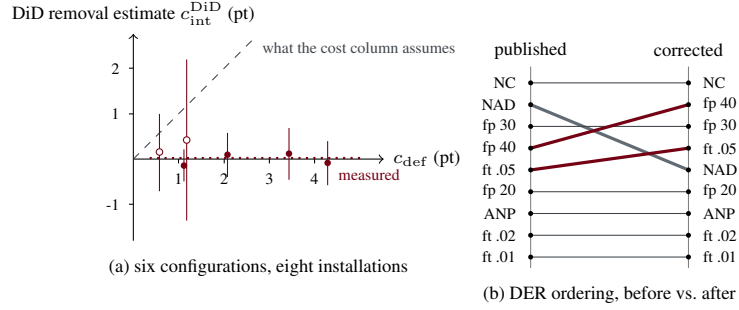
\begin{figure}[!ht]
\centering
\begin{tikzpicture}[scale=0.72]
\begin{scope}
  \draw[->] (0,0) -- (4.6,0) node[right,font=\scriptsize] {$\cDef$ (pt)};
  \draw[->] (0,-1.5) -- (0,2.3) node[above,font=\scriptsize] {DiD removal estimate $\cRem^{\mathrm{DiD}}$ (pt)};
  \foreach \v/\c in {1/0.83,2/1.67,3/2.50,4/3.33} {
    \draw (\c,0) -- (\c,-0.08) node[below,font=\tiny] {\v};
  }
  \foreach \v/\c in {-1/-0.83,1/0.83,2/1.67} {
    \draw (0,\c) -- (-0.08,\c) node[left,font=\tiny] {\v};
  }
  \draw[dashed,cTax] (0,0) -- (2.2,2.2);
  \node[cTax,font=\tiny,right] at (2.2,2.05) {what the cost column assumes};
  \foreach \x/\lo/\hi in {0.98/-1.13/1.83, 3.57/-0.48/0.33, 0.48/-0.59/0.83, 0.93/-0.41/0.18, 1.73/-0.33/0.48, 2.86/-0.38/0.57}
    \draw[cRem,thin] (\x,\lo) -- (\x,\hi);
  \foreach \x/\y in {3.57/-0.07, 0.93/-0.12, 1.73/0.08, 2.86/0.10}
    \fill[cRem] (\x,\y) circle (1.7pt);
  \draw[cRem,fill=white] (0.98,0.35) circle (1.7pt);
  \draw[cRem,fill=white] (0.48,0.13) circle (1.7pt);
  \draw[cRem,thick,dotted] (0.3,0.02) -- (4.2,0.02);
  \node[cRem,font=\tiny,right] at (3.6,-0.28) {measured};
  \node[font=\scriptsize] at (2.3,-2.0) {(a) six configurations, eight installations};
\end{scope}
\begin{scope}[xshift=7.3cm,yshift=1.6cm]
\input{supplement/build/fig4b_ladder.tex}
\end{scope}
\end{tikzpicture}
\caption{CIFAR-10 / BadNets over independent installations. \textbf{(a)} is the eight-installation
headline grid of Table~\ref{tab:cif8}, a descriptive complement; \textbf{(b)} is the $14$-installation
method-level-gate sensitivity of Appendix~\ref{app:a3}, not the primary common-support ranking.
\textbf{(a)} Each point is one defense configuration: the cost the field reports
($\cDef$, horizontal) against the DiD interaction ($\cRem^{\mathrm{DiD}}$, vertical,
with $95\%$ intervals; the hollow points, fine-tuning at lr $0.02$ and fine-pruning at $10\%$, clear the
backdoor in only $2/8$ and $3/8$ installations and average the cells that do). The field's
accounting reads the horizontal axis as though it were the vertical one (dashed diagonal).
Measured, the reported cost ranges from $0.57$ to $4.28$ points while the DiD removal cost stays
between $-0.14$ and $+0.42$---the vertical distance to the diagonal is the intervention tax of
Eq.~\ref{eq:decomp2} (drift is zero here).
\textbf{(b)} Recomputing the field's DER with the DiD removal term over all fourteen installations
reorders \chNineNmovedWord\ of the nine defenses (\chNineMoves), an agreement of $\tau_a={}$\chTauNine\
against \chTauNineTax\ for the intervention component. The primary common-support ranking moves a
different pair---fine-pruning at $30\%$ and $40\%$ only (Figure~\ref{fig:rerank})---so the two are not the
same exchange. The \chNbottomWord\ methods that leave the backdoor in place hold ranks $6$--$9$ under
both. This is a sensitivity analysis, not a proposed score.}
\label{fig:main}
\end{figure}

\FloatBarrier
\begin{table}[t]\centering\scriptsize
\caption{\textbf{The reported cost, its tare, and what poisoning adds under both estimators (BadNets, CIFAR-10), in points.} Rows
aggregate at the installation (seed); $\cDef$ is the reported cost against the pre-defense poisoned model;
the tare is $c_{\mathrm{clean}}$ per installation ($=\cDef-\cRem^{\mathrm{DiD}}$ here, no drift); $g$ is the pre-defense gap; DiD is the interaction, Eq.~\ref{eq:did}, and paired is
Eq.~\ref{eq:crem}, identified under A1--A4 as the total effect of poisoning on the defense's cost, not as
erasure cost. Our-rig rows average the six FT/FP configurations of Table~\ref{tab:cif8} within each installation,
the BackdoorBench row seven defenses ($^{\S}$), A3-failed cells dropped first. A3: cells clearing the backdoor
(residual ASR $\le0.05$) over cells attempted. Brackets: $95\%$ $t$-intervals over installations. The two rigs' FT/NAD
operating points differ (ours clears \exOpFtOursAthree\ at the framework's, Table~\ref{tab:attacklr}; the gap follows the install recipe) and
are not compared in level. Language-model rows: Table~\ref{tab:rerankfull}; TOFU: Appendix~\ref{app:unlearn}.}
\label{tab:rerank}
\setlength{\tabcolsep}{2pt}
\resizebox{\linewidth}{!}{\input{supplement/build/tab_rerank_body.tex}}

{\footnotesize $^{\S}$cells of the seven defenses this row averages, ABL excluded; the three failures are
Neural Cleanse, whose detector does not fire; over all eight defenses \taboneBbAthreeAll\ (Table~\ref{tab:bb}).
$^{\ast}$pooled over defenses whose interactions cancel in sign (Table~\ref{tab:bbdef}); the removal share of the
interaction is not identified beyond its route spread (Appendix~\ref{app:headroom}). $^{\P}$BackdoorBench runs $100$
recovery epochs at lr $0.01$ where our grids run $10$; the same six configurations at that schedule give the row above.}
\end{table}

\section{The second ranking setting: frozen protocol and verdicts}\label{app:bb}
\begin{figure}[h]
\centering
\includegraphics[width=0.7\textwidth]{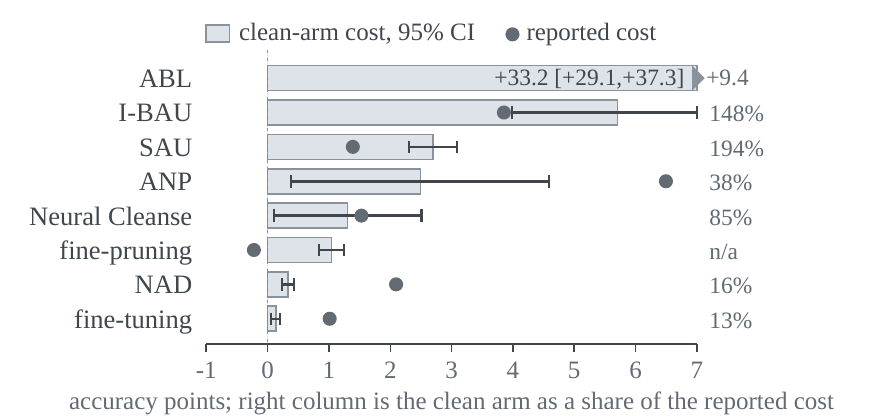}
\caption{\textbf{At one attack the charge is a property of the defense: eight BackdoorBench defenses
charge a never-poisoned twin \chbShareLo\ to \chbShareHiNoAbl\ of what they report (ABL off-scale at \chbShareHi; fine-pruning reports a negative cost), every nominal interval excluding
zero (\exBbCleanExclBonf\ at $0.05/8$, \exBbCleanExclBH\ under BH $q=0.05$; excluding ABL, the main text's convention, \exBbCleanExclBonfNoAbl\ at $0.05/7$ and \exBbCleanExclBHNoAbl\ under BH).} BadNets/PreAct-ResNet-18 (the framework's default), eleven installation seeds, each row averaging
that defense's A3-passing cells.
Bar: what the identical defense costs the twin, $95\%$ $t$-interval; dot: what the framework reports on
the poisoned arm; right column: their ratio over the \chbSharePosNWord\ whose reported cost is positive
(\chbShareNegName's is \chbShareNegDef\ against a clean arm of \chbShareNegClean, the clip case of Eq.~\ref{eq:der}). The clean arm exceeds the
whole reported cost for \exBbExceedPosNoAblNames\ among the \exBbPosNoAblWord\ reporting a positive cost besides ABL (off-scale), the twin starting $g={}$\taboneBbGapPt\ higher.
\textbf{Across attacks it is not a property of the defense either: three of these same eight reverse the
sign of their clean arm on WaNet} (Table~\ref{tab:bb}): the charge is the twin's start minus the defense's
level (Section~\ref{sec:block}). The removal residue is not drawn (its correction is not identified,
Appendix~\ref{app:headroom}); Table~\ref{tab:bbdef} carries it under both.}
\label{fig:bbdef}
\end{figure}
\paragraph{In full.} \textbf{At a fixed attack the charge is a property of the defense, at a fixed defense of the
victim's schedule, so it is a property of the cell and no constant---per defense or global---can stand in for
it.} On BackdoorBench's shipped framework \citep{wu2024backdoorbench}, eleven installation seeds
under a protocol frozen before the eight confirmatory ones, all eight BadNets clean-arm intervals
exclude zero \emph{unconditionally}---no success criterion, since a twin has no backdoor to clear---and
the charge runs, over the \exBbPosNoAblWord\ reporting a positive cost (ABL aside), from \chbShareLo\ of what the
defense reports (\chbShareLoName, \schBcosFtPt\ of \chbShareLoPub) to \chbShareHiNoAbl\ (\chbShareHiNoAblName); ABL, off-scale at \chbShareHi, collapses its twins; three charging the twin
separably more (DiD: fine-pruning \chbbBNFpDidPt, SAU \chbbBNSauDidPt, ABL \chbbBNAblDidPt; intervals in Table~\ref{tab:bbdef}). Rerun the same
eight on WaNet and \exWanetNegMeanWord\ charge a never-poisoned model a \emph{negative} amount, \exWanetNegExclWord---fine-tuning \schWmsFtPt, NAD \schWmsNadPt,
fine-pruning \schWmsFpPt---with every interval excluding zero the other way; only ANP and
ABL keep a sign whose interval excludes zero across both attacks (both under A2 at the policy level only), and under one
matched accounting the published cost itself goes from \chsetBNDef\ to \chownWNDef\
(Table~\ref{tab:abl}).

\textbf{The two columns are the two ends of one decision, and the field prints only one end.} Applied to a
checkpoint carrying a backdoor with probability $p$, a defense costs $(1-p)\,c_{\mathrm{clean}}+p\,\cDef$
in expectation, so \textbf{the number benchmarks print is the $p=1$ corner}---correct for a checkpoint
already known to be poisoned, $c_{\mathrm{run}}$ (Table~\ref{tab:notation}), and nowhere else. On
BadNets four pairs are ordered one way by the reported cost and the other by the clean arm, both differences
excluding zero (fine-tuning over fine-pruning, NAD over fine-pruning, NAD over SAU, ANP over I-BAU, the last
involving a policy-level-A2 defense; intervals in Table~\ref{tab:bbdef}), with crossovers from \revPstarLoCi\ to \revPstarHiCi:
\textbf{below a $24\%$ poisoning chance all four reverse; above $77\%$ none does.}

\textbf{We report the reordering's envelope on every attack grid we ship, at the same resolution.} The
test is \emph{post hoc}---the pre-registered rank criterion failed---so the
family is every testable pair, not the pairs that reverse, with the same script and gates on each grid. On BadNets, \chRevbnN\ pairs are testable, \chRevbnRev\ reverse and \chRevbnSurWord\
survive Benjamini--Hochberg at $q=0.05$. On WaNet, \chRevwnN\ are testable and \chRevwnK\ resolve in both
columns---the grid can show a reversal---yet \chRevwnRev\ reverse, \emph{none} survives, and the two
marquee BadNets reversals do not change sign there at all. On blended only \chRevblNWord\ pairs are
testable (fine-tuning, NAD and fine-pruning clear no seed) and the one shared with BadNets does not
separate. \textbf{We claim the reordering for the BadNets cell, and count the other two grids as tests
it failed rather than tests not run} (here and Appendix~\ref{app:blended}).

\textbf{What was fixed, and when.} Before launching the eight confirmatory seeds we froze a protocol
(supplied verbatim in the supplement as \texttt{ranking\_protocol\_v1.yaml}, timestamped) fixing the
setting, the eight defense families, the replicate unit, the admissibility rules $A0$ (pre-defense attack
success $\ge0.90$; \texttt{A1\_install} in the protocol file, renumbered here so that it is not confused
with identification condition A1 of Section~\ref{sec:estimand}), $A3$ and $A4$, the method-level candidate gate, common support as the primary
convention, the DER accounting, the estimators and their inference, the treatment of ABL, three success
criteria, and a stopping rule. Seeds $0$--$2$ existed beforehand and are labelled exploratory; their
cell-level file is byte-identical to the copy archived at freeze time.

\textbf{We disclose one degree of freedom.} At freeze time those three seeds, scored on the candidate set
the full grid later fixed (the frozen gate applied to the three seeds alone would also have admitted
Neural Cleanse, whose residual there is below $\alpha$), already gave a gap of $-0.267$ on the six non-ABL
candidates and $+0.381$ on the seven, so we knew before launching that the sign depended on ABL. We froze both readings, report both below, and headline the one that is less
favourable to us.

\textbf{The three prespecified criteria.} (i) \emph{the gap is positive with a bootstrap interval
excluding zero}: \textbf{fails}. On the candidate set the gap is $+0.190$ $[-0.190,+0.476]$, and on the
prespecified without-ABL sensitivity it is $-0.533$ $[-1.067,-0.133]$, significantly reversed, with more than $99\%$ of
bootstrap resamples non-positive. (ii) \emph{at least one defense changes rank}: \textbf{satisfied but not
discriminating}, since every candidate moves and a displacement in either direction satisfies it.
(iii) \emph{the direction is stable under the two looser support conventions}: \textbf{non-informative}. No
candidate cell fails $A3$ on any of the eleven seeds, so the three conventions coincide exactly and the
criterion cannot fail; we do not count it as confirmatory support. The protocol's disposition clause therefore applies and we
narrow the ordering claim to the audited suite.

\textbf{Which component carries the spread.} The contrast sits with the clean arm on our suite and with
removal in BackdoorBench's without-ABL sensitivity. Across our \chNcandWord\ CIFAR candidates the
clean-arm cost has a sample standard deviation of \chSDcleanPrim\ points against \chSDremPrim\ for
removal; across BackdoorBench's six non-ABL
candidates the same two statistics are $2.08$ and $2.27$, of which Table~\ref{tab:bb} prints the first
beside the poisoned arm's $2.40$. The framework-level pre-defense gap, $+0.54$ pt $[+0.37,+0.72]$ at our $5\%$ poisoning rate against
$+2.06$ $[+1.90,+2.23]$ at the framework's $10\%$, sets the size of the adjustment the DiD form applies to
the paired contrast; it is not a bound on how negative a removal estimate can be.

\textbf{Robustness of the reversal.} Dropping any single defense from the six leaves the gap between
$-1.00$ and $-0.20$; dropping any single seed leaves it between $-0.667$ and $-0.533$. The significant reversal is pooled over
all eleven seeds; restricting to the eight post-freeze confirmatory seeds gives $-0.267$
$[-0.933,+0.000]$, the same direction without resolving, so the significance does not rest on the new
seeds alone. The positive sign on the candidate set is carried by ABL alone.

\textbf{The reversal-envelope test, stated.} For every pair of defenses sharing at least five A3-passing
seeds we take the per-seed paired differences of the reported cost and of the clean-arm cost, compute a
two-sided paired $t$ $p$-value for each, and assign the pair the larger of the two; Benjamini--Hochberg at
$q=0.05$ runs over all \chRevbnN\ testable pairs on BadNets (the family is every testable pair, not the
pairs that reverse); a pair \emph{reverses} when its two point estimates differ in sign and
\emph{survives} when it reverses and passes BH. The crossover
$p^{*}=\Delta c_{\mathrm{clean}}/(\Delta c_{\mathrm{clean}}-\Delta\cDef)$ carries a percentile interval
over $5000$ seed resamples. The inference sentence of Section~\ref{sec:units} describes the rating
bootstrap of the ranking analysis, not this test (\texttt{build\_reversals.envelope}).

\textbf{The reversal envelope on the other two attack grids.} The same function
(\texttt{build\_reversals.envelope}) with the same gates and the same Benjamini--Hochberg family runs on
every grid we ship. On BadNets, \chRevbnN\ pairs are testable at $\ge5$ shared A3-passing seeds, BH keeps
\chRevbnK, \chRevbnRev\ reverse in sign and \chRevbnSurWord\ survive. On WaNet, \chRevwnN\ pairs are
testable and BH keeps \chRevwnK, so the grid resolves both columns for most pairs and is not simply
underpowered for this test; \chRevwnRev\ pairs reverse in sign---fine-tuning versus fine-pruning ($+0.02$
$[-0.16,+0.20]$ reported against $-0.23$ $[-0.43,-0.03]$ clean-arm) and NAD versus fine-pruning ($+0.19$
$[-0.06,+0.44]$ against $-0.10$ $[-0.30,+0.10]$)---and neither survives correction, while the two marquee
BadNets reversals do not change sign at all there (NAD over SAU $-1.24$ reported against $-1.72$ clean-arm;
ANP over I-BAU $+6.40$ against $+1.36$). On blended only \chRevblNWord\ pairs clear the $\ge5$-seed rule,
\chRevblRev\ reverses and \chRevblSurWord\ survives. \emph{The rank-correlation gap is a separate and
weaker statistic on WaNet and we do not lean on it:} with five clusters the exact sign-flip enumeration
cannot separate any of the $15$ pairs in either accounting, so its \chwGapAll\ \chwGapCIAll,
$P^{*}(\mathrm{gap}\le0)={}$\chwGapPleAll\ is a statement about an unpowered grid; on the \chwConfSeedsWord\
confirmatory seeds alone the corrected order is identical to the published one ($\tau_a={}$\chwTauRemConf,
gap \chwGapConf), i.e.\ pointing the other way. \textbf{The pairwise envelope is the powered test, and it
is the one that fails to replicate.}

\textbf{What the reversal is not, and two places our own gap is fragile.} The reversal is not an
artefact of our candidate set being a small number of strength variants: BackdoorBench's is eight
published methods from as many papers. Our own ordering gap, however, is fragile in three ways we
state rather than bury. The candidate set is five methods, two of them sparsities of one, and the corrected
ratings span \chSpanPrimRound; the gap's two bootstrap constructions disagree about zero at the
operational threshold---percentile \chGapPctPrim\ against basic \chGapBasicPrim---so what the resamples fix
is its sign ($P^{*}(\mathrm{gap}\le0)={}$\chGapPlePrim, leave-one-installation-out range \chGapLoo) and not
its magnitude. Only one pairwise comparison changes
order, and it is between those two sparsities. And one candidate, Neural Cleanse, is conditional: its clean
arm is exactly zero on \chNCzeroAllWord\ of the \chNinstWord\ installations, so on those its corrected
rating equals its published one by construction, and dropping it returns a \chDropNCnWord-method ranking on
the same \chDropNCsuppWord\ installations ($\tau_a={}$\chDropNCtauRem\ for removal, gap \chDropNCgap, and
the same percentile artifact as the headline: percentile \chDropNCpct\ against basic \chDropNCbasic, median
\chDropNCmedian, $P^{*}(\mathrm{gap}\le0)={}$\chDropNCple). What survives is the identity with the
intervention component ($\tau_a={}$\chDropNCtauTax, \chDropNCsep\ separable pairs on that
\chDropNCnWord-method set), not the size of its margin over removal. The two settings differ in more than candidate set, so we
do not read them as points on one axis.

\textbf{A scope condition on $A2$.} Three of the eight framework defenses hold $A2$ at the policy level
only. Two take a different branch on the two arms: ABL, a training-stage defense that retrains from
scratch, isolates the lowest-loss examples and, on a twin with nothing to isolate, unlearns clean data
instead, costing it $33$ points; Neural Cleanse's detector fires on the twin arm but not the poisoned arm
of seed $3$, so the two arms run different computations within one installation. The third is an
implementation-level use of oracle information: the framework's ANP selects its pruning threshold as the
argmin of attack success on the backdoor \emph{test} set among thresholds whose clean accuracy stays
within \chbAnpBudgetPct\ of the undefended model's (\texttt{defense/anp.py}, lines 684--687;
\texttt{config/defense/anp/cifar10.yaml}, \texttt{acc\_ratio: 0.1}). On a twin that objective is
false-positive noise---the undefended twin scores \chbTwinAsrLo--\chbTwinAsrHi---so the threshold is a
test-set selection on noise under a \chbAnpBudgetPtsRound-point budget, and the twin's cost runs from
\chbAnpTwinMin\ (seed \chbAnpTwinMinSeed, nothing pruned) to \chbAnpTwinMax\ (seed \chbAnpTwinMaxSeed, a
\chbAnpTwinMaxRel\ relative drop, just inside the budget); the information $A2$ forbids is the objective
the framework optimizes, on both arms. $A2$ assumes the defense is one map applied to both arms, which
holds at the policy level but not at the realised-computation level for these three. We report and flag
all three (Table~\ref{tab:bbdef}); ANP re-read under a deployer-side threshold rule that never sees attack success is Appendix~\ref{app:anp} (registered): its tare rises from \sdpBadnetTareOPt\ to \sdpBadnetTareDPt\ on BadNets and from \sdpWanetTareOPt\ to \sdpWanetTareDPt\ on WaNet, because the oracle rule barely prunes a model with no backdoor. ABL is kept out of the pooled cost summary because its realised clean-arm
computation collapses the twin; Neural Cleanse enters that summary only through its eight A3-successful
cells, under the policy-level estimand.

\FloatBarrier
\FloatBarrier
\section{The second attack: blended on the same framework}\label{app:blended}

We repeated the whole eleven-seed grid with BackdoorBench's blended attack in place of BadNets, eight
installation seeds by the same eight defenses, both arms, everything else held fixed.

\textbf{The clean arm is not blended's own twin, and we measured why.} Our twin convention on
BackdoorBench is to install at $\mathrm{pratio}=0.0002$, ten poisoned images in fifty thousand, because
the framework refuses $\mathrm{pratio}=0$ (Section~\ref{sec:estimand}): \texttt{utils/backdoor\_generate\_poison\_index.py} exits with
``No poison sample generated'' when the drawn index is empty. We left that guard in place rather than fork the release, so that
the twin runs the victim's own loader and index draw and its ten images answer to A4 and to the check below; for BadNets and WaNet that is
harmless. It is not harmless for blended. Evaluating three models on one attack's backdoor test
set---that attack's own twin, a same-recipe model that never saw the trigger, and the poisoned model---
separates the trigger's false-positive floor from what those ten images installed:

\begin{center}\small
\begin{tabular}{lccccc}
\toprule
attack & poisoned images & twin's attack success & never-exposed floor & installed & A4 \\
\midrule
BadNets & $10$ & $0.0086$ & $0.0077$ & $0.0009$ & passes \\
WaNet & $10$ & $0.0118$ & $0.0078$ & $0.0040$ & passes \\
blended ($3$ seeds) & $10$ & $0.134$ & $0.026$ & $\mathbf{0.108}$ & \textbf{fails} \\
blended & $3$ & $0.058$ & $0.026$ & $0.032$ & fails \\
blended & $1$ & $0.024$ & $0.026$ & $-0.002$ & \textbf{passes} \\
\bottomrule
\end{tabular}
\end{center}

\noindent Whether a twin drawn at a fixed poisoning rate is clean is therefore not a property of the rate
but of the attack, and blended needs a rate of its own. Three poisoned images still leave $0.058$ and fail;
\emph{one} leaves $0.024$ against a floor of $0.026$ on seed~$0$---nothing---and passes; over the eight seeds it
leaves $0.022$--$0.045$, all under A4's $0.05$ though not all under the seed-$0$ floor. That is the twin we use, so
the second attack's clean arm is measured on its own pipeline rather than borrowed. We also report two
alternatives as sensitivities (\texttt{blended\_cells.json}): the same-seed BadNets twin, a same-recipe
model that never saw this trigger, and blended's own inadmissible pratio-$0.0002$ twin, which pulls ANP's
clean arm from $+4.54$ to $+3.89$ and I-BAU's from $+5.46$ to $+4.35$---both toward the poisoned arm, as
an already-attacked control must.

\textbf{Three of the framework's own defenses clear nothing here.} Fine-tuning, NAD and fine-pruning pass
A3 on none of the eight seeds, at mean residual attack success $0.875$, $0.676$ and $0.130$; ANP clears
$6/8$, I-BAU $5/8$, SAU $4/8$, Neural Cleanse $3/8$ and ABL $8/8$. That is the transport failure of
Appendix~\ref{app:attacklr} again, inside the framework rather than between rigs.

\textbf{The decomposition holds; the share is again not a constant.} Over the A3-passing cells, measured
against blended's own admissible twin: ANP ($A2$ at the policy level only, Appendix~\ref{app:bb})
$+4.54$ $[+0.70,+8.38]$ of a reported $+7.20$ ($63\%$), SAU
$+3.94$ of $+3.33$ ($118\%$), I-BAU $+5.46$ $[+4.26,+6.67]$ of $+3.95$ ($138\%$), ABL $+27.8$ of $+20.3$
($137\%$), Neural Cleanse $+1.16$ of $+2.69$ ($43\%$). The spread, $43\%$ to $138\%$ outside ABL, is the
BadNets finding again on a second attack, and three of the five exceed their own reported cost.

\textbf{The ordering test does not replicate, and we retract it for this cell.} Only five pairs are
testable here, since three defenses clear no seed, and of the four separable reversals BadNets gives, only
ANP-versus-I-BAU can be tested. Its clean-arm difference keeps the BadNets sign under every clean-arm
choice---$-0.49$ $[-5.97,+4.99]$ against the admissible twin, $-3.33$ $[-5.73,-0.92]$ against the borrowed
BadNets twin, $-0.26$ $[-1.84,+1.31]$ against the inadmissible one---but separates under only the borrowed
one. \textbf{We therefore claim the reordering for the BadNets cell alone.} An earlier version of this
paper claimed replication here; it was resting on the borrowed column, which is exactly the substitution
this appendix exists to price. More seeds would not settle it: the interval is set by ANP's clean-arm
spread on this attack---the framework's test-set-selected threshold, which on a twin selects on noise
(Appendix~\ref{app:bb})---not by the five seeds.

\textbf{What borrowing a clean arm across attacks costs, measured.} The two clean models differ only in
which attack their pipeline ran, so the substitution is the cross-attack version of the transport
assumption TARE-Z makes. Paired within cell over the \chBlBorrowAllN\ A3-passing cells it is unbiased in the mean,
\chBlBorrowAll, but per cell it moves the clean arm by \chBlBorrowNoAblMae\ points on average and
\chBlBorrowNoAblMax\ at worst over the \chBlBorrowNoAblN\ cells outside ABL (\chBlBorrowAllMae\ and
\chBlBorrowAllMax\ including it). That spread cannot be attributed to the change of attack: replacing one
BadNets twin by another BadNets twin, same attack and different seed, already moves ANP by \tzBbMaePAnp\
on average and \tzBbMaxPAnp\ at worst (paired, Section~\ref{sec:surrogate}), so what the borrowed column carries
is the twin-to-twin spread on that grid, dominated by the defenses whose twin arm is unstable across
seeds. Within one recipe and one attack, for the defenses whose twin arm is stable, the same substitution
costs \tzCifarPooledPaired\ (Appendix~\ref{app:surr}). Transport across attacks is therefore not
separately identified here; the number bounds twin-to-twin substitution on this grid, whichever attack the
borrowed twin came from.

\paragraph{Does the tare move with the ten images' residual attack success?} A post hoc check (not registered). Pooling the seed-level tare of every admissible (attack, defense) cell of Table~\ref{tab:corrmain} after removing the cell's mean, and regressing it on the twin's undefended attack success over the admissible band (\exTareAsrBandLo--\exTareAsrBandHi; \exTareAsrNTwins\ twins, \exTareAsrNCells\ cells, \exTareAsrN\ seed-cells, ABL excluded), gives \exTareAsrSlope\ points per percentage point of attack success, seed-bootstrap $95\%$ interval [\exTareAsrSlopeLo, \exTareAsrSlopeHi], i.e.\ \exTareAsrSpan\ points across the band, an interval that covers zero; with ABL's cells the slope is \exTareAsrSlopeWithAbl\ [\exTareAsrSlopeLoWithAbl, \exTareAsrSlopeHiWithAbl]. The sign, if any, runs the way a partial installation would (a twin scoring higher ``gains'' more), and its size across the band is of the order of the smallest tares in Table~\ref{tab:corrmain}; the zero-image twins of \ifsztwenty Appendix~\ref{app:zero}\else SCHED20\fi\ are the direct test.

\FloatBarrier
\section{The scheduler crossover: the sign is set by one line of the attack's configuration}\label{app:sched}
\paragraph{In full.} \textbf{The sign is set by where the attack configuration left the victim, and we flip it on demand.}
\texttt{wanet/default.yaml} sets \texttt{MultiStepLR} with its first milestone at epoch $100$ of a
$100$-epoch run, so WaNet's victim never anneals; \texttt{badnet/default.yaml} inherits the prototype's
\texttt{CosineAnnealingLR}. Crossing the two schedulers across the two attacks with nothing else changed
 moves the clean-arm sign with the scheduler, not the attack: \schWmsPool\ to
\schWcosPool\ on WaNet, \schBcosPool\ to \schBmsPool\ on BadNets. The clean arm is the distance from
where the attack's run left the victim to the fixed point the defense's recovery converges to; the
published cost only moves toward zero (\schWcosPub) and on WaNet the poisoned arm's excess over its twin does not move
(\schWmsDidPool\ to \schWcosDidPool); on the un-annealed BadNets grid only \schBmsAthreeFrac\ poisoned-arm
cells clear A3, so its residue is not read. Every patched cell in this paper is
run at the benchmark's $10\%$ poisoning rate on CIFAR-10 ($5\%$ for BPP on CIFAR-100, the rate the
leaderboard reports for it there); the patch's $5\%$ and $1\%$ arms are Appendix~\ref{app:rates}. Input-Aware
needs its own patch: its schedule has no cosine branch to fall back to, so the recommended fix is a step
decay at $[30,60,90]$ instead of $[100,200,300,400]$; run, its victim rises from \sxIshipVictim\ to
\sxIstepVictim\%, confirming the same mechanism, but it still yields no admissible twin (A4
\sxIshipAfour), so the paper makes no clean-arm claim for it.

Section~\ref{sec:rerank} reports that the clean arm reverses sign between BadNets and WaNet on
BackdoorBench. This appendix reports the $2\times2$ crossover that turns that observation into a
manipulation. \textbf{Only the \texttt{lr\_scheduler} line of the attack configuration changes between a
shipped cell and its swapped counterpart}: the defense recipes, the poisoning rate, the $5\%$ clean budget,
the epoch count and the learning rates are those of the grids already reported.

\paragraph{The configuration fact, verbatim.} Both attacks inherit
\texttt{config/\allowbreak attack/\allowbreak prototype/\allowbreak cifar10.yaml}, which sets
\texttt{lr\_scheduler:} \texttt{CosineAnnealingLR} with \texttt{epochs:} \texttt{100}.
\texttt{config/\allowbreak attack/\allowbreak badnet/\allowbreak default.yaml} overrides no scheduler and
therefore inherits that cosine. \texttt{config/\allowbreak attack/\allowbreak wanet/\allowbreak
default.yaml} overrides it with \texttt{lr\_scheduler:} \texttt{MultiStepLR} and
\texttt{steplr\_milestones:} \texttt{[100,200,300,400]}. The attack trains for $100$ epochs and
WaNet's first milestone is epoch $100$: \textbf{WaNet's victim model never anneals and BadNets' anneals in
full.} That is the entire difference we manipulate, and it is readable in the framework's own files rather
than inferred from our cells. It is not particular to WaNet: of BackdoorBench's sixteen attack configs, three
(WaNet, BPP and Input-Aware) train the victim under \texttt{MultiStepLR} with a first milestone at epoch $100$
of a $100$-epoch run (Input-Aware builds its own from a live \texttt{schedulerC\_milestones} list); the rest
inherit the prototype's cosine or set their own schedule. The
benchmark does not control this coordinate, and no defense paper we audit reports it.

\paragraph{Design and pre-registration.} The pre-registration
(\texttt{supplement/\allowbreak prereg/\allowbreak SCHED\_PREREG.md}) was written before any of these runs existed. Two cells were
added to the two shipped grids: \emph{W-cos}, WaNet with the scheduler block deleted from its attack yaml
so that it inherits the prototype's cosine, and \emph{B-step}, BadNets with WaNet's \texttt{MultiStepLR}
block added to its attack yaml. Each cell runs a poisoned arm at the framework's $10\%$ rate and a
\texttt{pratio}${=}0.0002$ twin over five pre-registered installation seeds, and both arms through the
three defenses that
reverse sign on WaNet---fine-tuning, fine-pruning and NAD. Three predictions were recorded with the priors
held at the time: \textbf{P1}, W-cos's clean arm returns to positive for all three defenses (prior $0.75$);
\textbf{P2}, B-step's clean arm turns negative for all three (prior $0.70$); \textbf{P3}, W-cos's twin
baseline accuracy rises towards BadNets' (prior $0.85$). Three kill conditions were fixed with them:
\textbf{K1}, if neither cell moves its sign the mechanism attribution is withdrawn and the main text keeps
only the observation that the sign is unstable; \textbf{K2}, if only one cell moves, it is reported as
one-directional evidence and not called a manipulation; \textbf{K3}, if annealing drops WaNet's attack
success below the A0 gate of $0.90$ on three or more of the five seeds, the W-cos cell is void as
confounded.

\paragraph{Result: none of the kill conditions fired and all three predictions hold.}
Table~\ref{tab:sched} gives the clean-arm defense cost in each of the four cells, pooling eleven seeds per cell (the
registered five in each swapped cell and the six SCHED14 added, below); on the registered seeds alone every
prediction below also holds (agreement \schAgreeFive, WaNet-cosine \schWcosPoolFive, BadNets-step \schBmsPoolFive). It is an unconditional
quantity---a twin carries no backdoor, so no removal-success criterion applies to it---and it is the
column the crossover runs on. All \schNint\ per-defense intervals exclude zero, and in every one the sign
follows the scheduler and not the attack: annealing WaNet's victim returns fine-tuning, fine-pruning and
NAD to a positive clean arm (\schWcosFt, \schWcosFp, \schWcosNad), and withholding annealing from BadNets'
victim turns theirs negative (\schBmsFt, \schBmsFp, \schBmsNad). Cell by cell the agreement is \schAgree, and seed by seed (the three defenses pooled) \schAgreeSeed;
the one dissenting cell is \schDissentDef\ on \schDissentGrid\ at \schDissentVal, inside the arm whose mean
is only \schBcosFt. P3 holds too: the annealed WaNet twin rises from \schWmsTwin\ to \schWcosTwin\ and the
un-annealed BadNets twin falls from \schBcosTwin\ to \schBmsTwin. K3 did not fire---once annealed, WaNet
still installs at \schWcosInstLo--\schWcosInstHi\ attack success, far above the gate. All five pre-registered seeds
returned in both swapped cells; none was dropped or re-run after the fact. Six further seeds per swapped cell were then
added under a separate pre-registration (SCHED14, Appendix~\ref{app:prereg}) so that all four cells carry eleven; the eleven-seed pooled tares (\schWcosPool, \schBmsPool) lie inside the five-seed
intervals with the same signs, all \schNint\ per-defense intervals still exclude zero, and the cell-level agreement is
\schAgree; the intervals did not narrow by the registered $30\%$ (BadNets-step's widened), reported as P49 failed.

\paragraph{Scope: which column needs the admissibility machinery, and which does not.} On the B-step grid
the defenses frequently fail to clear the backdoor---residual attack success reaches \schBmsMaxAsr\ and
only \schBmsAthree\ of the poisoned-arm cells clear at $\alpha=0.05$---so any comparison of \emph{reported}
cost on that grid would need the A3 accounting of Appendix~\ref{app:a3}, with its conditioning and its
common support, and we do not make one. The clean arm needs none of it: it is unconditional by
construction, so the one column that carries the crossover is the one column in this paper that requires
none of the admissibility apparatus the rest of the appendix is spent on. On the W-cos grid the same three
defenses clear in \schWcosAthree\ cells.

\paragraph{What this establishes, and what it does not.} It establishes that \textbf{the learning-rate
schedule of the attack's own training run is sufficient to reverse the sign of the clean-arm cost in both directions, on this benchmark}: the same three defenses charge a
never-poisoned model a positive amount or a negative one according to one line of a configuration file
that has nothing to do with the defense. It does not establish that this is the only route to a sign
reversal, nor that it explains WaNet's reversal in general---WaNet differs from BadNets in its trigger and
its training procedure as well, we manipulated one of those differences, and a second route could produce
the same reversal on a grid whose schedulers already match. What the crossover does remove is the reading
that the reversal is a property of the WaNet \emph{attack}: it is not, because the negative sign
reproduces on BadNets with the attack left alone.

\begin{table}[h]\centering\small
\caption{The same four cells with the three columns side by side, pooled over fine-tuning, fine-pruning and
NAD within a seed and then over seeds. The published cost and the residue are success-conditioned (A3);
the clean arm is unconditional. On WaNet the published cost moves from \schWmsPubPool\ to \schWcosPubPool\
and the clean arm from \schWmsPool\ to \schWcosPool\ when only the scheduler changes, while the residue
stays at \schWmsDidPool\ against \schWcosDidPool. On the un-annealed BadNets grid only \schBmsAthreeFrac\
poisoned-arm cells clear, so its published and residue intervals are wide and we draw no conclusion from
them. Printed by \texttt{supplement/analysis/build\_scheduler.py}.}
\label{tab:schedcols}
\resizebox{\linewidth}{!}{\input{supplement/build/tab_sched_cols_body.tex}}
\end{table}

\section{Decomposing a published LLM defense}\label{app:fmix}

The two language-model defenses of Section~\ref{sec:setup} are our own operating points. To decompose a
defense the field actually publishes, we implemented Fine-mixing \citep{zhang2022finemixing}, whose
two-stage recipe is designed for exactly this setting: mix the backdoored fine-tuned weights back toward
the pre-trained weights, then fine-tune the mixture on a small clean set. We interpolate parameterwise,
$\theta_{\mathrm{mix}}=(1-\lambda)\,\theta_{\mathrm{bd}}+\lambda\,\theta_{\mathrm{pre}}$, and run stage
two with the identical budget, optimiser and pool as our other LLM defenses ($150$ steps, Adafactor at
$4\times10^{-5}$, $2000$ Alpaca examples, a $3.87\%$ clean-data budget), so the arms stay comparable. The
mixing ratio is the method's own strength dial and we sweep it. Both arms of the Qwen2.5-7B pair are run
at every ratio; utility is MMLU over $2000$ items.

\textbf{Every ratio clears the backdoor, and the decomposition bounces.} Table~\ref{tab:fmix} sweeps
$\lambda$ on a single defender draw. Residual attack success is $0.00$ at every ratio, so A3 holds
throughout, yet the reported cost moves between $-0.45$ and $+5.00$ points and the DiD term changes sign
twice, ending negative at both ends of the sweep.

\textbf{The reported cost is dominated by which clean examples the defender happens to draw.} We
replicated $\lambda=0.5$ on both arms at three defender pool draws (Table~\ref{tab:fmixrep}). All three
cleared the backdoor, but the reported cost came out $-0.45$, $+8.60$ and $-0.70$ points: a standard
deviation of $5.30$, which is $2.1$ times the mean it is a deviation from. The removal term inherits that
spread, $-0.45$ $[-10.86,+9.96]$, and so does the tax, $+2.93$ $[-2.18,+8.05]$. A single-run cost figure
for Fine-mixing at this budget is therefore not a measurement of the defense; the clean arm is what makes
that visible, because it shows the same instability on a model that never carried a backdoor. The
instability is visible in training as well: three of the fourteen stage-two runs spike above a loss of $6$
at a single logging step before recovering, and they are spread across both arms and two mixing ratios
rather than concentrated in one cell (peak losses are a column of the supplementary cell file). We report
every run; none is excluded.

\begin{table}[!ht]\centering\small
\caption{Fine-mixing \citep{zhang2022finemixing} on Qwen2.5-7B, one defender draw, mixing ratio swept.
MMLU over $2000$ items; the pre-defense poisoned model scores $0.6775$ and its twin $0.6850$, so the
pre-defense gap is $+0.75$ points. Residual attack success is $0.00$ everywhere.}
\label{tab:fmix}
\begin{tabular}{lccccc}
\toprule
mixing ratio $\lambda$ & poisoned MMLU & twin MMLU & reported $\cDef$ & tax & removal (DiD) \\
\midrule
$0.1$ & $0.6640$ & $0.6520$ & $+1.35$ & $+3.30$ & $-1.95$ \\
$0.2$ & $0.6590$ & $0.6745$ & $+1.85$ & $+1.05$ & $+0.80$ \\
$0.3$ & $0.6275$ & $0.6575$ & $+5.00$ & $+2.75$ & $+2.25$ \\
$0.5$ & $0.6820$ & $0.6480$ & $-0.45$ & $+3.70$ & $-4.15$ \\
$0.7$ & $0.6560$ & $0.6520$ & $+2.15$ & $+3.30$ & $-1.15$ \\
\bottomrule
\end{tabular}
\end{table}

\begin{table}[!ht]\centering\small
\caption{Fine-mixing at $\lambda=0.5$, both arms, three independent defender pool draws from the same
Alpaca corpus at the same $3.87\%$ budget. All six runs cleared the backdoor. Intervals are $95\%$
$t$-intervals over the three draws.}
\label{tab:fmixrep}
\begin{tabular}{lcccc}
\toprule
 & draw $2$ & draw $3$ & draw $4$ & mean $[95\%]$ \\
\midrule
reported $\cDef$ (pt) & $-0.45$ & $+8.60$ & $-0.70$ & $+2.48\ \scriptstyle[-10.68,+15.65]$ \\
intervention tax (pt) & $+3.70$ & $+4.50$ & $+0.60$ & $+2.93\ \scriptstyle[-2.18,+8.05]$ \\
interaction, DiD (pt) & $-4.15$ & $+4.10$ & $-1.30$ & $-0.45\ \scriptstyle[-10.86,+9.96]$ \\
\bottomrule
\end{tabular}
\end{table}

\FloatBarrier
\section{Extension to machine unlearning}\label{app:unlearn}
\textbf{Machine unlearning has the analogous baseline-accounting problem}, and there the reference
backdoor benchmarks omit is already standard. On TOFU \citep{maini2024tofu} the clean twin is
\emph{exact}, a from-scratch retrain on the retain split, and it removes the forget set at $+0.82$ points
of utility ($[-1.17,+2.81]$, five seeds), similar across forget-set sizes ($1$--$10\%$;
Table~\ref{tab:tofu}), while the utility drops that published TOFU reproductions report for algorithms
scored against that same gold span $-2$ to $62$ points \citep{fan2025simnpo,mekala2025altpo}: a
cross-study span read off those papers, not a quantity we recompute.

\FloatBarrier
\section{Per-defense re-ranking}\label{app:zoo}
Per-defense detail behind the re-ranking of Section~\ref{sec:corrected} (the slope diagram is
Figure~\ref{fig:main}(b)).

\begin{figure}[!ht]
\centering
\includegraphics[width=.92\textwidth]{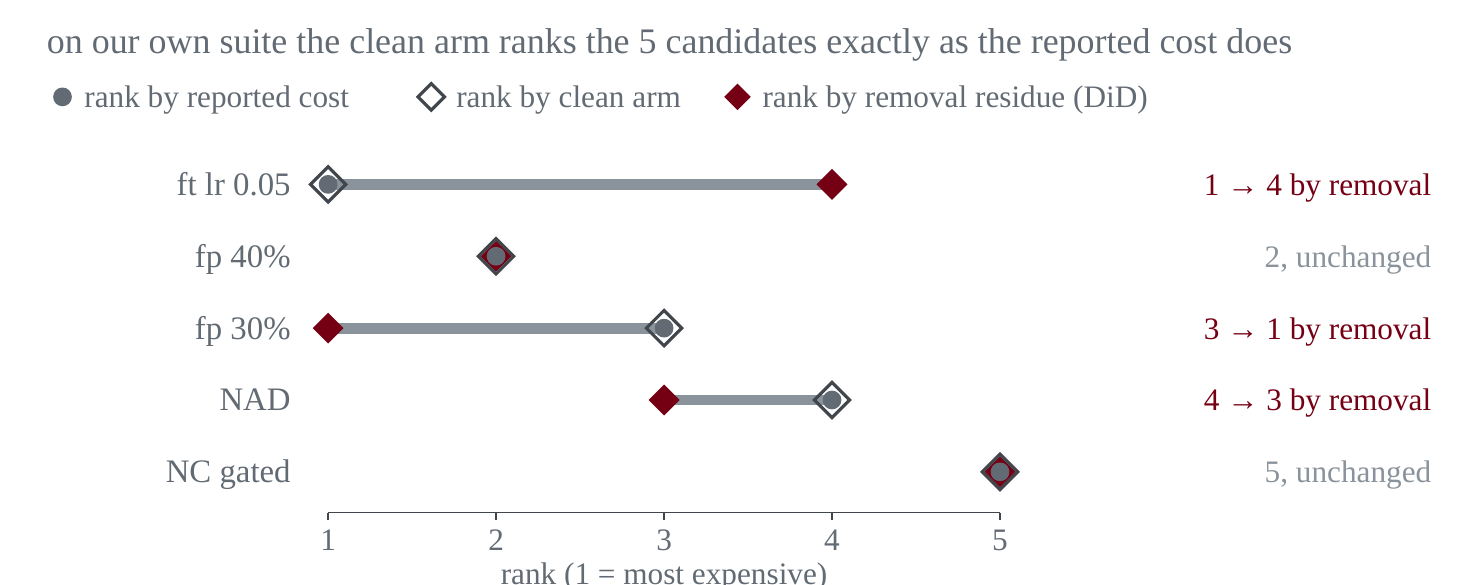}
\caption{\textbf{On our own suite the clean arm ranks the candidates exactly as the reported cost does
($\tau_a={}$\chTauTaxPrim, \chSepPub\ pairs separable), which is the identity of Eq.~\ref{eq:decomp2} made
visible; only the removal residue moves anything.} The \chNcandWord\ candidate defenses on
CIFAR-10/BadNets over the thirteen
installations of the common support, where every candidate clears the backdoor and keeps an admissible
twin. Grey dot: rank by reported cost. Hollow diamond: rank by the clean arm---it coincides with the grey
dot on every row. Crimson diamond: rank by the DiD removal residue. \textbf{The rows are drawn in order
of cost magnitude, not of the DER rating every $\tau_a$ we quote is computed on}: on the drawn rule the
published-versus-removal agreement is \chFigTauRemRaw\ with a gap of \chFigGapRaw\ and
\chFigNmovedWordRaw\ of the \chNcandWord\ rows moving, while on the DER rule of Table~\ref{tab:csper} it is
\chFigTauRemDer\ with a gap of \chFigGapDer\ (percentile \chGapPctPrim, basic \chGapBasicPrim---the two
constructions disagree about zero, Appendix~\ref{app:a3}), \chSepDid\ pairs separable and one adjacent
transposition. The corrected
ratings all lie within \chSpanPrimRound\ of one another, so the corrected rank intervals are wide and we report the
ordering as a consequence of the identity rather than a result; the separable reversals that are a result
are on BackdoorBench (\S\ref{sec:abl}). NAD enters this suite only in this version: its teacher had been
built at a learning rate that leaves the backdoor intact, and at a teacher that loses it NAD clears
\chNADAthree\ (Appendix~\ref{sec:selfcorrect}). All rank correlations are Kendall $\tau_a$ on unrounded scores;
rank intervals are conditional on the selected candidate set and the realized common support, both adopted
during internal review and reported with the alternatives they replaced (Table~\ref{tab:choices}), and
that selection uncertainty is not propagated by the bootstrap. The \chNbottomWord\ methods excluded by the
candidate gate are tabulated in Appendix~\ref{app:a3} and below.}
\label{fig:rerank}
\end{figure}

\begin{table}[!ht]\centering\small
\caption{\textbf{On the community's own benchmark the decomposition replicates and the ordering result
does not.} BackdoorBench's shipped attack$+$defense framework, BadNets on PreAct-ResNet-18 (the
framework's default), eleven independent
installation seeds under the protocol of Appendix~\ref{app:bb}, frozen before the eight confirmatory
seeds. Cost rows average the seven defenses other than ABL, a training-stage defense retrained from
scratch, whose clean-twin drop exceeds $30$ points, dropping A3-failed cells first as elsewhere: Neural Cleanse contributes eight of its eleven cells, the rest all eleven. The ABL row and the $21$ pairwise comparisons include ABL. The ranking rows use different sets: the candidate set is the seven
methods that pass the method-level gate, which excludes Neural Cleanse and includes ABL, and its
alignment gap is $+0.190$ $[-0.190,+0.476]$; the rank-correlation row below is the prespecified
without-ABL sensitivity over six methods; the pairwise comparisons run over the seven gate-passing candidates. The entries close at the point-estimate level as
Eq.~\ref{eq:decomp2} requires: $+2.42-2.06=+0.36$ (paired minus gap) and $+0.36+1.98=+2.34$ (removal plus
clean arm). WaNet is a third attack on the same framework under the same frozen protocol, eleven
installation seeds: three fail A0 (attack success \chwAoneRange\ against $0.94$--$0.99$ elsewhere),
leaving a success-conditioned common support of \chwNsuppWord\ seeds over the \chwNcandWord\ candidate
defenses (pre-defense gap $+1.43$ $[+0.48,+2.39]$). Its ranking gap is \chwGapAll, bootstrap \chwGapCIAll\
with $P^{*}(\mathrm{gap}\le0)={}$\chwGapPleAll, and the exact sign-flip enumeration separates \chwSepPubAll\
pairs in \emph{both} accountings, so that zero is an unpowered grid rather than a null. On the
\chwConfSeedsWord\ confirmatory seeds the corrected order is \emph{identical} to the published one
($\tau_a={}$\chwTauRemConf) but the intervention ordering is not (\chwTauTaxConf), so the gap there is
\chwGapConf\ \chwGapCIConf\ with $P^{*}(\mathrm{gap}\le0)={}$\chwGapPleConf: on those seeds it runs the other
way. The powered test on this grid is the pairwise envelope (Appendix~\ref{app:bb}). $^{\dagger}$ANP
holds $A2$ at the policy level only: the framework selects its threshold by test-set attack success
(Appendix~\ref{app:bb}).}
\label{tab:bb}
\setlength{\tabcolsep}{4pt}
\resizebox{\linewidth}{!}{%
\begin{tabular}{lcc}
\toprule
 & published accounting & twin-corrected \\
\midrule
clean-accuracy cost term (pt, $11$ seeds) $\downarrow$ & $+2.34\ \scriptstyle[+2.10,+2.57]$ & $+0.36\ \scriptstyle[-0.26,+0.98]$ (DiD) \\
\hspace{1em}pre-defense gap, twin $-$ poisoned (pt) & \multicolumn{2}{c}{$+2.06\ \scriptstyle[+1.90,+2.23]$} \\
\hspace{1em}paired post-defense contrast (pt) & --- & $+2.42\ \scriptstyle[+1.81,+3.04]$ \\
\hspace{1em}clean-arm defense cost (pt) & --- & $+1.98\ \scriptstyle[+1.50,+2.46]$ \\
\hspace{1em}with ABL, eight defenses (pt) & $+3.25\ \scriptstyle[+3.01,+3.49]$ & $-2.79\ \scriptstyle[-3.69,-1.88]$ (DiD) \\
attack-success drop (\%) & \multicolumn{2}{c}{$93$--$95$ (near-constant)} \\
SD across the six non-ABL defenses (pt) & $2.40$ (poisoned arm) & $2.08$ (backdoor-free twin) \\
rank correlation (without-ABL, $6$) & --- & $+0.733$ rem., $+0.200$ interv. \\
separable pairs (BH, $q{=}0.05$) & $20/21$ & $18/21$ \\
\midrule
third attack (WaNet): mean, $n{=}5$ cond. & \chownWNDef & \chownWNDid\ (DiD) \\
\hspace{1em}clean-arm defense cost, $n{=}5$ cond. & --- & \chownWNClean \\
\hspace{1em}clean-arm defense cost, $n{=}11$ uncond. & --- & \chuncWNClean \\
\hspace{1em}paired post-defense contrast, $n{=}11$ (Table~\ref{tab:bb}'s convention) & --- & \chbbWNPaired \\
\hspace{1em}fine-tuning alone, paired, $n{=}11$, BadNets / WaNet & \chbbBNFtPaired & \chbbWNFtPaired \\
\hspace{1em}\emph{clean arms of reversed sign}, $n{=}11$ & \multicolumn{2}{c}{$3/8$ excl.\ zero: ft \schWmsFtPt, fp \schWmsFpPt, NAD \schWmsNadPt} \\
\hspace{1em}\emph{negative} reported cost, $n{=}5$ cond. & \multicolumn{2}{c}{\chwCandNeg\ candidates, every interval excl.\ zero} \\
\hspace{1em}largest (ANP$^{\dagger}$), $n{=}5$ cond. & $+4.37\ \scriptstyle[+0.46,+8.28]$ & $+2.28\ \scriptstyle[-2.07,+6.64]$ \\
\hspace{1em}seeds installing below the A0 gate & \multicolumn{2}{c}{\chwAoneFailN\ (success \chwAoneRange)} \\
\bottomrule
\end{tabular}}
\end{table}

\begin{table}[!ht]\centering\small
\caption{\textbf{At this attack the non-removal share is a property of the defense, not a constant; on
WaNet the same eight do not even agree on its sign (Table~\ref{tab:bb}), so it is a property of the
cell.} Per-defense decomposition on BackdoorBench over the eleven installation seeds, $95\%$
$t$-intervals. Every defense
levies a clean-arm cost whose nominal interval excludes zero (\exBbCleanExclBonf\ at $0.05/8$, \exBbCleanExclBH\ under BH $q=0.05$; excluding ABL, \exBbCleanExclBonfNoAbl\ at $0.05/7$ and \exBbCleanExclBHNoAbl\ under BH, the main text's convention). Four of the seven unconditional methods have
non-positive interaction (DiD) point estimates: the identical defense costs
the never-poisoned twin more than it costs the poisoned model, which is what a rating that charges the
whole drop to removal cannot represent; in paired form (A0 seeds, Table~\ref{tab:corrmain}) only ABL is
non-positive (fine-pruning \steBadnetFpPaired, SAU \steBadnetSauPaired, I-BAU \steBadnetIbauPaired), the two forms
differing by $g={}$\taboneBbGapPt, and which part of either is removal is not identified (Section~\ref{sec:estimand}). Every row averages that defense's A3-passing cells, so Neural
Cleanse's row is over its eight successes rather than all eleven seeds; an earlier version printed its row
over all eleven, the only row in the table that used a different convention. Neural Cleanse is
listed for completeness; the framework's detector does not fire on three of its eleven seeds, so the
candidate gate excludes it there at a mean residual attack success of $0.267$. In our own harness, with the
trigger reverse-engineered in pixel space, the same method fires on all fourteen and clears every one
(Appendix~\ref{app:cond}); the two rigs disagree about this defense as they do about several others.}
\label{tab:bbdef}
\setlength{\tabcolsep}{3pt}
\resizebox{\linewidth}{!}{\begin{tabular}{lccccc}
\toprule
Defense & reported $\cDef$ (pt) & clean arm (pt) & interaction, DiD (pt) & $+$noise & $+$shortage \\
\midrule
\input{supplement/build/tab_bbdef_rows.tex}
pooled, $7$ excl.\ ABL & \multicolumn{2}{c}{share of $\cDef$ left for the interaction} & \textbf{\exShareUncorr} & \exShareNoise & \exShareShort \\
\bottomrule
\end{tabular}}

{\footnotesize The last two columns apply the headroom correction of Appendix~\ref{app:headroom} to every
row or to none, once for each degradation route: an arm degraded by label noise implies \exNoiseAtG\ at this
benchmark's gap, an arm degraded by training on less data \exShortAtG. The routes disagree in sign, and the
poisoned arm's own deficit resists these defenses the way the second route's does, so we apply neither and
report the uncorrected column in bold as the measurement. The clean arm needs no correction. $^{\dagger}$$A2$ at the policy level only: ABL and Neural
Cleanse run different computations on the two arms, and ANP selects its pruning threshold by test-set
attack success under a \chbAnpBudgetPct\ accuracy budget, on a twin a selection on false-positive noise
(Appendix~\ref{app:bb}). ABL is excluded from the
pooled row; Neural Cleanse enters it through its eight A3-successful cells. $^{\ddagger}$conditional;
excluded from the ranking by the method-level gate.}
\end{table}

\begin{table}[!ht]
\centering
\caption{Sensitivity of the field's ordering to how its cost term is defined. Nine defenses,
CIFAR-10 / BadNets, $5\%$ clean budget, all fourteen independent installations, corrected defender
draw. The last column gives, within the five candidate rows, the rank
under the published, intervention-component and DiD-removal DER ($\tau_a={}$\chTauTaxAllFourteen\ and
\chTauRemAllFourteen); primary inference uses the \chNsuppWord-installation common support, not these
fourteen, which carry the rank column only. $^{\P}$ANP is excluded on A3, not on A4: under its own
accuracy budget it prunes too little to clear this attack (Table~\ref{tab:bbmain},
Appendix~\ref{app:a3}). A sensitivity analysis, not a
proposed score. $^{\ddagger}$Neural Cleanse is conditional and gated as its
authors specify; its clean arm no-ops in \chNCzeroAllWord\ of the \chNinstWord\ installations, so on those
its two columns coincide by construction, and on the other four it charges a never-poisoned model
\chNCfireLo--\chNCfireHi\ points
(Appendix~\ref{app:cond}). Over all nine rows the published-versus-corrected
agreement is $\tau_a={}$\chTauNine. These are the \chNinst-installation values under the candidate gate;
the primary common-support analysis, over the \chNcandWord\ gate-passing methods on the \chNsuppWord\
installations they all clear, is Table~\ref{tab:csper}. \chTieB\ and \chTieA\ agree to five decimals on the
published score (\chTieScoreB\ against \chTieScoreA, a margin of \chTieMargin) and are ordered on the
unrounded value.
$^{\mathsection}$marks methods whose \emph{mean} residual attack success exceeds the $5\%$
method-level gate. They are not universally inadmissible---A3 is an installation-level event and
some of their cells clear it (Appendix~\ref{app:a3})---but they are excluded from the candidate
set for cross-method ranking and enter no primary figure.}
\label{tab:zoo}
\scriptsize
\input{supplement/build/tab_zoo_body.tex}
\end{table}

\FloatBarrier
\section{Independently re-trained language-model seeds}\label{app:indep}

Table~\ref{tab:indep} gives the per-defense detail behind the pooled language-model figures
of Section~\ref{sec:results}. Each seed is a full end-to-end re-run: poisoning, clean twin,
and both defenses retrained from the base model.

\begin{table}[!ht]
\centering
\caption{Language-model costs under eight independently re-trained installation seeds
(Qwen2.5-7B-Instruct, MMLU over $2000$ items; each seed is a full re-run of poisoning, twin, and
both defenses with the corrected, independent defender draw; installation-level A3 successes are
$7/8$ for each defense, fine-tuning missing it in installation $15$ at residual ASR $0.09$ and
fine-pruning in installation $17$ at $0.25$; restricting to the six installations where both succeed
gives a complete-case removal aggregate of $+1.60$ $[+0.79,+2.42]$ against a reported cost of $+12.05$ on
those same six, so the headline does not depend on variable-composition averaging). Per-defense intervals are $95\%$ over
installation seeds; the aggregate averages the admissible defenses within each seed ($\mathrm{df}=7$).
Pre-defense gap $g=+0.04$ $[-0.19,+0.27]$.}
\label{tab:indep}
\begin{tabular}{lccc}
\toprule
Defense & $\cDef$ (pt) & $\cRem$ (pt) & $\cRem^{\mathrm{DiD}}$ (pt) \\
\midrule
full fine-tune lr $4\!\times\!10^{-5}$ & $+8.74\ \scriptstyle[+5.52,+11.97]$   & $+2.78\ \scriptstyle[-0.19,+5.75]$ & $+2.77\ \scriptstyle[-0.17,+5.71]$ \\
fine-prune $20\%$                      & $+14.68\ \scriptstyle[+11.41,+17.95]$ & $+0.25\ \scriptstyle[-1.94,+2.44]$ & $+0.17\ \scriptstyle[-2.11,+2.46]$ \\
\midrule
install level ($n{=}8$) & $+11.46\ \scriptstyle[+8.94,+13.97]$ & $+1.42\ \scriptstyle[+0.69,+2.14]$ & $+1.37\ \scriptstyle[+0.52,+2.23]$ \\
\bottomrule
\end{tabular}
\end{table}

\FloatBarrier
\section{Conditional defenses}\label{app:cond}

\textbf{Policy-level consistency (A2).} A conditional defense is one function at the policy
level---detection, then repair only if the detector fires---so the twin arm is well-defined and
simply takes the no-op branch on a clean model. With the trigger reverse-engineered in pixel space
(Appendix~\ref{app:hyper}) Neural Cleanse \citep{wang2019neuralcleanse} fires on all fourteen poisoned
models, names the true target class in each, clears the backdoor in all of them (mean residual attack
success $0.008$), and charges $+0.78$ $[+0.60,+0.96]$ points. Its gated twin arm no-ops in ten of the
fourteen installations, so on those the corrected and reported costs coincide by construction.

\textbf{The gate does not make the repair cheap; it makes it rare.} That reading is testable, and we
tested it: forcing the repair branch on every never-poisoned twin costs \ncForcedOursCi\ points,
more than four times what the same defense charges the poisoned models it does fire on. The price is a
property of the repair, not of our patching schedule: at \citeauthor{wang2019neuralcleanse}'s own recipe,
one epoch over $10\%$ of the training set with a fifth of each batch patched, the forced repair costs
\ncForcedPubCi. The asymmetry has
a mechanism---on a poisoned model the search recovers a small, real trigger and unlearning it is cheap,
while on a clean model it must invent a large diffuse pattern and unlearning that damages the network---and
it is invisible without the twin, because the gated clean arm reads exactly zero. The detector also
false-positives on four of the fourteen twins, naming a wrong class each time at an anomaly index just
over the threshold of $2$, and on those it charges a never-poisoned model $2.8$--$3.7$ points. \textbf{The
confound we identify is a property of unconditional defenses and of false-positive repairs}; for a gated
policy it is priced by the false-positive rate times the forced-repair cost, and a benchmark scoring
gated and ungated defenses under one rating should say which it is scoring.

\FloatBarrier
\section{The surrogate's two tiers}\label{app:tiers}

The estimator has two tiers, and we report both. \textbf{Within the tested architecture and
recipe} its error is a clean-versus-clean difference: on Qwen2.5-7B, four independently trained
clean models used one at a time give DiD-form surrogate-minus-oracle errors from $-0.5$ to $+3.0$
points, mean $+1.7$, against a twin-arm cost of $8.9$ (Table~\ref{tab:qsurr}); averaging them brings
the error to the mean, a fifth of the clean-arm cost it replaces. On vision the instability is concentrated in ANP (Table~\ref{tab:surr}).
\textbf{Across architectures} it is qualitative: Llama-3.2-3B as a surrogate for Qwen2.5-7B
(different family, $2.3\times$ scale gap) carries a $-1.29$-point systematic bias (range $-2.47$
to $-0.42$ over five fixed defense configurations, all same-signed; Table~\ref{tab:xsurr}; we
report no interval because the configurations are design points, not replicates), which supports the qualitative verdict that the reported cost is
overstated, but not a quantitative decomposition. The
boundary is set by a third architecture---Mistral-7B catastrophically forgets under the same
recipe (a clean Mistral loses $26.0$ MMLU points to the identical defense, lr $4\times10^{-5}$ for $150$
steps, against $7.7$ for a clean Qwen; both drops are within-model, each arm scored on its own $2{,}000$-item
pool, \texttt{supplement/data/llm\_arm\_evaluations.json}), an instance of the path-dependence formalized by
\citet{impossibility2025retrain}---and a same-recipe degradation probe identifies
that catastrophic case without an oracle. The probe does not by itself certify exchangeability or
guarantee surrogate validity; it screens out the regime in which the surrogate is known to fail.

\textbf{Deployment scenario.} A model provider receives a suspected or externally supplied checkpoint
and can test whether a defense suppresses a known or independently discovered trigger, but does not
hold the poisoned training examples needed to reconstruct a matched twin. It does hold a clean model
of the same architecture and training recipe---a prior production replica or an independently
trained release candidate. TARE-Z uses that model to estimate the clean-arm correction at the cost of
one defense run; the degradation probe screens out the catastrophic regime observed here, and the
removal-success evidence must still be obtained independently.

\begin{table}[!ht]\centering\small
\caption{Same-recipe surrogates on Qwen2.5-7B-Instruct: continued fine-tuning at lr
$4\times10^{-5}$ for $250$ steps applied to the paired twin and to four independently trained clean
models (MMLU over $1{,}000$ items, in points). The DiD-form surrogate error is the twin's drop minus the surrogate's drop; it does not involve the poisoned arm,
so these rows are unaffected by the defender-draw correction of Appendix~\ref{sec:selfcorrect}. The
poisoned-arm reference row an earlier version carried was affected and is omitted.}
\label{tab:qsurr}
\begin{tabular}{lcccc}
\toprule
clean model & $Q(\theta)$ & $Q(\mathcal{D}_b\theta)$ & drop & error vs.\ twin \\
\midrule
paired twin (oracle) & $70.7$ & $61.8$ & $8.9$ & --- \\
clean seed 1 & $71.2$ & $65.3$ & $5.9$ & $+3.0$ \\
clean seed 2 & $70.4$ & $64.0$ & $6.4$ & $+2.5$ \\
clean seed 3 & $70.3$ & $63.1$ & $7.2$ & $+1.7$ \\
clean seed 4 & $70.9$ & $61.5$ & $9.4$ & $-0.5$ \\
mean of the four & $70.7$ & $63.5$ & $7.2$ & $+1.7$ \\
\bottomrule
\end{tabular}
\end{table}

\begin{table}[!ht]\centering\small
\caption{Cross-architecture surrogate: Llama-3.2-3B's clean-arm defense drop substituted for
Qwen2.5-7B's, per defense configuration (points; every value, both arms and both models, is MMLU over $2{,}000$ items; $n$ is the number of defense seeds averaged). The error, Qwen drop minus Llama
drop, is the error of the DiD-form surrogate estimate: mean $-1.29$, range $-2.47$ to
$-0.42$, all five same-signed; the configurations are fixed design points, so no interval is
attached.}
\label{tab:xsurr}
\begin{tabular}{lccc}
\toprule
configuration & Qwen twin drop ($n$) & Llama twin drop ($n$) & error \\
\midrule
fine-tune, lr $3\times10^{-5}$ & $5.22$ ($3$) & $6.60$ ($1$) & $-1.38$ \\
fine-tune, lr $4\times10^{-5}$ & $7.20$ ($3$) & $8.80$ ($3$) & $-1.60$ \\
fine-tune, lr $5\times10^{-5}$ & $9.83$ ($3$) & $12.30$ ($3$) & $-2.47$ \\
fine-prune $20\%$ & $12.40$ ($2$) & $13.00$ ($3$) & $-0.60$ \\
fine-prune $40\%$ & $23.65$ ($2$) & $24.07$ ($3$) & $-0.42$ \\
\bottomrule
\end{tabular}
\end{table}

\FloatBarrier
\section{Surrogate calibration, per defense}\label{app:surr}

\begin{figure}[!ht]
\centering
\includegraphics[width=\textwidth]{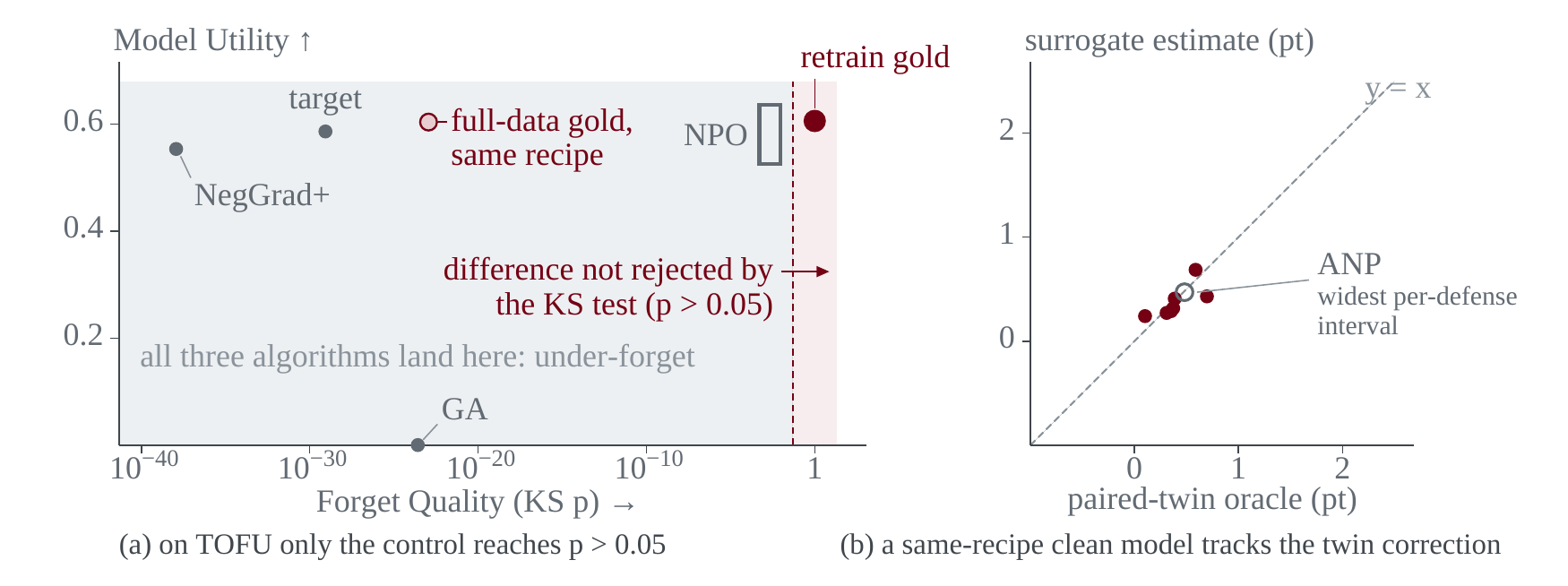}
\caption{\textbf{Two checks on the control arm.} \textbf{(a)} On TOFU the from-scratch retrain
(\textcolor{cRem}{crimson}) is the only point outside the shaded under-forgetting region: it
forgets perfectly by construction---it \emph{is} the reference the KS test compares against, so its
$p=1$ is a definition, not a finding---at a utility that does not resolve from the full-data model's,
while none of the three algorithms we run reaches that forget quality, and harder NPO settings
lower utility without reaching it. GA collapses entirely; NPO is drawn as a box over seeds and
harder settings; the open marker is the full-data gold under the same recipe, which has forgotten
nothing and lands with the target at $p\approx10^{-23}$, the anchor at the far end of the axis.
Every $p$ drawn here is recomputed under one definition by
\texttt{analysis/tofu\_fq.py}, which reproduces the three values the eval dumps carried
(one seed shown; five-seed means in Table~\ref{tab:tofu}). \textbf{(b)} Pairing is not essential for the mean correction in this suite: an
independently trained clean model of the same recipe (Eq.~\ref{eq:surr}) puts all eight unconditional
defenses within $0.3$ pt of $y=x$ in the mean (pooled error \tzCifarPooledPaired\ pt, largest single
defense-by-installation error $0.88$ pt). The open marker is ANP, whose \emph{oracle} estimate is the most
installation-unstable of the eight, so its wider per-defense interval cannot be attributed
cleanly to either side---it marks the boundary of the estimator's precision, not an outlier we
discard.
CIFAR-10, eight defenses over four installations; each point is a defense's mean over the four.}
\label{fig:controls}
\end{figure}

Table~\ref{tab:surr} gives the per-defense detail behind the pooled figures of
Section~\ref{sec:surrogate}.
On these four installations the per-defense oracle values outside ANP span only $0.6$ points, so an
ordering comparison is uninformative there (Kendall $\tau_a=+0.857$ over all eight, $+0.905$ without ANP); the
calibration claim is the error column, not an ordering.

\begin{table}[!ht]
\centering
\caption{The surrogate agrees with the oracle on average and tracks all eight unconditional defenses
closely.
CIFAR-10, $5\%$ clean budget, four installations; surrogates are independently trained clean
models with no pairing to the poisoned run. Neural Cleanse is excluded because its gated clean
arm usually no-ops, leaving no clean-arm intervention correction for the surrogate to estimate.
The last column is what a practitioner using the surrogate instead of the true twin would be wrong
by; per-defense rows give the paired form, and the four pooled rows give both the paired and the
DiD form. All intervals are $t$-intervals over the four installations, pooled rows over
the four installation means. Two independently
trained clean models serve as surrogates for each installation and their estimates are averaged; both arms
use the identical defense implementation, including the BN-masked fine-pruning. Fine-tuning at lr $0.01$
and $0.02$ do not clear the backdoor on these installations (NAD, on its corrected teacher, clears three of
four), so their oracle column is a
post-defense contrast rather than an identified removal cost; the surrogate's error is a clean-versus-clean
quantity either way. ANP no longer needs a separate reading: with its reference mask semantics and accuracy-guarded
threshold its surrogate error is $-0.01$ $[-1.02,+0.99]$ like the rest, and the pooled error is the same
with and without it. An earlier implementation of ANP made it the one outlier and dominated the pooled
row (Appendix~\ref{sec:selfcorrect}).}
\label{tab:surr}
\begin{tabular}{lccc}
\toprule
Defense & oracle $\cRem$ (pt) & surrogate $\hat{c}_{\mathrm{int}}$ (pt) & difference (pt) \\
\midrule
fine-tune lr $0.01$ & $+0.39$ & $+0.41$ & $+0.02\ \scriptstyle[-0.38,+0.42]$ \\
fine-tune lr $0.02$ & $+0.59$ & $+0.69$ & $+0.10\ \scriptstyle[-0.14,+0.34]$ \\
fine-tune lr $0.05$ & $+0.70$ & $+0.43$ & $-0.27\ \scriptstyle[-0.65,+0.12]$ \\
fine-prune $20\%$   & $+0.10$ & $+0.24$ & $+0.14\ \scriptstyle[-0.05,+0.32]$ \\
fine-prune $30\%$   & $+0.37$ & $+0.32$ & $-0.06\ \scriptstyle[-0.20,+0.09]$ \\
fine-prune $40\%$   & $+0.31$ & $+0.27$ & $-0.04\ \scriptstyle[-0.35,+0.27]$ \\
NAD                 & $+0.35$ & $+0.29$ & $-0.06\ \scriptstyle[-0.57,+0.45]$ \\
ANP                 & $+0.48$ & $+0.47$ & $-0.01\ \scriptstyle[-1.02,+0.99]$ \\
\midrule
pooled paired, all eight   & --- & --- & \tzCifarPooledPaired \\
pooled paired, excl.\ ANP  & --- & --- & \tzCifarPooledPairedNoAnp \\
pooled DiD, all eight      & --- & --- & \tzCifarPooledDid \\
pooled DiD, excl.\ ANP     & --- & --- & \tzCifarPooledDidNoAnp \\
\bottomrule
\end{tabular}
\end{table}

\FloatBarrier
\section{Three cost estimators, side by side}\label{app:did}
\paragraph{In full.} \textbf{What the twin attributes to the interaction is estimator- and defense-dependent under either form.}
Table~\ref{tab:rerank} gives every setting under both, and matched denominators put none above $55\%$
(Appendix~\ref{app:lim}); on CIFAR-10 a reported $+2.51$ leaves $+0.02$ $[-0.31,+0.34]$ by
difference-in-differences and $+0.43$ $[+0.30,+0.55]$ paired. Over the \gdNinst\ CIFAR installations
the paired estimate's slope in $g$ is \gdPairedSlopeBoot\ (exact permutation $p={}$\gdPairedPermP\ over
\gdPermN\ orderings); since $\cRem^{\mathrm{DiD}}=\cRem-g$ cell by cell, the DiD slope is that minus one,
\gdDidSlopeBoot, an identity rather than an independent finding. Qwen's $+1.37$ of $+11.46$ is the only
all-installation estimate above zero (BackdoorBench's confirmatory seeds alone give \taboneBbConfDid) and the one we hold most lightly; its operating points are expensive because nothing
cheaper works there (Appendix~\ref{app:hyper}).

Table~\ref{tab:cif8} reports all three estimators of Section~\ref{sec:protocol} on the CIFAR
defense families over eight independent installations. The difference-in-differences
correction subtracts the pre-defense gap ($+0.41$ pt, CI $[+0.15,+0.68]$ over the eight
installations), so it shifts every $\cRem$ down by roughly that amount; on this grid that is the
difference between a paired contrast that resolves ($+0.43$ $[+0.30,+0.55]$) and a DiD that does not
($+0.02$ $[-0.31,+0.34]$); Appendix~\ref{app:headroom} reports the slope of each estimator in $g$ over the
eight installations---the paired estimate's is \gdPairedSlopeBoot, and the DiD's is that minus one by
identity, \gdDidSlopeBoot. The qualitative picture is unchanged under all three: the reported cost $\cDef$ grows with defense
strength while the removal cost does not.

On the language model the two estimators nearly coincide on the eight-installation data
($\cRem=+1.42$ and $\cRem^{\mathrm{DiD}}=+1.37$ pt at the installation level; pre-defense gap $+0.04$
pt). No conclusion in the paper depends on which estimator is used; Table~\ref{tab:rerank} prints both,
and Section~\ref{sec:estimand} states why neither is corrected into a removal cost.

\paragraph{An attempt to manipulate $g$ experimentally, reported as inconclusive.} Because the two
estimators differ by $g$, the natural design is to vary $g$ at the source. We installed BadNets at $1\%$
and $10\%$ poisoning instead of the headline $5\%$, three installations each, and ran four defenses on
each (supplementary \texttt{cifar\_poisonrate\_cells.csv}). The manipulation did not move the manipulated
variable: $g$ is $+0.94$ $[-0.22,+2.10]$ at $1\%$ and $+1.15$ $[-0.29,+2.58]$ at $10\%$, against $+0.53$
$[-0.18,+1.24]$ at $5\%$ on the same seeds, three overlapping intervals; pre-defense attack success is
$0.94$ and $0.97$. What the poisoning rate does change is removal success, which falls at the low rate:
at $1\%$ only fine-pruning at $40\%$ clears the backdoor in all three installations, and fine-tuning at lr
$0.02$ in none. A second reason not to lean on this design: the $1\%$ and $10\%$ arms were installed for
$30$ epochs where the headline installations run $40$, so their pre-defense gaps mix the poisoning rate
with a training-budget difference and are not comparable to the headline $g$ on equal terms. We therefore
report the slope in $g$ of Appendix~\ref{app:headroom} from the variation the installations supply on their
own ($g$ from \gdGLo\ to \gdGHi), and record this design as inconclusive rather than as evidence.

\FloatBarrier
\section{Self-correction details}\label{app:selfcorr}\label{sec:selfcorrect}
\paragraph{In full.} \textbf{Whether a named defense removes the backdoor is a property of the cell, not the method.} At
BackdoorBench's own operating point its fine-tuning leaves a mean residual attack success of $0.015$ on
its installations and $0.369$ on ours, and on a second attack inside that framework its own fine-tuning,
NAD and fine-pruning clear none of eight seeds. One such gap was ours, not the installation's: our NAD
cleared $0/14$ until we rebuilt its teacher, which we had been fine-tuning at a rate that leaves the
backdoor intact; at a teacher that loses it NAD clears $13/14$, and its clean arm of $+1.64$ exceeds its
entire reported cost of $+1.49$. That a defense's verdict depends on how intensely the model was trained
on the poison is established \citep{zhu2025rethinkdetect,min2023fst,zhang2023ftsam}; we cite it as the
explanation rather than offer the failure as a finding
(Appendices~\ref{app:attacklr},~\ref{app:blended},~\ref{sec:selfcorrect}).

\paragraph{The defender's clean data was not independent of the attacker's poison draw.}
In the vision harness the poisoned indices and the defender's clean subset were drawn from the same
seeded stream with the same call, so for every installation the two index sets were bit-identical: the
defender fine-tuned on exactly the images whose labels the attacker had flipped, now correctly labelled.
The language-model harness had the same defect, the defender's pool being the installation pool. The bias
runs in the direction of our own thesis, because an oracle repair set makes removal on the poisoned arm
unusually cheap and unusually reliable. We found it when an internal code review read the shipped code, fixed the draw
to an independent stream, and re-ran every affected cell: the entire CIFAR grid, the eight Qwen
installations, which had to be rebuilt from scratch because their weights had been deleted, the Llama
supporting row, the ranking grid and the Fine-mixing decomposition. Three results moved materially.
Success rates fell, from $47/48$ to $37/48$ on the CIFAR headline grid and from $18/32$ to $0/32$ for
last-convolution fine-pruning, which under a realistic defender never reaches the $5\%$ residual-attack
threshold at any sparsity we tried. The CIFAR removal estimate fell from $+1.05$ $[+0.78,+1.33]$ to
$+0.02$ $[-0.31,+0.34]$. The Qwen estimate rose and changed sign, from $-0.24$ $[-1.45,+0.98]$ to $+1.37$
$[+0.52,+2.23]$. Everything reported here is post-fix; the pre-fix numbers appear only in this paragraph.
BackdoorBench and TOFU were unaffected, since they use their own data pipelines.

\paragraph{Three defects in our own fine-pruning, found by re-reading the reference method.}
Fine-pruning ranks channels by activation and then recovers by fine-tuning. Our first implementation
zeroed the selected channels once and then ran the recovery loop, so gradient descent refilled them: by
the end of recovery the ``pruned'' network was a fine-tuned dense one. We now re-apply the channel mask
after every optimizer step. Second, the last-convolution scope selected \texttt{layer4.1.conv2}, whose
output the residual identity path routes around, so masking it changed almost nothing; the scope now
ranks and masks the post-residual \texttt{layer4} output the classifier actually sees, and at $95\%$
sparsity $26$ of $512$ output channels remain non-zero. Third, zeroing a convolution's weights does not
silence its channel, because the following BatchNorm adds $\beta$: on the masked channels the post-norm
output had mean $0.331$, not zero. We now zero the matching BatchNorm $\gamma$ and $\beta$ as well, and
all $153$ masked channels measure silent. The three fixes together move the CIFAR headline from $+2.81$ to
$+2.51$ points reported and leave the removal term at $+0.02$ $[-0.31,+0.34]$ against $+0.04$ before
(Table~\ref{tab:abl}); they do not move the paper's conclusion, and we report them because the first
version would have let a reader reproduce a number our own description did not describe.

\paragraph{A bookkeeping error the rebuild caught.}
The framework does not run the defended clean arm in any of its shipped attack--defense pipelines, so every
difference-in-differences here is assembled from its per-cell summaries by our own scripts. An earlier
internal version of this paper read one such assembled quantity as a paired post-defense contrast and
subtracted the pre-defense gap a second time, reporting a removal component of
$-2.23$ and a clean-arm cost of $4.48$ points where the correct three-seed values were $-0.09$ and
$+2.34$. Every BackdoorBench quantity here is rebuilt from the framework's own per-cell summaries; the
eleven-seed values are $+0.36$ $[-0.26,+0.98]$ for removal and $+1.98$ $[+1.50,+2.46]$ for the clean arm
(Table~\ref{tab:bb}). The error was ours, not the framework's, and it inflated the correction in our
favour, which is why we state it here rather than in a changelog.

\paragraph{Clean-data budget is the controlling variable.}
Holding the model (ResNet-18), the poisoning run, and the defense (fine-pruning at $30\%$)
fixed, and varying only the fraction of clean data available to the defender, the reported
cost moves monotonically to zero while the removal term stays within half a point of it; the full seven-point sweep is Table~\ref{tab:cr},
Appendix~\ref{app:cr}. The same parameter count spans four accuracy points as the budget
alone changes, which is why we treat the defender's clean-data budget as a protocol
coordinate and report all vision results at the field-standard $5\%$.

\paragraph{Clean--clean variability calibration.}
The placebo arm pairs two never-poisoned \emph{vision} models differing only in seed and
applies the identical defense to both. Over four such pairs the difference in defense cost is $+0.06$ pt with a
$95\%$ $t$-half-width of $0.74$ pt (an earlier version quoted $0.39$, which was $1.96\times$ the
standard error rather than the $t$ interval this paper otherwise uses). The band bounds the
\emph{vision} claim only: the DiD estimate, $+0.02$ $[-0.31,+0.34]$, sits inside it, and so does
the paired contrast, $+0.43$ $[+0.30,+0.55]$, which resolves by its own interval.
It does \emph{not} transfer to the language model, which is a different instrument; an
earlier version of this paper applied it there, and Appendix~\ref{sec:selfcorrect} records
the correction. The language-model claim is stated at the installation level ($n{=}8$, $\mathrm{df}=7$): a
$+1.37$ pt removal cost with $95\%$ CI $[+0.52,+2.23]$ (Table~\ref{tab:indep}). A direct
language-model placebo---five independently trained clean models of the same recipe, each put
through the identical defense---pays $5.9$--$9.4$ pt, and the ten pairwise differences among
those five costs have one standard deviation of $\pm2.1$ pt. Repeating the construction on the
eight headline installations' own twin arms gives $\pm3.2$ pt; we report the smaller $\pm2.1$
descriptive scale, whose upper edge the removal interval just crosses ($+2.23$ against $+2.10$). We draw no equivalence
conclusion from the band: the language-model claim in Table~\ref{tab:rerankfull} is the interval itself, whose
upper end $+2.23$ is $19\%$ of the reported $11.46$.

\begin{figure}[!ht]
\centering
\includegraphics[width=.9\textwidth]{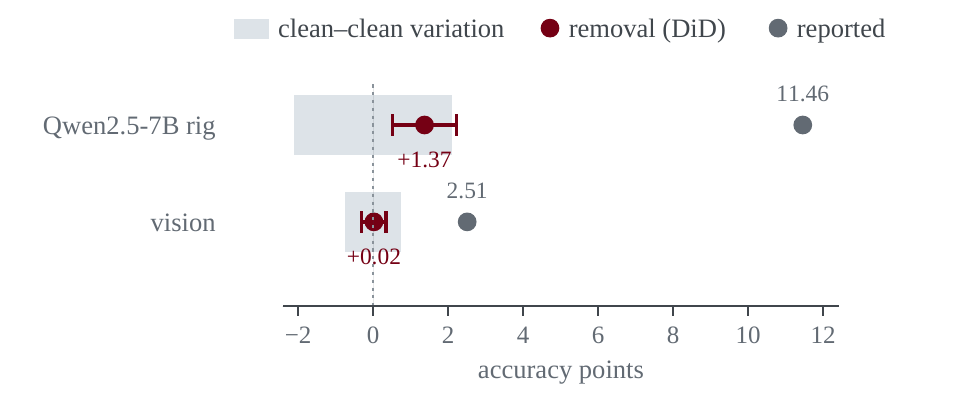}
\caption{\textbf{Both reported costs sit far outside the clean-versus-clean variation of their own
rig; both interaction estimates sit near it.} DiD estimates in accuracy points with $95\%$
intervals over installations ($n{=}8$ in both settings). The shaded band is clean-versus-clean
variation, shown only as a descriptive, rig-specific reference: $\pm0.74$ pt on vision (the $95\%$
$t$-half-width of four clean--clean pair differences) and $\pm2.1$ pt on the language model (one
standard deviation of five clean models' pairwise cost differences). These are different
statistics on different instruments and are not compared across rigs. The statistical conclusions
rest on the intervals themselves: the Qwen interval excludes zero and its upper endpoint, $+2.23$ points,
is $19\%$ of the reported-cost point estimate, while the vision interval includes zero.}
\label{fig:floor}
\end{figure}

\paragraph{The threshold metric hides no margin.}
\texttt{asr50} is thresholded, so it could saturate while an unthresholded margin kept
growing. It does not: over a $20\times$ dose sweep the unbounded trigger-vs-clean KL
divergence slightly \emph{falls} rather than rises, and thresholds $50/90/99$ agree throughout
(Appendix~\ref{app:dose}).

\paragraph{A conditional defense, first run unconditionally, then given a trigger it could reach.}
Neural Cleanse repairs only when its anomaly index exceeds $2$; our first implementation
repaired regardless. We caught that only because a reader asked why a defense would damage a model it had
declared clean. Gating it exposed a second defect. The trigger search composed its candidate pattern on
the loader's already-normalised tensors and squashed it through $\tanh$, so the pattern could not exceed
$1.0$ in those units while the BadNets white patch sits at $(1-0.4914)/0.2470=2.06$; the optimiser
compensated with a larger mask, which inflated every $\ell_1$ norm and flattened the outlier score the
detector reads. The weight on that norm was also a fixed constant rather than the published dynamic
schedule. Reverse-engineering in $[0,1]$ pixel space with that schedule, the detector fires on
$14/14$ poisoned models against $9/14$ before, names the true target class every time, and clears the
backdoor everywhere (mean residual attack success $0.008$ against $0.368$) at $+0.78$ points against
$+4.24$ over the cells that cleared before. It becomes a ranking candidate, bringing the candidate set to
\chNcandWord, and it is from a family that is not a strength variant of any other member. The scope condition the gate exposes---the confound is a property of unconditional
defenses and of false-positive repairs---is stated in Section~\ref{sec:admiss} and now priced in
Appendix~\ref{app:cond}.

\paragraph{An objective we had transcribed backwards, and a threshold rule we had not implemented.}
Adversarial Neuron Pruning trades a natural loss against a loss under adversarial neuron noise. We had put
the mixing weight $\alpha$ on the perturbed term rather than the natural one, drew the neuron noise once
instead of redrawing it each round, multiplied the whole batch-norm output by the mask where the reference
multiplies only the scale---so our ``pruned'' neuron could not even emit its bias, the mirror image of the
BatchNorm defect above---and pruned a fixed global fraction where the method selects its threshold under a
clean-accuracy budget. All four are fixed. The correction runs against us in one direction and for us in
another: ANP no longer destroys anything, so the twin arm is admissible in all fourteen installations
rather than nine, but under its own two-point accuracy budget it prunes too little to clear this attack
(\chANPAthree, mean residual attack success \chANPResTwo) and is excluded on removal success instead. It also removes
the one outlier from the surrogate table: the pooled TARE-Z error over all eight defenses is now
\tzCifarPooledPaired, against $-0.52$ $[-5.86,+4.82]$ with the earlier implementation.

\paragraph{An error the twin caught.}
Our first interpolation-vs-fine-tuning comparison reported a fivefold, sign-reversed gap---an
unfair baseline that charged fine-tuning's own cost to removal on one side only ($\cDef$ for
training-side, $\cRem$ for interpolation). The twin makes it immediate: the same fine-tuning
costs \emph{more} on the twin. This is precisely the error the paper attributes to the field,
committed once by us and caught by the control we advocate.

\paragraph{Defense grids.}
The complete grids for both modalities are in Appendix~\ref{app:hyper}. The language model is run at a
single fine-pruning operating point ($20\%$ of MLP channels); an earlier version reported a
$\{10,20,30,40\}\%$ sweep from a single installation, which the defender-draw correction invalidated and
which we did not rebuild.

\input{appendix_sched.tex}

\end{document}

%% file: iclr2027/math_commands.tex
\usepackage{amsmath,amsfonts,bm}

\def\eqref#1{equation~\ref{#1}}

\def\1{\bm{1}}

\DeclareMathAlphabet{\mathsfit}{\encodingdefault}{\sfdefault}{m}{sl}
\SetMathAlphabet{\mathsfit}{bold}{\encodingdefault}{\sfdefault}{bx}{n}



%% file: supplement/build/chain_numbers.tex
\newcommand{\chANPAthree}{$3/14$}

\newcommand{\chANPResTwo}{$0.51$}

\newcommand{\chAllFourteenMoves}{fine-pruning at $40\%$ 4 to 2; fine-tuning at lr $0.05$ 5 to 4; NAD 2 to 5}

\newcommand{\chClipCellsPrim}{$65$}

\newcommand{\chClipListPrim}{NAD in $10$ of $13$, fine-tuning at lr $0.05$ in $9$ of $13$, fine-pruning at $40\%$ in $7$ of $13$, fine-pruning at $30\%$ in $7$ of $13$, Neural Cleanse in $4$ of $13$}

\newcommand{\chClipTotPrim}{$37$}
\newcommand{\chDropNCbasic}{$[-0.333,+0.333]$}
\newcommand{\chDropNCgap}{$+0.333$}
\newcommand{\chDropNCmedian}{$+0.667$}

\newcommand{\chDropNCnWord}{four}
\newcommand{\chDropNCpct}{$[+0.333,+1.000]$}

\newcommand{\chDropNCple}{$0.014$}
\newcommand{\chDropNCsep}{$6/6$}

\newcommand{\chDropNCsuppWord}{thirteen}
\newcommand{\chDropNCtauRem}{$+0.667$}
\newcommand{\chDropNCtauTax}{$+1.000$}
\newcommand{\chFigGapDer}{$+0.200$}
\newcommand{\chFigGapRaw}{$+0.800$}

\newcommand{\chFigNmovedWordRaw}{three}
\newcommand{\chFigTauRemDer}{$+0.800$}
\newcommand{\chFigTauRemRaw}{$+0.200$}

\newcommand{\chGapAFive}{$+0.200$}
\newcommand{\chGapATen}{$+0.667$}

\newcommand{\chGapATwo}{$+0.333$}
\newcommand{\chGapBasicAFive}{$[-0.200,+0.200]$}

\newcommand{\chGapBasicPrim}{$[-0.200,+0.200]$}
\newcommand{\chGapBiasPrim}{$+0.181$}
\newcommand{\chGapClip}{$+0.200$}
\newcommand{\chGapLoo}{$[+0.200,+0.400]$}
\newcommand{\chGapMeanPrim}{$+0.381$}

\newcommand{\chGapMedianPrim}{$+0.400$}
\newcommand{\chGapNboot}{$5{,}000$}
\newcommand{\chGapNoClip}{$+0.400$}
\newcommand{\chGapPctAFive}{$[+0.200,+0.600]$}
\newcommand{\chGapPctATen}{$[+0.400,+1.200]$}

\newcommand{\chGapPctATwo}{$[+0.000,+0.667]$}
\newcommand{\chGapPctPrim}{$[+0.200,+0.600]$}
\newcommand{\chGapPctilePrim}{$0.9$}

\newcommand{\chGapPlePrim}{$0.0086$}
\newcommand{\chGapPlePrimPct}{$0.9\%$}

\newcommand{\chGapStep}{$0.200$}

\newcommand{\chMissRes}{$0.138$}

\newcommand{\chNADAthree}{$13/14$}

\newcommand{\chNCfireHi}{$3.67$}
\newcommand{\chNCfireLo}{$2.75$}

\newcommand{\chNCzeroAllWord}{ten}

\newcommand{\chNbottomWord}{four}

\newcommand{\chNcandAFiveWord}{five}

\newcommand{\chNcandATwoWord}{four}
\newcommand{\chNcandWord}{five}

\newcommand{\chNexclBasicWord}{two}

\newcommand{\chNexclPctWord}{three}
\newcommand{\chNineMoves}{NAD 2 to 5; fine-pruning at $40\%$ 4 to 2; fine-tuning at lr $0.05$ 5 to 4}

\newcommand{\chNineNmovedWord}{three}
\newcommand{\chNinst}{$14$}
\newcommand{\chNinstWord}{fourteen}
\newcommand{\chNoClipMoves}{fp 40 4$\to$3, fp 30 3$\to$5, ft .05 5$\to$4}

\newcommand{\chNsuppAFiveWord}{thirteen}

\newcommand{\chNsuppATenWord}{nine}

\newcommand{\chNsuppATwoWord}{fourteen}
\newcommand{\chNsuppWord}{thirteen}

\newcommand{\chNthrRowsWord}{four}

\newcommand{\chSDcleanPrim}{$1.38$}

\newcommand{\chSDremPrim}{$0.11$}
\newcommand{\chSepDid}{$3/10$}
\newcommand{\chSepPub}{$10/10$}
\newcommand{\chSignAssign}{$8{,}192$}

\newcommand{\chSpanPrimRound}{$0.003$}

\newcommand{\chTauNine}{$+0.778$}
\newcommand{\chTauNineTax}{$+0.944$}

\newcommand{\chTauRemAllFourteen}{$+0.200$}
\newcommand{\chTauRemClip}{$+0.800$}
\newcommand{\chTauRemNoClip}{$+0.600$}
\newcommand{\chTauRemPrim}{$+0.800$}

\newcommand{\chTauTaxAllFourteen}{$+0.800$}
\newcommand{\chTauTaxPrim}{$+1.000$}
\newcommand{\chTieA}{fine-pruning at $30\%$}
\newcommand{\chTieB}{NAD}
\newcommand{\chTieMargin}{$3.6\times 10^{-6}$}
\newcommand{\chTieScoreA}{$0.9633964$}
\newcommand{\chTieScoreB}{$0.9634000$}

%% file: supplement/build/attack_chain.tex
\newcommand{\chBlBorrowAll}{$+1.50\ \scriptstyle[-2.81,+5.80]$}
\newcommand{\chBlBorrowAllMae}{$6.57$}
\newcommand{\chBlBorrowAllMax}{$30.23$}
\newcommand{\chBlBorrowAllN}{$26$}

\newcommand{\chBlBorrowNoAblMae}{$2.65$}
\newcommand{\chBlBorrowNoAblMax}{$8.84$}
\newcommand{\chBlBorrowNoAblN}{$18$}

\newcommand{\chRevblNWord}{three}
\newcommand{\chRevblRev}{$1$}

\newcommand{\chRevblSurWord}{zero}
\newcommand{\chRevbnK}{$20$}
\newcommand{\chRevbnN}{$28$}

\newcommand{\chRevbnRev}{$7$}

\newcommand{\chRevbnSurWord}{four}

\newcommand{\chRevwnK}{$13$}
\newcommand{\chRevwnN}{$20$}

\newcommand{\chRevwnRev}{$2$}
\newcommand{\chRevwnSur}{$0$}
\newcommand{\chRevwnSurWord}{zero}
\newcommand{\chbAnpBudgetPct}{$10\%$}

\newcommand{\chbAnpBudgetPtsRound}{$9$}

\newcommand{\chbAnpDefAthreePt}{$+6.50$}

\newcommand{\chbAnpTwinMax}{$8.84$}
\newcommand{\chbAnpTwinMaxRel}{$9.4\%$}
\newcommand{\chbAnpTwinMaxSeed}{$0$}
\newcommand{\chbAnpTwinMin}{$0.00$}
\newcommand{\chbAnpTwinMinSeed}{$10$}

\newcommand{\chbShareHi}{$354\%$}

\newcommand{\chbShareHiNoAbl}{$194\%$}
\newcommand{\chbShareHiNoAblClean}{$+2.70$}
\newcommand{\chbShareHiNoAblName}{SAU}
\newcommand{\chbShareHiNoAblPub}{$+1.39$}
\newcommand{\chbShareLo}{$13\%$}
\newcommand{\chbShareLoName}{fine-tuning}
\newcommand{\chbShareLoPub}{$+1.01$}

\newcommand{\chbShareNegClean}{$+1.05$}
\newcommand{\chbShareNegDef}{$-0.22$}
\newcommand{\chbShareNegName}{fine-pruning}

\newcommand{\chbSharePosNWord}{seven}
\newcommand{\chbTwinAsrHi}{$0.90\%$}
\newcommand{\chbTwinAsrLo}{$0.66\%$}

\newcommand{\chbbBNAblDidPt}{$-23.82$}

\newcommand{\chbbBNClean}{$+1.98\ \scriptstyle[+1.50,+2.46]$}

\newcommand{\chbbBNDefPt}{$+2.34$}

\newcommand{\chbbBNFpDidPt}{$-1.27$}

\newcommand{\chbbBNFtPaired}{$+2.94\ \scriptstyle[+2.75,+3.14]$}

\newcommand{\chbbBNSauDidPt}{$-1.31$}
\newcommand{\chbbWNClean}{$-0.55\ \scriptstyle[-0.90,-0.20]$}

\newcommand{\chbbWNDef}{$-1.00\ \scriptstyle[-1.53,-0.47]$}

\newcommand{\chbbWNDid}{$-0.45\ \scriptstyle[-1.03,+0.13]$}

\newcommand{\chbbWNFtPaired}{$+0.34\ \scriptstyle[+0.18,+0.50]$}

\newcommand{\chbbWNPaired}{$+1.00\ \scriptstyle[+0.55,+1.45]$}

\newcommand{\chownWNClean}{$-0.69\ \scriptstyle[-1.46,+0.09]$}

\newcommand{\chownWNDef}{$-1.32\ \scriptstyle[-2.18,-0.46]$}

\newcommand{\chownWNDid}{$-0.63\ \scriptstyle[-1.94,+0.68]$}

\newcommand{\chsetBNDef}{$+2.44\ \scriptstyle[+2.18,+2.70]$}

\newcommand{\chuncWNClean}{$-0.61\ \scriptstyle[-0.93,-0.29]$}

\newcommand{\chwAfour}{$11/11$}
\newcommand{\chwAoneFail}{$3, 8, 10$}
\newcommand{\chwAoneFailN}{$3/11$}

\newcommand{\chwAoneRange}{$0.37$--$0.75$}
\newcommand{\chwCandNeg}{$5/6$}
\newcommand{\chwCandShareHi}{$67\%$}
\newcommand{\chwCandShareLo}{$-20\%$}

\newcommand{\chwConfSeedsWord}{four}
\newcommand{\chwGapAll}{$+0.000$}
\newcommand{\chwGapCIAll}{$[-0.667,+0.400]$}
\newcommand{\chwGapCIConf}{$[-0.667,+0.133]$}
\newcommand{\chwGapConf}{$-0.400$}
\newcommand{\chwGapPleAll}{$0.7842$}
\newcommand{\chwGapPleConf}{$0.9382$}

\newcommand{\chwNcandWord}{six}

\newcommand{\chwNseedsWord}{eleven}

\newcommand{\chwNsuppWord}{five}
\newcommand{\chwSeedZeroGap}{$-0.05$}

\newcommand{\chwSepPubAll}{$0/15$}

\newcommand{\chwTauRemConf}{$+1.000$}

\newcommand{\chwTauTaxConf}{$+0.600$}
\newcommand{\chwTwinMean}{$0.0135$}
\newcommand{\chwTwinRange}{$0.0073$--$0.0178$}

%% file: supplement/build/sched_numbers.tex
\newcommand{\schAgreeFive}{$95/96$}
\newcommand{\schAgreeSeed}{$44/44$}
\newcommand{\schAgree}{$131/132$}

\newcommand{\schBcosDidPool}{$+0.46\ \scriptstyle[+0.27,+0.65]$}
\newcommand{\schBcosFpPt}{$+1.05$}

\newcommand{\schBcosFtPt}{$+0.13$}
\newcommand{\schBcosFt}{$+0.13\ \scriptstyle[+0.06,+0.21]$}

\newcommand{\schBcosNadPt}{$+0.33$}

\newcommand{\schBcosPool}{$+0.50\ \scriptstyle[+0.42,+0.59]$}

\newcommand{\schBcosTwin}{$0.9371$}

\newcommand{\schBmsAthreeFrac}{$15/33$}
\newcommand{\schBmsAthree}{$15/33$}
\newcommand{\schBmsDidPool}{$-0.74\ \scriptstyle[-1.62,+0.13]$}

\newcommand{\schBmsFp}{$-1.95\ \scriptstyle[-2.36,-1.53]$}

\newcommand{\schBmsFt}{$-2.29\ \scriptstyle[-2.61,-1.96]$}
\newcommand{\schBmsHalfChange}{$22\%$ wider}
\newcommand{\schBmsHalfEleven}{$0.32$}
\newcommand{\schBmsHalfFive}{$0.26$}

\newcommand{\schBmsMaxAsr}{$0.655$}

\newcommand{\schBmsNad}{$-2.00\ \scriptstyle[-2.30,-1.71]$}

\newcommand{\schBmsPoolFive}{$-2.03\ \scriptstyle[-2.28,-1.77]$}

\newcommand{\schBmsPool}{$-2.08\ \scriptstyle[-2.40,-1.76]$}

\newcommand{\schBmsTwin}{$0.9109$}

\newcommand{\schDidWcosFt}{$-0.94\ \scriptstyle[-1.38,-0.49]$}
\newcommand{\schDidWcosNad}{$-1.00\ \scriptstyle[-1.46,-0.55]$}

\newcommand{\schDidWmsFt}{$-1.11\ \scriptstyle[-1.64,-0.58]$}
\newcommand{\schDidWmsNad}{$-1.09\ \scriptstyle[-1.57,-0.62]$}
\newcommand{\schDissentDef}{fine-tuning}
\newcommand{\schDissentGrid}{BadNets+Cosine}
\newcommand{\schDissentVal}{$-0.03$}

\newcommand{\schFixFpRange}{$0.66$}

\newcommand{\schFixFtRange}{$0.30$}
\newcommand{\schFixMaxRange}{$0.66$}

\newcommand{\schFixNadRange}{$0.29$}
\newcommand{\schNint}{$12$}
\newcommand{\schPfortyeight}{\checkmark}
\newcommand{\schPfortynine}{$\times$}

\newcommand{\schUndefRange}{$3.10$}

\newcommand{\schWcosAthree}{$33/33$}
\newcommand{\schWcosDidPool}{$-1.03\ \scriptstyle[-1.49,-0.58]$}

\newcommand{\schWcosFp}{$+0.86\ \scriptstyle[+0.74,+0.97]$}

\newcommand{\schWcosFt}{$+0.51\ \scriptstyle[+0.43,+0.59]$}
\newcommand{\schWcosHalfChange}{$15\%$ narrower}
\newcommand{\schWcosHalfEleven}{$0.09$}
\newcommand{\schWcosHalfFive}{$0.10$}
\newcommand{\schWcosInstHi}{$0.995$}
\newcommand{\schWcosInstLo}{$0.979$}
\newcommand{\schWcosInstN}{$11/11$}

\newcommand{\schWcosNad}{$+0.99\ \scriptstyle[+0.86,+1.12]$}

\newcommand{\schWcosPoolFive}{$+0.82\ \scriptstyle[+0.72,+0.92]$}

\newcommand{\schWcosPool}{$+0.78\ \scriptstyle[+0.70,+0.87]$}
\newcommand{\schWcosPubPool}{$-0.25\ \scriptstyle[-0.66,+0.16]$}
\newcommand{\schWcosPub}{$-0.25\ \scriptstyle[-0.66,+0.16]$}
\newcommand{\schWcosTwin}{$0.9418$}
\newcommand{\schWcosVictim}{$92.69$}

\newcommand{\schWmsDidPool}{$-1.17\ \scriptstyle[-1.70,-0.64]$}
\newcommand{\schWmsFpPt}{$-1.68$}
\newcommand{\schWmsFp}{$-1.68\ \scriptstyle[-2.12,-1.24]$}
\newcommand{\schWmsFtPt}{$-1.99$}
\newcommand{\schWmsFt}{$-1.99\ \scriptstyle[-2.30,-1.68]$}
\newcommand{\schWmsInstN}{$8/11$}
\newcommand{\schWmsNadPt}{$-1.85$}
\newcommand{\schWmsNad}{$-1.85\ \scriptstyle[-2.14,-1.56]$}

\newcommand{\schWmsPool}{$-1.84\ \scriptstyle[-2.18,-1.50]$}
\newcommand{\schWmsPubPool}{$-3.05\ \scriptstyle[-3.37,-2.73]$}
\newcommand{\schWmsTwin}{$0.9150$}
\newcommand{\schWmsVictim}{$90.05$}

%% file: supplement/build/gdep.tex
\newcommand{\gdDidSlopeBoot}{$-1.16\ \scriptstyle[-1.46,-0.90]$}

\newcommand{\gdGHi}{$+0.83$}
\newcommand{\gdGLo}{$-0.06$}

\newcommand{\gdNinst}{$8$}

\newcommand{\gdPairedPermP}{$0.3950$}

\newcommand{\gdPairedSlopeBoot}{$-0.16\ \scriptstyle[-0.46,+0.10]$}

\newcommand{\gdPermN}{$40320$}

%% file: supplement/build/table1.tex
\newcommand{\taboneBbAthreeAll}{$85/88$}

\newcommand{\taboneBbConfDid}{$+0.52\ \scriptstyle[+0.06,+0.99]$}

\newcommand{\taboneBbDefPt}{$+2.34$}
\newcommand{\taboneBbDid}{$+0.36\ \scriptstyle[-0.26,+0.98]$}

\newcommand{\taboneBbGap}{$+2.06\ \scriptstyle[+1.90,+2.23]$}

\newcommand{\taboneBbGapPt}{$+2.06$}

\newcommand{\taboneBbPaired}{$+2.42\ \scriptstyle[+1.81,+3.04]$}

\newcommand{\taboneQwenBothN}{$6$}

\newcommand{\taboneQwenDid}{$+1.37\ \scriptstyle[+0.52,+2.23]$}

\newcommand{\taboneQwenFpDid}{$+0.17\ \scriptstyle[-2.11,+2.46]$}

\newcommand{\taboneQwenFtDid}{$+2.77\ \scriptstyle[-0.17,+5.71]$}

\newcommand{\taboneQwenIndepCI}{$[-0.31,+3.05]$}

\newcommand{\taboneQwenR}{$-0.90$}

%% file: supplement/build/tarez_bb.tex
\newcommand{\tzBbLargeMaePHi}{$6.79$}
\newcommand{\tzBbLargeMaePLo}{$1.58$}

\newcommand{\tzBbLargeSet}{Neural Cleanse, I-BAU, ANP, ABL}

\newcommand{\tzBbMaePAnp}{$3.20$}

\newcommand{\tzBbMaxPAnp}{$8.97$}

\newcommand{\tzBbPooledPaired}{$+0.00\ \scriptstyle[-0.51,+0.51]$}

\newcommand{\tzBbSmallMaePHi}{$0.68$}
\newcommand{\tzBbSmallMaePLo}{$0.12$}

\newcommand{\tzBbSmallSet}{fine-tuning, NAD, fine-pruning, SAU}

\newcommand{\tzCifarNdef}{$8$}
\newcommand{\tzCifarNinst}{$4$}
\newcommand{\tzCifarPooledDid}{$-0.03\ \scriptstyle[-0.48,+0.42]$}
\newcommand{\tzCifarPooledDidNoAnp}{$-0.03\ \scriptstyle[-0.52,+0.46]$}
\newcommand{\tzCifarPooledDidPt}{$-0.03$}
\newcommand{\tzCifarPooledPaired}{$-0.02\ \scriptstyle[-0.17,+0.12]$}

\newcommand{\tzCifarPooledPairedNoAnp}{$-0.02\ \scriptstyle[-0.10,+0.05]$}
\newcommand{\tzCifarPooledPairedPt}{$-0.02$}

%% file: supplement/build/extras.tex
\newcommand{\exBbCleanExclBH}{$8/8$}
\newcommand{\exBbCleanExclBHNoAbl}{$7/7$}

\newcommand{\exBbCleanExclBonf}{$6/8$}
\newcommand{\exBbCleanExclBonfNoAbl}{$5/7$}

\newcommand{\exBbCleanMaxName}{I-BAU}
\newcommand{\exBbCleanMaxPt}{$+5.70$}

\newcommand{\exBbCleanMinName}{fine-tuning}
\newcommand{\exBbCleanMinPt}{$+0.13$}

\newcommand{\exBbExceedPosNoAblNames}{SAU, I-BAU}

\newcommand{\exBbPosNoAblWord}{six}

\newcommand{\exBetaNoise}{$0.45$}
\newcommand{\exBetaShort}{$1.2$}

\newcommand{\exBetaStableAb}{FT, NAD and ANP}

\newcommand{\exBetaUnstableAb}{NC, FP, SAU and I-BAU}
\newcommand{\exG}{$+2.06$}
\newcommand{\exNoiseAtG}{$-1.14$}
\newcommand{\exOpFtBb}{$0.015$}
\newcommand{\exOpFtOurs}{$0.369$}
\newcommand{\exOpFtOursAthree}{$1/8$}
\newcommand{\exOpFtOursTwo}{$0.37$}
\newcommand{\exOpNadBb}{$0.014$}
\newcommand{\exOpNadOurs}{$0.48$}
\newcommand{\exShareNoise}{$64\%$}
\newcommand{\exShareShort}{$-3\%$}
\newcommand{\exShareUncorr}{$15\%$}
\newcommand{\exShortAtG}{$+0.41$}
\newcommand{\exSlopeNoise}{$-0.55$}
\newcommand{\exSlopeNoiseTwoPt}{$-0.44$}

\newcommand{\exTareAsrBandHi}{$4.47\%$}
\newcommand{\exTareAsrBandLo}{$0.59\%$}
\newcommand{\exTareAsrN}{$350$}

\newcommand{\exTareAsrNCells}{$49$}
\newcommand{\exTareAsrNTwins}{$50$}

\newcommand{\exTareAsrSlope}{$-0.28$}
\newcommand{\exTareAsrSlopeHi}{$+0.03$}
\newcommand{\exTareAsrSlopeHiWithAbl}{$+1.01$}
\newcommand{\exTareAsrSlopeLo}{$-0.58$}
\newcommand{\exTareAsrSlopeLoWithAbl}{$-1.46$}
\newcommand{\exTareAsrSlopeWithAbl}{$-0.36$}
\newcommand{\exTareAsrSpan}{$-1.08$}
\newcommand{\exWanetAblAthree}{$3/11$}
\newcommand{\exWanetNegExclWord}{three}
\newcommand{\exWanetNegMeanWord}{five}
\newcommand{\ncForcedOurs}{$+3.40$}
\newcommand{\ncForcedOursCi}{$+3.40$ $[+3.21,+3.59]$}

\newcommand{\ncForcedPubCi}{$+3.51$ $[+2.79,+4.23]$}
\newcommand{\pairRbb}{$-0.27$}
\newcommand{\pairRbbN}{$11$}
\newcommand{\pairRcifar}{$+0.19$}
\newcommand{\pairRcifarN}{$8$}
\newcommand{\pairRpostHi}{$+0.43$}
\newcommand{\pairRpostLo}{$-0.74$}
\newcommand{\pairRwanet}{$-0.38$}
\newcommand{\pairRwanetN}{$11$}

%% file: supplement/build/reversal_numbers.tex
\newcommand{\revPstarLo}{$0.24$}
\newcommand{\revPstarLoCi}{$0.24$ $[0.20,0.28]$}

\newcommand{\revPstarHiCi}{$0.77$ $[0.72,0.82]$}

%% file: supplement/build/budget_numbers.tex
\newcommand{\crFpOne}{$3.79$ $[3.25,4.33]$}
\newcommand{\crFpHundred}{$0.28$ $[0.09,0.47]$}

%% file: supplement/build/llm_ops.tex
\newcommand{\qwOpN}{$17$}
\newcommand{\qwOpCheapest}{$+5.43$}
\newcommand{\qwOpBelowN}{$6$}

\newcommand{\qwOpBelowCostLo}{$+2.88$}
\newcommand{\qwOpBelowCostHi}{$+4.38$}
\newcommand{\qwOpBelowAsrLo}{$0.44$}
\newcommand{\qwOpBelowAsrHi}{$0.96$}
\newcommand{\qwOpClearN}{$9$}

%% file: supplement/sched/build/csched_numbers.tex
\newcommand{\cschBcosAthreeFrac}{$33/33$}

\newcommand{\cschBcosProtoRef}{$+3.03\ \scriptstyle[+2.89,+3.16]$}

\newcommand{\cschBmsAthreeFrac}{$15/33$}

\newcommand{\cschBmsProtoRef}{$+1.28\ \scriptstyle[+1.13,+1.43]$}

\newcommand{\cschDissentDef}{fine-tuning}

\newcommand{\cschDissentVal}{$-0.03$}

\newcommand{\cschWcosAthreeFrac}{$33/33$}

\newcommand{\cschWcosCleanDerHalf}{$-0.39$}

\newcommand{\cschWcosDidPool}{$-1.03\ \scriptstyle[-1.49,-0.58]$}

\newcommand{\cschWcosMaxAsr}{$0.038$}

\newcommand{\cschWcosPool}{$+0.78\ \scriptstyle[+0.70,+0.87]$}

\newcommand{\cschWcosProtoRef}{$+0.77\ \scriptstyle[+0.66,+0.88]$}
\newcommand{\cschWcosPubPool}{$-0.25\ \scriptstyle[-0.66,+0.16]$}

\newcommand{\cschWmsAthreeFrac}{$21/24$}

\newcommand{\cschWmsDidPool}{$-1.30\ \scriptstyle[-2.00,-0.61]$}

\newcommand{\cschWmsProtoRef}{$+0.59\ \scriptstyle[+0.44,+0.74]$}

\newcommand{\cschWmsPubPool}{$-3.07\ \scriptstyle[-3.42,-2.73]$}

\newcommand{\cschWmsUnclipDer}{$+1.54$}

%% file: supplement/sched/build/leaderboard_numbers.tex
\newcommand{\lbAblBadnetCost}{$+8.00$}
\newcommand{\lbCifarLow}{$30/31$}
\newcommand{\lbClipAllAnn}{$16/40$}

\newcommand{\lbDbrCaBadnet}{$29.41$}
\newcommand{\lbDbrCaBlended}{$14.77$}
\newcommand{\lbDbrCaBpp}{$83.11$}
\newcommand{\lbDbrCaLf}{$11.23$}
\newcommand{\lbDbrCaTrojan}{$12.03$}
\newcommand{\lbDbrCaWanet}{$61.70$}
\newcommand{\lbDerClipAnn}{$0/25$}
\newcommand{\lbDerClipDefs}{FT, FP, NAD, NC, I-BAU}

\newcommand{\lbDerClipNever}{$13/15$}

\newcommand{\lbGapChundPre}{$+6.3$}

\newcommand{\lbGapCtenPre}{$+2.1$}
\newcommand{\lbGapCtenVgg}{$+2.6$}

\newcommand{\lbGapTinyVit}{$+11.3$}

%% file: supplement/sched/build/fixedpoint_lb_numbers.tex
\newcommand{\fpLbBadnetTenUndef}{$91.32$}
\newcommand{\fpLbCellsFp}{$0/9$}
\newcommand{\fpLbCellsFt}{$18/41$}
\newcommand{\fpLbCellsNad}{$11/39$}

\newcommand{\fpLbCommonFtSlope}{$0.18$}

\newcommand{\fpLbCommonNadSlope}{$0.22$}

\newcommand{\fpLbFtTenDefHi}{$93.5$}
\newcommand{\fpLbFtTenDefLo}{$90.0$}
\newcommand{\fpLbFtTenDef}{$3.5$}

\newcommand{\fpLbFtTenSlope}{$0.01$}

\newcommand{\fpLbFtTenUndefLo}{$62.5$}
\newcommand{\fpLbFtTenUndef}{$31.0$}

\newcommand{\fpLbHealthyFtDef}{$2.82$}
\newcommand{\fpLbHealthyFtN}{$5$}
\newcommand{\fpLbHealthyFtSlopeNoBn}{$+0.86$}
\newcommand{\fpLbHealthyFtSlopeSe}{$0.10$}
\newcommand{\fpLbHealthyFtSlope}{$+1.26$}
\newcommand{\fpLbHealthyFtUndefNoBn}{$0.59$}
\newcommand{\fpLbHealthyFtUndefTwo}{$2.15$}
\newcommand{\fpLbHealthyFtUndef}{$2.1$}
\newcommand{\fpLbHealthyList}{BadNets, Blended, LF, SSBA, Trojan}

\newcommand{\fpLbHealthyNadSlopeNoBn}{$+0.14$}
\newcommand{\fpLbHealthyNadSlopeSe}{$0.20$}
\newcommand{\fpLbHealthyNadSlope}{$+1.08$}

\newcommand{\fpLbLooFtSlope}{$0.06$}

\newcommand{\fpLbLooNadSlope}{$0.07$}

\newcommand{\fpLbNadTenSlope}{$0.01$}

\newcommand{\lbHealthyCutHi}{$91.3$}

\newcommand{\lbHealthyCutLo}{$84.6$}
\newcommand{\lbSlopeFtAllAnnThree}{$-0.002$}
\newcommand{\lbSlopeFtAllAnn}{$+0.00$}
\newcommand{\lbSlopeNadAllAnnThree}{$-0.001$}

%% file: supplement/sched/build/paper_extras.tex
\newcommand{\exBcosTwinPct}{$93.71$}

\newcommand{\exBmsTwinPct}{$91.09$}

\newcommand{\exBppTwinGap}{$+2.79$}

\newcommand{\exFixFigN}{$162$}

\newcommand{\exGpuCorr}{118}
\newcommand{\exGpuDose}{32}

\newcommand{\exGpuFour}{14}

\newcommand{\exGpuTotal}{328}

\newcommand{\exWanetTwinGap}{$+2.68$}

\newcommand{\exWcosTwinPct}{$94.18$}

\newcommand{\exWmsTwinPct}{$91.50$}

%% file: supplement/sched/build/audit_numbers.tex
\newcommand{\auTsbdNd}{$91.34$}
\newcommand{\auTsbdAvg}{$91.70$}
\newcommand{\auTsbdMargin}{$+0.36$}
\newcommand{\auTsbdIa}{$+2.39$}
\newcommand{\auTsbdWanet}{$+2.01$}
\newcommand{\auTsbdSig}{$+5.93$}
\newcommand{\auTsbdNdSix}{$91.46$}
\newcommand{\auTsbdSix}{$91.21$}
\newcommand{\auTsbdAdjTwin}{$91.24$}
\newcommand{\auTsbdAdjDeficit}{$91.04$}
\newcommand{\auCleanGain}{$1.84$}
\newcommand{\auDeficit}{$2.64$}
\newcommand{\auUnobtainedJournal}{$11$}

\newcommand{\auCiting}{$252$}
\newcommand{\auFulltext}{$48$}
\newcommand{\auReuse}{$13$}
\newcommand{\auCopied}{$7$}
\newcommand{\auLoadBearing}{$2$}

%% file: supplement/sched/build/rerank_lb_numbers.tex
\newcommand{\rkLbHundQclean}{$70.75$}

\newcommand{\rkLbTenOrderProto}{ANP $>$ ABL $>$ i-bau $>$ FT $>$ NAD $>$ FP $>$ NC $>$ CLP $>$ spectral $>$ d-br $>$ AC $>$ DBD}
\newcommand{\rkLbTenOrder}{ANP $>$ ABL $>$ i-bau $>$ FT $>$ NC $>$ FP $>$ NAD $>$ CLP $>$ spectral $>$ d-br $>$ DBD $>$ AC}

%% file: supplement/sched/build/sched2_numbers.tex
\newcommand{\sxBcosLegacyAfour}{$1/5$}
\newcommand{\sxBcosLegacyFp}{$+0.94\ \scriptstyle[+0.75,+1.13]$}
\newcommand{\sxBcosLegacyFt}{$+0.58\ \scriptstyle[+0.24,+0.91]$}
\newcommand{\sxBcosLegacyNad}{$+0.96\ \scriptstyle[+0.66,+1.27]$}
\newcommand{\sxBcosOneAfour}{$5/5$}

\newcommand{\sxBcosOneFp}{$+0.97\ \scriptstyle[+0.76,+1.17]$}

\newcommand{\sxBcosOneFt}{$+0.55\ \scriptstyle[+0.30,+0.80]$}

\newcommand{\sxBcosOneNad}{$+1.11\ \scriptstyle[+0.92,+1.30]$}

\newcommand{\sxBcosOnePool}{$+0.87\ \scriptstyle[+0.69,+1.06]$}
\newcommand{\sxBcosOneProtoRef}{$+0.89\ \scriptstyle[+0.55,+1.23]$}

\newcommand{\sxBshipOneAfour}{$5/5$}

\newcommand{\sxBshipOnePool}{$-2.01\ \scriptstyle[-2.63,-1.40]$}
\newcommand{\sxBshipOneProtoRef}{$+0.57\ \scriptstyle[+0.29,+0.85]$}

\newcommand{\sxIshipAfour}{$0/5$}

\newcommand{\sxIshipVictim}{$91.21$}

\newcommand{\sxIstepVictim}{$93.09$}

%% file: supplement/sched/build/sched3_numbers.tex
\cifarhundredtrue
\newif\ifcbpp
\cbpptrue

\newcommand{\syBcosFp}{$+6.20\ \scriptstyle[+5.23,+7.16]$}

\newcommand{\syBcosFt}{$+0.54\ \scriptstyle[+0.12,+0.97]$}

\newcommand{\syBcosNad}{$+0.62\ \scriptstyle[+0.21,+1.02]$}

\newcommand{\syBlshipPool}{$+2.76\ \scriptstyle[+2.50,+3.03]$}

\newcommand{\syBlshipTwin}{$70.87$}

\newcommand{\syBlstepPool}{$-5.23\ \scriptstyle[-7.60,-2.86]$}

\newcommand{\syBlstepTwin}{$63.36$}

\newcommand{\syBppTwinGap}{$+6.75$}

\newcommand{\syBshipFp}{$-3.28\ \scriptstyle[-5.54,-1.01]$}

\newcommand{\syBshipFt}{$-4.89\ \scriptstyle[-6.65,-3.13]$}

\newcommand{\syBshipNad}{$-4.62\ \scriptstyle[-6.46,-2.79]$}

\newcommand{\syBshipTwin}{$65.02$}

\newcommand{\syFixSixFpRange}{$5.01$}
\newcommand{\syFixSixFtHi}{$71.23$}
\newcommand{\syFixSixFtLo}{$69.36$}
\newcommand{\syFixSixFtRange}{$1.86$}

\newcommand{\syFixSixNadRange}{$2.07$}
\newcommand{\syFixSixUndef}{$8.41$}

\newcommand{\syPeight}{\checkmark}
\newcommand{\syPeleven}{$\times$}
\newcommand{\syPfive}{\checkmark}
\newcommand{\syPfour}{\checkmark}
\newcommand{\syPnine}{\checkmark}
\newcommand{\syPone}{\checkmark}
\newcommand{\syPsevenAll}{failed}

\newcommand{\syPsixAll}{held}

\newcommand{\syPten}{\checkmark}
\newcommand{\syPthree}{\checkmark}
\newcommand{\syPtwo}{\checkmark}

\newcommand{\syPublicBship}{$63.96$}
\newcommand{\syPublicWship}{$64.05$}

\newcommand{\syWanetTwinGap}{$+6.01$}

\newcommand{\syWcosAzero}{$5/5$}

\newcommand{\syWcosFp}{$+4.41\ \scriptstyle[+3.68,+5.15]$}

\newcommand{\syWcosFt}{$+0.34\ \scriptstyle[-0.15,+0.83]$}

\newcommand{\syWcosNad}{$+0.45\ \scriptstyle[-0.10,+1.00]$}

\newcommand{\syWcosTwin}{$71.39$}

\newcommand{\syWshipAzero}{$1/5$}
\newcommand{\syWshipDidAll}{$-1.37\ \scriptstyle[-4.29,+1.56]$}

\newcommand{\syWshipFp}{$-2.95\ \scriptstyle[-4.41,-1.50]$}

\newcommand{\syWshipFt}{$-4.54\ \scriptstyle[-5.74,-3.34]$}

\newcommand{\syWshipNad}{$-4.11\ \scriptstyle[-5.17,-3.04]$}

\newcommand{\syWshipPoisAsrLo}{$0.767$}

\newcommand{\syWshipPubAll}{$-5.24\ \scriptstyle[-7.48,-2.99]$}

\newcommand{\syWshipTwin}{$65.38$}
\newcommand{\syWshipVictimAll}{$63.12$}

%% file: supplement/sched/build/sched4_numbers.tex
\newif\ifcbppfour
\cbppfourtrue

\newcommand{\sqBcosIbau}{$+4.09\ \scriptstyle[+2.70,+5.49]$}

\newcommand{\sqBcosSauAllN}{$5$}
\newcommand{\sqBcosSauAll}{$+10.67\ \scriptstyle[-12.25,+33.59]$}

\newcommand{\sqBcosSau}{$+2.41\ \scriptstyle[+1.50,+3.33]$}

\newcommand{\sqBnshipFtsam}{$+0.88\ \scriptstyle[+0.76,+1.00]$}

\newcommand{\sqBnshipIbauDefTwin}{$88.01$}

\newcommand{\sqBnshipIbau}{$+5.70\ \scriptstyle[+3.99,+7.42]$}

\newcommand{\sqBnshipSau}{$+2.70\ \scriptstyle[+2.31,+3.08]$}

\newcommand{\sqBnstepFtsam}{$-2.45\ \scriptstyle[-2.84,-2.06]$}

\newcommand{\sqBnstepIbauDefTwin}{$90.90$}

\newcommand{\sqBnstepIbau}{$+0.26\ \scriptstyle[-0.51,+1.03]$}

\newcommand{\sqBnstepSau}{$-0.58\ \scriptstyle[-0.91,-0.25]$}

\newcommand{\sqBshipIbau}{$+0.82\ \scriptstyle[-0.13,+1.78]$}

\newcommand{\sqCollapsedAcc}{$50.6$}

\newcommand{\sqFixFtRange}{$0.21$}

\newcommand{\sqFixFtsamRange}{$0.81$}

\newcommand{\sqFixIbauRange}{$3.26$}

\newcommand{\sqFixNadRange}{$0.23$}

\newcommand{\sqFixSauMean}{$91.4$}
\newcommand{\sqFixSauRange}{$0.73$}

\newcommand{\sqFixTrioHi}{$93.7$}
\newcommand{\sqFixTrioLo}{$92.8$}
\newcommand{\sqFixUndefRange}{$3.07$}
\newcommand{\sqFtsamVsFt}{$0.75$}
\newcommand{\sqIbauSwingBadnet}{$+5.44$}
\newcommand{\sqIbauSwingWanet}{$+4.77$}
\newcommand{\sqPfifteen}{\checkmark}
\newcommand{\sqPfourteen}{\checkmark}

\newcommand{\sqPthirteen}{\checkmark}
\newcommand{\sqPtwelve}{\checkmark}

\newcommand{\sqWcosFtsam}{$+1.09\ \scriptstyle[+0.92,+1.26]$}

\newcommand{\sqWcosIbauDefTwin}{$89.23$}

\newcommand{\sqWcosIbau}{$+5.00\ \scriptstyle[+3.68,+6.31]$}

\newcommand{\sqWcosSau}{$+2.88\ \scriptstyle[+1.93,+3.82]$}

\newcommand{\sqWshipFtsam}{$-2.14\ \scriptstyle[-2.46,-1.81]$}

\newcommand{\sqWshipIbauDefTwin}{$91.27$}

\newcommand{\sqWshipIbau}{$+0.23\ \scriptstyle[-0.30,+0.75]$}

\newcommand{\sqWshipSau}{$-0.15\ \scriptstyle[-0.77,+0.47]$}

%% file: supplement/sched/build/sched5_numbers.tex
\sfivetrue
\newcommand{\sqvAthreeShort}{$0/20$}

\newcommand{\sqvBstar}{$5$}

\newcommand{\sqvCleanEfiftyLrAWcos}{$+0.45\ \scriptstyle[+0.30,+0.60]$}
\newcommand{\sqvCleanEfiftyLrAWship}{$-0.95\ \scriptstyle[-1.47,-0.43]$}

\newcommand{\sqvCleanEfiftyLrCBnship}{$+0.13\ \scriptstyle[-0.04,+0.29]$}

\newcommand{\sqvCleanEfiveLrAWship}{$-0.26\ \scriptstyle[-0.81,+0.28]$}

\newcommand{\sqvLrAtFifty}{$0.00516$}
\newcommand{\sqvLrAtFive}{$0.00996$}
\newcommand{\sqvLrAtTwenty}{$0.00914$}
\newcommand{\sqvPeighteen}{$\times$}
\newcommand{\sqvPnineteen}{\checkmark}
\newcommand{\sqvPseventeenExist}{\checkmark}
\newcommand{\sqvPseventeen}{$\times$}
\newcommand{\sqvPsixteen}{$\times$}
\newcommand{\sqvRangeEfiftyLrA}{$1.39$}

\newcommand{\sqvRangeEfiveLrC}{$0.71$}

\newcommand{\sqvRangeEtwentyLrC}{$0.27$}

\newcommand{\sqvWcosDropFive}{$-1.04$}
\newcommand{\sqvWcosDropHundred}{$-0.56$}
\newcommand{\sqvWshipRiseFiveShare}{$76\%$}
\newcommand{\sqvWshipRiseFive}{$+1.52$}
\newcommand{\sqvWshipRiseHundred}{$+2.01$}
\newcommand{\sqvWshipRiseTail}{$+0.49$}

%% file: supplement/sched/build/sched6_numbers.tex
\ssixtrue
\newcommand{\svFixMax}{$1.72$}
\newcommand{\svPtwentyone}{\checkmark}
\newcommand{\svPtwentytwo}{$\times$}
\newcommand{\svPtwenty}{\checkmark}
\newcommand{\svTwinGap}{$+3.12$}
\newcommand{\svVcosAfour}{$5/5$}
\newcommand{\svVcosAzero}{$5/5$}

\newcommand{\svVcosFp}{$+1.86\ \scriptstyle[+1.69,+2.03]$}
\newcommand{\svVcosFtAthree}{$4/5$}
\newcommand{\svVcosFtDefTwin}{$90.44$}

\newcommand{\svVcosFtPub}{$-0.76\ \scriptstyle[-5.02,+3.49]$}
\newcommand{\svVcosFt}{$+2.09\ \scriptstyle[+1.59,+2.59]$}

\newcommand{\svVcosNad}{$+2.92\ \scriptstyle[+2.26,+3.58]$}
\newcommand{\svVcosTwin}{$92.53$}

\newcommand{\svVshipAfour}{$5/5$}
\newcommand{\svVshipAzero}{$4/5$}

\newcommand{\svVshipFp}{$-2.18\ \scriptstyle[-2.58,-1.78]$}
\newcommand{\svVshipFtAthree}{$2/4$}
\newcommand{\svVshipFtDefTwin}{$91.72$}

\newcommand{\svVshipFtPub}{$-8.44\ \scriptstyle[-20.58,+3.69]$}
\newcommand{\svVshipFt}{$-2.30\ \scriptstyle[-2.72,-1.89]$}

\newcommand{\svVshipNad}{$-1.92\ \scriptstyle[-2.40,-1.44]$}
\newcommand{\svVshipTwin}{$89.41$}
\newcommand{\svVshipVictimHi}{$87.9$}
\newcommand{\svVshipVictimLo}{$71.8$}

%% file: supplement/sched/build/sched7_numbers.tex
\sseventrue
\newif\ifstcgd
\stcgdtrue

\newcommand{\stCgdShare}{$66\%$}

\newcommand{\stCgdWcosClean}{$+0.41\ \scriptstyle[+0.26,+0.57]$}

\newcommand{\stCgdWcosPub}{$-1.00\ \scriptstyle[-2.10,+0.10]$}

\newcommand{\stCgdWshipClean}{$-2.43\ \scriptstyle[-3.00,-1.86]$}

\newcommand{\stCgdWshipPub}{$-3.69\ \scriptstyle[-4.22,-3.16]$}
\newcommand{\stPtwentyfive}{\checkmark}
\newcommand{\stPtwentyfour}{\checkmark}
\newcommand{\stPtwentyseven}{\checkmark}
\newcommand{\stPtwentysix}{\checkmark}
\newcommand{\stPtwentythree}{\checkmark}
\newcommand{\stTsbdFixRange}{$0.92$}
\newcommand{\stTsbdShare}{$38\%$}
\newcommand{\stTsbdStageOneWcosTwin}{$60.4$}
\newcommand{\stTsbdStageOneWcosVictim}{$79.3$}
\newcommand{\stTsbdStageOneWshipTwin}{$45.7$}
\newcommand{\stTsbdStageOneWshipVictim}{$78.0$}
\newcommand{\stTsbdSweepBigMinCost}{$+2.9$}

\newcommand{\stTsbdSweepNmax}{$0.9$}
\newcommand{\stTsbdSweepNmin}{$0.01$}
\newcommand{\stTsbdSweepN}{$11$}
\newcommand{\stTsbdSweepOrder}{$11$}
\newcommand{\stTsbdSweepSign}{$9$}
\newcommand{\stTsbdSweepTwinOrder}{$11$}
\newcommand{\stTsbdSwingThird}{$0.93$}
\newcommand{\stTsbdTwinSwing}{$+2.78$}

\newcommand{\stTsbdWcosClean}{$+2.59\ \scriptstyle[+1.37,+3.82]$}

\newcommand{\stTsbdWcosPub}{$+0.43\ \scriptstyle[-0.61,+1.47]$}

\newcommand{\stTsbdWshipClean}{$-1.11\ \scriptstyle[-1.86,-0.35]$}

\newcommand{\stTsbdWshipPub}{$-2.90\ \scriptstyle[-3.49,-2.32]$}

%% file: supplement/sched/build/sched9_numbers.tex
\sninetrue
\newcommand{\sgAfour}{$1/5$}
\newcommand{\sgAsrHi}{$0.989$}
\newcommand{\sgAsrLo}{$0.980$}
\newcommand{\sgAsrTwinHi}{$0.213$}

\newcommand{\sgAzero}{$5/5$}

\newcommand{\sgFpClean}{$+1.23\ \scriptstyle[+0.79,+1.66]$}

\newcommand{\sgFpPub}{$-5.89\ \scriptstyle[-6.25,-5.53]$}
\newcommand{\sgFtAthree}{$4/5$}
\newcommand{\sgFtClean}{$+0.07\ \scriptstyle[-0.12,+0.25]$}
\newcommand{\sgFtDefTwin}{$93.76$}
\newcommand{\sgFtDefVictim}{$90.74$}
\newcommand{\sgFtFixGap}{$3.03$}
\newcommand{\sgFtPub}{$-6.16\ \scriptstyle[-6.40,-5.91]$}
\newcommand{\sgGap}{$+9.25$}

\newcommand{\sgNadClean}{$+0.31\ \scriptstyle[+0.04,+0.59]$}

\newcommand{\sgNadFixGap}{$3.72$}
\newcommand{\sgNadPub}{$-5.22\ \scriptstyle[-5.38,-5.06]$}
\newcommand{\sgPthirtyone}{$\times$}
\newcommand{\sgPthirtytwo}{\checkmark}
\newcommand{\sgPthirty}{$\times$}
\newcommand{\sgPtwentyeight}{\checkmark}
\newcommand{\sgPtwentynine}{\checkmark}
\newcommand{\sgTwin}{$93.83$}
\newcommand{\sgVictimHi}{$84.7$}
\newcommand{\sgVictimLo}{$84.4$}
\newcommand{\sgVictim}{$84.58$}

%% file: supplement/sched/build/sched10_numbers.tex
\stentrue
\newcommand{\steAblFloorAdmitN}{$7$}
\newcommand{\steAblFloorN}{$14$}
\newcommand{\steAblTwinCollapseAdmit}{$41/50$}
\newcommand{\steAblTwinCollapse}{$50/60$}
\newcommand{\steAdmitList}{BadNets, Blended, LF, FTrojanNN, Refool, WaNet, BPP}
\newcommand{\steAdmitN}{$7$}

\newcommand{\steAnnAdmitN}{$5$}
\newcommand{\steAnnPos}{$35/35$}
\newcommand{\steAnpIbauGapAllReg}{$0.06$}

\newcommand{\steAttList}{BadNets, Blended, LF, FTrojanNN, Refool, Trojan, SIG, WaNet, BPP}

\newcommand{\steBadnetAblPub}{$+9.39\ \scriptstyle[+8.12,+10.65]$}

\newcommand{\steBadnetFpPaired}{$+0.80$}

\newcommand{\steBadnetIbauPaired}{$+0.21$}

\newcommand{\steBadnetSauPaired}{$+0.76$}

\newcommand{\steBppAfour}{$5/5$}

\newcommand{\steBppNcFire}{$0/5$}

\newcommand{\steBppNewShare}{$1/5$}

\newcommand{\steCellsBonfAdmit}{$15$}
\newcommand{\steCellsBonf}{$20$}
\newcommand{\steCellsKAdmit}{$49$}
\newcommand{\steCellsK}{$63$}

\newcommand{\steCellsNominalAdmit}{$44$}
\newcommand{\steCellsNominal}{$56$}
\newcommand{\steCleanOrderAll}{FT $<$ NAD $<$ FP $<$ NC $<$ SAU $<$ ANP $<$ I-BAU $<$ ABL}
\newcommand{\steCleanOrderReg}{FT $<$ NAD $<$ FP $<$ NC $<$ SAU $<$ ANP $<$ I-BAU (ABL off-scale, last)}
\newcommand{\steCleanOrder}{FT $<$ NAD $<$ FP $<$ NC $<$ SAU $<$ ANP $<$ I-BAU (ABL off-scale, last)}
\newcommand{\steClipDamaged}{$22/25$}
\newcommand{\steClipDefs}{FT, FP, NAD, NC, I-BAU}

\newcommand{\steClipHealthyFpList}{BadNets $11/11$, Refool $2/5$}
\newcommand{\steClipHealthy}{$13/195$}
\newcommand{\steClipNever}{$58/65$}
\newcommand{\steCollapsedTwinList}{SIG seed 4 SAU (defended twin 54.26\%)}

\newcommand{\steDerFlipsReg}{$0/28$}
\newcommand{\steDerFlips}{$0/28$}
\newcommand{\steDerOrderProto}{SAU $>$ ANP $>$ I-BAU $>$ NAD $>$ FP $>$ FT $>$ NC $>$ ABL}
\newcommand{\steDerOrderPub}{SAU $>$ ANP $>$ I-BAU $>$ NAD $>$ FP $>$ FT $>$ NC $>$ ABL}
\newcommand{\steDerShiftHi}{$1.1$}
\newcommand{\steDerShiftLo}{$0.5$}
\newcommand{\steDescList}{Trojan, SIG}

\newcommand{\steFloorExcludedList}{Trojan/SAU seed 3, SIG/SAU seed 2}
\newcommand{\steFloorExcludedN}{$2$}
\newcommand{\steFtfamClearOut}{$1$ of $69$}

\newcommand{\steLfFtClean}{$+0.11\ \scriptstyle[-0.03,+0.25]$}

\newcommand{\steNeverAnpPos}{$2/2$}

\newcommand{\steNeverBonf}{$3/8$}

\newcommand{\steNeverFtFamNeg}{$6/6$}

\newcommand{\steNeverSauIbauZero}{$4/4$}

\newcommand{\steNewAfour}{FTrojanNN $5/5$, Refool $5/5$}
\newcommand{\steNewFtFamExcl}{$6/6$}

\newcommand{\steNewGapRefool}{$1.67$}

\newcommand{\stePfiftyone}{\checkmark}
\newcommand{\stePfiftytwo}{\checkmark}
\newcommand{\stePfifty}{\checkmark}

\newcommand{\stePthirtyfive}{\checkmark}
\newcommand{\stePthirtyfourAdmit}{\checkmark}
\newcommand{\stePthirtyfour}{$\times$}
\newcommand{\stePthirtysix}{$\times$}
\newcommand{\stePthirtythree}{$\times$}

\newcommand{\steRefoolFpClean}{$+0.91\ \scriptstyle[+0.62,+1.21]$}

\newcommand{\steRefoolFpPub}{$+0.08\ \scriptstyle[-0.25,+0.40]$}

\newcommand{\steSigAfour}{$1/5$}

\newcommand{\steSigNewShare}{$5/5$}

\newcommand{\steTrojanAfour}{$0/5$}

\newcommand{\steTwinAsrHi}{$4.5$}
\newcommand{\steTwinAsrLo}{$0.6$}

\newcommand{\steWanetAblClean}{$+27.77\ \scriptstyle[+21.27,+34.26]$}

\newcommand{\steWanetAnpClean}{$+1.78\ \scriptstyle[+0.52,+3.04]$}

\newcommand{\steWanetNcCond}{$-1.30$}

\newcommand{\steWanetNcFire}{$3/11$}

%% file: supplement/sched/build/sched11_numbers.tex
\seleventrue

\newcommand{\selCleanOrderReg}{MCR $<$ FT $<$ NAD $<$ FP $<$ NC $<$ CLP $<$ SAU $<$ ANP $<$ I-BAU $<$ NPD $<$ RNP (ABL, D-BR off-scale, last)}

\newcommand{\selDbrTwinCollapseReg}{$35/50$}
\newcommand{\selDbrTwinCollapse}{$45/60$}

\newcommand{\selDerFlipList}{FT--NPD}
\newcommand{\selDerFlipMaxGap}{$0.03$}
\newcommand{\selDerFlipsReg}{$0/78$}
\newcommand{\selDerFlips}{$1/78$}

\newcommand{\selFloorDbrN}{$18$}

\newcommand{\selFloorRnpN}{$48$}

\newcommand{\selMcrClears}{BadNets, WaNet, BPP}
\newcommand{\selMcrLeaves}{Blended, LF, FTrojanNN, Refool}

\newcommand{\selNpdTrigList}{BadNets $0.99$, $0.71$; FTrojanNN $0.64$}

\newcommand{\selNpdTrigN}{$3$}

\newcommand{\selPfortyone}{$\times$}
\newcommand{\selPforty}{\checkmark}
\newcommand{\selPthirtyeight}{$\times$}
\newcommand{\selPthirtynine}{$\times$}
\newcommand{\selPthirtysevenExcl}{$3/4$}
\newcommand{\selPthirtyseven}{\checkmark}

\newcommand{\selRnpMaxDrop}{$74$}
\newcommand{\selRnpMedianDrop}{$8.4$}

\newcommand{\selRnpTwinCollapseReg}{$10/50$}
\newcommand{\selRnpTwinCollapse}{$11/60$}
\newcommand{\selRnpWithinOne}{$23/60$}

\newcommand{\selWanetMcrCleanPt}{$-1.78$}

%% file: supplement/sched/build/sched12_numbers.tex
\stwelvetrue
\newcommand{\shsFpDefRange}{$1.24$}

\newcommand{\shsFpSlope}{$+0.16$}
\newcommand{\shsFtDefRange}{$1.78$}

\newcommand{\shsFtSlope}{$+0.32$}

\newcommand{\shsHundredAzero}{$3/3$}

\newcommand{\shsHundredFpLevel}{$92.37$}

\newcommand{\shsHundredFtLevel}{$93.63$}
\newcommand{\shsHundredFtPub}{$+0.98$}

\newcommand{\shsHundredNadLevel}{$93.44$}

\newcommand{\shsNadDefRange}{$1.72$}

\newcommand{\shsNadSlope}{$+0.31$}
\newcommand{\shsPfortyfour}{\checkmark}
\newcommand{\shsPfortythree}{$\times$}
\newcommand{\shsPfortytwo}{\checkmark}

\newcommand{\shsSixtyAzero}{$3/3$}

\newcommand{\shsStartGap}{$4.41$}

\newcommand{\shsThirtyAzero}{$3/3$}

\newcommand{\shsThirtyFpLevel}{$91.87$}

\newcommand{\shsThirtyFtLevel}{$92.18$}
\newcommand{\shsThirtyFtPub}{$-1.96$}

\newcommand{\shsThirtyNadLevel}{$91.97$}

\newcommand{\shsUndefRange}{$4.83$}
\newcommand{\shsUndefThird}{$1.61$}

%% file: supplement/sched/build/sched13_numbers.tex
\sthirteentrue
\newcommand{\sdrAzeroFailRates}{0.5\%, 1\%}

\newcommand{\sdrFloorList}{SAU at 5\%, seed 2}
\newcommand{\sdrFloorN}{$1$}
\newcommand{\sdrFtPubDropN}{$4$}
\newcommand{\sdrFtPubDropPt}{$-0.82$}
\newcommand{\sdrFtPubDrop}{$-0.82\ \scriptstyle[-1.43,-0.21]$}
\newcommand{\sdrGapHiRate}{$20\%$}
\newcommand{\sdrGapHi}{$+3.58$}
\newcommand{\sdrGapLoRate}{$0.5\%$}
\newcommand{\sdrGapLo}{$-0.09$}

\newcommand{\sdrGapRisePt}{$+2.33$}
\newcommand{\sdrGapRise}{$+2.33\ \scriptstyle[+1.50,+3.15]$}

\newcommand{\sdrHalfAzero}{$0/5$}

\newcommand{\sdrHiRate}{$20\%$}

\newcommand{\sdrLoRate}{$5\%$}

\newcommand{\sdrOneAzero}{$0/5$}

\newcommand{\sdrPfortyfiveNote}{untestable as registered: K20 voids the 0.5\% and 1\% poisoned arms its five-rate test needs; read there anyway, $\rho=1.00$ over all five rates and $1.00$ over the A0-passing ones}
\newcommand{\sdrPfortyfive}{$\times$}
\newcommand{\sdrPfortysevenNote}{untestable: K20 voided every rate $\le1\%$, counted as failed}
\newcommand{\sdrPfortyseven}{$\times$}
\newcommand{\sdrPfortysixRatio}{untestable (K20 voided the $1\%$ arm), counted as failed}
\newcommand{\sdrPfortysixRise}{\checkmark}
\newcommand{\sdrPfortysix}{$\times$}
\newcommand{\sdrPubFellN}{$8/8$}

\newcommand{\sdrRatesAll}{0.5\%, 1\%, 5\%, 10\%, 20\%}

\newcommand{\sdrRhoGapBoth}{$\rho=1.00$ over all five rates, $1.00$ over the A0-passing ones}

\newcommand{\sdrRhoPairedFp}{$1.00$}
\newcommand{\sdrRhoPairedFt}{$1.00$}
\newcommand{\sdrRhoPairedNad}{$1.00$}
\newcommand{\sdrRhoPubFp}{$-1.00$}
\newcommand{\sdrRhoPubFt}{$-1.00$}
\newcommand{\sdrRhoPubNad}{$-1.00$}

%% file: supplement/sched/build/sched16_numbers.tex
\ssixteentrue

\newcommand{\sdbBlFtPubChange}{$+1.06\ \scriptstyle[+0.67,+1.44]$}
\newcommand{\sdbBlGapRisePt}{$+1.02$}
\newcommand{\sdbBlGapRise}{$+1.02\ \scriptstyle[+0.78,+1.27]$}

\newcommand{\sdbBlKtwenty}{none}

\newcommand{\sdbFloorDefs}{ABL}
\newcommand{\sdbFloorN}{$7$}

\newcommand{\sdbLfFiveGapThree}{$+0.562$}

\newcommand{\sdbLfFtPubChange}{$+0.96\ \scriptstyle[+0.67,+1.24]$}
\newcommand{\sdbLfGapRisePt}{$+0.50$}
\newcommand{\sdbLfGapRise}{$+0.50\ \scriptstyle[+0.17,+0.82]$}

\newcommand{\sdbLfKtwenty}{1\%}

\newcommand{\sdbLfOneAzero}{$0/5$}

\newcommand{\sdbLfTenGapThree}{$+0.560$}

\newcommand{\sdbPfiftyfiveN}{$0/4$}
\newcommand{\sdbPfiftyfive}{$\times$}
\newcommand{\sdbPfiftyfour}{$\times$}
\newcommand{\sdbPfiftysixN}{$4/4$}
\newcommand{\sdbPfiftysix}{\checkmark}
\newcommand{\sdbPfiftythree}{$\times$}

\newcommand{\sdbRepSummary}{registered for Blended and LF, where the start moves by \sdbBlGapRisePt\ and \sdbLfGapRisePt, it reversed (\sdbPfiftyfiveN\ pairs fall; a failed registered prediction)}
\newcommand{\sdbRhoGapBl}{$1.00$}
\newcommand{\sdbRhoGapLf}{$0.50$}

\newcommand{\sdbRhoPubBlFt}{$+1.00$}
\newcommand{\sdbRhoPubBlNad}{$+1.00$}
\newcommand{\sdbRhoPubLfFt}{$+1.00$}
\newcommand{\sdbRhoPubLfNad}{$+1.00$}
\newcommand{\sdbScorecardRows}{SCHED16 (Blended and LF dose, \cref{app:bddose}) & P53 A0 holds at every rate (1, 5, 10, 20\%) on both attacks & 0.75 & \sdbPfiftythree{} (void by K20: Blended \sdbBlKtwenty; LF \sdbLfKtwenty) \\
 & P54 the starting gap rises with the rate, $\rho\ge0.8$, each attack & 0.60 & \sdbPfiftyfour{} (\sdbRhoGapBl, \sdbRhoGapLf) \\
 & P55 FT and NAD published cost falls with the rate, $\rho\le-0.8$, in $\ge3$ of the $4$ (attack, defense) pairs & 0.45 & \sdbPfiftyfive{} (\sdbPfiftyfiveN) \\
 & P56 FT and NAD paired contrast rises, $\rho\ge0.8$, in $\ge3$ of $4$ & 0.55 & \sdbPfiftysix{} (\sdbPfiftysixN) \\}

%% file: supplement/sched/build/sched17_numbers.tex
\sseventeentrue
\newcommand{\sbxNAzero}{$5/5$}
\newcommand{\sbxNAfour}{$4/5$}

\newcommand{\sbxNTwin}{82.83}
\newcommand{\sbxNTwinAsrMax}{0.089}
\newcommand{\sbxNPois}{79.64}

\newcommand{\sbxAAzero}{$5/5$}
\newcommand{\sbxAAfour}{$5/5$}

\newcommand{\sbxATwin}{91.44}

\newcommand{\sbxAPois}{90.38}

\newcommand{\sbxFtNTare}{$-6.10$ {\scriptsize[$-7.67$, $-4.52$]}}
\newcommand{\sbxFtNTarePt}{$-6.10$}

\newcommand{\sbxFtNPub}{$-8.24$ {\scriptsize[$-10.59$, $-5.90$]}}

\newcommand{\sbxFtATare}{$+0.01$ {\scriptsize[$-0.03$, $+0.04$]}}
\newcommand{\sbxFtATarePt}{$+0.01$}

\newcommand{\sbxFtAPub}{$+0.19$ {\scriptsize[$+0.02$, $+0.37$]}}

\newcommand{\sbxFtoneNTare}{$-7.12$ {\scriptsize[$-8.68$, $-5.57$]}}
\newcommand{\sbxFtoneNTarePt}{$-7.12$}

\newcommand{\sbxFtoneATare}{$-0.01$ {\scriptsize[$-0.10$, $+0.08$]}}
\newcommand{\sbxFtoneATarePt}{$-0.01$}

\newcommand{\sbxNadNTare}{$-7.23$ {\scriptsize[$-8.82$, $-5.63$]}}
\newcommand{\sbxNadNTarePt}{$-7.23$}

\newcommand{\sbxNadATare}{$+0.81$ {\scriptsize[$+0.67$, $+0.95$]}}
\newcommand{\sbxNadATarePt}{$+0.81$}

\newcommand{\sbxNadSwing}{$+8.03$ {\scriptsize[$+6.54$, $+9.53$]}}
\newcommand{\sbxPruneNTare}{$+3.44$ {\scriptsize[$+0.83$, $+6.06$]}}
\newcommand{\sbxPruneNTarePt}{$+3.44$}

\newcommand{\sbxPruneATare}{$+1.69$ {\scriptsize[$+0.51$, $+2.87$]}}
\newcommand{\sbxPruneATarePt}{$+1.69$}

\newcommand{\sbxAthreeAll}{$0/40$}
\newcommand{\sbxTwinGap}{8.61}

\newcommand{\sbxBnNAzero}{$5/5$}
\newcommand{\sbxBnNAfour}{$4/5$}

\newcommand{\sbxBnNTwinAsrMax}{0.051}

\newcommand{\sbxBnAAzero}{$5/5$}
\newcommand{\sbxBnAAfour}{$5/5$}

\newcommand{\sbxBnFtNTare}{$-5.91$ {\scriptsize[$-7.92$, $-3.91$]}}
\newcommand{\sbxBnFtNTarePt}{$-5.91$}

\newcommand{\sbxBnFtATare}{$+0.09$ {\scriptsize[$-0.05$, $+0.24$]}}

\newcommand{\sbxBnFtoneNTarePt}{$-7.10$}

\newcommand{\sbxBnFtoneATarePt}{$+0.05$}

\newcommand{\sbxBnNadNTare}{$-7.34$ {\scriptsize[$-9.35$, $-5.33$]}}

\newcommand{\sbxBnNadATare}{$+1.15$ {\scriptsize[$+0.51$, $+1.79$]}}

\newcommand{\sbxBnNadSwing}{$+8.49$ {\scriptsize[$+6.55$, $+10.43$]}}
\newcommand{\sbxBnPruneNTare}{$+3.46$ {\scriptsize[$+1.86$, $+5.05$]}}
\newcommand{\sbxBnPruneNTarePt}{$+3.46$}

\newcommand{\sbxBnPruneATare}{$+1.80$ {\scriptsize[$+0.68$, $+2.91$]}}
\newcommand{\sbxBnPruneATarePt}{$+1.80$}

\newcommand{\sbxBnAthreeAll}{$5/40$}
\newcommand{\sbxBnTwinGap}{8.94}

\newcommand{\sbxPfiftyseven}{\checkmark}
\newcommand{\sbxPfiftyeight}{\checkmark}
\newcommand{\sbxPfiftynine}{$\times$}
\newcommand{\sbxPsixty}{$\times$}
\newcommand{\sbxPsixtyone}{\checkmark}
\newcommand{\sbxPsixtytwo}{\checkmark}
\newcommand{\sbxPsixtythree}{\checkmark}
\newcommand{\sbxPsixtyfour}{\checkmark}
\newcommand{\sbxPsixtyfive}{$\times$}
\newcommand{\sbxPsixtysix}{\checkmark}
\newcommand{\sbxPsixtyseven}{\checkmark}
\newcommand{\sbxPsixtyeight}{\checkmark}
\newcommand{\sbxHeldN}{5}
\newcommand{\sbxBnHeldN}{4}
\newcommand{\sbxGpuH}{21.7}

%% file: supplement/sched/build/sched20_numbers.tex
\sztwentytrue

\newcommand{\szrBshipFtTareZero}{$+0.07$ {\scriptsize[$-0.05$, $+0.20$]}}

\newcommand{\szrPsixtynine}{\checkmark}
\newcommand{\szrPseventy}{\checkmark}
\newcommand{\szrPseventyone}{\checkmark}
\newcommand{\szrPseventyN}{$8/9$}

\newcommand{\szrStatus}{final}
\newcommand{\szrAfourAll}{$15/15$}
\newcommand{\szrAccDiffMaxAbs}{0.24}
\newcommand{\szrTareDiffMaxAbs}{0.35}

\newcommand{\szrKthirtyN}{$0$}

\newcommand{\szrKsixteenN}{$0$}

\newcommand{\szrRunsN}{$60/60$}

\newcommand{\szrRcFailN}{$0$}

\newcommand{\szrGpuH}{5.9}

%% file: supplement/sched/build/sched21_numbers.tex
\srtwentyonetrue

\newcommand{\srtFiveWshipAzero}{$4/5$}

\newcommand{\srtFiveWcosAzero}{$5/5$}

\newcommand{\srtFiveBshipAzero}{$4/5$}

\newcommand{\srtFiveBstepAzero}{$4/5$}

\newcommand{\srtOneWshipGate}{void (K31: A0 fails in $3/5$)}

\newcommand{\srtOneWcosGate}{counted}

\newcommand{\srtOneBshipGate}{void (K31: A0 fails in $5/5$)}

\newcommand{\srtOneBstepGate}{void (K31: A0 fails in $5/5$)}

\newcommand{\srtPseventytwo}{$\times$}
\newcommand{\srtPseventythree}{$\times$}
\newcommand{\srtPseventyfour}{$\times$}
\newcommand{\srtPseventyfive}{$\times$}
\newcommand{\srtPseventytwoN}{$2/4$}
\newcommand{\srtPseventythreeN}{$3/6$}
\newcommand{\srtPseventyfourN}{$3/6$}

\newcommand{\srtStatus}{final}

\newcommand{\srtKthirtytwoN}{$0$}

\newcommand{\srtVictimsN}{$40/40$}

\newcommand{\srtAzeroAll}{$23/40$}
\newcommand{\srtAthreeAll}{$29/69$}

\newcommand{\srtGpuH}{15.1}

%% file: supplement/sched/build/sched22_numbers.tex
\sdptwentytwotrue

\newcommand{\sdpBadnetTareOPt}{$+2.49$}

\newcommand{\sdpBadnetTareD}{$+6.44$ {\scriptsize[$+5.11$, $+7.77$]}}
\newcommand{\sdpBadnetTareDPt}{$+6.44$}

\newcommand{\sdpBadnetTareDiff}{$+3.95$ {\scriptsize[$+2.35$, $+5.55$]}}
\newcommand{\sdpBadnetTareDiffPt}{$+3.95$}

\newcommand{\sdpWanetTareO}{$+1.78$ {\scriptsize[$+0.52$, $+3.04$]}}
\newcommand{\sdpWanetTareOPt}{$+1.78$}

\newcommand{\sdpWanetPairedO}{$+4.95$ {\scriptsize[$+1.96$, $+7.93$]}}

\newcommand{\sdpWanetTareD}{$+6.79$ {\scriptsize[$+5.88$, $+7.71$]}}
\newcommand{\sdpWanetTareDPt}{$+6.79$}

\newcommand{\sdpWanetPairedD}{$-0.11$ {\scriptsize[$-2.05$, $+1.82$]}}

\newcommand{\sdpPeighty}{\checkmark}
\newcommand{\sdpPeightyone}{$\times$}
\newcommand{\sdpPeightytwo}{\checkmark}

\newcommand{\sdpSweepMissingN}{$0$}

\newcommand{\sdpAssertN}{$225/225$}

\newcommand{\sdpSameThrPois}{$97/110$}
\newcommand{\sdpSameThrTwin}{$35/110$}

%% file: supplement/sched/build/sched23_numbers.tex
\sedtwentythreetrue

\newcommand{\sedPseventysix}{\checkmark}
\newcommand{\sedPseventyseven}{\checkmark}
\newcommand{\sedPseventyeight}{$\times$}
\newcommand{\sedPseventynine}{\checkmark}

\newcommand{\sedStatus}{harvest before the deadline, every registered arm complete}
\newcommand{\sedPseventysevenCounts}{WaNet never $7/11$; WaNet cosine $7/11$; BPP never $3/5$; BPP cosine $5/5$}
\newcommand{\sedPseventyeightFire}{$6/11$}
\newcommand{\sedPseventyeightCond}{$+3.31$}

\newcommand{\sedPseventynineFlipList}{FP/I-BAU, FP/SAU}
\newcommand{\sedPseventynineFlipsAll}{$3/28$}
\newcommand{\sedPseventynineFlipListAll}{FT/NAD, FP/I-BAU, FP/SAU}
\newcommand{\sedFlipSepShipN}{$0/3$}

\newcommand{\sedKthirtythreeN}{$0$}

\newcommand{\sedKthirtyfourN}{$0$}

\newcommand{\sedKsixteenN}{$2$}
\newcommand{\sedKsixteenList}{WaNet cosine SAU: s5; BPP cosine SAU: s2}

\newcommand{\sedNewRunsN}{$146/146$}

\newcommand{\sedGpuH}{27.1}

%% file: authors.tex
\author{Ruizhi Xu \\
School of Mathematics \\
University of Edinburgh \\
\texttt{s2479881@ed.ac.uk}
\And
Wei Xu \\
Beijing Jinhe Technology Co., Ltd. \\
\texttt{xuwei@hedechina.com}
\And
Sibo Zhu \\
Inno Asset Management \\
\texttt{zhusibo@innoam.com}
}
\hypersetup{pdfauthor={Ruizhi Xu, Wei Xu, Sibo Zhu}}

%% file: supplement/build/tab_bbdef_main_body.tex
\begin{tabular}{lccccc}
\toprule
Defense & reported $\cDef$ (pt) & tare $c_{\mathrm{clean}}$ (pt) & paired $Q(\mathcal{D}\theta_{\mathrm{twin}})-Q(\mathcal{D}\theta_{\mathrm{pois}})$ (pt) & interaction, DiD (pt) & $A3$ ($k/n$) \\
\midrule
fine-tuning                        & $+1.01\ \scriptstyle[+0.84,+1.19]$ & $+0.13\ \scriptstyle[+0.06,+0.21]$ & $+2.94\ \scriptstyle[+2.75,+3.14]$ & $+0.88\ \scriptstyle[+0.68,+1.08]$ & $11/11$ \\
NAD                                & $+2.10\ \scriptstyle[+1.89,+2.30]$ & $+0.33\ \scriptstyle[+0.24,+0.42]$ & $+3.83\ \scriptstyle[+3.59,+4.07]$ & $+1.77\ \scriptstyle[+1.52,+2.02]$ & $11/11$ \\
ANP$^{\dagger}$                    & $+6.50\ \scriptstyle[+5.73,+7.27]$ & $+2.49\ \scriptstyle[+0.39,+4.60]$ & $+6.07\ \scriptstyle[+3.75,+8.39]$ & $+4.01\ \scriptstyle[+1.69,+6.32]$ & $11/11$ \\
\midrule
fine-pruning                       & $-0.22\ \scriptstyle[-0.36,-0.08]$ & $+1.05\ \scriptstyle[+0.84,+1.25]$ & $+0.80\ \scriptstyle[+0.50,+1.09]$ & $-1.27\ \scriptstyle[-1.60,-0.93]$ & $11/11$ \\
SAU                                & $+1.39\ \scriptstyle[+1.11,+1.67]$ & $+2.70\ \scriptstyle[+2.31,+3.08]$ & $+0.76\ \scriptstyle[+0.26,+1.25]$ & $-1.31\ \scriptstyle[-1.80,-0.82]$ & $11/11$ \\
I-BAU                              & $+3.85\ \scriptstyle[+3.05,+4.66]$ & $+5.70\ \scriptstyle[+3.99,+7.42]$ & $+0.21\ \scriptstyle[-1.84,+2.27]$ & $-1.85\ \scriptstyle[-3.93,+0.23]$ & $11/11$ \\
ABL$^{\dagger}$                    & $+9.39\ \scriptstyle[+8.12,+10.7]$ & $+33.2\ \scriptstyle[+29.1,+37.3]$ & $-21.8\ \scriptstyle[-26.3,-17.2]$ & $-23.8\ \scriptstyle[-28.3,-19.3]$ & $11/11$ \\
\midrule
Neural Cleanse$^{\dagger\ddagger}$ & $+1.53\ \scriptstyle[+0.99,+2.07]$ & $+1.31\ \scriptstyle[+0.10,+2.51]$ & $+2.28\ \scriptstyle[+1.16,+3.39]$ & $+0.22\ \scriptstyle[-0.85,+1.29]$ & $8/11$ \\
\midrule
pooled, $7$ excl.\ ABL             & $+2.34\ \scriptstyle[+2.10,+2.57]$ & $+1.98\ \scriptstyle[+1.50,+2.46]$ & $+2.42\ \scriptstyle[+1.81,+3.04]$ & $+0.36\ \scriptstyle[-0.26,+0.98]$ & $74/77$ \\
\bottomrule
\end{tabular}

%% file: supplement/build/tab_sched_main_body.tex
\begin{tabular}{llcccccrc}
\toprule
attack & victim's scheduler & seeds & fine-tuning & fine-pruning & NAD & pooled & twin (\%) & cells \\
\midrule
WaNet & step, never fires (shipped) & $11$ & $-1.99$ {\tiny$\pm0.31$} & $-1.68$ {\tiny$\pm0.44$} & $-1.85$ {\tiny$\pm0.29$} & $-1.84$ {\tiny$\pm0.34$} & $91.50$ & $33/33$ \\
WaNet & cosine (swapped) & $11$ & $+0.51$ {\tiny$\pm0.08$} & $+0.86$ {\tiny$\pm0.12$} & $+0.99$ {\tiny$\pm0.13$} & $+0.78$ {\tiny$\pm0.09$} & $94.18$ & $33/33$ \\
\midrule
BadNets & cosine (shipped) & $11$ & $+0.13$ {\tiny$\pm0.07$} & $+1.05$ {\tiny$\pm0.20$} & $+0.33$ {\tiny$\pm0.09$} & $+0.50$ {\tiny$\pm0.09$} & $93.71$ & $32/33$ \\
BadNets & step, never fires (swapped) & $11$ & $-2.29$ {\tiny$\pm0.32$} & $-1.95$ {\tiny$\pm0.41$} & $-2.00$ {\tiny$\pm0.30$} & $-2.08$ {\tiny$\pm0.32$} & $91.09$ & $33/33$ \\
\bottomrule
\end{tabular}

%% file: supplement/sched/build/tab_lb_fourds_body.tex
\begin{tabular}{lccccc}
\toprule
dataset & PreAct-R18 & VGG19-BN & ConvNeXt-T & ViT-B/16 & lowest three \\
\midrule
CIFAR-10 & $+2.1$ & $+2.6$ & $+4.1$ & $+4.9$ & $15/16$ \\
CIFAR-100 & $+6.3$ & $+5.9$ & $+9.6$ & $+10.0$ & $15/15$ \\
Tiny-ImageNet$^\dagger$ & --- & $-2.4$ & $+10.5$ & $+11.3$ & $4/10$ \\
GTSRB (saturated) & $+0.3$ & $+0.9$ & $+0.4$ & $-7.3$ & $2/16$ \\
\bottomrule
\end{tabular}

%% file: supplement/sched/build/tab_sched10_main_body.tex
\begin{tabular}{lccccccc}
\toprule
defense & BadNets & Blended & LF & FTrojanNN & Refool & WaNet & BPP \\
{\tiny victim schedule} & {\tiny cosine} & {\tiny cosine} & {\tiny cosine} & {\tiny cosine} & {\tiny cosine} & {\tiny never} & {\tiny never} \\
{\tiny undefended twin} & {\tiny $93.71$} & {\tiny $93.84$} & {\tiny $93.92$} & {\tiny $93.84$} & {\tiny $93.74$} & {\tiny $91.50$} & {\tiny $91.48$} \\
{\tiny A4: clean twins} & {\tiny $11/11$} & {\tiny $8/8$} & {\tiny $5/5$} & {\tiny $5/5$} & {\tiny $5/5$} & {\tiny $11/11$} & {\tiny $5/5$} \\
\midrule
FT & $+0.13$ {\tiny$\pm0.07$} & $+0.20$ {\tiny$\pm0.11$} & $+0.11$ {\tiny$\pm0.14$} & $+0.24$ {\tiny$\pm0.15$} & $+0.16$ {\tiny$\pm0.15$} & $-1.99$ {\tiny$\pm0.31$} & $-2.07$ {\tiny$\pm0.75$} \\
\;\emph{paired} & $+2.94$ & $+0.86^{\ddagger}$ & $+1.32^{\ddagger}$ & $+0.84^{\ddagger}$ & $+1.69^{\ddagger}$ & $+0.30$ & $+0.22$ \\
FP & $+1.05$ {\tiny$\pm0.20$} & $+1.12$ {\tiny$\pm0.23$} & $+1.08$ {\tiny$\pm0.46$} & $+1.04$ {\tiny$\pm0.36$} & $+0.91$ {\tiny$\pm0.30$} & $-1.68$ {\tiny$\pm0.44$} & $-1.82$ {\tiny$\pm1.07$} \\
\;\emph{paired} & $+0.80$ & $+0.28^{\ddagger}$ & $+0.39^{\ddagger}$ & $+0.15^{\ddagger}$ & $+0.84^{\ddagger}$ & $-0.06$ & $+0.25^{\ddagger}$ \\
NAD & $+0.33$ {\tiny$\pm0.09$} & $+0.36$ {\tiny$\pm0.15$} & $+0.22$ {\tiny$\pm0.18$} & $+0.47$ {\tiny$\pm0.25$} & $+0.36$ {\tiny$\pm0.16$} & $-1.85$ {\tiny$\pm0.29$} & $-1.84$ {\tiny$\pm0.64$} \\
\;\emph{paired} & $+3.83$ & $+1.65^{\ddagger}$ & $+1.96^{\ddagger}$ & $+1.27^{\ddagger}$ & $+2.00^{\ddagger}$ & $+0.37$ & $+0.33$ \\
NC$^{\dagger}$ & $5/11$\,\textbar\,$+2.71$ & $3/8$\,\textbar\,$+3.06$ & $1/5$\,\textbar\,$+3.22$ & $3/5$\,\textbar\,$+2.86$ & $1/5$\,\textbar\,$+2.79$ & $3/11$\,\textbar\,$-1.30$ & $0/5$\,\textbar\,--- \\
\;\emph{paired} & $+1.94$ & $+0.57^{\ddagger}$ & $+2.35$ & $+0.57$ & $+1.12^{\ddagger}$ & $+0.46$ & $-1.22$ \\
ANP$^{\dagger}$ & $+2.49$ {\tiny$\pm2.10$} & $+3.86$ {\tiny$\pm2.86$} & $+3.25$ {\tiny$\pm3.08$} & $+1.43$ {\tiny$\pm1.01$} & $+4.72$ {\tiny$\pm3.62$} & $+1.78$ {\tiny$\pm1.26$} & $+4.28$ {\tiny$\pm2.48$} \\
\;\emph{paired} & $+6.07$ & $+3.61$ & $+4.09$ & $+5.80$ & $+3.28$ & $+4.95$ & $+2.51$ \\
ABL$^{\dagger\natural}$ & $+33.20$ {\tiny$\pm4.06$} & $+27.83$ {\tiny$\pm11.98$} & $+31.28$ {\tiny$\pm15.75$} & $+32.36$ {\tiny$\pm12.73$} & $+35.91$ {\tiny$\pm11.28$} & $+27.77$ {\tiny$\pm6.49$} & $+29.83$ {\tiny$\pm12.18$} \\
\;\emph{paired} & $-21.75$ & $-7.08$ & $-6.32^{\ddagger}$ & $-15.48$ & $+4.90^{\ddagger}$ & $-6.27^{\ddagger}$ & $-2.05^{\ddagger}$ \\
I-BAU & $+5.70$ {\tiny$\pm1.71$} & $+4.95$ {\tiny$\pm1.02$} & $+5.98$ {\tiny$\pm2.26$} & $+6.06$ {\tiny$\pm2.37$} & $+5.66$ {\tiny$\pm1.68$} & $+0.23$ {\tiny$\pm0.53$} & $+0.71$ {\tiny$\pm1.67$} \\
\;\emph{paired} & $+0.21$ & $-0.42$ & $-0.61^{\ddagger}$ & $-1.32^{\ddagger}$ & $+0.26^{\ddagger}$ & $+0.20$ & $+0.06$ \\
SAU & $+2.70$ {\tiny$\pm0.39$} & $+3.06$ {\tiny$\pm1.76$} & $+2.83$ {\tiny$\pm0.80$} & $+2.92$ {\tiny$\pm0.51$} & $+3.71$ {\tiny$\pm1.95$} & $-0.15$ {\tiny$\pm0.62$} & $-0.60$ {\tiny$\pm0.96$} \\
\;\emph{paired} & $+0.76$ & $+1.50$ & $+0.25$ & $-0.35$ & $-0.28^{\ddagger}$ & $-0.10$ & $+0.56$ \\
\bottomrule
\end{tabular}

%% file: supplement/build/tab_rerank_full_body.tex
\begin{tabular}{lcccccc}
\toprule
Setting (unit) & A3 & $\cDef$ & tare $c_{\mathrm{clean}}$ & $g$ & DiD & paired \\
\midrule
BackdoorBench, $11$ seeds$^{\ast}$ & $74/77^{\S}$ & $+2.34\ \scriptstyle[+2.10,+2.57]$ & $+1.98\ \scriptstyle[+1.50,+2.46]$ & $+2.06$ & $+0.36\ \scriptstyle[-0.26,+0.98]$ & $+2.42\ \scriptstyle[+1.81,+3.04]$ \\
our rig, $8$ installations & $37/48$ & $+2.51\ \scriptstyle[+2.25,+2.77]$ & $+2.49\ \scriptstyle[+2.03,+2.96]$ & $+0.41$ & $+0.02\ \scriptstyle[-0.31,+0.34]$ & $+0.43\ \scriptstyle[+0.30,+0.55]$ \\
\hspace{1em}framework recovery schedule$^{\P}$ & $42/48$ & $+2.20\ \scriptstyle[+2.04,+2.36]$ & $+2.24\ \scriptstyle[+1.84,+2.63]$ & $+0.41$ & $-0.04\ \scriptstyle[-0.30,+0.23]$ & $+0.38\ \scriptstyle[+0.22,+0.53]$ \\
Qwen2.5-7B ($8$ installations) & $14/16$ & $+11.46\ \scriptstyle[+8.94,+13.97]$ & $+10.08\ \scriptstyle[+7.61,+12.56]$ & $+0.04$ & $+1.37\ \scriptstyle[+0.52,+2.23]^{\diamond}$ & $+1.42\ \scriptstyle[+0.69,+2.14]$ \\
\bottomrule
\end{tabular}

%% file: supplement/build/tab_a3succ_body.tex
\begin{tabular}{lcccc}
\toprule
Defense & successes (A3) & defended twin clears & mean residual $A$ & max residual $A$ \\
\midrule
Neural Cleanse & $14/14$ & $14/14$ & $0.008$ & $0.012$ \\
fine-prune $40\%$ & $14/14$ & $14/14$ & $0.014$ & $0.019$ \\
fine-prune $30\%$ & $14/14$ & $14/14$ & $0.016$ & $0.045$ \\
fine-tune lr $0.05$ & $14/14$ & $14/14$ & $0.018$ & $0.021$ \\
NAD$^{\S}$ & $13/14$ & $14/14$ & $0.022$ & $0.138$ \\
\midrule
fine-prune $20\%$ & $10/14$ & $14/14$ & $0.062$ & $0.318$ \\
ANP$^{\P}$ & $3/14$ & $14/14$ & $0.509$ & $0.965$ \\
fine-tune lr $0.02$ & $1/14$ & $14/14$ & $0.613$ & $0.914$ \\
fine-tune lr $0.01$ & $0/14$ & $14/14$ & $0.856$ & $0.940$ \\
\bottomrule
\end{tabular}

%% file: supplement/build/tab_a3thr_body.tex
\begin{tabular}{lccccccc}
\toprule
$\alpha$ & methods & common support & $\tau_a$(published, tax) & $\tau_a$(published, removal) & gap & percentile $95\%$ & basic $95\%$ \\
\midrule
$0.02$ & $4$ & $14$ & $+1.000$ & $+0.667$ & $+0.333$ & $\scriptstyle[+0.000,+0.667]$ & $\scriptstyle[+0.000,+0.667]$ \\
$0.05$ & $5$ & $13$ & $+1.000$ & $+0.800$ & $+0.200$ & $\scriptstyle[+0.200,+0.600]$ & $\scriptstyle[-0.200,+0.200]$ \\
$0.10$ & $6$ & $9$ & $+1.000$ & $+0.333$ & $+0.667$ & $\scriptstyle[+0.400,+1.200]$ & $\scriptstyle[+0.133,+0.933]$ \\
$0.25$ & $6$ & $9$ & $+1.000$ & $+0.333$ & $+0.667$ & $\scriptstyle[+0.400,+1.200]$ & $\scriptstyle[+0.133,+0.933]$ \\
\bottomrule
\end{tabular}

%% file: supplement/build/tab_csper_body.tex
\begin{tabular}{lcccc}
\toprule
Defense & DER published & DER intervention & DER removal (DiD) & rank pub./interv./removal \\
\midrule
Neural Cleanse & $0.9741$ & $0.9731$ & $0.9755$ & $1/1/1$ \\
NAD & $0.9679$ & $0.9671$ & $0.9750$ & $2/2/2$ \\
fine-prune $30\%$ & $0.9632$ & $0.9634$ & $0.9728$ & $3/3/4$ \\
fine-prune $40\%$ & $0.9582$ & $0.9579$ & $0.9744$ & $4/4/3$ \\
fine-tune lr $0.05$ & $0.9522$ & $0.9512$ & $0.9727$ & $5/5/5$ \\
\bottomrule
\end{tabular}

%% file: supplement/build/tab_a3supp_body.tex
\begin{tabular}{lcccc}
\toprule
convention & installations & $\tau_a$(published, tax) & $\tau_a$(published, removal) & fine-pruning $30\%$'s rank \\
\midrule
common support (primary) & $13$ & $+1.000$ & $+0.800$ & $3\to4$ \\
drop failed cells & $14$, unbalanced & $+1.000$ & $+0.800$ & $3\to4$ \\
method-level mean-ASR gate & $14$ & $+0.800$ & $+0.200$ & $3\to3$ \\
\bottomrule
\end{tabular}

%% file: supplement/build/tab_qwen_ops_body.tex
\begin{tabular}{lrrrrrc}
\toprule
configuration & lr & steps & residual ASR & $\cDef$ (pt) & $c_{\mathrm{clean}}$ (pt) & A3 \\
\midrule
\texttt{pois\_s200} & 0.0002 & 200 & $0.96$ & $+2.88$ & $+3.05$ & no \\
\texttt{pois\_full2e-5\_s4} & 2e-05 & 150 & $0.64$ & $+3.18$ & $+3.15$ & no \\
\texttt{pois\_full2e-5} & 2e-05 & 150 & $0.44$ & $+3.43$ & $+4.55$ & no \\
\texttt{pois\_hi5e-4} & 0.0005 & 300 & $0.96$ & $+3.88$ & $+3.05$ & no \\
\texttt{pois\_s600} & 0.0002 & 600 & $0.94$ & $+3.88$ & $+4.55$ & no \\
\texttt{pois\_hi1e-3} & 0.001 & 300 & $0.94$ & $+4.38$ & $+4.55$ & no \\
\texttt{pois\_full4e-5\_s4} & 4e-05 & 150 & $0.00$ & $+5.43$ & $+5.65$ & yes \\
\texttt{pois\_full3e-5} & 3e-05 & 150 & $0.10$ & $+5.88$ & $+5.90$ & no \\
\texttt{pois2\_full2e-5} & 2e-05 & 150 & $0.14$ & $+7.88$ & $+4.55$ & no \\
\texttt{pois\_full3e-5\_s5} & 3e-05 & 150 & $0.00$ & $+7.98$ & $+6.20$ & yes \\
\texttt{pois\_full4e-5} & 4e-05 & 150 & $0.00$ & $+8.03$ & $+7.65$ & yes \\
\texttt{pois\_full5e-5\_s4} & 5e-05 & 150 & $0.00$ & $+8.03$ & $+8.50$ & yes \\
\texttt{pois\_full5e-5\_s5} & 5e-05 & 150 & $0.00$ & $+8.33$ & $+11.60$ & yes \\
\texttt{pois\_full4e-5\_s5} & 4e-05 & 150 & $0.00$ & $+9.13$ & $+8.30$ & yes \\
\texttt{pois2\_full4e-5} & 4e-05 & 150 & $0.00$ & $+10.13$ & $+7.65$ & yes \\
\texttt{pois\_full5e-5} & 5e-05 & 150 & $0.00$ & $+10.93$ & $+9.40$ & yes \\
\texttt{pois2\_full5e-5} & 5e-05 & 150 & $0.00$ & $+11.88$ & $+9.40$ & yes \\
\bottomrule
\end{tabular}

%% file: supplement/build/tab_cr_body.tex
\begin{tabular}{lcccccccc}
\toprule
 & \multicolumn{4}{c}{fine-prune $30\%$} & \multicolumn{4}{c}{fine-tune lr $0.02$} \\
\cmidrule(lr){2-5}\cmidrule(lr){6-9}
clean budget & $c^{0}_{\mathrm{def}}$ & $\cDef$ & $\cRem^{\mathrm{DiD}}$ & A3 & $c^{0}_{\mathrm{def}}$ & $\cDef$ & $\cRem^{\mathrm{DiD}}$ & A3 \\
\midrule
$1\%$ & $+3.79\ \scriptstyle[+3.25,+4.33]$ & $+3.80$ & $+0.01\ \scriptstyle[-0.45,+0.46]$ & $9/10$ & $+0.28\ \scriptstyle[+0.15,+0.41]$ & \textcolor{cRep}{$+0.59$} & \textcolor{cRep}{$+0.31\ \scriptstyle[+0.05,+0.57]$} & $0/10$ \\
$2\%$ & $+3.19\ \scriptstyle[+2.73,+3.66]$ & $+3.10$ & $-0.10\ \scriptstyle[-0.56,+0.36]$ & $10/10$ & $+0.57\ \scriptstyle[+0.41,+0.73]$ & \textcolor{cRep}{$+0.93$} & \textcolor{cRep}{$+0.36\ \scriptstyle[+0.13,+0.59]$} & $1/10$ \\
$2.5\%$ & $+2.88\ \scriptstyle[+2.50,+3.25]$ & $+2.72$ & $-0.16\ \scriptstyle[-0.56,+0.24]$ & $10/10$ & $+0.67\ \scriptstyle[+0.50,+0.84]$ & \textcolor{cRep}{$+1.01$} & \textcolor{cRep}{$+0.34\ \scriptstyle[+0.16,+0.52]$} & $1/10$ \\
$5\%$ & $+2.17\ \scriptstyle[+1.82,+2.51]$ & $+2.27$ & $+0.10\ \scriptstyle[-0.24,+0.44]$ & $9/9$ & $+0.97\ \scriptstyle[+0.76,+1.18]$ & \textcolor{cRep}{$+1.16$} & \textcolor{cRep}{$+0.19\ \scriptstyle[-0.01,+0.39]$} & $1/10$ \\
$10\%$ & $+1.72\ \scriptstyle[+1.49,+1.94]$ & $+1.62$ & $-0.09\ \scriptstyle[-0.32,+0.14]$ & $9/9$ & $+1.06\ \scriptstyle[+0.94,+1.19]$ & \textcolor{cRep}{$+1.04$} & \textcolor{cRep}{$-0.02\ \scriptstyle[-0.19,+0.15]$} & $1/10$ \\
$20\%$ & $+1.21\ \scriptstyle[+1.02,+1.39]$ & $+1.04$ & $-0.17\ \scriptstyle[-0.38,+0.05]$ & $10/10$ & $+0.72\ \scriptstyle[+0.58,+0.86]$ & $+0.69$ & $-0.03\ \scriptstyle[-0.27,+0.21]$ & $6/10$ \\
$100\%$ & $+0.28\ \scriptstyle[+0.09,+0.47]$ & $-0.05$ & $-0.33\ \scriptstyle[-0.59,-0.07]$ & $9/9$ & $-0.08\ \scriptstyle[-0.25,+0.08]$ & $-0.40$ & $-0.31\ \scriptstyle[-0.51,-0.12]$ & $10/10$ \\
\bottomrule
\end{tabular}

%% file: supplement/build/fig4b_ladder.tex
  \node[font=\scriptsize] at (0,0.4) {published};
  \node[font=\scriptsize] at (2.9,0.4) {corrected};
  \draw[gray] (0,0.1) -- (0,-3.6);  \draw[gray] (2.9,0.1) -- (2.9,-3.6);
  \draw[cTax] (0,-0.2) -- (2.9,-0.2);      %
  \draw[cRep,very thick] (0,-0.6) -- (2.9,-1.8);      %
  \draw[cTax] (0,-1.0) -- (2.9,-1.0);      %
  \draw[cUp,very thick] (0,-1.4) -- (2.9,-0.6);      %
  \draw[cUp,very thick] (0,-1.8) -- (2.9,-1.4);      %
  \draw[cTax] (0,-2.2) -- (2.9,-2.2);      %
  \draw[cTax] (0,-2.6) -- (2.9,-2.6);      %
  \draw[cTax] (0,-3.0) -- (2.9,-3.0);      %
  \draw[cTax] (0,-3.4) -- (2.9,-3.4);      %
  \foreach \y in {-0.2,-0.6,-1.0,-1.4,-1.8,-2.2,-2.6,-3.0,-3.4} {
    \fill (0,\y) circle (1.3pt); \fill (2.9,\y) circle (1.3pt); }
  \node[left,font=\tiny] at (-0.1,-0.2) {NC};
  \node[left,font=\tiny] at (-0.1,-0.6) {NAD};
  \node[left,font=\tiny] at (-0.1,-1.0) {fp 30};
  \node[left,font=\tiny] at (-0.1,-1.4) {fp 40};
  \node[left,font=\tiny] at (-0.1,-1.8) {ft .05};
  \node[left,font=\tiny] at (-0.1,-2.2) {fp 20};
  \node[left,font=\tiny] at (-0.1,-2.6) {ANP};
  \node[left,font=\tiny] at (-0.1,-3.0) {ft .02};
  \node[left,font=\tiny] at (-0.1,-3.4) {ft .01};
  \node[right,font=\tiny] at (3.0,-0.2) {NC};
  \node[right,font=\tiny] at (3.0,-0.6) {fp 40};
  \node[right,font=\tiny] at (3.0,-1.0) {fp 30};
  \node[right,font=\tiny] at (3.0,-1.4) {ft .05};
  \node[right,font=\tiny] at (3.0,-1.8) {NAD};
  \node[right,font=\tiny] at (3.0,-2.2) {fp 20};
  \node[right,font=\tiny] at (3.0,-2.6) {ANP};
  \node[right,font=\tiny] at (3.0,-3.0) {ft .02};
  \node[right,font=\tiny] at (3.0,-3.4) {ft .01};
  \node[font=\scriptsize] at (1.45,-4.1) {(b) DER ordering, before vs.\ after};

%% file: supplement/build/tab_rerank_body.tex
\begin{tabular}{lcccccc}
\toprule
Setting (unit) & A3 & $\cDef$ & tare $c_{\mathrm{clean}}$ & $g$ & DiD & paired \\
\midrule
BackdoorBench, $11$ seeds$^{\ast}$ & $74/77^{\S}$ & $+2.34\ \scriptstyle[+2.10,+2.57]$ & $+1.98\ \scriptstyle[+1.50,+2.46]$ & $+2.06$ & $+0.36\ \scriptstyle[-0.26,+0.98]$ & $+2.42\ \scriptstyle[+1.81,+3.04]$ \\
our rig, $8$ installations & $37/48$ & $+2.51\ \scriptstyle[+2.25,+2.77]$ & $+2.49\ \scriptstyle[+2.03,+2.96]$ & $+0.41$ & $+0.02\ \scriptstyle[-0.31,+0.34]$ & $+0.43\ \scriptstyle[+0.30,+0.55]$ \\
\hspace{1em}framework recovery schedule$^{\P}$ & $42/48$ & $+2.20\ \scriptstyle[+2.04,+2.36]$ & $+2.24\ \scriptstyle[+1.84,+2.63]$ & $+0.41$ & $-0.04\ \scriptstyle[-0.30,+0.23]$ & $+0.38\ \scriptstyle[+0.22,+0.53]$ \\
\bottomrule
\end{tabular}

%% file: supplement/build/tab_sched_cols_body.tex
\begin{tabular}{llcccc}
\toprule
attack & victim's \texttt{lr\_scheduler} & published cost & clean arm & removal residue (DiD) & A3 clears \\
\midrule
WaNet & \texttt{MultiStepLR} & $-3.05\ \scriptstyle[-3.37,-2.73]$ & $-1.84\ \scriptstyle[-2.18,-1.50]$ & $-1.17\ \scriptstyle[-1.70,-0.64]$ & $30/33$ \\
WaNet & \texttt{CosineAnnealingLR} & $-0.25\ \scriptstyle[-0.66,+0.16]$ & $+0.78\ \scriptstyle[+0.70,+0.87]$ & $-1.03\ \scriptstyle[-1.49,-0.58]$ & $33/33$ \\
\midrule
BadNets & \texttt{CosineAnnealingLR} & $+0.96\ \scriptstyle[+0.83,+1.09]$ & $+0.50\ \scriptstyle[+0.42,+0.59]$ & $+0.46\ \scriptstyle[+0.27,+0.65]$ & $33/33$ \\
BadNets & \texttt{MultiStepLR} & $-2.78\ \scriptstyle[-3.59,-1.96]$ & $-2.08\ \scriptstyle[-2.40,-1.76]$ & $-0.74\ \scriptstyle[-1.62,+0.13]$ & $15/33$ \\
\bottomrule
\end{tabular}

%% file: supplement/build/tab_bbdef_rows.tex
fine-tuning                        & $+1.01\ \scriptstyle[+0.84,+1.19]$ & $+0.13\ \scriptstyle[+0.06,+0.21]$ & $+0.88\ \scriptstyle[+0.68,+1.08]$ & $+2.02$ & $+0.47$ \\
NAD                                & $+2.10\ \scriptstyle[+1.89,+2.30]$ & $+0.33\ \scriptstyle[+0.24,+0.42]$ & $+1.77\ \scriptstyle[+1.52,+2.02]$ & $+2.90$ & $+1.35$ \\
ANP$^{\dagger}$                    & $+6.50\ \scriptstyle[+5.73,+7.27]$ & $+2.49\ \scriptstyle[+0.39,+4.60]$ & $+4.01\ \scriptstyle[+1.69,+6.32]$ & $+5.14$ & $+3.59$ \\
\midrule
fine-pruning                       & $-0.22\ \scriptstyle[-0.36,-0.08]$ & $+1.05\ \scriptstyle[+0.84,+1.25]$ & $-1.27\ \scriptstyle[-1.60,-0.93]$ & $-0.13$ & $-1.68$ \\
SAU                                & $+1.39\ \scriptstyle[+1.11,+1.67]$ & $+2.70\ \scriptstyle[+2.31,+3.08]$ & $-1.31\ \scriptstyle[-1.80,-0.82]$ & $-0.17$ & $-1.72$ \\
I-BAU                              & $+3.85\ \scriptstyle[+3.05,+4.66]$ & $+5.70\ \scriptstyle[+3.99,+7.42]$ & $-1.85\ \scriptstyle[-3.93,+0.23]$ & $-0.71$ & $-2.26$ \\
ABL$^{\dagger}$                    & $+9.39\ \scriptstyle[+8.12,+10.7]$ & $+33.2\ \scriptstyle[+29.1,+37.3]$ & $-23.8\ \scriptstyle[-28.3,-19.3]$ & $-22.7$ & $-24.2$ \\
\midrule
Neural Cleanse$^{\dagger\ddagger}$ & $+1.53\ \scriptstyle[+0.99,+2.07]$ & $+1.31\ \scriptstyle[+0.10,+2.51]$ & $+0.22\ \scriptstyle[-0.85,+1.29]$ & $+1.36$ & $-0.19$ \\
\midrule

%% file: supplement/build/tab_zoo_body.tex
\begin{tabular}{lcccc}
\toprule
Defense & $A$ after & DER published & DER corrected & rank pub./interv./DiD \\
\midrule
Neural Cleanse & $0.008$ & $0.9741$ & $0.9754$ & $1/1/1$ \\
NAD & $0.022$ & $0.96340$ & $0.9705$ & $2/3/5$ \\
fine-prune $30\%$ & $0.016$ & $0.96340$ & $0.9731$ & $3/2/3$ \\
fine-prune $40\%$ & $0.014$ & $0.9582$ & $0.9745$ & $4/4/2$ \\
fine-tune lr $0.05$ & $0.018$ & $0.9520$ & $0.9725$ & $5/5/4$ \\
\midrule
fine-prune $20\%$$^{\mathsection}$ & $0.062$ & $0.9454$ & $0.9507$ & --- \\
ANP$^{\mathsection}$ & $0.509$ & $0.7183$ & $0.7250$ & --- \\
fine-tune lr $0.02$$^{\mathsection}$ & $0.613$ & $0.6698$ & $0.6746$ & --- \\
fine-tune lr $0.01$$^{\mathsection}$ & $0.856$ & $0.5522$ & $0.5535$ & --- \\
\bottomrule
\end{tabular}

%% file: appendix_sched.tex
\section{CIFAR-100: the crossover at twice the swing}\label{app:c100}
The leaderboard gap doubles from CIFAR-10 to CIFAR-100 (\cref{tab:fourds}). Two further pre-registrations (\cref{app:prereg}; \cref{tab:c100} in \cref{app:ext}) ask whether the causal swing does too, on CIFAR-100 / PreAct-ResNet18 with the benchmark's own CIFAR-100 defense configurations: the same WaNet swap; BPP with the same swap at $5\%$ poisoning (the rate the leaderboard reports for it there); and Blended~\citep{chen2017blended} with the step block \emph{added}, as the annealed control (BadNets installs below A0 on CIFAR-100 on the leaderboard).

\textbf{The swing doubles, on both attacks.} Shipped, WaNet's clean twin on CIFAR-100 ``gains'' \syWshipFt{} (FT), \syWshipFp{} (FP) and \syWshipNad\pt{} (NAD) from the defenses; with the cosine schedule FT and NAD show no measurable cost (\syWcosFt{} and \syWcosNad\pt) and FP costs \syWcosFp\pt{} (\cref{tab:c100}). BPP is the same picture, with all six annealed intervals excluding zero: \syBshipFt, \syBshipFp, \syBshipNad{} shipped, \syBcosFt, \syBcosFp, \syBcosNad{} annealed. The undefended twins move \syWanetTwinGap{} (WaNet) and \syBppTwinGap\pt{} (BPP) between schedules, more than twice the CIFAR-10 swing of \exWanetTwinGap{} and \exBppTwinGap\pt; the shipped twins (\syWshipTwin, \syBshipTwin) sit within $1.4$\pt{} of the benchmark's own never-annealed entries (\syPublicWship, \syPublicBship), so this is the leaderboard's CIFAR-100 gap of \lbGapChundPre\pt{} reproduced causally. Blended, the annealed control, shows the mirror image: its clean-arm cost, pooled over the three defenses, is \syBlshipPool\pt{} with its shipped cosine schedule and \syBlstepPool\pt{} with the step block added (\cref{tab:c100}, ``clean arm (pooled)''), and its undefended twin drops from \syBlshipTwin{} to \syBlstepTwin. The effect follows the three configuration keys, not the attack.

\textbf{The fixed point holds for the fine-tuning defenses, not for pruning.} Over the six CIFAR-100 cells the undefended twins span \syFixSixUndef\pt; after fine-tuning they span \syFixSixFtRange\pt{} (\syFixSixFtLo--\syFixSixFtHi), after NAD \syFixSixNadRange\pt. Fine-pruning does not converge to one level: \syFixSixFpRange\pt{} across the six cells, with the pruned annealed twins ending \emph{below} the pruned never-annealed ones. Our pre-registered fixed-point test (P11: the widest per-defense range under a third of the undefended range) therefore fails on CIFAR-100 because of FP, and we report it as failed; FT and NAD, the two defenses that re-run a full schedule, satisfy it on their own.

\textbf{The line also moves the attack.} On CIFAR-10 the swap left installation essentially untouched (\cref{tab:schedcols}). On CIFAR-100 it does not: the shipped, never-annealed WaNet victim installs at ASR \syWshipPoisAsrLo--$0.91$ across seeds (A0 \syWshipAzero), the annealed one at $\ge0.97$ (A0 \syWcosAzero). The benchmark's published CIFAR-100 WaNet entry at $10\%$ (\texttt{results.csv}: accuracy \syPublicWship, ASR $0.977$) has the clean accuracy of the shipped victim---our shipped twin sits at \syWshipTwin, the annealed one at \syWcosTwin---so it was produced under the shipped schedule; its ASR exceeds all five of our shipped seeds, and we could not reproduce that installation rate. Under-training therefore weakens both columns on the harder dataset, and K3 voids the shipped cell's published cost rather than letting us report a cost on an attack that did not install; P6 (the published cost itself) and P7 (interaction invariance) rest on that void cell, so we record both as failed and draw nothing from them. Removal is harder here: A3 clears in a minority of the fine-pruning cells and in none of the Blended cells (\cref{tab:c100}); no clean-arm number depends on A3.

\section{FT-SAM, SAU and I-BAU on the crossover cells}\label{app:defenses}
\begin{table}[h]
\caption{The scheduler crossover on CIFAR-10 / PreAct-ResNet18 for six defenses: clean-arm cost per defense (pt, $t$-intervals over seeds; negative is a ``gain''), the undefended twin, and the A3 row over A0 seeds. The shipped WaNet and BadNets cells use $11$ seeds and are the same cells as Table~\ref{tab:sched}; shipped BPP and the three swapped cells use their five pre-registered seeds for all six defenses (SCHED14's eleven-seed FT, FP and NAD swapped cells are in Table~\ref{tab:sched}, so those rows differ slightly). $^{\dagger}$: SAU on annealed BPP excludes one seed whose defended twin collapsed to \sqCollapsedAcc\% (post hoc, \cref{app:prereg}); with all five seeds it is \sqBcosSauAll, $n={}$\sqBcosSauAllN.}
\label{tab:defenses}
\begin{center}\resizebox{\textwidth}{!}{\input{supplement/sched/build/tab_sched4_body.tex}}\end{center}
\end{table}

A fourth pre-registration (\cref{app:prereg}) ran three defenses downstream papers compare against---FT-SAM~\citep{zhang2023ftsam}, SAU~\citep{wei2023sau}, i-BAU---on the same six cells with the benchmark's configurations (\exGpuFour{} GPU-hours; the shipped grids' SAU/i-BAU readings predate the pre-registration, which says so). 

\textbf{The flip is not a property of the three old defenses.} FT-SAM behaves as FT (\sqWshipFtsam{} to \sqWcosFtsam\pt{} on WaNet, \sqBnshipFtsam{} to \sqBnstepFtsam\pt{} on BadNets), with defended twins within \sqFtsamVsFt\pt{} of FT's. SAU moves at a different level: within a point of zero on the never-annealed twins (\sqWshipSau, \sqBnstepSau\pt), a cost of \sqWcosSau{} and \sqBnshipSau\pt{} on the annealed ones. i-BAU moves the same way without reversing sign: the never-annealed cells straddle zero (\sqWshipIbau, \sqBnstepIbau, \sqBshipIbau\pt) and the annealed ones cost \sqWcosIbau, \sqBnshipIbau{} and \sqBcosIbau\pt. The BPP cells replicate all three (\cref{tab:defenses}; SAU's annealed-BPP cell excludes one collapsed seed, post hoc, with the $n=5$ value \sqBcosSauAll\pt{} in the caption).

\begin{figure}[h]
\centering
\begin{minipage}[c]{0.66\textwidth}\centering
\includegraphics[width=\linewidth]{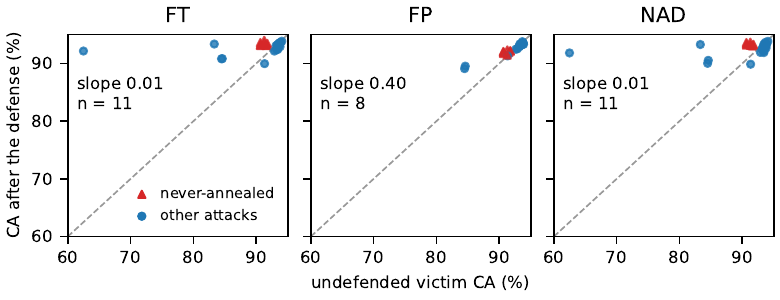}\\[-2pt]{\small (a)}
\end{minipage}\hfill
\begin{minipage}[c]{0.30\textwidth}\centering
\includegraphics[width=\linewidth]{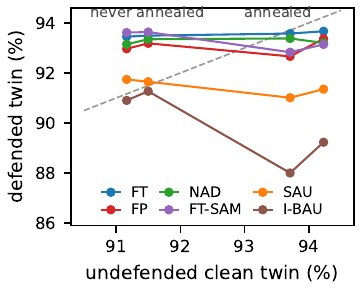}\\[-2pt]{\small (b)}
\end{minipage}
\caption{The level per defense. \textbf{(a)} Every attack at every rate on the public CIFAR-10 / PreAct-ResNet18 leaderboard (the site's \texttt{results.csv}): accuracy after FT, NAD and FP against before; slopes at $10\%$: \fpLbFtTenSlope\ and \fpLbNadTenSlope\ over all eleven attacks, \fpLbCommonFtSlope/\fpLbCommonNadSlope\ on the eight all three cover, against \fpLbHealthyFtSlope\ and \fpLbHealthyNadSlope\ on the healthy annealed five (grouped after reading the leaderboard). Triangles: the never-annealed three. \textbf{(b)} The same on the never-poisoned twins we trained, one line per defense over the four WaNet/BadNets cells: FT, NAD and FT-SAM return every twin to \sqFixTrioLo--\sqFixTrioHi\%, SAU to near \sqFixSauMean\%; I-BAU has no level of its own.}
\label{fig:levels}
\end{figure}

\textbf{Each defense has its own fixed point, and the cost is the distance to it.} Across the four WaNet/BadNets cells the undefended twins span \sqFixUndefRange\pt; after FT, NAD and FT-SAM the defended twins span \sqFixFtRange, \sqFixNadRange{} and \sqFixFtsamRange\pt{} near $93.5\%$, and after SAU \sqFixSauRange\pt{} near \sqFixSauMean\%---a fixed point too, but a lower one. It is not a missing anneal: plain fine-tuning at SAU's own $10^{-4}$ costs an annealed twin \ifsfive\sqvCleanEfiftyLrAWcos\pt{} over fifty epochs (\cref{app:budget})\else{}under a point\fi, so the \sqWcosSau\pt{} SAU costs it is the price of its adversarial unlearning, which happens to bring it to about where the never-annealed twins start. That is why SAU is free on those victims and costs two to three points on annealed ones: for SAU as for FT, the published cost is the victim's start minus the defense's level plus what the poison leaves at it. i-BAU has no fixed point (defended range \sqFixIbauRange\pt) and runs against the start: the never-annealed twins finish at \sqWshipIbauDefTwin{} and \sqBnstepIbauDefTwin\%, the annealed ones at \sqWcosIbauDefTwin{} and \sqBnshipIbauDefTwin\% (\cref{fig:levels}); its cost moves with the block by \sqIbauSwingWanet{} and \sqIbauSwingBadnet\pt, about twice what the block moves the start, so the fixed-point account explains at most half of that swing and the rest we measure and do not explain. The fixed-point statement of \cref{sec:bias} holds for FT, NAD, FT-SAM and SAU, fails for i-BAU and partly for fine-pruning; the sign reversal of \cref{sec:block} holds with both intervals excluding zero on all twelve CIFAR-10 pairs for FT, FP, NAD and FT-SAM and on the BadNets pair for SAU (\sqBnshipSau{} to \sqBnstepSau\pt), and on CIFAR-100 (\cref{tab:c100}) for fine-pruning on all three pairs and for FT and NAD on two; for SAU and i-BAU the block changes the size of the cost by two and a half to five and a half points (BPP's SAU swing rests on the $\dagger$ cell), not, in general, its sign. The methods of \cref{sec:downstream} beat SAU, FT-SAM and i-BAU as baselines: on a never-annealed victim SAU and i-BAU report near-zero cost, so a method that fine-tunes to its own level reports against them a gain that is the block, not the method.

\section{Where the milestones come from, and the block in a second toolkit}\label{app:toolkits}
\paragraph{Seven code bases.} \textbf{The attacks' own releases anneal; BackdoorBench kept their milestones and cut the run to
$100$ epochs.} Table~\ref{tab:toolkits} reads the victim's learning-rate schedule off each code base. The WaNet, Input-Aware and
BppAttack releases step a \texttt{MultiStepLR} at $[100,200,300,400]$ once per epoch over $1000$, $600$ and $1000$ epochs, so their
victims first decay at epoch $100$ and spend most of training annealed. BackdoorBench carries the same milestones into a
$100$-epoch run, where the first one would fire only after the last epoch. BackdoorBox's examples (WaNet, BadNets and Blended at $[150,180]$ of
$200$ epochs; Input-Aware at the release's milestones over $600$) and backdoor-toolbox ($[50,75]$ of $100$ on CIFAR-10) anneal.
TrojanZoo's documented attack recipe fine-tunes a pretrained model for $50$ epochs at a constant lr $0.01$: its scheduler is
opt-in (\texttt{--lr\_scheduler}, off by default), so its poisoning phase ends unannealed too. The coordinate is set three
ways across seven code bases, and none of the defense papers we audit reports it (Appendix~\ref{app:lit}).

\begin{table}[!ht]\centering\small
\caption{The victim's learning-rate schedule, read from each code base at the commit given ($\gamma=0.1$ throughout; the scheduler
steps once per epoch in every code base listed). ``Decays'': whether the rate falls before the last epoch of the run as shipped.}
\label{tab:toolkits}
\resizebox{\linewidth}{!}{\begin{tabular}{llcc}
\toprule
Code base (commit), attack & Victim schedule & Epochs & Decays \\
\midrule
WaNet release (\texttt{45e8c33}) & \texttt{MultiStepLR} $[100,200,300,400]$ & $1000$ & yes \\
Input-Aware release (\texttt{1750e4f}), classifier & \texttt{MultiStepLR} $[100,200,300,400]$ & $600$ & yes \\
BppAttack release (\texttt{f6b6580}) & \texttt{MultiStepLR} $[100,200,300,400]$ & $1000$ & yes \\
BackdoorBench (\texttt{f02e353}), WaNet, BPP, Input-Aware & \texttt{MultiStepLR} $[100,200,300,400]$ & $100$ & \textbf{no} \\
BackdoorBench (\texttt{f02e353}), LIRA & \texttt{MultiStepLR} $[50,100,150,200]$ & $100$ & yes \\
BackdoorBench (\texttt{f02e353}), the other twelve & prototype \texttt{CosineAnnealingLR} & $100$ & yes \\
BackdoorBox (\texttt{af3afd1c}), WaNet, BadNets, Blended & $[150,180]$ & $200$ & yes \\
BackdoorBox (\texttt{af3afd1c}), Input-Aware & $[100,200,300,400]$ & $600$ & yes \\
backdoor-toolbox (\texttt{9d4d909}), CIFAR-10 & \texttt{MultiStepLR} $[50,75]$ & $100$ & yes \\
TrojanZoo (\texttt{45d680c}), README attack recipe & none (opt-in), lr $0.01$ from a pretrained model & $50$ & \textbf{no} \\
\bottomrule
\end{tabular}}
\end{table}

\ifsseventeen
\paragraph{The block in BackdoorBox (three registrations).} \textbf{Cutting BackdoorBox's own WaNet run to $100$ epochs under its own
milestones reproduces the gain with none of BackdoorBench's code.} SCHED17 (\cref{app:prereg}) trains BackdoorBox's
WaNet, ResNet-18 and training loop (commit \texttt{af3afd1c}; only the package \texttt{\_\_init\_\_} files trimmed to skip modules
with missing dependencies) on CIFAR-10 with its example's settings (lr $0.1$, batch $128$, $100$ epochs), under N, its milestones
$[150,180]$ left in place so no decay fires, or A, milestones $[50,75]$; five seeds, a poisoned arm at $10\%$ and a twin with ten
poisoned images sharing its seed's warping grid. The undefended twin sits at \sbxNTwin\% under N and \sbxATwin\% under A, and the
poisoned victim at \sbxNPois\% and \sbxAPois\% (A0 \sbxNAzero\ and \sbxAAzero; A4 \sbxNAfour\ and \sbxAAfour, the one N twin above
$0.05$ at \sbxNTwinAsrMax, below K23's two). Defended by BackdoorBox's own FineTuning (full layers, its default lr $0.001$, $10$
epochs on a $10\%$ clean split), the twin's tare is \sbxFtNTare\ under N and \sbxFtATare\ under A. SCHED18, registered while
SCHED17's first runs were at epoch $16$, adds three BackdoorBox defenses on the same checkpoints: FineTuning at lr $0.01$
(\sbxFtoneNTare\ and \sbxFtoneATare), NAD (\sbxNadNTare\ and \sbxNadATare, the one sign reversal with both intervals excluding
zero) and Pruning, the registered negative control with no training phase, which charges both schedules, both intervals excluding zero
(\sbxPruneNTare\ and \sbxPruneATare; Table~\ref{tab:bbox}). \textbf{The poisoned arm shows the leaderboard's artifact in a second
toolkit:} BackdoorBox's defaults remove its WaNet in no cell (A3 \sbxAthreeAll\ over the four defenses and two schedules), yet on the never-decayed
victim FineTuning reports a published cost of \sbxFtNPub, an accuracy ``improvement'' of the kind \cref{sec:downstream}'s papers
report, and under A \sbxFtAPub. Predictions P57--P63 (scorecard in \cref{app:prereg}): \sbxHeldN\ of seven hold; P59 and P60
fail on their A halves, where fine-tuning at either rate leaves an annealed twin where it was rather than charging it.
\ifdefined\sbxPsixtyfour{}SCHED19 repeats both registrations on BackdoorBox's own BadNets ($3\times3$ patch, $5\%$, its example's trigger; \cref{tab:bbox}, lower block) and reads the same way: the twin starts \sbxBnTwinGap\ points lower under N; FineTuning's tare is \sbxBnFtNTare\ under N and \sbxBnFtATare\ under A, NAD's \sbxBnNadNTare\ and \sbxBnNadATare\ (the sign reversal again), Pruning's \sbxBnPruneNTare\ and \sbxBnPruneATare; A0 \sbxBnNAzero\ and \sbxBnAAzero, A4 \sbxBnNAfour\ and \sbxBnAAfour\ (the one N twin at \sbxBnNTwinAsrMax); the defaults remove BadNets in \sbxBnAthreeAll\ poisoned cells, Pruning's under N; \sbxBnHeldN\ of its five predictions hold, P65 failing on its A half as P60 did.\fi{}
\sbxGpuH\ GPU-hours.

\begin{table}[!ht]\centering\small
\caption{The scheduler block in BackdoorBox (SCHED17/SCHED18\ifdefined\sbxPsixtyfour; BadNets block: SCHED19\fi): BackdoorBox's WaNet\ifdefined\sbxPsixtyfour{} and BadNets\fi{} on CIFAR-10 / ResNet-18 under N (milestones
$[150,180]$ in a $100$-epoch run: never decays) or A ($[50,75]$), five seeds; tare and published cost in points with $t$-intervals
(negative is a ``gain''); defended twin: clean accuracy after the defense.}
\label{tab:bbox}
\resizebox{\linewidth}{!}{\input{supplement/sched/build/tab_sched17_body.tex}}
\end{table}
\fi

\ifsztwenty
\FloatBarrier
\section{Zero-image twins}\label{app:zero}
Registered as SCHED20 (Appendix~\ref{app:prereg}) before any run started. BackdoorBench refuses a zero poisoning rate (Section~\ref{sec:estimand}, Appendix~\ref{app:blended}); for this experiment only, the guard in \texttt{utils/backdoor\_generate\_poison\_index.py} is skipped when the environment variable \texttt{BB\_ALLOW\_ZERO\_POISON} is set, and a plotting helper that crashed on an epoch with no poisoned sample skips \texttt{None} entries (both recorded in \texttt{LOCAL\_CHANGES.md}); nothing else changes, so a pratio-$0$ run is the attack's own pipeline with an all-zero poison index (WaNet poisons per batch and draws zero). Three cells, five seeds each, the same seed, recipe and schedule as the paper's ten-image twins: BadNets under the shipped block (cosine), WaNet under the shipped block (never fires) and WaNet with the block removed (cosine); then FT, FP and NAD as everywhere else, and the ten-image twin of the same seed for comparison (Table~\ref{tab:zero}). Registered predictions: P69, the zero-minus-ten undefended accuracy has $|\Delta|\le0.3$ points and an interval covering zero in all three cells: \szrPsixtynine; P70, at least $8$ of the $9$ (cell, defense) tare differences have an interval covering zero and $|\Delta|\le0.5$: \szrPseventy; P71, all $15$ zero-image twins pass A4: \szrPseventyone. Kill lines: K29 (a one-epoch smoke test, crash only) did not trigger; K30 (the deadline, 2026-09-25 20:00 UTC+8) voided \szrKthirtyN\ cells; K16 (a twin losing more than $20$ points) applied to \szrKsixteenN. \szrGpuH\ GPU-hours; \szrRunsN\ runs\ifdefined\szrRcFailN, \szrRcFailN\ with a non-zero exit code\fi. Status: \szrStatus.
\begin{table}[h]
\caption{SCHED20: zero-image twins against the paper's ten-image twins of the same seed (five seeds per cell; ASR$_{\max}$: the largest attack success any zero-image twin scores, the A4 gate being $0.05$). Tare per defense on each twin, and the seed-paired difference with its $95\%$ $t$-interval.}
\label{tab:zero}
\centering\small
\resizebox{\linewidth}{!}{\input{supplement/sched/build/tab_sched20_body.tex}}
\end{table}
\fi
\ifsrtwentyone
\FloatBarrier
\section{The crossover at $1\%$ and $5\%$ poisoning}\label{app:rates}
Registered as SCHED21 (Appendix~\ref{app:prereg}) before any run started. The $2\times2$ of Section~\ref{sec:block} was run at the benchmark's $10\%$ rate; here the victims are re-trained at pratio $0.05$ and $0.01$, five seeds each, under the shipped block and under the swapped one (WaNet: the shipped \ms\ that never fires against \cosA; BadNets: the shipped \cosA\ against a \ms\ whose first milestone is epoch $100$), then FT, FP and NAD as everywhere else; a twin has no rate, so the tare is the paper's ten-image twin of the same seed and schedule, not re-run (Table~\ref{tab:rates}). Registered predictions: P72, the annealed victim starts at least one point higher in at least $4$ of $5$ seeds for each (attack, rate): \srtPseventytwo; P73, on WaNet the published cost of FT, FP and NAD is negative under the shipped block and higher under the swapped one, the difference's interval excluding zero, in at least $5$ of $6$ (defense, rate) pairs: \srtPseventythree; P74, the same on BadNets: \srtPseventyfour; P75, WaNet at $1\%$ installs (A0 in at least $3$ of $5$ seeds) under both schedules: \srtPseventyfive. Kill lines: K31 (A0 failing in $3$ or more of $5$ seeds voids that victim cell): WaNet at $1\%$ under the shipped block, \srtOneWshipGate; under the swapped block, \srtOneWcosGate; BadNets at $1\%$ under the shipped block, \srtOneBshipGate; under the swapped block, \srtOneBstepGate; every $5\%$ cell counted (A0 \srtFiveWshipAzero, \srtFiveWcosAzero, \srtFiveBshipAzero, \srtFiveBstepAzero). K32 (the deadline, 2026-09-25 20:00 UTC+8) voided \srtKthirtytwoN\ cells. A prediction whose test arm a kill line voids counts as failed, so the four predictions, each spanning both rates, are recorded as failed although their $5\%$ halves hold (P72 \srtPseventytwoN, P73 \srtPseventythreeN, P74 \srtPseventyfourN). A0 over all victims \srtAzeroAll, A3 over all defended victims \srtAthreeAll; \srtVictimsN\ victims, \srtGpuH\ GPU-hours. Status: \srtStatus.
\begin{sidewaystable}
\caption{SCHED21: the scheduler crossover at $5\%$ and $1\%$ poisoning. Per (attack, rate, schedule): A0 (seeds installing), the undefended victim's accuracy, the published cost of FT, FP and NAD (mean over A0 seeds, $95\%$ $t$-interval) and the tare on the same-seed ten-image twin; the third line of each block is the annealed-minus-not-annealed difference.}
\label{tab:rates}
\centering\small
\resizebox{\textheight}{!}{\input{supplement/sched/build/tab_sched21_body.tex}}
\end{sidewaystable}
\fi
\ifsdptwentytwo
\FloatBarrier
\section{ANP under a deployer-side threshold rule}\label{app:anp}
Registered as SCHED22 (Appendix~\ref{app:prereg}), an analysis-only registration written before the analysis ran. BackdoorBench's ANP scans pruning thresholds and keeps the one with the smallest \emph{test-set} attack success within a clean-accuracy budget ($|Q-Q_{\mathrm{ori}}|/Q_{\mathrm{ori}}<0.1$; \texttt{defense/anp.py}), an oracle rule the paper discloses at every ANP cell ($^{\dagger}$). Every ANP run in the archive saved its full sweep (\texttt{threshold\_df.csv}), so each cell can be re-read under a rule that never sees attack success: \emph{rule D}, the largest threshold (the most pruning) whose clean accuracy stays within the same budget. Rule O, the framework's, is reproduced from the sweeps in \sdpAssertN\ runs to $10^{-6}$ (the sweep's row $0$ is not the unpruned model---neurons whose mask prints as $0.0000$ are already pruned there---so rule O takes the unpruned accuracy from \texttt{attack\_df\_summary}, as the framework does; rule D keeps row $0$ as registered, and a variant on the framework's budget changes no verdict). Registered predictions: P80, BadNets tare under rule D positive with interval excluding zero: \sdpPeighty\ (\sdpBadnetTareD); P81, the two rules' BadNets tares within one point: \sdpPeightyone\ (D $-$ O \sdpBadnetTareDiff); P82, WaNet tare of the same sign under both rules: \sdpPeightytwo\ (O \sdpWanetTareO, D \sdpWanetTareD). The mechanism is visible in the thresholds: on the poisoned arm the two rules pick the same threshold in \sdpSameThrPois\ runs, on the twin arm in \sdpSameThrTwin---on a model with no backdoor the oracle rule picks the minimum of attack-success noise and often barely prunes, so it makes ANP look cheap on the twin; the paired contrast on WaNet moves from \sdpWanetPairedO\ to \sdpWanetPairedD. Of the seven other attacks, five have a higher tare under rule D and two (Trojan, SIG) the same threshold under both rules (Table~\ref{tab:anp}); the threshold counts run over the poisoned and twin arms of the registered and descriptive cells and of the dose-response cells that share a twin; runs without a sweep at collection time (\sdpSweepMissingN) are not counted.
\begin{table}[h]
\caption{SCHED22: ANP's tare under the framework's threshold rule (O: smallest test-set attack success within the budget) and under a deployer-side rule (D: most pruning within the budget, attack success unseen), seed-paired, $95\%$ $t$-intervals; A3 under each rule; $n$ seeds. The two registered cells first, the other Table~\ref{tab:corrmain} attacks descriptive.}
\label{tab:anp}
\centering\small
\resizebox{\linewidth}{!}{\input{supplement/sched/build/tab_sched22_body.tex}}
\end{table}
\fi
\ifsedtwentythree
\FloatBarrier
\section{The eight-defense crossover}\label{app:eight}
Registered as SCHED23 (Appendix~\ref{app:prereg}) before any run started; no new victim or twin was trained. The defenses missing from the swapped and shipped arms of the crossover (NC, ANP, ABL on WaNet swapped and BadNets swapped; six on BPP swapped) were run on the existing arms so that WaNet ($11$ seeds), BadNets (swapped: $5$) and BPP ($5$) carry all eight BackdoorBench defenses under both schedules (Table~\ref{tab:eight}). Registered predictions: P76, ANP's tare on WaNet positive under both schedules: \sedPseventysix; P77, ABL destroys most twins under both schedules (\sedPseventysevenCounts): \sedPseventyseven; P78, NC's detector fires on at most half of the cosine WaNet twins (\sedPseventyeightFire) and its repair costs them a positive amount where it fires (\sedPseventyeightCond): \sedPseventyeight; P79, the benchmark's own DER ordering of the eight defenses on WaNet changes between the shipped and the patched victim in at least one (fine-tuning-family, other) pair: \sedPseventynine\ (\sedPseventynineFlipList; Table~\ref{tab:eightder}). Kill lines: K33 voided \sedKthirtythreeN\ (cell, defense) pairs; K34 (the deadline, 2026-09-26 09:00 UTC+8) voided \sedKthirtyfourN; K16 applied to \sedKsixteenN\ non-ABL twins (\sedKsixteenList), excluded as failure records. \sedNewRunsN\ registered runs plus the make-up runs of FT-SAM, I-BAU and SAU on six WaNet-cosine seeds the registration wrongly assumed present (its outcome record), \sedGpuH\ GPU-hours. Status: \sedStatus.
\begin{table}[h]
\caption{SCHED23: the tare of all eight defenses on each arm of the crossover (mean over seeds, $95\%$ $t$-interval; WaNet $11$ seeds, BadNets swapped and BPP $5$; $^{\dagger}$ policy-level A2; $^{\natural}$ ABL is a failure record, $\flat k/n$ twins destroyed; NC printed as twins fired $|$ cost where fired; --: not harvested by the deadline).}
\label{tab:eight}
\centering\small
\resizebox{\linewidth}{!}{\input{supplement/sched/build/tab_sched23_body.tex}}
\end{table}
\begin{table}[h]
\caption{SCHED23: the benchmark's DER (victim-referenced, clipped; mean over the A0 seeds harvested, which needs only the poisoned arm) for the eight defenses on the shipped and the patched victims, with the rank in parentheses; --: not harvested by the deadline.}
\label{tab:eightder}
\centering\small
\resizebox{\linewidth}{!}{\input{supplement/sched/build/tab_sched23_der_body.tex}}
\end{table}
\fi
\section{Sources and hashes for the leaderboard analysis}
\label{app:lbsources}
The under-convergence gap per (dataset, backbone) is Table~\ref{tab:fourds} in the main text.

The rendered CIFAR-10 leaderboard (S1), the site's previous and current \yaml{results.csv} (S2, S3), Tables 3/4/13 of the 2022 paper (S4) and Table III of the 2025 paper (S5) were downloaded on 2026-09-12; SHA-256 sums are in the supplement. All runs use BackdoorBench at commit \texttt{f02e353} (2025-07-04) with seven local changes, listed in \texttt{supplement/sched/LOCAL\_CHANGES.md}: six NumPy-2 compatibility renames (\texttt{np.round\_}, \texttt{np.infty}; \texttt{np.float} in the patch transform and in the Trojan attack, made for SCHED10; \texttt{np.float} in the FTrojanNN transform and \texttt{np.lib.pad} in the Refool attack, made for SCHED15) and a label-transform option no run behind this paper selects; none alters a number in this paper. We do not know the commit or seed behind the public CIFAR-100 WaNet entry, whose ASR ($0.977$) we could not reproduce under the shipped configuration (\cref{app:c100}). S1 lists ten attacks on CIFAR-10, eight at $10\%$ (LC and SIG have no entry there) and carries BPP; S3 lacks BPP at $10\%$ and carries LIRA, Blind, LC and SIG. The sign counts, permutation tests and the $13/15$ vs.\ $0/25$ DER count of \cref{sec:leaderboard} use S1; \cref{fig:level} and \cref{app:fplb}. S2 and S3 agree with S1 in $226$ of $252$ CIFAR-10 / PreAct-ResNet18 cells; the $26$ mismatches are LC cells re-run for the current site and one LF/ABL cell at $0.5\%$, none in a cell used in \cref{sec:leaderboard}. \Cref{tab:fourds} excludes the $10\%$ cells because LC and SIG are annealed but heavily damaged at that rate, which was stated before the four-dataset table was computed.

\section{Leaderboard fixed point, all rates}
\label{app:fplb}
For each defense and poisoning rate on CIFAR-10 and CIFAR-100 / PreAct-ResNet18, the number of attacks with a public entry, the least-squares slope of after-defense accuracy on before-defense accuracy across those attacks, and the two ranges. Entries recorded as $-100$ (failed runs) are dropped. The last three columns restrict the $10\%$ CIFAR-10 cells to the eight attacks all three defenses cover, to the ten attacks without the LIRA victim (which alone spans most of the undefended range), and to the healthy annealed attacks (neither never-annealed nor damaged by their own poisoning: \fpLbHealthyFtN\ attacks spanning \fpLbHealthyFtUndef\ points). On that last subset the slope is \fpLbHealthyFtSlope\ (SE \fpLbHealthyFtSlopeSe) for FT and \fpLbHealthyNadSlope\ (SE \fpLbHealthyNadSlopeSe) for NAD, and the defended spread (\fpLbHealthyFtDef\ points under FT) is not smaller than the undefended one (\fpLbHealthyFtUndefTwo): on healthy annealed victims the defense leaves the start near where it was (the healthy/damaged split was made after reading the leaderboard). The slope on this subset leans on BadNets, its lowest start: without it the other four span \fpLbHealthyFtUndefNoBn\ points and the slopes are \fpLbHealthyFtSlopeNoBn\ (FT) and \fpLbHealthyNadSlopeNoBn\ (NAD), too narrow a spread to pin a slope. The fixed point is a property of the victims that start low, and the main text says so wherever it is stated; the twin-arm evidence of \cref{sec:bias} and \cref{fig:level} (four cells \sqFixUndefRange\ points apart returned to within \sqFixFtRange\ by FT) is its direct measurement. Over every (dataset, backbone, poisoning-rate) cell with at least four attacks outside GTSRB, the compression criterion ($|\text{slope}|<0.25$ and a defended range under a third of the undefended one) holds in \fpLbCellsFt\ FT, \fpLbCellsNad\ NAD and \fpLbCellsFp\ fine-pruning cells; it can also fail where the victims start close together, since a third of a small range is within seed-to-seed noise.
\begin{center}\resizebox{\textwidth}{!}{\input{supplement/sched/build/tab_fixedpoint_lb_body.tex}}\end{center}
\paragraph{Grouping sensitivity (post hoc).} The healthy annealed five of \cref{fig:level} were grouped after reading the leaderboard; any cutoff on undefended accuracy between \lbHealthyCutLo\ and \lbHealthyCutHi\% selects the same five. Over every annealed attack at $10\%$ on the \yaml{results.csv} roster, the damaged four (SIG, LC, LIRA, Blind) included, the FT and NAD slopes are \lbSlopeFtAllAnnThree\ and \lbSlopeNadAllAnnThree: those four start low and are returned to the level as the never-annealed victims are, so the healthy-five contrast is a statement about victims that already start at the level, not about annealing as such; the clip then fires in \lbClipAllAnn\ annealed (attack, defense) cells on that roster, none of them a healthy victim's (the rendered site's own $10\%$ page carries no damaged attack, hence its \lbDerClipAnn).

\section{The corrected reference}
\label{app:lbreference}
Removing the clip from its accuracy term is not enough: $-\Delta Q$ without the $\max(0,\cdot)$ would reward a negative cost rather than expose it---on the shipped WaNet grid (A0-gated published cost \cschWmsPubPool\pt; \schWmsPubPool{} A3-conditioned in \cref{tab:schedcols}) the unclipped term adds \cschWmsUnclipDer{} DER points, where the clean-arm cost of \cschWcosPool\pt{} would subtract \cschWcosCleanDerHalf. The clip hides the artifact; unclipping alone amplifies it in the headline number---a nonlinearity manufacturing a qualitative claim in either direction~\citep{schaeffer2023emergent}. DER's accuracy term should instead be referenced to a clean model trained from the prototype recipe---one per (dataset, backbone, seed), sharing no scheduler key with any attack---as $-[\Q(\theta_{\rm clean})-\Q(D\theta_{\rm pois})]$. This reference does not depend on the victim's schedule, so a never-annealed victim can no longer be booked as a gain, and the defense's own cost on a clean model stays in the score: i-BAU's \sqWcosIbau{} and SAU's \sqWcosSau\pt{} on annealed twins (\cref{tab:defenses}) are paid by every model those defenses touch, and a defended-clean reference $-[\Q(D\theta_{\rm clean})-\Q(D\theta_{\rm pois})]$ would cancel them (that bracket is the published cost minus the clean-arm cost plus the start gap). With the prototype reference (here the same seed's BadNets-cosine twin) the bracket is \cschWmsProtoRef{} shipped and \cschWcosProtoRef\pt{} annealed on WaNet, \sxBshipOneProtoRef{} and \sxBcosOneProtoRef\pt{} on BPP---within half a point across schedules, a non-rejection at five seeds---and \cschBcosProtoRef\pt{} on BadNets with cosine, \cschBmsProtoRef\pt$^\ddagger$ with the block. The prototype clean model serves every attack, including Input-Aware; where the fixed point fails (i-BAU, fine-pruning on CIFAR-100) the per-attack twin remains the right flag. $\Q(D\theta_{\rm clean})$ and the signed, unclipped $\Delta Q$ belong beside the score as flag columns (the clean-arm cost, and the twin-referenced bracket of \cref{tab:schedcols}), so that a reader can tell the three parts of the cost apart while the score that gets copied keeps all three.

\section{Leaderboard re-ranking under the corrected reference}
\label{app:rerank}
For each defense with public PreAct-ResNet18 entries, mean DER over the attacks shared by all defenses in the panel (CIFAR-10: every defense with at least five entries at $10\%$, then the fine-tuning family at $10\%$ and $5\%$; CIFAR-100 (C100): every defense with at least four entries, prototype level \rkLbHundQclean\%, the mean $0.1\%$-poisoning accuracy of the healthy annealed attacks), under the benchmark's DER and under a DER whose accuracy term is referenced to the benchmark's clean PreAct-ResNet18 model ($93.65\%$, \citealp{wu2024backdoorbench}; the $93.9\%$ of the 2022 paper gives the same order). Original order at $10\%$: \rkLbTenOrder; prototype-referenced: \rkLbTenOrderProto.
\begin{center}\small\input{supplement/sched/build/tab_rerank_lb_body.tex}\end{center}

\section{Leaderboard cost table}
\label{app:lbcost}
Per-(attack, defense) $\Delta Q$ at $10\%$ and $5\%$ poisoning, with sign counts by group and exact permutation tests, are printed by \yaml{leaderboard/leaderboard_artifact_test.py}; the markdown report it writes is included verbatim in the supplement.

\section{Pre-registrations}
\label{app:prereg}
All \preregWord{} reported pre-registrations are included verbatim (for double-blind review, zone labels, the GPU model name and one host path in the scheduler study's files were normalized after registration, to ``UTC+8'', ``GPU'' and ``$<$server$>$''; no other text changed)---the scheduler study's (\texttt{supplement/sched/prereg}; SCHED3b as an addendum inside the SCHED3 file), the eleven-seed WaNet grid's (\texttt{supplement/prereg}) and the ranking protocol (\texttt{supplement/configs})---with timestamps, predictions with priors and kill lines; the timestamps are those typed in each file, and for SCHED11 onward also our file system's record of when the file was written, which the supplement's zip (modification times only) does not carry; none is an external registry, and one arm of the CIFAR-10 crossover was already known when it was written (stated below). Files: the CIFAR-10 crossover (WaNet $\times$ BadNets); the BPP / Input-Aware extension; the CIFAR-100 crossover (WaNet, with Blended as the annealed control); and its BPP addendum (SCHED3b, at the poisoning rate the leaderboard reports for BPP on CIFAR-100); and SCHED4, the three further defenses on the existing records (\cref{app:defenses}), whose one deviation is a single SAU run (BPP cosine, seed 4) whose defended twin collapsed to \sqCollapsedAcc\% and is excluded as a failed run---a post-hoc judgement, not a pre-registered rule, so the $\dagger$ note of the \cref{tab:defenses} caption gives the interval with it included (\sqBcosSauAll, $n={}$\sqBcosSauAllN) next to the $n=4$ interval printed in the cell. Every deviation is logged in the file it belongs to and stated in the text (one exception: the SCHED9 and SCHED10 header-time corrections, stated only below): the BPP twin reset from ten images to one on CIFAR-10 (prescribed by SCHED2's K3, \cref{app:twin}); rule K3, registered before any CIFAR-100 run as voiding a whole cell whose attack installs in a minority of seeds, and its disposition, written after seed 0 of the shipped WaNet cell returned a low installation rate and before seeds 1--4 existed: void the poisoned arm and report the clean arm as a declared deviation; the failed predictions P6 and P7 (published cost and interaction invariance of the K3-voided CIFAR-100 cell) and P11 (four-cell fixed point on CIFAR-100), reported as failed in \cref{app:c100}; and the one-image twin for Blended, which follows the convention established for BPP on CIFAR-10. Six decisions were taken after data existed. Two are rules written before their outcome was determined: the BPP twin reset (before the one-image twins were run) and K3's disposition (after seed 0 of the shipped CIFAR-100 WaNet cell returned an ASR below A0, before seeds 1--4, whose installation decided whether it fired; timestamps in the SCHED3 file). The third was taken after seeing its outcome: the exclusion of the SAU run on the BPP-cosine twin of seed 4, implemented in \texttt{build\_sched4.py} as ``a defended twin that loses more than $20$\pt{} is a failed run''; the threshold was written after the run had been seen, it affects only that twin (the seed's poisoned arm is kept), and it is the only reason the SAU cell for annealed BPP reads \sqBcosSau{} ($n=4$) rather than \sqBcosSauAll{} ($n=5$). Every sentence that leans on that cell is marked $\dagger$. The fourth and fifth were also taken after their data: TSBD's operating point, fixed at the paper's printed default after the first harvest had taken the sweep's last value (\cref{app:tsbd}), and SIG's twin, left unreset after it failed A4 although SCHED9's K14 allowed one one-image reset (\cref{app:sig}). No other number in the tables was chosen after seeing data; the sixth decision was a grouping: Figure~\ref{fig:level}'s healthy/damaged split of the leaderboard's annealed attacks (\cref{app:fplb}). The eleven-seed shipped grids and their SAU / i-BAU readings predate the pre-registrations that use them, which state so: the CIFAR-10 crossover and SCHED4 were pre-registered with one arm already known, SCHED2, SCHED3 and SCHED3b with both arms unrun\ifsthirteen, and SCHED13 reuses three seed-0 BadNets victims (at $1$, $5$ and $20\%$, with their FT and NAD runs) that predate it, as its file states; the $5\%$ one is the seed that fails A0 there\fi. The eleven-seed shipped grids predate the pre-registrations (they were run for an earlier study); the admissibility thresholds (A0 $0.90$, A3/A4 $0.05$) and the ten-image twin convention were fixed with them. \ifsfive SCHED5 (the FT budget sweep, \cref{app:budget})\ifssix, \else{} and \fi\fi\ifssix SCHED6 (the VGG19-BN crossover, \cref{app:vgg})\ifsseven{} and \fi\fi\ifsseven SCHED7 (TSBD\ifstcgd{} and CGD\fi{} measured in the WaNet $2\times2$, \cref{app:tsbd})\fi\ifsfive{} were added in that order, each written before its runs\else\ifssix{} were added in that order, each written before its runs\else\ifsseven{} was added, written before its runs\fi\fi\fi. SCHED9 (SIG) and SCHED10 onward followed, each written before its new runs. For SCHED11--SCHED15 the evidence is the file system's write time, which precedes each launcher's first run (their typed header times are $4$ to $17$ minutes later and are not the evidence)\ifssixteen; SCHED16's file was written at 01:34 UTC on 2026-09-23, the minute its header states, before its launcher\fi\ifsseventeen; SCHED17's at 03:02 UTC on 2026-09-24, six minutes before its launcher started, and SCHED18's and SCHED19's at 03:14 and 03:35 UTC, before SCHED17's first readout at 03:39 (SCHED18's header time was corrected by two minutes at 03:15, recorded in its file)\fi; for SCHED9 and SCHED10 it is only our command log, not in the supplement, which shows each file written seconds and three minutes before its first run, while their typed header times, corrected after launch, read $1$ and $12$ minutes earlier than the files were written. One exception to ``before its new runs'': seven of SCHED11's seed-0 BadNets cells predate it---CLP and NPD on the poisoned arm are runs of 2026-08-28 under the same configuration, and the five defenses on the twin come from the timing smoke test whose RNP and D-BR readings P39's prior quotes. One deviation: D-ST's registered override (adding only \texttt{client\_optimizer}) failed on its first arm, which under SCHED11 drops D-ST; we then widened the override to the fine-tuning defaults and ran it twice on the seed-0 BadNets twin (a trial stopped at epoch 13 and a launcher run stopped by hand at epoch 115 of its 200, about $2.1$ GPU-hours in all) before dropping it for cost; no D-ST number is printed. SCHED15 was registered as adding two leaderboard attacks; FTrojanNN and Refool are BackdoorBench attacks the public leaderboard does not list, and the paper says so. \ifssixteen\else A further registration, SCHED16 (the dose fall of \cref{app:bddose} registered for Blended and LF), is included as well; it had not completed at submission and none of its numbers is printed. \fi A prediction whose test arm a kill line voids is counted as failed (SCHED2's P3, P4 and the Input-Aware half of P5; SCHED3's P1, P3 and P4 under K3 as registered, which held as its declared deviation; P6; P7; P45; P46's ratio clause; P47). The scorecard below covers the scheduler study. The ranking protocol's three criteria are scored in \cref{app:bb} (criterion (i) fails), and the eleven-seed WaNet grid's three predictions all held: P1, its A3-passing published cost negative with the interval excluding zero (\chbbWNDef, \cref{tab:bb}'s convention); P2, no reordering (\chRevwnRev{} of \chRevwnN{} testable pairs reverse, \chRevwnSur{} survives, \cref{app:bb}); P3, ABL failing A3 on most seeds (it clears \exWanetAblAthree). The scheduler scorecard:

\begin{center}\footnotesize
\begin{tabular}{p{1.7cm}p{5.3cm}p{1.2cm}p{3.8cm}}
\toprule
pre-registration & prediction & prior & outcome \\
\midrule
CIFAR-10 $2\times2$ & P1 WaNet clean arm reverses with cosine; P2 BadNets reverses with the block & 0.75, 0.70 & both held (\cref{tab:xover}) \\
 & P3 the annealed WaNet twin rises toward BadNets' & 0.85 & \checkmark{} (\schWmsTwin$\to$\schWcosTwin) \\
BPP / Input-Aware & P1--P2 BPP flips both ways; P3--P4 Input-Aware flips; P5 shipped twins $\ge1.5$\pt{} lower & 0.65, 0.70, 0.55, 0.55, 0.75 & P1, P2 and P5's BPP half held (twin gap \exBppTwinGap\pt); P3, P4 and P5's Input-Aware half $\times$: untestable, SCHED2's K3 (a twin failing A4 is reset; failing again voids the attack) voided Input-Aware's clean arm (its ten-image twin fails A4 and its one-image reset diverges in training); its published cost moved in the predicted direction \\
CIFAR-100 (WaNet, Blended) & P1 shipped WaNet clean arm negative & 0.75 & $\times$ as registered (K3 voids the cell); \syPone{} as the declared deviation \\
 & P2 cosine WaNet not significantly negative & 0.70 & \syPtwo \\
 & P3 undefended twin moves $\ge4.0$\pt & 0.60 & $\times$ as registered (K3); \syPthree{} as the declared deviation (\syWanetTwinGap) \\
 & P4 fixed point, WaNet cells & 0.65 & $\times$ as registered (K3); \syPfour{} as the declared deviation \\
 & P5 Blended reverses when given the block & 0.65 & \syPfive \\
 & P6 shipped WaNet published cost negative & 0.75 & $\times$: untestable, K3 voids its test arm (attack installed in 1/5 seeds; on all five seeds it would have \syPsixAll, \cref{tab:c100}) \\
 & P7 interaction invariant ($<0.5$\pt) & 0.55 & $\times$: untestable, K3 voids its shipped side (on all five seeds it would have \syPsevenAll) \\
CIFAR-100 (BPP addendum) & P8 shipped BPP clean arm negative & 0.70 & \syPeight \\
 & P9 cosine BPP not negative & 0.65 & \syPnine \\
 & P10 twin moves $\ge4.0$\pt & 0.60 & \syPten{} (\syBppTwinGap) \\
  & P11 four-cell fixed point (all three defenses) & 0.60 & \syPeleven{} (fine-pruning) \\
SCHED4 (FT-SAM, SAU, i-BAU) & P12 FT-SAM flips as FT; defended twin within $1$\pt{} of FT's & 0.75 & \sqPtwelve{} (\sqFtsamVsFt\pt) \\
 & P13 SAU flips; four-cell fixed point & 0.65 & \sqPthirteen{} as registered (cosine WaNet \sqWcosSau; BadNets+block \sqBnstepSau); strict two-sided reversal on BadNets only---on WaNet and BPP the never-annealed interval straddles zero \\
 & P14 i-BAU moves in the predicted direction (shipped$\to$swapped) & 0.55 & \sqPfourteen{} (direction only; no sign change) \\
 & P15 published WaNet cost rises with cosine for all three & 0.70 & \sqPfifteen \\
\bottomrule
\end{tabular}
\end{center}
\newif\iflate\latefalse\ifsfive\latetrue\fi\ifssix\latetrue\fi\ifsseven\latetrue\fi\ifsnine\latetrue\fi\ifsten\latetrue\fi
\iflate
\begin{center}\footnotesize
\begin{tabular}{p{1.7cm}p{5.3cm}p{1.2cm}p{3.8cm}}
\toprule
pre-registration & prediction & prior & outcome \\
\midrule
\ifsfive SCHED5 (FT budget) & P16 defended-twin range shrinks monotonically with budget; $\ge1.5$\pt{} at $5$ epochs of $10^{-2}$ and at $50$ of $10^{-4}$ & 0.70 & \sqvPsixteen{} (monotone, but \sqvRangeEfiveLrC{} and \sqvRangeEfiftyLrA\pt: the fixed point forms faster than predicted) \\
 & P17 a budget $B^*$ exists above which the range is $<1.0$\pt; predicted $20$--$50$ epochs at $10^{-2}$ & 0.60 & \sqvPseventeen{} as registered: $B^*$ exists (\sqvPseventeenExist) but is $\le$\sqvBstar{} epochs, the smallest budget run, against the predicted $20$--$50$ \\
 & P18 clean-arm sign reversal at every $10^{-2}$ budget $\ge10$ epochs, all four intervals excluding zero; none at $10^{-4}$ & 0.60 & \sqvPeighteen{} as registered (at $50$ epochs the BadNets-cosine interval is \sqvCleanEfiftyLrCBnship); both schedule contrasts reverse at every $10^{-2}$ budget $\ge10$, none at $10^{-4}$ \\
 & P19 A3 fails for most poisoned arms at $5$ epochs of $10^{-4}$ & 0.70 & \sqvPnineteen{} (\sqvAthreeShort) \\
\fi\ifssix SCHED6 (VGG19-BN) & P20 clean arm negative shipped, positive cosine (FT, NAD), intervals excluding zero & 0.70 & \svPtwenty{} (FT \svVshipFt{} / \svVcosFt) \\
 & P21 twin swing $\ge2.0$\pt & 0.65 & \svPtwentyone{} (\svTwinGap) \\
 & P22 FT/NAD defended twins within a third of the swing & 0.60 & \svPtwentytwo{} (\svFixMax\pt) \\
\fi\ifsseven SCHED7 (TSBD\ifstcgd, CGD\fi) & P23 TSBD gains on the shipped victim; cost $\ge0$ or straddles zero on cosine & 0.70 & \stPtwentythree{} (\stTsbdWshipPub, \stTsbdWcosPub) \\
 & P24 twin ``gain'' $\ge1.0$\pt{} shipped; cost $\ge0$ cosine & 0.65 & \stPtwentyfour{} (\stTsbdWshipClean, \stTsbdWcosClean) \\
 & P25 defended twins within a third of the twin swing & 0.50 & \stPtwentyfive{} (\stTsbdFixRange\pt{} against \stTsbdSwingThird, a third of the five-seed swing \stTsbdTwinSwing) \\
\ifstcgd & P26 CGD gains on the shipped victim; cost $\ge-1.0$ or straddles zero on cosine & 0.65 & \stPtwentysix{} (\stCgdWshipPub, \stCgdWcosPub) \\
 & P27 CGD twin ``gain'' $\ge1.0$\pt{} shipped & 0.60 & \stPtwentyseven{} (\stCgdWshipClean) \\
\fi\fi
\ifsnine SCHED9 (SIG) & P28 damaged annealed victim starts $\ge5$\pt{} below its twin, A0 holds & 0.75 & \sgPtwentyeight{} (\sgGap) \\
 & P29 FT and NAD published cost negative, intervals excluding zero & 0.80 & \sgPtwentynine{} (\sgFtPub) \\
 & P30 defended victim within $1$\pt{} of the defended twin (FT, NAD) & 0.55 & \sgPthirty{} (\sgFtFixGap\pt) \\
 & P31 twin's clean-arm cost positive (FT, NAD) & 0.70 & \sgPthirtyone{} (FT \sgFtClean; twins fail A4, \cref{app:sig}) \\
 & P32 DER's clip fires on the cell & 0.80 & \sgPthirtytwo \\
\fi
\ifsten SCHED10 (tare column for the leaderboard) & P33 LF and Trojan: $|g|<1.5$\pt, FT/NAD clean arm positive with intervals excluding zero & 0.70 & \stePthirtythree{} as registered (LF FT \steLfFtClean; Trojan A4 \steTrojanAfour) \\
 & P34 by mean tare, FT cheapest and ABL, I-BAU the two most expensive & 0.60 & \stePthirtyfour{} as registered over all seven attacks, where ANP and I-BAU swap (\steAnpIbauGapAllReg\pt{} apart); \stePthirtyfourAdmit{} on the A4-passing twins: \steCleanOrderReg \\
 & P35 prototype-referenced DER flips $\le4/28$ pairs, all near-ties & 0.60 & \stePthirtyfive{} (\steDerFlipsReg) \\
 & P36 on SIG and BPP at least 3 of the 5 added defenses share FT's tare sign & 0.65 & \stePthirtysix{} (SIG \steSigNewShare, BPP \steBppNewShare\ share fine-tuning's sign) \\
\fi
\bottomrule
\end{tabular}
\end{center}
\begin{center}\footnotesize
\begin{tabular}{p{1.7cm}p{5.3cm}p{1.2cm}p{3.8cm}}
\toprule
pre-registration & prediction & prior & outcome \\
\midrule
\ifseleven SCHED11 (five further defenses, \cref{app:corrected}) & P37 CLP and NPD tare $>0$ on all seven attacks; on WaNet and BPP intervals excluding zero in $\ge3/4$ & 0.70 & \selPthirtyseven{} (\selPthirtysevenExcl) \\
 & P38 MCR $|$tare$|<0.5$\pt{} on annealed and $<1$\pt{} on never-annealed victims & 0.50 & \selPthirtyeight{} as registered (WaNet \selWanetMcrCleanPt) \\
 & P39 RNP and D-BR destroy $\ge$ half their twins & 0.60 & \selPthirtynine{} as registered (D-BR \selDbrTwinCollapseReg, RNP \selRnpTwinCollapseReg) \\
 & P40 prototype-referenced DER flips $\le10\%$ of pairs, near-ties only & 0.60 & \selPforty{} on the seven registered attacks (\selDerFlipsReg; \selDerFlips\ on all nine) \\
 & P41 FT cheapest by mean tare, CLP between FT and FP & 0.55 & \selPfortyone{} as registered: \selCleanOrderReg \\
\fi
\ifstwelve SCHED12 (healthy start moved by training length, \cref{app:hstart}) & P42 the 30-epoch twin starts $\ge1$\pt{} below the 100-epoch twin & 0.65 & \shsPfortytwo{} (\shsStartGap\pt) \\
 & P43 after FT and NAD the defended twins' range $\le$ a third of the undefended range, slope $\le0.35$; FP slope $\ge0.5$ & 0.60 & \shsPfortythree{} as registered: slopes FT \shsFtSlope, NAD \shsNadSlope, FP \shsFpSlope; ranges \shsFtDefRange, \shsNadDefRange, \shsFpDefRange\pt{} against a third of \shsUndefRange, \shsUndefThird \\
 & P44 FT's published cost $\ge1$\pt{} lower on the 30-epoch victim & 0.60 & \shsPfortyfour{} (\shsThirtyFtPub\ vs \shsHundredFtPub) \\
\fi
\bottomrule
\end{tabular}
\end{center}
\begin{center}\footnotesize
\begin{tabular}{p{1.7cm}p{5.3cm}p{1.2cm}p{3.8cm}}
\toprule
pre-registration & prediction & prior & outcome \\
\midrule
\ifsthirteen SCHED13 (BadNets dose, \cref{app:bddose}) & P45 the starting gap $g$ rises with the rate, Spearman $\rho\ge0.8$ over the five rates & 0.65 & \sdrPfortyfive: \sdrPfortyfiveNote \\
 & P46 FT, NAD and FP paired contrast rises, $\rho\ge0.8$; the $20\%$ paired $\ge2\times$ the $1\%$ & 0.60 & \sdrPfortysix{} as registered: the rise \sdrPfortysixRise{} (\sdrRhoPairedFt, \sdrRhoPairedNad, \sdrRhoPairedFp); ratio clause \sdrPfortysixRatio \\
 & P47 at $\le1\%$ FT and NAD published cost within $0.5$\pt{} of the tare & 0.55 & \sdrPfortyseven: \sdrPfortysevenNote{} (A0 \sdrHalfAzero, \sdrOneAzero) \\
\fi
SCHED14 (swapped cells to eleven seeds) & P48 eleven-seed pooled tare inside the five-seed interval, same sign & 0.75 & \schPfortyeight{} (W-cos \schWcosPool\ vs \schWcosPoolFive; B-step \schBmsPool\ vs \schBmsPoolFive) \\
 & P49 interval half-width shrinks $\ge30\%$; $\ge95\%$ of cells keep the sign & 0.60 & \schPfortynine{} as registered (half-width W-cos \schWcosHalfFive$\to$\schWcosHalfEleven, \schWcosHalfChange; B-step \schBmsHalfFive$\to$\schBmsHalfEleven, \schBmsHalfChange; agreement \schAgree) \\
\ifdefined\stePfifty SCHED15 (FTrojanNN and Refool, \cref{app:corrected}) & P50 both twins pass A4 in $\ge4/5$ seeds & 0.65 & \stePfifty{} (\steNewAfour) \\
 & P51 FT, FP and NAD tare positive with intervals excluding zero; published cost positive & 0.70 & \stePfiftyone{} (\steNewFtFamExcl\ intervals) \\
 & P52 prototype-referenced DER flips $\le10\%$ of pairs, near-ties only & 0.60 & \stePfiftytwo{} (\steDerFlips; \selDerFlips\ with the SCHED11 rows) \\
\fi
\ifssixteen\sdbScorecardRows\fi
\bottomrule
\end{tabular}
\end{center}
\ifsseventeen
\begin{center}\footnotesize
\begin{tabular}{p{1.7cm}p{5.3cm}p{1.2cm}p{3.8cm}}
\toprule
pre-registration & prediction & prior & outcome \\
\midrule
SCHED17 (BackdoorBox, \cref{app:toolkits}) & P57 undefended twin $\ge1$\pt{} higher under A than under N & 0.85 & \sbxPfiftyseven{} (\sbxTwinGap\pt) \\
 & P58 FineTuning (lr $0.001$) tare negative under N, interval excluding zero & 0.75 & \sbxPfiftyeight{} (\sbxFtNTarePt) \\
 & P59 FineTuning (lr $0.001$) tare positive under A, interval excluding zero & 0.50 & \sbxPfiftynine{} (\sbxFtATarePt) \\
SCHED18 (three more BackdoorBox defenses) & P60 FineTuning at lr $0.01$: tare negative under N and positive under A, both intervals excluding zero & 0.75 & \sbxPsixty{} (\sbxFtoneNTarePt, \sbxFtoneATarePt) \\
 & P61 NAD tare higher under A than under N, seed-paired interval excluding zero & 0.65 & \sbxPsixtyone{} (\sbxNadSwing) \\
 & P62 NAD tare negative under N, interval excluding zero & 0.55 & \sbxPsixtytwo{} (\sbxNadNTarePt) \\
 & P63 Pruning, no training phase: tare $\ge0$ under both schedules & 0.65 & \sbxPsixtythree{} (\sbxPruneNTarePt, \sbxPruneATarePt) \\
\ifdefined\sbxPsixtyfour SCHED19 (BackdoorBox BadNets) & P64 undefended twin $\ge1$\pt{} higher under A than under N & 0.85 & \sbxPsixtyfour{} (\sbxBnTwinGap\pt) \\
 & P65 FineTuning at lr $0.01$: tare negative under N and positive under A, both intervals excluding zero & 0.75 & \sbxPsixtyfive{} (\sbxBnFtoneNTarePt, \sbxBnFtoneATarePt) \\
 & P66 FineTuning (lr $0.001$) tare negative under N, interval excluding zero & 0.70 & \sbxPsixtysix{} (\sbxBnFtNTarePt) \\
 & P67 NAD tare higher under A than under N, seed-paired interval excluding zero & 0.65 & \sbxPsixtyseven{} (\sbxBnNadSwing) \\
 & P68 Pruning, no training phase: tare $\ge0$ under both schedules & 0.60 & \sbxPsixtyeight{} (\sbxBnPruneNTarePt, \sbxBnPruneATarePt) \\
\fi
\bottomrule
\end{tabular}
\end{center}
\begin{center}\footnotesize
\begin{tabular}{p{1.7cm}p{5.3cm}p{1.2cm}p{3.8cm}}
\toprule
pre-registration & prediction & prior & outcome \\
\midrule
\ifsztwenty SCHED20 (zero-image twins, \cref{app:zero}) & P69 zero-minus-ten undefended accuracy $|\Delta|\le0.3$\pt{} with interval covering zero, all three cells & 0.85 & \szrPsixtynine{} (max $|\Delta|$ \szrAccDiffMaxAbs) \\
 & P70 $\ge8$ of $9$ (cell, defense) tare differences with interval covering zero and $|\Delta|\le0.5$\pt & 0.80 & \szrPseventy{} (\szrPseventyN) \\
 & P71 all $15$ zero-image twins pass A4 & 0.85 & \szrPseventyone{} (\szrAfourAll) \\
\fi
\ifsrtwentyone SCHED21 (crossover at $1\%$/$5\%$, \cref{app:rates}) & P72 annealed victim starts $\ge1$\pt{} higher in $\ge4/5$ seeds, every (attack, rate) & 0.85 & \srtPseventytwo{} (\srtPseventytwoN) \\
 & P73 WaNet: FT/FP/NAD published cost negative under the shipped block and higher under the swapped one, interval excluding zero, $\ge5/6$ & 0.75 & \srtPseventythree{} (\srtPseventythreeN) \\
 & P74 BadNets: the same & 0.75 & \srtPseventyfour{} (\srtPseventyfourN) \\
 & P75 WaNet at $1\%$ installs (A0 $\ge3/5$) under both schedules & 0.60 & \srtPseventyfive \\
\fi
\ifsdptwentytwo SCHED22 (ANP deployer-side rule, \cref{app:anp}; analysis only) & P80 BadNets tare under rule D positive, interval excluding zero & 0.70 & \sdpPeighty{} (\sdpBadnetTareDPt) \\
 & P81 rule D and rule O BadNets tares within one point & 0.60 & \sdpPeightyone{} (D $-$ O \sdpBadnetTareDiffPt) \\
 & P82 WaNet tare of the same sign under both rules & 0.60 & \sdpPeightytwo{} (\sdpWanetTareOPt, \sdpWanetTareDPt) \\
\fi
\ifsedtwentythree SCHED23 (eight-defense crossover, \cref{app:eight}) & P76 ANP tare on WaNet positive under both schedules & 0.60 & \sedPseventysix \\
 & P77 ABL destroys most twins under both schedules on WaNet and BPP & 0.80 & \sedPseventyseven{} (\sedPseventysevenCounts) \\
 & P78 NC fires on $\le$ half the cosine WaNet twins and costs them a positive amount where it fires & 0.65 & \sedPseventyeight{} (\sedPseventyeightFire; \sedPseventyeightCond) \\
 & P79 the DER ordering of the eight on WaNet changes in a (fine-tuning-family, other) pair & 0.60 & \sedPseventynine{} (\sedPseventynineFlipList) \\
\fi
\bottomrule
\end{tabular}
\end{center}
\fi
\fi
A ``P7-type'' invariance claim would need an equivalence bound; at five seeds we report it as a non-rejection (\cref{app:lim}).

\section{The twin fixed point, all cells}
\label{app:fixfig}
\begin{figure}[h]
\centering
\includegraphics[width=0.5\textwidth]{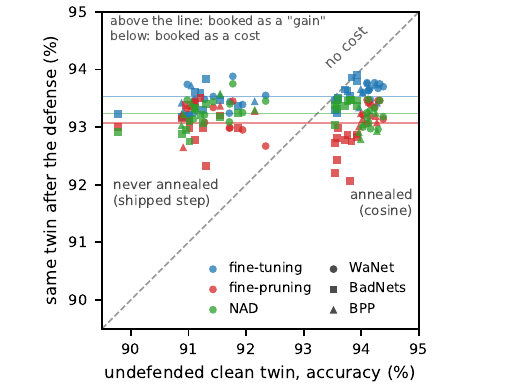}
\caption{The per-defense fixed point. Each point is one clean twin from one of the six admissible cells (WaNet, BadNets, BPP $\times$ two schedulers; \exFixFigN\ twin--defense pairs): its accuracy before the defense ($x$) and after ($y$). Never-annealed twins start near $91$, annealed ones near $94$; every defense returns both to the same $93$--$93.5$. Above the diagonal the benchmark books a ``gain'', below it a cost. The scheduler decides which side a victim starts on.}
\label{fig:fixedall}
\end{figure}

\Cref{fig:fixedall} plots all \exFixFigN{} twin--defense pairs of the six admissible cells.

\section{Crossover and extension tables}
\label{app:ext}
\Cref{tab:c100} gives the CIFAR-100 cells, \cref{tab:xover} the CIFAR-10 $2\times2$ crossover and \cref{tab:ext} the BPP/Input-Aware extension.
\begin{sidewaystable}
\caption{CIFAR-100 / PreAct-ResNet18, $5$ seeds per cell. Clean-arm cost per defense, pooled published cost, undefended twin accuracy, and A3 (residual ASR $\le0.05$) per defense. The shipped WaNet victim installs the attack in only \syWshipAzero{} seeds (ASR \syWshipPoisAsrLo{} and up), so by rule K3 (registered before any CIFAR-100 run; its disposition, a timestamped amendment written after seed 0 and before seeds 1--4, voids the poisoned arm and reports the clean arm as a deviation, \cref{app:prereg}) its poisoned-arm quantities are void. For the record, over all five seeds ignoring A0 the published cost is \syWshipPubAll\pt, the interaction \syWshipDidAll\pt{} and the victim base \syWshipVictimAll; evaluated on these, P6 would have \syPsixAll{} and P7 would have \syPsevenAll. Its clean arm does not depend on installation and is reported. Twins: ten images for WaNet, one image for BPP and Blended (\cref{app:twin}).}
\label{tab:c100}
\centering\small
\resizebox{\textheight}{!}{\input{supplement/sched/build/tab_sched3_body.tex}}
\end{sidewaystable}

\begin{sidewaystable}
\caption{The $2\times2$ scheduler crossover on CIFAR-10 / PreAct-ResNet18: clean-arm cost $c_{\rm clean}$ (pt, positive = the defense cost accuracy on a model with no backdoor), $t$-interval over seeds. Only the victim's scheduler keys differ between a shipped row and its swapped partner. ``twin base'': the undefended twin's accuracy; ``sign agree'': (seed $\times$ defense) cells whose cost has the predicted sign (the one dissent is BadNets + cosine, \cschDissentDef, \cschDissentVal\pt). Every twin passes A4; A3 is in \cref{tab:schedcols} and the WaNet victim's A0 in \cref{sec:block}.}
\label{tab:xover}
\centering\small
\input{supplement/sched/build/tab_sched_body.tex}
\end{sidewaystable}

\begin{sidewaystable}
\caption{Extension to BPP and Input-Aware, CIFAR-10 / PreAct-ResNet18, $5$ seeds. Clean-arm cost per defense and pooled (per-seed mean over the three defenses, $t$-interval over seeds), and pooled published cost. A4: twin admissible (ASR $\le0.05$); the attack installs in every seed (A0 $5/5$ throughout); $\ddagger$ marks a published number on which fewer than half the cells clear A3. BPP's twin is one poisoned image ($2\times10^{-5}$) because ten images already install a $5$--$7\%$ ASR on an annealed BPP victim (\cref{app:twin}). Input-Aware's twin fails A4 at ten images and diverges at one; its clean-arm row is shown, not claimed.}
\label{tab:ext}
\centering\small
\resizebox{\textheight}{!}{\input{supplement/sched/build/tab_sched2_body.tex}}
\end{sidewaystable}

\FloatBarrier
\section{The twin convention}
\label{app:twin}
A twin is the attack's own pipeline run at a poisoning rate too small to install the attack. Ten images ($2\times10^{-4}$) is clean for WaNet and BadNets (max twin ASR \cschWcosMaxAsr{} and below). On BPP with the cosine schedule ten images install a $0.045$--$0.068$ ASR in $4/5$ seeds (A4 \sxBcosLegacyAfour); the pre-registration's disposition for this case was to reset to one image, which passes A4 on both schedules (\sxBshipOneAfour, \sxBcosOneAfour); Blended on CIFAR-100 uses the one-image twin from the start, following BPP and gives readings within $0.15$\pt{} of the ten-image ones (\sxBcosLegacyFt/\sxBcosLegacyFp/\sxBcosLegacyNad{} vs.\ \sxBcosOneFt/\sxBcosOneFp/\sxBcosOneNad). On Input-Aware ten images carry $0.17$--$0.46$ ASR and one image makes the generator's joint training diverge (NaN loss at epoch $26$--$28$ in all ten runs); no further rate was tried. On BPP's shipped schedule the ten-image twin passes A4 (\steBppAfour) and is the twin Table~\ref{tab:corrmain} reports; Tables~\ref{tab:defenses} and~\ref{tab:ext} use the one-image reset for both BPP rows; on shipped BPP, where both twins exist, they read within $0.16$\pt{} of each other for FT, FP, NAD and I-BAU and $0.53$\pt{} apart for SAU.

\paragraph{What the matched twin buys.} The recipe match, not variance reduction. Across installations the poisoned and
twin arms' pre-defense accuracies correlate at $r={}$\pairRbb\ (BackdoorBench BadNets, $n={}$\pairRbbN), \pairRwanet\
(WaNet, $n={}$\pairRwanetN) and \pairRcifar\ (our CIFAR-10 headline grid, $n={}$\pairRcifarN), and after each defense on
the two BackdoorBench grids at \pairRpostLo\ to \pairRpostHi, so pairing by seed does not in general narrow the paired contrast's
interval against an unpaired one; what the twin controls is the recipe, schedule and
data order the defense starts from, which is where the tare comes from.

\ifsfive
\section{The fixed point as a function of budget}
\label{app:budget}
Pre-registered as SCHED5 (\cref{app:prereg}): FT with the benchmark's configuration except \texttt{--epochs} and \texttt{--lr}, on all $40$ models of the four CIFAR-10 cells ($480$ runs), twelve budgets plus the benchmark's own. \Cref{fig:budget} and \cref{tab:budget} give the defended clean-twin level per cell and the four-cell range. One fact about the benchmark's trainer, found after the sweep had run and recorded in the pre-registration as a post-hoc note: it builds \cosA{} with $T_{\max}=100$ regardless of \texttt{--epochs} (\path{utils/aggregate_block/train_settings_generate.py}, line 67; no FT configuration sets \texttt{cos\_t\_max}), so a budget of $E$ epochs is the first $E$ epochs of a $100$-epoch cosine, not a shorter complete one---the learning rate at the last epoch is \sqvLrAtFive{} at five, \sqvLrAtTwenty{} at twenty and \sqvLrAtFifty{} at fifty. Two regimes: at $10^{-2}$ the range is already \sqvRangeEfiveLrC\pt{} after five epochs, before any decay, and \sqvRangeEtwentyLrC\pt{} after twenty (K8, range $\ge1.5$ at $20$ epochs, did not fire); at $10^{-4}$ the twins keep their order and the range is \sqvRangeEfiftyLrA\pt{} after fifty epochs. Of the shipped WaNet twin's \sqvWshipRiseHundred\pt{} rise over the full schedule, \sqvWshipRiseFive\pt{} arrives in the first five epochs at $10^{-2}$ and \sqvWshipRiseTail\pt{} over the ninety-five that contain the decay; the annealed twin first drops \sqvWcosDropFive\pt{} in those five epochs and climbs back to \sqvWcosDropHundred\pt{} by the end. So the level the defenses share is set mostly by the fine-tuner's own regime (learning rate $10^{-2}$, batch $256$ against the victim's $128$, the $5\%$ clean subset) and only in its last half-point by the decay: ``the defense completes the anneal'' (our first phrasing) is about a quarter of the story, and the main text now says ``return a low start to a level of their own.'' In the first regime the clean-arm cost is the start-to-fixed-point distance of \cref{sec:bias} at every budget; in the second it is small for every cell (shipped WaNet twin \sqvCleanEfiveLrAWship\pt{} at five epochs, \sqvCleanEfiftyLrAWship\pt{} at fifty) and the poisoned arm is not cleaned (A3 \sqvAthreeShort{} at five epochs of $10^{-4}$), so the published cost then reports neither the backdoor nor the start. Two of the four predictions failed on their thresholds, both in the same direction: the fixed point forms faster than we predicted (P16, P17's $B^*$ of \sqvBstar{} epochs against a predicted $20$--$50$), and P18's four-interval criterion fails at $50$ epochs of $10^{-2}$ because the BadNets-cosine cell, whose cost is near zero by construction, straddles it (\sqvCleanEfiftyLrCBnship); the two schedule contrasts themselves reverse at every $10^{-2}$ budget from ten epochs up and at none of the $10^{-4}$ budgets.
\begin{figure}[h]
\centering
\includegraphics[width=0.95\textwidth]{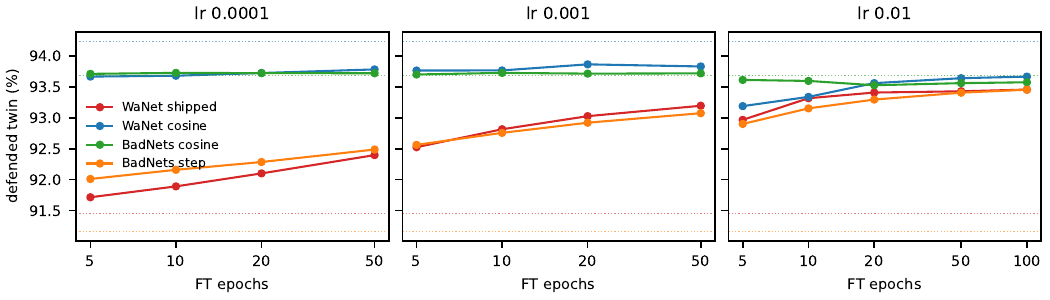}
\caption{Defended clean-twin accuracy against FT epochs, one panel per learning rate, one line per cell; dotted lines are the undefended twins. At $10^{-2}$ the four lines meet within a point from five epochs on; at $10^{-4}$ they keep their starting order through fifty.}
\label{fig:budget}
\end{figure}
\begin{table}[h]
\caption{Defended clean-twin accuracy (\%, mean of $5$ seeds) per cell and budget, and the four-cell range (pt).}
\label{tab:budget}
\begin{center}\small\input{supplement/sched/build/tab_sched5_body.tex}\end{center}
\end{table}
\fi
\ifssix
\section{The WaNet pair on VGG19-BN}
\label{app:vgg}
Pre-registered as SCHED6 (\cref{app:prereg}): the WaNet crossover of \cref{sec:block} with \texttt{--model vgg19\_bn}, five seeds per cell, FT, FP and NAD (\cref{tab:vgg}). The defenses run with AMP off on this backbone (the benchmark's AMP path aborts with a floating-point exception on VGG19-BN on our machine, so all thirty defense calls of the first launch produced nothing; the attack side trains without AMP by default), identically under both schedules, recorded as a timestamped amendment before any VGG19-BN defense result existed. Undefended twins: \svVshipTwin\% shipped, \svVcosTwin\% cosine, a swing of \svTwinGap\pt{} (the leaderboard's VGG19-BN gap is \lbGapCtenVgg\pt); A4 \svVshipAfour{} and \svVcosAfour, A0 \svVshipAzero{} and \svVcosAzero. The reversal is as clean as on PreAct-ResNet18: clean-arm cost \svVshipFt\pt{} shipped against \svVcosFt\pt{} cosine under FT, \svVshipFp{} against \svVcosFp{} under FP, \svVshipNad{} against \svVcosNad{} under NAD, every interval excluding zero (P20 \svPtwenty, P21 \svPtwentyone). The fixed point is not: the defended twins land at \svVshipFtDefTwin\% from the shipped start and \svVcosFtDefTwin\% from the cosine start under FT (largest FT/NAD gap \svFixMax\pt, against a third of the swing), so P22 fails---on this backbone the benchmark's fine-tuning takes a well-trained clean model down by two points and an under-trained one up by two, and the two do not meet. The published cost on the poisoned arm is a start-point story with wide intervals: the shipped victims start between \svVshipVictimLo{} and \svVshipVictimHi\% (WaNet at $10\%$ damages VGG19-BN unevenly across seeds), and FT books \svVshipFtPub\pt{} against \svVcosFtPub\pt{} on the cosine victims; A3 is \svVshipFtAthree{} shipped and \svVcosFtAthree{} cosine under FT. What transfers across backbones is therefore the sign of the clean-arm cost and its dependence on the three keys; the common level that all defenses return to is a property of PreAct-ResNet18 on these datasets, and \cref{sec:bias} should be read with that scope.
\begin{table}[h]
\caption{VGG19-BN / CIFAR-10 WaNet cells, $5$ seeds, $t$-intervals; costs in pt, negative is a ``gain''.}
\label{tab:vgg}
\begin{center}\small\resizebox{\textwidth}{!}{\input{supplement/sched/build/tab_sched6_body.tex}}\end{center}
\end{table}
\fi

\ifsnine
\section{A second cause of a low start: SIG at $10\%$}\label{app:sig}
Pre-registered as SCHED9 (\cref{app:prereg}) after the VGG19-BN cell, to test the leaderboard's reading that
the clip fires wherever the victim starts below the defense's level, whatever made it low. SIG is a
clean-label attack that leaves every scheduler key alone---its victim anneals under the prototype's cosine
schedule---and at the benchmark's $10\%$ rate its poisoning damages the victim outright: five seeds start at
\sgVictim\% (\sgVictimLo--\sgVictimHi, ASR \sgAsrLo--\sgAsrHi, A0 \sgAzero) against a same-recipe twin at
\sgTwin\%, a gap of \sgGap\ points with no scheduler involved. Fine-tuning books the rise as a gain
(published cost \sgFtPub; NAD \sgNadPub, fine-pruning \sgFpPub), so DER's clip fires on this cell (P32),
while on the twin the same defenses cost \sgFtClean, \sgNadClean\ and \sgFpClean. The defended arms do not
meet: \sgFtDefVictim\% against \sgFtDefTwin\% under fine-tuning (\sgFtFixGap\ points; NAD \sgNadFixGap),
so P30 fails as registered---the damage is erased only in part, and the fixed-point accounting of
\cref{sec:bias} holds here for the sign of the published cost and not for its size. Two things this cell
cannot say. Its twins fail A4 (\sgAfour; twin ASR up to \sgAsrTwinHi) for a reason the protocol's rule does
not distinguish from installation: the SIG signal itself moves a clean model's predictions (the
leaderboard's own $0.1\%$ SIG victim has an ASR of $41\%$), a one-image reset would not change that, and we
made none; the twin's clean-arm numbers are therefore descriptive, and P31 (fine-tuning's clean-arm cost
positive with its interval excluding zero) fails on that arm. And A3 clears in \sgFtAthree\ of the
poisoned arms under fine-tuning.
\begin{table}[h]
\caption{SIG at $10\%$ on CIFAR-10 / PreAct-ResNet18, five seeds, $t$-intervals; costs in pt, negative is a
``gain''; the twin columns are descriptive (A4 \sgAfour).}
\label{tab:sig}
\begin{center}\small\input{supplement/sched/build/tab_sched9_body.tex}\end{center}
\end{table}
\fi

\ifstwelve
\section{A healthy start moved by training length}\label{app:hstart}
Pre-registered as SCHED12 (\cref{app:prereg}) to test whether the level of \cref{sec:bias} belongs to the defense when a healthy,
fully annealed victim's start is moved by something other than the scheduler block: BadNets under the prototype's cosine
schedule trained for $30$ and $60$ epochs (the $100$-epoch cells are seeds 0--2 of the shipped grid), three installation seeds,
poisoned arm and twin, then FT, NAD and FP at their shipped configurations (\cref{tab:hstart}). The start moved
(\shsStartGap\pt{} between the 30- and 100-epoch twins, P42 \shsPfortytwo) and every victim installed (A0 \shsThirtyAzero,
\shsSixtyAzero, \shsHundredAzero). The three defenses compress that spread to \shsFtDefRange, \shsNadDefRange\ and
\shsFpDefRange\pt{} (slopes of defended on undefended twin over the nine twin arms: \shsFtSlope, \shsNadSlope, \shsFpSlope), each
to a level of its own (FT \shsThirtyFtLevel--\shsHundredFtLevel, NAD \shsThirtyNadLevel--\shsHundredNadLevel, FP
\shsThirtyFpLevel--\shsHundredFpLevel\%), so the accounting holds approximately and not as an identity on a healthy victim:
P43 fails as registered (the range criterion by \shsFtDefRange\ against a third of \shsUndefRange, and FP's slope is below,
not above, $0.5$). What the block-free manipulation reproduces is the sign: FT's published cost on the 30-epoch victim is
\shsThirtyFtPub\ against \shsHundredFtPub\ on the 100-epoch one (P44 \shsPfortyfour), so a fully annealed victim whose start
sits below the defense's level is booked as a gain and clipped by DER exactly as a never-annealed one is.
\begin{table}[h]
\caption{SCHED12: BadNets under the prototype cosine schedule at three training lengths, $3$ seeds each. Undefended twin and
victim (A0), the tare per defense ($t$-intervals over seeds) and the defended twin's level.}
\label{tab:hstart}
\begin{center}\small\resizebox{\textwidth}{!}{\input{supplement/sched/build/tab_sched12_body.tex}}\end{center}
\end{table}
\fi
\ifsthirteen
\section{More poison, a lower start, a cheaper-looking removal}\label{app:bddose}
Pre-registered as SCHED13 (\cref{app:prereg}): BadNets on CIFAR-10 / PreAct-ResNet18 under the prototype's cosine schedule at
poisoning rates \sdrRatesAll{} (the $10\%$ cells are seeds 0--4 of the shipped grid; the seed-0 victims at $1$, $5$ and $20\%$,
with their FT and NAD runs, predate the registration, which says so), five installation seeds per rate, the
eight defenses at their shipped configurations (\exGpuDose\ GPU-hours). The twin does not depend on the rate, so one twin per
seed serves every rate and the tare is the same number in every row by construction. K20 voids the poisoned arm at
\sdrAzeroFailRates{}, where the attack does not install (A0 \sdrHalfAzero{} and \sdrOneAzero); P47, which needed those rates, and
P46's ratio clause, which needed $1\%$, are untestable and counted as failed. The starting gap rises over all five rates (\sdrGapLo{} at \sdrGapLoRate{} to
\sdrGapHi{} at \sdrGapHiRate). P45 was registered over the five rates, but K20 reports only A0 and the twin at a voided
rate, so at \sdrAzeroFailRates\ its registered test is untestable and counted as failed; read there anyway, \sdrRhoGapBoth. On the
A0-passing rates the paired contrast rises for FT, NAD and FP (Spearman \sdrRhoPairedFt, \sdrRhoPairedNad, \sdrRhoPairedFp)
and the published cost falls: for all eight defenses between \sdrLoRate{} and \sdrHiRate{} (\sdrPubFellN, the utility floor
applied; Spearman \sdrRhoPubFt, \sdrRhoPubNad, \sdrRhoPubFp{} for FT, NAD, FP). Per seed, over the \sdrFtPubDropN{} seeds whose
poisoned victim installs at both rates, fine-tuning's published cost changes by \sdrFtPubDrop{} while the starting gap grows
\sdrGapRise{} (the $g$ column of \cref{tab:bddose} averages all five seeds). SCHED13's file had named the opposite, a published
cost rising with the rate, as half of its falsifiable signature; that half failed, and the fall was read off this
table\ifssixteen{} before SCHED16 registered it for Blended and LF (\cref{tab:bddose2})\fi. The identity of \cref{sec:bias}
(published cost = tare + paired contrast $-$ starting gap) does not fix its sign; the data do: the start drops by as much as the starting gap grows (the twin is fixed) while the defended victim's level
drops by less, by as much as the paired contrast rises, so the defense recovers part of what the extra poison took and books
the recovery as a smaller cost. A defense whose level did not depend on
the start at all (\cref{sec:bias}'s fixed point) would have cut the published cost one for one; fine-tuning's changes by about a third as much
(\sdrFtPubDropPt\ against a starting-gap rise of \sdrGapRisePt\ per seed): its defended level follows the start part of the way.
\begin{table}[h]
\caption{SCHED13: BadNets dose--response, five seeds per rate. A0 per rate, the starting gap $g$ (twin minus victim), and for
FT, NAD and FP the tare (one twin per seed for every rate), the published cost and the paired contrast (A0-passing seeds;
--- where K20 voids the poisoned arm).}
\label{tab:bddose}
\begin{center}\small\resizebox{\textwidth}{!}{\input{supplement/sched/build/tab_sched13_body.tex}}\end{center}
\end{table}
\begin{figure}[h]
\centering\includegraphics[width=\linewidth]{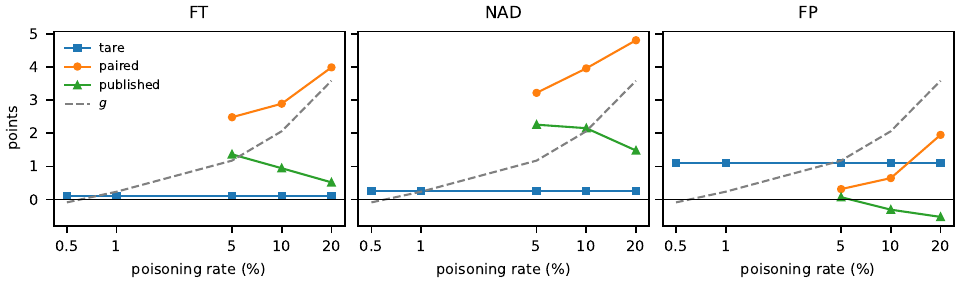}
\caption{SCHED13, the rows of \cref{tab:bddose} against the poisoning rate (log scale): the tare is flat by construction, the
starting gap $g$ and the paired contrast rise, the published cost falls.}
\label{fig:bddose}
\end{figure}
\ifssixteen
\paragraph{The registered replication on Blended and LF (SCHED16).} Registered before any run (\cref{app:prereg}) to test
the fall on two further annealed CIFAR-10 attacks whose $10\%$ installs reach $0.99$, Blended and LF, at $1$, $5$, $10$ and
$20\%$ with five seeds each, the eight defenses and the same twins (\cref{tab:bddose2}). A0 holds at every rate on Blended and
fails at $1\%$ on LF (\sdbLfOneAzero), so P53 fails. The start gap rises with the rate on Blended (Spearman \sdbRhoGapBl) and
not monotonically on LF (\sdbRhoGapLf: its $5\%$ and $10\%$ seed means, \sdbLfFiveGapThree{} and \sdbLfTenGapThree, differ below the printed precision; P54 fails as registered). The paired contrast rises for FT and NAD on both attacks
(\sdbPfiftysixN\ pairs; P56 holds), and the published cost rises with it: \sdbPfiftyfiveN\ of the four (attack, defense) pairs
fall (Spearman \sdbRhoPubBlFt, \sdbRhoPubBlNad, \sdbRhoPubLfFt, \sdbRhoPubLfNad), so P55, the primary prediction, fails in the
opposite direction. Per seed between the lowest and highest installing rate, fine-tuning's published cost changes by
\sdbBlFtPubChange{} on Blended and \sdbLfFtPubChange{} on LF while the start gap grows by \sdbBlGapRise{} and \sdbLfGapRise,
against \sdrGapRise{} on BadNets. The identity of \cref{sec:bias} reads the three attacks together---with the twin fixed, the
published cost falls where the victim's start drops by more than its defended level (BadNets) and rises where it drops by less
(Blended, LF)---but that reading is after the fact: SCHED13's fall is a BadNets observation, not a replicated one.
\begin{table}[h]
\caption{SCHED16: Blended and LF dose--response, five seeds per rate (the $10\%$ rows are the shipped seeds 0--4), laid out as
\cref{tab:bddose}; --- where A0 fails in at least three of five seeds (K20).}
\label{tab:bddose2}
\begin{center}\small\resizebox{\textwidth}{!}{\input{supplement/sched/build/tab_sched16_body.tex}}\end{center}
\end{table}
\fi
\fi
\ifsten
\section{The CIFAR-10 leaderboard at $10\%$ with its tare column: published costs, DER and seeds}\label{app:corrected}
Pre-registered as SCHED10 (\cref{app:prereg}) and extended by SCHED11 (five further defenses) and SCHED15 (FTrojanNN, Refool). For every attack of the
benchmark's CIFAR-10 / $10\%$ roster that admits a never-poisoned twin under this protocol---\steAttList---and the eight
defenses of Table~\ref{tab:bb}, \cref{tab:corrclean} prints the tare (the clean-arm cost on the twin, unconditional) and
\cref{tab:corrpub} the published cost on the victim (A0-gated; $\ddagger$ where fewer than half the cells clear A3).
Twins are the attack's own pipeline at pratio $0.0002$ (ten images; one image for Blended, whose ten-image twin fails A4). Seed by seed on our grids (\steClipDefs; Table~\ref{tab:corrpub} prints the cell means), DER's clip fires in \steClipNever\ never-annealed, \steClipDamaged\ SIG (damaged by its own poison) and \steClipHealthy\ healthy seed-cells, all fine-pruning's (\steClipHealthyFpList).
Two attacks' twins fail A4 and their tare columns are descriptive: SIG's signal moves a clean model's predictions
(A4 \steSigAfour, \cref{app:sig}), and Trojan's twins answer its trigger (A4 \steTrojanAfour); whether its optimised trigger
fires on any same-recipe model or ten images install it we did not test. Attacks without an admissible twin under this protocol
are not printed: Input-Aware (its ten-image twin fails A4 too, its one-image twin diverges, and only FT, FP and NAD were run
on it, \cref{app:ext}), LIRA and Blind (no per-sample poisoning rate), SSBA and LC (the
benchmark does not ship the encoded or adversarial data they need, and we download nothing), and Poison Ink and CTRL (a trigger generator or a self-supervised pipeline would have to be trained first). LF and Trojan are new here, and FTrojanNN and Refool (SCHED15; the checkout ships their frequency trigger and reflection images; neither is on the public leaderboard) were added last;
Trojan's trigger optimisation needs a pretrained clean model the benchmark does not ship, and we used the seed-0 BadNets
twin (the prototype recipe plus ten poisoned images) for it. The table's sign pattern is the paper's accounting read
across a roster; the sign counts and the ordering below use only the \steAdmitN\ attacks whose twins pass A4
(\steAdmitList), the \steDescList\ rows being descriptive, and leave ABL out: it destroys \steAblTwinCollapse\ twins,
so under the registered kill rule K16 (a defended twin that loses more than $20$ points is a failed run) its row is a
failure record, printed and marked $^{\natural}$ but outside every sign count and the ordering. Outside ABL's row no run in Table~\ref{tab:corrmain} is excluded as a failed run (a defended twin that lost more than $20$ points) or by A3's utility floor (\cref{app:a3}), which keeps \steAblFloorAdmitN\ of ABL's poisoned-arm runs out of its A3 count. Over all nine attacks, among these eight defenses K16 fired once outside ABL
(\steCollapsedTwinList; that cell is printed over the remaining seeds, marked $^{\flat}$), and the utility floor of
A3 (\cref{app:a3}) excludes \steFloorExcludedN\ poisoned-arm runs outside ABL (\steFloorExcludedList) from A3 and from the paired
contrast, and \steAblFloorN\ of ABL's from its A3 count only. On the \steAnnAdmitN\ admissible annealed victims every tare is positive (\steAnnPos\ cells); on Refool, whose victim starts \steNewGapRefool\ points below its twin, fine-pruning's published cost (\steRefoolFpPub) already sits below its tare (\steRefoolFpClean). On the two
never-annealed victims FT, FP and NAD ``gain'' in \steNeverFtFamNeg\ cells and ANP charges the twin in
\steNeverAnpPos, every interval excluding zero; SAU, which re-runs a full schedule but lands at a level of its own
(\cref{sec:replic}), and I-BAU, which has none, sit at zero in \steNeverSauIbauZero\ of their cells; Neural Cleanse's
detector fires on \steWanetNcFire\ WaNet and \steBppNcFire\ BPP twins, and where it fires its repair, the benchmark's
own $100$-epoch fine-tune, ``gains'' \steWanetNcCond\ as fine-tuning does. This pattern was read off the table after
P36 failed on BPP, not predicted. Ordering the defenses by mean tare over the admissible attacks gives
\steCleanOrder\ (over all nine attacks, \steCleanOrderAll). A prototype-referenced DER (\cref{sec:corrected}; the same seed's BadNets twin as
reference) shifts every defense's mean DER down by \steDerShiftLo\ to \steDerShiftHi\ points and re-ranks
\steDerFlips\ of the defense pairs over all nine attacks (\cref{tab:corrder}; on the public twelve-defense panel of \cref{app:rerank}
it re-ranks near-ties only): the reference changes attribution, not order. Scorecard, judged on the seven attacks registered: P33
\stePthirtythree\ (LF's fine-tuning interval touches zero, \steLfFtClean; Trojan's twins fail A4), P34 \stePthirtyfour\
(over all seven registered attacks ANP and I-BAU swap; on their A4-passing twins \stePthirtyfourAdmit), P35 \stePthirtyfive, P36
\stePthirtysix\ (on SIG \steSigNewShare\ of the five added defenses share fine-tuning's sign; on BPP \steBppNewShare, SAU alone, the
pattern above).\ifdefined\stePfifty{} SCHED15, for the two attacks added last: P50 \stePfifty\ (A4 \steNewAfour), P51 \stePfiftyone\
(\steNewFtFamExcl\ fine-tuning-family tares positive with intervals excluding zero), P52 \stePfiftytwo\ (\steDerFlips).\fi
\begin{table}[h]
\caption{The tare per (attack, defense) cell: clean-arm cost on the never-poisoned twin (pt; negative is a ``gain''), mean over seeds with its $t$-interval; the victim's schedule, the undefended twin and A4 beside it. $^{\flat}$: one K16 run excluded (\steCollapsedTwinList), $n-1$ seeds; $^{\natural}$: ABL, a K16 failure record (\steAblTwinCollapse\ twins destroyed), outside every sign count and the ordering. $\dagger$: A2 holds at the policy level only (Table~\ref{tab:bb}, Appendix~\ref{app:bb}); ANP's threshold is chosen by test-set attack success within an accuracy budget, which on a twin selects on false-positive noise. Neural Cleanse's cell is printed as the fraction of twins on which its detector fired \textbar\ the cost conditional on firing (the others are returned unchanged). Intervals are per-cell and nominal: of the \steCellsK\ cells printed with one, \steCellsNominal\ exclude zero at $0.05$ and \steCellsBonf\ at the Bonferroni level $0.05/K$, $K={}$\steCellsK\ (\steCellsNominalAdmit\ and \steCellsBonfAdmit\ of the \steCellsKAdmit\ admissible-attack cells); the sign counts in the text are read from the cell means, not from these intervals.}
\label{tab:corrclean}
\begin{center}\small\resizebox{\textwidth}{!}{\input{supplement/sched/build/tab_sched10_clean_body.tex}}\end{center}
\end{table}
\begin{table}[h]
\caption{The published cost per (attack, defense) cell on the poisoned victim (pt; A0 seeds), with the undefended victim and A0 beside it; $\ddagger$: fewer than half the cells clear A3 (residual attack success $\le0.05$ and the utility floor of \cref{app:a3}).}
\label{tab:corrpub}
\begin{center}\small\resizebox{\textwidth}{!}{\input{supplement/sched/build/tab_sched10_pub_body.tex}}\end{center}
\end{table}
\ifseleven
\begin{table}[h]
\caption{SCHED11: the five further defenses that run unchanged on the same arms (\cref{app:prereg}; EP, BNP and NAB do not run in this checkout, \texttt{LOCAL\_CHANGES.md}; D-ST failed under its registered override and is dropped, \cref{app:prereg}). Tare per cell as in \cref{tab:corrclean}; $^{\flat}$ marks a cell with K16 exclusions (RNP: \selRnpTwinCollapse\ twins lost more than $20$ points, printed over the rest), $^{\natural}$ a failure record (D-BR destroys \selDbrTwinCollapse\ twins), $^{\dagger}$ RNP's threshold is chosen by test-set attack success. RNP leaves \selRnpWithinOne\ twins within a point of the undefended one and cuts others by up to \selRnpMaxDrop\ points (median over all runs \selRnpMedianDrop); NPD makes \selNpdTrigN\ never-poisoned twins answer the trigger (attack success \selNpdTrigList).}
\label{tab:selclean}
\begin{center}\small\resizebox{\textwidth}{!}{\input{supplement/sched/build/tab_sched11_clean_body.tex}}\end{center}
\end{table}
\begin{table}[h]
\caption{SCHED11: the published cost per cell on the poisoned victim for the same five defenses (A0 seeds; $^{\ddagger}$: fewer than half the cells clear A3, which is most of them: CLP and NPD leave the backdoor in place on most attacks at their shipped configurations; among the admissible attacks MCR clears it on \selMcrClears{} and not on \selMcrLeaves). The public leaderboard prints the same shape at $10\%$ on PreAct-ResNet18: D-BR leaves \lbDbrCaBadnet\ (BadNets), \lbDbrCaBlended\ (Blended), \lbDbrCaLf\ (LF) and \lbDbrCaTrojan\ (TrojanNN) clean accuracy, \lbDbrCaWanet\ on WaNet and \lbDbrCaBpp\ on BPP; ABL's published BadNets cost there is \lbAblBadnetCost\ against \steBadnetAblPub\ here; RNP has no public CIFAR-10 cell.}
\label{tab:selpub}
\begin{center}\small\resizebox{\textwidth}{!}{\input{supplement/sched/build/tab_sched11_pub_body.tex}}\end{center}
\end{table}
\fi
\begin{table}[h]
\caption{DER per defense, mean over the attacks above: the benchmark's clipped, victim-referenced form and the prototype-referenced form of \cref{sec:corrected} (reference: the same seed's BadNets twin). Published order \steDerOrderPub; prototype order \steDerOrderProto\ifseleven; with the SCHED11 defenses added, \selDerFlips\ of the pairs re-rank (\selDerFlipList, \selDerFlipMaxGap\ DER points apart)\fi.}
\label{tab:corrder}
\begin{center}\small\input{supplement/sched/build/tab_sched10_der_body.tex}\end{center}
\end{table}
\fi
\ifsseven
\section{TSBD\ifstcgd{} and CGD\fi{} measured in the WaNet $2\times2$}
\label{app:tsbd}
Pre-registered as SCHED7 (\cref{app:prereg}). TSBD's released code was dropped into our BackdoorBench checkout unchanged except for device and record paths; its clean reference model is the same-seed BadNets-cosine twin's weights (no new model was trained). We used the paper's printed defaults ($n=0.15$, $m=0.7$, $20$ fine-tuning epochs at $10^{-2}$, $5\%$ clean data). One disclosure: the released code sweeps the re-initialisation ratio over \stTsbdSweepN{} values from \stTsbdSweepNmin{} to \stTsbdSweepNmax{} and never selects one, so our first harvest took the last row ($n=0.9$), which destroys the network; we fixed the operating point to the printed default before anything entered the paper (timestamped in the pre-registration) and record the whole sweep in \cref{tab:tsbdsweep}. The sign pattern of \cref{tab:tsbd}---shipped victim gains, cosine twin loses---holds at \stTsbdSweepSign{} of the \stTsbdSweepN{} ratios; the shipped victim gains more than the cosine victim at all \stTsbdSweepOrder, and the cosine twin loses more than the shipped twin at all \stTsbdSweepTwinOrder; at $n\ge0.5$ the twins lose \stTsbdSweepBigMinCost\pt{} or more. Stage 1 (clean unlearning, then zero re-initialisation) leaves the shipped victim at \stTsbdStageOneWshipVictim\% and the shipped twin at \stTsbdStageOneWshipTwin\% (cosine: \stTsbdStageOneWcosVictim\% and \stTsbdStageOneWcosTwin\%); stage 2, plain fine-tuning with a gradient-norm regulariser, is what carries all four to the defended values in \cref{tab:tsbd}. P23--P25: \stPtwentythree, \stPtwentyfour, \stPtwentyfive{} (P25 by \stTsbdFixRange{} against \stTsbdSwingThird\pt, a third of the five-seed swing---a near miss, not a margin); K11 did not fire.\ifstcgd{} CGD (P26 \stPtwentysix{} on its interval clause---the cosine-victim point estimate sits on the $-1.0$ line---, P27 \stPtwentyseven) used its released two-step code: CLIP ViT-B/32 separation, then $10$ epochs of fine-tuning on the retained subset; our victims are at the benchmark's $10\%$ rate, the CGD paper's at $5\%$.\fi
\begin{table}[h]
\caption{TSBD\ifstcgd{} and CGD\fi{} on the WaNet victims and their clean twins, CIFAR-10 / PreAct-ResNet18, $t$-intervals over seeds, A0-gated. Published cost $=\Q(\theta_{\mathrm{pois}})-\Q(D\theta_{\mathrm{pois}})$; clean-arm cost likewise on the twin; negative is a ``gain''.}
\label{tab:tsbd}
\begin{center}\small\resizebox{\textwidth}{!}{\input{supplement/sched/build/tab_sched7_body.tex}}\end{center}
\end{table}
\begin{table}[h]
\caption{TSBD across the released code's re-initialisation sweep: mean cost (pt) per cell at the final fine-tuning epoch. The paper's printed default is the operating point of \cref{tab:tsbd}.}
\label{tab:tsbdsweep}
\begin{center}\small\input{supplement/sched/build/tab_sched7_sweep_body.tex}\end{center}
\end{table}
\fi

\section{Citation audit}
\label{app:audit}
The full audit table (paper, arXiv id, venue, victim provenance, verbatim clean-accuracy claim, where it appears, whether load-bearing, whether it rests on the three cells), the search queries with hit counts, and the \auUnobtainedJournal\ journal-only papers that could not be obtained (\texttt{AUDIT.md}, \S1D) are in the supplement.

\paragraph{The three claims in detail.} The main text compresses the following.

\paragraph{Roster.}
We pulled the $252$ papers citing either BackdoorBench paper (Semantic Scholar, deduplicated), read the $48$ that propose a defense against installed backdoors and could be obtained in full, and asked of each: does it report WaNet, BPP or Input-Aware on a BackdoorBench victim, does it make a clean-accuracy claim, and does that claim survive removing those cells? Thirteen reuse the benchmark's victims---their ``No Defense'' rows reproduce the leaderboard to two decimals---and their baseline columns reproduce the artifact: FT, FP, NAD and NC ``gaining'' two points on WaNet and Input-Aware appears, unremarked, in the tables of at least seven.

\paragraph{Three claims rest on it.}
\emph{TSBD}~\citep{lin2024unveiling} opens its CIFAR-10 results with an average accuracy ``higher than the original No Defense models,'' attributed to ``removing the backdoor effect with the least cost'': \auTsbdAvg\% against \auTsbdNd\% over eight attacks (\auTsbdMargin\pt, no interval reported). The margin is carried by the two never-annealed cells (Input-Aware \auTsbdIa, WaNet \auTsbdWanet\pt) and by SIG (\auTsbdSig\pt), a damaged annealed victim. Dropping the two never-annealed cells, the average is \auTsbdSix{} against \auTsbdNdSix; subtracting from each of the two the \auCleanGain\pt{} that the benchmark's fine-tuning family (FT, FP and NAD pooled) gains on the clean twin leaves \auTsbdAdjTwin{} against \auTsbdNd; subtracting the victim's \auDeficit\pt{} scheduler deficit leaves \auTsbdAdjDeficit. The sign reverses under each accounting. \ifsseven TSBD itself, run in the WaNet $2\times2$, confirms it (\cref{app:tsbd}).\else We have not run TSBD itself on an annealed victim or a clean twin.\fi{} \emph{CGD}~\citep{xu2025cgd} reports that it ``can identify and unlearn poisoned samples, improving CA over the No Defense baseline by 3.4\% for WaNet and 3.6\% for IAB''---the sign survives, the attribution does not. CGD fine-tunes the backdoored model on a CLIP-separated clean subset at $5\%$ poisoning, a shorter schedule than the benchmark's; what transfers from our measurements is the victim's deficit (\lbGapCtenPre\pt{} at that rate, \cref{tab:fourds}), not a value, \ifstcgd{}and CGD itself, run in the same $2\times2$ (\cref{app:tsbd}), gives the value: published cost \stCgdWshipPub\pt{} on the shipped victim and \stCgdWcosPub\pt{} on the cosine victim; the clean twin gains \stCgdWshipClean\pt{} under the shipped schedule and loses \stCgdWcosClean\pt{} under cosine, so \stCgdShare{} of the shipped-victim gain is what a never-poisoned model gets from the same fine-tuning. \else{}so ``roughly two of the $3.4$ points are the anneal'' is an estimate from the benchmark's own fine-tuning (\cref{tab:schedcols}), not a measurement on CGD; measuring it costs the one run of \cref{sec:surrogate}.\fi{} The sentence attributes all of them to unlearning. \emph{PDB}~\citep{wei2024pdb} is a training-time defense: it does not fine-tune the victim but trains its own model, so the mechanism there is not a defense completing the anneal but a comparison against a No-Defense baseline that the attack's schedule left under-annealed. Its Tiny-ImageNet/ViT-B/16 table has a positive average accuracy change that is produced entirely by WaNet ($+11.9$) and BPP ($+10.2$); the other four attacks are all negative, and the same two cells carry the table's highest DER. The two gains match the \lbGapTinyVit\pt{} leaderboard gap on that backbone (\cref{tab:fourds}); but Tiny-ImageNet is the once-firing contrast on which we ran no swap and on which VGG19-BN reverses (\cref{sec:leaderboard}), so for PDB the attribution to the scheduler is correlational, causal only for TSBD and CGD. Two further papers~\citep{min2024ulrl,min2023fst} have load-bearing no-cost statements that these cells buffer without reversing, and two more~\citep{wei2023sau,wei2025d3} report DER values that the same cells lift through the clip. \emph{None} of the $48$, nor either benchmark paper, notes that the three victims are under-trained; the closest is FST's appendix, which lowers WaNet's learning rate ``since a large initial learning rate would make the attack collapse'' without connecting the resulting gap to the fine-tuning gains it then reports. One author coded all $48$ papers; no second coder or inter-rater agreement was measured. The audit's criterion, written before coding (\texttt{AUDIT.md}), also defines a loose count that adds the load-bearing claims the three cells buffer without reversing (PDB, ULRL and FST) to the strict \auLoadBearing{} the text uses.

\paragraph{We are not saying these methods do not work.}
Every one of the three has other evidence. We are saying that a specific sentence in each depends on these cells. For TSBD and CGD we have shown causally that the victims in those cells start two to three points low because of the scheduler\ifsseven; for TSBD\ifstcgd{} and CGD\fi{} the anneal is measured on the method itself (\cref{app:tsbd})\ifstcgd\else, for CGD it is still inferred from the benchmark's fine-tuning defenses (\cref{tab:schedcols})\fi\else; we have not run either method on an annealed victim or a clean twin, so how much of their own gain is the anneal is inferred from the benchmark's fine-tuning defenses (\cref{tab:schedcols}), not measured\fi. For PDB even the victim's deficit is attributed only on the leaderboard. And a benchmark whose cost metric cannot go negative let those sentences be written.

%% file: supplement/sched/build/tab_sched4_body.tex
\begin{tabular}{llccccccc}
\toprule
attack & victim's scheduler & FT & FP & NAD & FT-SAM & SAU & i-BAU & twin base \\
\midrule
WaNet & step (shipped) & $-1.99\ \scriptstyle[-2.30,-1.68]$ & $-1.68\ \scriptstyle[-2.12,-1.24]$ & $-1.85\ \scriptstyle[-2.14,-1.56]$ & $-2.14\ \scriptstyle[-2.46,-1.81]$ & $-0.15\ \scriptstyle[-0.77,+0.47]$ & $+0.23\ \scriptstyle[-0.30,+0.75]$ & $91.50$ \\
WaNet & cosine (swapped) & $+0.56\ \scriptstyle[+0.44,+0.68]$ & $+0.86\ \scriptstyle[+0.77,+0.95]$ & $+1.03\ \scriptstyle[+0.76,+1.30]$ & $+1.09\ \scriptstyle[+0.92,+1.26]$ & $+2.88\ \scriptstyle[+1.93,+3.82]$ & $+5.00\ \scriptstyle[+3.68,+6.31]$ & $94.23$ \\
\midrule
BadNets & cosine (shipped) & $+0.13\ \scriptstyle[+0.06,+0.21]$ & $+1.05\ \scriptstyle[+0.84,+1.25]$ & $+0.33\ \scriptstyle[+0.24,+0.42]$ & $+0.88\ \scriptstyle[+0.76,+1.00]$ & $+2.70\ \scriptstyle[+2.31,+3.08]$ & $+5.70\ \scriptstyle[+3.99,+7.42]$ & $93.71$ \\
BadNets & step (swapped) & $-2.29\ \scriptstyle[-2.59,-1.99]$ & $-1.80\ \scriptstyle[-2.49,-1.12]$ & $-1.98\ \scriptstyle[-2.26,-1.71]$ & $-2.45\ \scriptstyle[-2.84,-2.06]$ & $-0.58\ \scriptstyle[-0.91,-0.25]$ & $+0.26\ \scriptstyle[-0.51,+1.03]$ & $91.16$ \\
\midrule
BPP & step (shipped) & $-2.18\ \scriptstyle[-2.86,-1.49]$ & $-1.87\ \scriptstyle[-2.48,-1.25]$ & $-2.00\ \scriptstyle[-2.62,-1.39]$ & $-2.30\ \scriptstyle[-3.12,-1.47]$ & $-0.07\ \scriptstyle[-0.80,+0.66]$ & $+0.82\ \scriptstyle[-0.13,+1.78]$ & $91.31$ \\
BPP & cosine (swapped) & $+0.55\ \scriptstyle[+0.30,+0.80]$ & $+0.97\ \scriptstyle[+0.76,+1.17]$ & $+1.11\ \scriptstyle[+0.92,+1.30]$ & $+1.10\ \scriptstyle[+0.78,+1.42]$ & $+2.41\ \scriptstyle[+1.50,+3.33]$$^{\dagger}$ & $+4.09\ \scriptstyle[+2.70,+5.49]$ & $94.10$ \\
\midrule
\multicolumn{2}{l}{A3 clears (poisoned arm, A0 seeds, six cells)} & $36/39$ & $23/39$ & $37/39$ & $35/39$ & $37/39$ & $31/39$ & --- \\
\multicolumn{2}{l}{defended twin, range over the four CIFAR-10 cells} & $0.21$ & $0.70$ & $0.23$ & $0.81$ & $0.73$ & $3.26$ & $3.07$ \\
\bottomrule
\end{tabular}

%% file: supplement/sched/build/tab_sched17_body.tex
\begin{tabular}{lcccccc}
\toprule
 & \multicolumn{2}{c}{tare (clean arm)} & \multicolumn{2}{c}{published cost} & \multicolumn{2}{c}{defended twin (\%)} \\
\cmidrule(lr){2-3}\cmidrule(lr){4-5}\cmidrule(lr){6-7}
BackdoorBox defense & N & A & N & A & N & A \\
\midrule
\multicolumn{7}{l}{\emph{WaNet (SCHED17, SCHED18)}} \\
FineTuning, lr $0.001$ (BackdoorBox default) & $-6.10$ {\scriptsize[$-7.67$, $-4.52$]} & $+0.01$ {\scriptsize[$-0.03$, $+0.04$]} & $-8.24$ {\scriptsize[$-10.59$, $-5.90$]} & $+0.19$ {\scriptsize[$+0.02$, $+0.37$]} & 88.93 & 91.43 \\
FineTuning, lr $0.01$ & $-7.12$ {\scriptsize[$-8.68$, $-5.57$]} & $-0.01$ {\scriptsize[$-0.10$, $+0.08$]} & $-9.46$ {\scriptsize[$-12.08$, $-6.85$]} & $-0.06$ {\scriptsize[$-0.16$, $+0.04$]} & 89.95 & 91.45 \\
NAD & $-7.23$ {\scriptsize[$-8.82$, $-5.63$]} & $+0.81$ {\scriptsize[$+0.67$, $+0.95$]} & $-9.46$ {\scriptsize[$-12.17$, $-6.75$]} & $+0.82$ {\scriptsize[$+0.42$, $+1.22$]} & 90.06 & 90.63 \\
Pruning (no training phase) & $+3.44$ {\scriptsize[$+0.83$, $+6.06$]} & $+1.69$ {\scriptsize[$+0.51$, $+2.87$]} & $+3.29$ {\scriptsize[$+1.76$, $+4.81$]} & $+2.80$ {\scriptsize[$+2.32$, $+3.28$]} & 79.39 & 89.75 \\
Undefended twin (\%) & 82.83 & 91.44 & & & & \\
\midrule
\multicolumn{7}{l}{\emph{BadNets (SCHED19)}} \\
FineTuning, lr $0.001$ (BackdoorBox default) & $-5.91$ {\scriptsize[$-7.92$, $-3.91$]} & $+0.09$ {\scriptsize[$-0.05$, $+0.24$]} & $-6.57$ {\scriptsize[$-9.15$, $-3.98$]} & $+0.08$ {\scriptsize[$-0.05$, $+0.20$]} & 88.68 & 91.61 \\
FineTuning, lr $0.01$ & $-7.10$ {\scriptsize[$-9.14$, $-5.06$]} & $+0.05$ {\scriptsize[$-0.13$, $+0.23$]} & $-7.79$ {\scriptsize[$-10.23$, $-5.35$]} & $+0.10$ {\scriptsize[$+0.03$, $+0.17$]} & 89.86 & 91.65 \\
NAD & $-7.34$ {\scriptsize[$-9.35$, $-5.33$]} & $+1.15$ {\scriptsize[$+0.51$, $+1.79$]} & $-7.97$ {\scriptsize[$-10.29$, $-5.65$]} & $+1.09$ {\scriptsize[$+0.69$, $+1.48$]} & 90.10 & 90.55 \\
Pruning (no training phase) & $+3.46$ {\scriptsize[$+1.86$, $+5.05$]} & $+1.80$ {\scriptsize[$+0.68$, $+2.91$]} & $+3.44$ {\scriptsize[$+1.88$, $+5.00$]} & $+1.37$ {\scriptsize[$+0.67$, $+2.07$]} & 79.31 & 89.91 \\
Undefended twin (\%) & 82.76 & 91.70 & & & & \\
\bottomrule
\end{tabular}

%% file: supplement/sched/build/tab_sched20_body.tex
\begin{tabular}{llcccc}
\toprule
cell & quantity & zero-image twin & ten-image twin & zero $-$ ten {\scriptsize[95\% $t$]} & $n$ \\
\midrule
BadNets, shipped block (cosine) & clean acc.\ (\%), ASR$_{\max}$ 0.009 & 93.82 & 93.69 & $+0.14$ {\scriptsize[$-0.12$, $+0.39$]} & 5 \\
 & tare, FT & $+0.07$ {\scriptsize[$-0.05$, $+0.20$]} & $+0.11$ {\scriptsize[$+0.00$, $+0.22$]} & $-0.04$ {\scriptsize[$-0.21$, $+0.13$]} & 5 \\
 & tare, FP & $+1.20$ {\scriptsize[$+0.85$, $+1.54$]} & $+1.11$ {\scriptsize[$+0.55$, $+1.67$]} & $+0.09$ {\scriptsize[$-0.51$, $+0.68$]} & 5 \\
 & tare, NAD & $+0.38$ {\scriptsize[$+0.12$, $+0.65$]} & $+0.26$ {\scriptsize[$+0.08$, $+0.43$]} & $+0.13$ {\scriptsize[$-0.19$, $+0.45$]} & 5 \\
\midrule
WaNet, shipped block (never fires) & clean acc.\ (\%), ASR$_{\max}$ 0.015 & 91.34 & 91.45 & $-0.11$ {\scriptsize[$-0.73$, $+0.51$]} & 5 \\
 & tare, FT & $-2.21$ {\scriptsize[$-2.84$, $-1.57$]} & $-2.01$ {\scriptsize[$-2.57$, $-1.44$]} & $-0.20$ {\scriptsize[$-0.79$, $+0.39$]} & 5 \\
 & tare, FP & $-1.89$ {\scriptsize[$-2.55$, $-1.23$]} & $-1.74$ {\scriptsize[$-2.77$, $-0.72$]} & $-0.14$ {\scriptsize[$-1.15$, $+0.86$]} & 5 \\
 & tare, NAD & $-1.93$ {\scriptsize[$-2.45$, $-1.41$]} & $-1.86$ {\scriptsize[$-2.40$, $-1.32$]} & $-0.06$ {\scriptsize[$-0.54$, $+0.42$]} & 5 \\
\midrule
WaNet, block removed (cosine) & clean acc.\ (\%), ASR$_{\max}$ 0.009 & 93.99 & 94.23 & $-0.24$ {\scriptsize[$-0.51$, $+0.04$]} & 5 \\
 & tare, FT & $+0.51$ {\scriptsize[$+0.30$, $+0.72$]} & $+0.56$ {\scriptsize[$+0.44$, $+0.68$]} & $-0.05$ {\scriptsize[$-0.21$, $+0.11$]} & 5 \\
 & tare, FP & $+1.21$ {\scriptsize[$+0.89$, $+1.52$]} & $+0.86$ {\scriptsize[$+0.77$, $+0.95$]} & $+0.35$ {\scriptsize[$+0.03$, $+0.66$]} & 5 \\
 & tare, NAD & $+0.88$ {\scriptsize[$+0.62$, $+1.15$]} & $+1.03$ {\scriptsize[$+0.76$, $+1.30$]} & $-0.15$ {\scriptsize[$-0.32$, $+0.03$]} & 5 \\
\bottomrule
\end{tabular}

%% file: supplement/sched/build/tab_sched21_body.tex
\begin{tabular}{lllcc ccc ccc}
\toprule
 & & & & undefended & \multicolumn{3}{c}{published cost {\scriptsize[95\% $t$]}} & \multicolumn{3}{c}{tare (same-seed twin)} \\
\cmidrule(lr){6-8}\cmidrule(lr){9-11}
attack & rate & schedule & A0 & acc.\ (\%) & FT & FP & NAD & FT & FP & NAD \\
\midrule
WaNet & 5\% & shipped (\ms, never fires) & $4/5$ & 89.50 & $-4.11$ {\scriptsize[$-7.05$, $-1.17$]} & $-4.12$ {\scriptsize[$-7.13$, $-1.11$]} & $-3.97$ {\scriptsize[$-6.91$, $-1.03$]} & $-2.01$ {\scriptsize[$-2.57$, $-1.44$]} & $-1.74$ {\scriptsize[$-2.77$, $-0.72$]} & $-1.86$ {\scriptsize[$-2.40$, $-1.32$]} \\
 &  & swapped (\cosA) & $5/5$ & 91.99 & $-1.45$ {\scriptsize[$-3.86$, $+0.96$]} & $-1.37$ {\scriptsize[$-3.62$, $+0.88$]} & $-1.00$ {\scriptsize[$-3.27$, $+1.27$]} & $+0.56$ {\scriptsize[$+0.44$, $+0.68$]} & $+0.86$ {\scriptsize[$+0.77$, $+0.95$]} & $+1.03$ {\scriptsize[$+0.76$, $+1.30$]} \\
 & & \emph{annealed $-$ not annealed} & & $+2.49$ {\scriptsize[$+0.65$, $+4.33$]} & $+2.77$ {\scriptsize[$+0.34$, $+5.20$]} & $+2.89$ {\scriptsize[$+0.36$, $+5.41$]} & $+3.07$ {\scriptsize[$+0.67$, $+5.48$]} & & & \\
\midrule
WaNet & 1\% & shipped (\ms, never fires) & $2/5$ & 90.32 & -- & -- & -- & $-2.01$ {\scriptsize[$-2.57$, $-1.44$]} & $-1.74$ {\scriptsize[$-2.77$, $-0.72$]} & $-1.86$ {\scriptsize[$-2.40$, $-1.32$]} \\
 &  & swapped (\cosA) & $4/5$ & 92.42 & $-1.22$ {\scriptsize[$-3.16$, $+0.73$]} & $-0.98$ {\scriptsize[$-2.86$, $+0.91$]} & $-0.77$ {\scriptsize[$-2.79$, $+1.24$]} & $+0.56$ {\scriptsize[$+0.44$, $+0.68$]} & $+0.86$ {\scriptsize[$+0.77$, $+0.95$]} & $+1.03$ {\scriptsize[$+0.76$, $+1.30$]} \\
 & & \emph{annealed $-$ not annealed} & & $+2.10$ {\scriptsize[$+0.94$, $+3.26$]} & $+2.71$ {\scriptsize[$+1.31$, $+4.11$]} & $+2.80$ {\scriptsize[$+2.55$, $+3.05$]} & $+3.07$ {\scriptsize[$+0.33$, $+5.80$]} & & & \\
\midrule
BadNets & 5\% & swapped (\ms, first milestone 100) & $4/5$ & 90.25 & $-2.70$ {\scriptsize[$-3.92$, $-1.49$]} & $-2.82$ {\scriptsize[$-4.01$, $-1.62$]} & $-2.46$ {\scriptsize[$-3.78$, $-1.13$]} & $-2.29$ {\scriptsize[$-2.59$, $-1.99$]} & $-1.80$ {\scriptsize[$-2.49$, $-1.12$]} & $-1.98$ {\scriptsize[$-2.26$, $-1.71$]} \\
 &  & shipped (\cosA) & $4/5$ & 92.51 & $+1.36$ {\scriptsize[$+1.11$, $+1.62$]} & $+0.07$ {\scriptsize[$-0.30$, $+0.44$]} & $+2.26$ {\scriptsize[$+1.76$, $+2.76$]} & $+0.11$ {\scriptsize[$+0.00$, $+0.22$]} & $+1.11$ {\scriptsize[$+0.55$, $+1.67$]} & $+0.26$ {\scriptsize[$+0.08$, $+0.43$]} \\
 & & \emph{annealed $-$ not annealed} & & $+2.26$ {\scriptsize[$+1.14$, $+3.38$]} & $+3.86$ {\scriptsize[$+1.71$, $+6.01$]} & $+2.77$ {\scriptsize[$+0.39$, $+5.15$]} & $+4.68$ {\scriptsize[$+1.84$, $+7.51$]} & & & \\
\midrule
BadNets & 1\% & swapped (\ms, first milestone 100) & $0/5$ & 91.01 & -- & -- & -- & $-2.29$ {\scriptsize[$-2.59$, $-1.99$]} & $-1.80$ {\scriptsize[$-2.49$, $-1.12$]} & $-1.98$ {\scriptsize[$-2.26$, $-1.71$]} \\
 &  & shipped (\cosA) & $0/5$ & 93.45 & -- & -- & -- & $+0.11$ {\scriptsize[$+0.00$, $+0.22$]} & $+1.11$ {\scriptsize[$+0.55$, $+1.67$]} & $+0.26$ {\scriptsize[$+0.08$, $+0.43$]} \\
 & & \emph{annealed $-$ not annealed} & & $+2.45$ {\scriptsize[$+1.77$, $+3.12$]} & -- & -- & -- & & & \\
\bottomrule
\end{tabular}

%% file: supplement/sched/build/tab_sched22_body.tex
\begin{tabular}{lcccccc}
\toprule
cell & tare, rule O {\scriptsize[95\% $t$]} & tare, rule D {\scriptsize[95\% $t$]} & D $-$ O {\scriptsize[95\% $t$]} & A3, O & A3, D & $n$ \\
\midrule
BadNets (registered) & $+2.49$ {\scriptsize[$+0.39$, $+4.60$]} & $+6.44$ {\scriptsize[$+5.11$, $+7.77$]} & $+3.95$ {\scriptsize[$+2.35$, $+5.55$]} & $11/11$ & $11/11$ & 11 \\
WaNet (registered) & $+1.78$ {\scriptsize[$+0.52$, $+3.04$]} & $+6.79$ {\scriptsize[$+5.88$, $+7.71$]} & $+5.01$ {\scriptsize[$+3.41$, $+6.61$]} & $8/8$ & $8/8$ & 11 \\
\midrule
Blended & $+3.86$ {\scriptsize[$+1.01$, $+6.72$]} & $+5.95$ {\scriptsize[$+4.01$, $+7.90$]} & $+2.09$ {\scriptsize[$-0.02$, $+4.20$]} & $6/8$ & $6/8$ & 8 \\
LF & $+3.25$ {\scriptsize[$+0.17$, $+6.33$]} & $+5.20$ {\scriptsize[$+3.76$, $+6.64$]} & $+1.95$ {\scriptsize[$-0.37$, $+4.27$]} & $5/5$ & $5/5$ & 5 \\
FTrojanNN & $+1.43$ {\scriptsize[$+0.42$, $+2.43$]} & $+5.78$ {\scriptsize[$+3.31$, $+8.25$]} & $+4.35$ {\scriptsize[$+1.71$, $+6.99$]} & $5/5$ & $5/5$ & 5 \\
Refool & $+4.72$ {\scriptsize[$+1.11$, $+8.34$]} & $+6.18$ {\scriptsize[$+3.54$, $+8.82$]} & $+1.45$ {\scriptsize[$-1.02$, $+3.93$]} & $5/5$ & $5/5$ & 5 \\
Trojan & $+6.53$ {\scriptsize[$+5.53$, $+7.52$]} & $+6.53$ {\scriptsize[$+5.53$, $+7.52$]} & $0.00$ {\scriptsize(same row, every seed)} & $5/5$ & $5/5$ & 5 \\
SIG & $+7.27$ {\scriptsize[$+5.52$, $+9.02$]} & $+7.27$ {\scriptsize[$+5.52$, $+9.02$]} & $0.00$ {\scriptsize(same row, every seed)} & $5/5$ & $5/5$ & 5 \\
BPP & $+4.28$ {\scriptsize[$+1.80$, $+6.75$]} & $+6.55$ {\scriptsize[$+4.61$, $+8.49$]} & $+2.27$ {\scriptsize[$+0.28$, $+4.26$]} & $5/5$ & $5/5$ & 5 \\
\bottomrule
\end{tabular}

%% file: supplement/sched/build/tab_sched23_body.tex
\begin{tabular}{llcccccccc}
\toprule
attack & schedule & FT & FP & NAD & NC$^{\dagger}$ & ANP$^{\dagger}$ & ABL$^{\dagger\natural}$ & I-BAU & SAU \\
\midrule
WaNet & never (shipped) & $-1.99$ {\tiny$\pm0.31$} & $-1.68$ {\tiny$\pm0.44$} & $-1.85$ {\tiny$\pm0.29$} & $3/11$\,\textbar\,$-1.30$ & $+1.78$ {\tiny$\pm1.26$} & $+27.77$ {\tiny$\pm6.49$}{\tiny$\flat7/11$ destroyed} & $+0.23$ {\tiny$\pm0.53$} & $-0.15$ {\tiny$\pm0.62$} \\
WaNet & cosine (swapped) & $+0.51$ {\tiny$\pm0.08$} & $+0.86$ {\tiny$\pm0.12$} & $+0.99$ {\tiny$\pm0.13$} & $6/11$\,\textbar\,$+3.31$ & $+3.96$ {\tiny$\pm1.74$} & $+30.45$ {\tiny$\pm6.47$}{\tiny$\flat7/11$ destroyed} & $+4.76$ {\tiny$\pm0.69$} & $+2.61$ {\tiny$\pm0.44$}{\tiny$\flat1/11$} \\
\midrule
BadNets & cosine (shipped) & $+0.13$ {\tiny$\pm0.07$} & $+1.05$ {\tiny$\pm0.20$} & $+0.33$ {\tiny$\pm0.09$} & $5/11$\,\textbar\,$+2.71$ & $+2.49$ {\tiny$\pm2.10$} & $+33.20$ {\tiny$\pm4.06$}{\tiny$\flat11/11$ destroyed} & $+5.70$ {\tiny$\pm1.71$} & $+2.70$ {\tiny$\pm0.39$} \\
BadNets & step (swapped) & $-2.29$ {\tiny$\pm0.30$} & $-1.80$ {\tiny$\pm0.69$} & $-1.98$ {\tiny$\pm0.28$} & $4/5$\,\textbar\,$-1.24$ & $+1.57$ {\tiny$\pm1.76$} & $+30.72$ {\tiny$\pm8.70$}{\tiny$\flat4/5$ destroyed} & $+0.26$ {\tiny$\pm0.77$} & $-0.58$ {\tiny$\pm0.33$} \\
\midrule
BPP & never (shipped) & $-2.07$ {\tiny$\pm0.75$} & $-1.82$ {\tiny$\pm1.07$} & $-1.84$ {\tiny$\pm0.64$} & $0/5$\,\textbar\,--- & $+4.28$ {\tiny$\pm2.48$} & $+29.83$ {\tiny$\pm12.18$}{\tiny$\flat3/5$ destroyed} & $+0.71$ {\tiny$\pm1.67$} & $-0.60$ {\tiny$\pm0.96$} \\
BPP & cosine (swapped) & $+0.58$ {\tiny$\pm0.33$} & $+0.94$ {\tiny$\pm0.19$} & $+0.96$ {\tiny$\pm0.31$} & $0/5$\,\textbar\,--- & $+4.17$ {\tiny$\pm2.19$} & $+32.46$ {\tiny$\pm12.42$}{\tiny$\flat5/5$ destroyed} & $+5.28$ {\tiny$\pm1.53$} & $+2.49$ {\tiny$\pm0.80$}{\tiny$\flat1/5$} \\
\bottomrule
\end{tabular}

%% file: supplement/sched/build/tab_sched23_der_body.tex
\begin{tabular}{llcccccccc}
\toprule
attack & schedule & FT & FP & NAD & NC$^{\dagger}$ & ANP$^{\dagger}$ & ABL$^{\dagger\natural}$ & I-BAU & SAU \\
\midrule
WaNet & never (shipped) & $97.5$ {\tiny(2)} & $95.8$ {\tiny(5)} & $97.7$ {\tiny(1)} & $80.0$ {\tiny(7)} & $95.6$ {\tiny(6)} & $67.8$ {\tiny(8)} & $96.5$ {\tiny(4)} & $97.2$ {\tiny(3)} \\
WaNet & cosine (swapped) & $99.1$ {\tiny(1)} & $98.6$ {\tiny(3)} & $99.0$ {\tiny(2)} & $85.4$ {\tiny(7)} & $96.7$ {\tiny(6)} & $68.3$ {\tiny(8)} & $97.0$ {\tiny(5)} & $98.1$ {\tiny(4)} \\
\midrule
BadNets & cosine (shipped) & $96.3$ {\tiny(3)} & $97.1$ {\tiny(1)} & $95.8$ {\tiny(4)} & $83.6$ {\tiny(8)} & $94.3$ {\tiny(6)} & $92.9$ {\tiny(7)} & $94.8$ {\tiny(5)} & $96.4$ {\tiny(2)} \\
BadNets & step (swapped) & $85.9$ {\tiny(8)} & $93.0$ {\tiny(5)} & $88.8$ {\tiny(6)} & $88.4$ {\tiny(7)} & $95.4$ {\tiny(3)} & $94.6$ {\tiny(4)} & $97.1$ {\tiny(2)} & $97.8$ {\tiny(1)} \\
\midrule
BPP & never (shipped) & $98.4$ {\tiny(2)} & $81.3$ {\tiny(7)} & $98.5$ {\tiny(1)} & $89.0$ {\tiny(6)} & $96.3$ {\tiny(4)} & $56.5$ {\tiny(8)} & $94.6$ {\tiny(5)} & $98.1$ {\tiny(3)} \\
BPP & cosine (swapped) & $98.2$ {\tiny(1)} & $80.5$ {\tiny(7)} & $98.1$ {\tiny(2)} & $97.8$ {\tiny(3)} & $96.6$ {\tiny(5)} & $55.2$ {\tiny(8)} & $94.4$ {\tiny(6)} & $97.0$ {\tiny(4)} \\
\bottomrule
\end{tabular}

%% file: supplement/sched/build/tab_fixedpoint_lb_body.tex
\begin{tabular}{llrrrrrrrrr}
\toprule
dataset & defense & rate & $n$ & slope & undefended range & defended range & common $n$ & common-set slope & slope w/o LIRA & healthy-annealed slope ($n$) \\
\midrule
CIFAR-10 & FT & 10\% & 11 & $+0.01$ & $31.0$ & $3.5$ & 8 & $+0.18$ & $+0.06$ & $+1.26$ (5) \\
CIFAR-10 & FT & 5\% & 5 & $-0.20$ & $2.3$ & $0.9$ & --- & --- & --- & --- \\
CIFAR-10 & FT & 1\% & 9 & $+0.01$ & $3.2$ & $0.5$ & --- & --- & --- & --- \\
CIFAR-10 & FT & 0.5\% & 5 & $-0.01$ & $3.2$ & $0.8$ & --- & --- & --- & --- \\
CIFAR-10 & FP & 10\% & 8 & $+0.40$ & $9.0$ & $4.1$ & 8 & $+0.40$ & $+0.40$ & $+0.85$ (4) \\
CIFAR-10 & FP & 5\% & 8 & $+0.70$ & $2.2$ & $1.6$ & --- & --- & --- & --- \\
CIFAR-10 & FP & 1\% & 8 & $+0.50$ & $3.2$ & $1.5$ & --- & --- & --- & --- \\
CIFAR-10 & FP & 0.5\% & 4 & $+0.50$ & $3.2$ & $1.8$ & --- & --- & --- & --- \\
CIFAR-10 & NAD & 10\% & 11 & $+0.01$ & $31.0$ & $3.4$ & 8 & $+0.22$ & $+0.07$ & $+1.08$ (5) \\
CIFAR-10 & NAD & 5\% & 5 & $-0.21$ & $2.3$ & $0.6$ & --- & --- & --- & --- \\
CIFAR-10 & NAD & 1\% & 9 & $-0.07$ & $3.2$ & $1.1$ & --- & --- & --- & --- \\
CIFAR-10 & NAD & 0.5\% & 5 & $-0.06$ & $3.2$ & $0.7$ & --- & --- & --- & --- \\
\midrule
CIFAR-100 & FT & 10\% & 7 & $+0.01$ & $65.3$ & $3.2$ & --- & --- & --- & --- \\
CIFAR-100 & FT & 5\% & 7 & $-0.16$ & $6.1$ & $1.8$ & --- & --- & --- & --- \\
CIFAR-100 & FT & 1\% & 6 & $+0.06$ & $7.4$ & $0.8$ & --- & --- & --- & --- \\
CIFAR-100 & FT & 0.5\% & 6 & $+0.15$ & $6.4$ & $1.2$ & --- & --- & --- & --- \\
CIFAR-100 & FT & 0.1\% & 6 & $+0.08$ & $6.1$ & $0.7$ & --- & --- & --- & --- \\
CIFAR-100 & FP & 10\% & 4 & $+0.45$ & $5.2$ & $2.4$ & --- & --- & --- & --- \\
CIFAR-100 & FP & 5\% & 6 & $+0.32$ & $6.1$ & $2.1$ & --- & --- & --- & --- \\
CIFAR-100 & FP & 1\% & 5 & $+0.48$ & $7.1$ & $3.7$ & --- & --- & --- & --- \\
CIFAR-100 & FP & 0.5\% & 5 & $+0.56$ & $6.4$ & $3.6$ & --- & --- & --- & --- \\
CIFAR-100 & FP & 0.1\% & 5 & $+0.62$ & $5.8$ & $3.7$ & --- & --- & --- & --- \\
CIFAR-100 & NAD & 10\% & 7 & $+0.01$ & $65.3$ & $3.8$ & --- & --- & --- & --- \\
CIFAR-100 & NAD & 5\% & 7 & $-0.10$ & $6.1$ & $1.5$ & --- & --- & --- & --- \\
CIFAR-100 & NAD & 1\% & 6 & $+0.12$ & $7.4$ & $1.6$ & --- & --- & --- & --- \\
CIFAR-100 & NAD & 0.5\% & 6 & $+0.17$ & $6.4$ & $1.3$ & --- & --- & --- & --- \\
CIFAR-100 & NAD & 0.1\% & 6 & $+0.15$ & $6.1$ & $1.2$ & --- & --- & --- & --- \\
\bottomrule
\end{tabular}

%% file: supplement/sched/build/tab_rerank_lb_body.tex
\begin{tabular}{lrrrrrr}
\toprule
defense & rate & attacks & DER (original) & rank & DER (prototype-referenced) & rank \\
\midrule
FT & 10\% & 4 & $74.25$ & 4 & $73.75$ & 4 \\
FP & 10\% & 4 & $71.89$ & 6 & $70.93$ & 6 \\
NAD & 10\% & 4 & $71.85$ & 7 & $71.22$ & 5 \\
NC & 10\% & 4 & $72.20$ & 5 & $70.70$ & 7 \\
ANP & 10\% & 4 & $94.69$ & 1 & $93.12$ & 1 \\
i-BAU & 10\% & 4 & $88.62$ & 3 & $87.20$ & 3 \\
ABL & 10\% & 4 & $90.26$ & 2 & $88.69$ & 2 \\
CLP & 10\% & 4 & $53.44$ & 8 & $51.88$ & 8 \\
DBD & 10\% & 4 & $49.50$ & 11 & $47.94$ & 12 \\
AC & 10\% & 4 & $49.45$ & 12 & $47.97$ & 11 \\
Spectral & 10\% & 4 & $50.70$ & 9 & $49.22$ & 9 \\
D-BR & 10\% & 4 & $50.28$ & 10 & $48.72$ & 10 \\
\midrule
FT (C100) & 10\% & 4 & $73.64$ & 6 & $72.55$ & 6 \\
FP (C100) & 10\% & 4 & $60.74$ & 7 & $59.16$ & 7 \\
NAD (C100) & 10\% & 4 & $73.99$ & 5 & $72.85$ & 5 \\
NC (C100) & 10\% & 4 & $96.23$ & 1 & $94.97$ & 1 \\
ANP (C100) & 10\% & 4 & $84.12$ & 4 & $82.45$ & 4 \\
i-BAU (C100) & 10\% & 4 & $86.24$ & 3 & $84.73$ & 3 \\
ABL (C100) & 10\% & 4 & $90.16$ & 2 & $88.49$ & 2 \\
CLP (C100) & 10\% & 4 & $46.50$ & 11 & $44.83$ & 11 \\
DBD (C100) & 10\% & 4 & $44.43$ & 12 & $42.76$ & 12 \\
AC (C100) & 10\% & 4 & $48.52$ & 10 & $46.85$ & 10 \\
Spectral (C100) & 10\% & 4 & $50.99$ & 8 & $49.62$ & 8 \\
D-BR (C100) & 10\% & 4 & $42.13$ & 13 & $40.46$ & 13 \\
D-ST (C100) & 10\% & 4 & $49.19$ & 9 & $47.52$ & 9 \\
\midrule
FT (C100) & 5\% & 4 & $64.41$ & 7 & $63.83$ & 7 \\
FP (C100) & 5\% & 4 & $63.98$ & 8 & $62.90$ & 8 \\
NAD (C100) & 5\% & 4 & $65.46$ & 6 & $64.81$ & 6 \\
NC (C100) & 5\% & 4 & $91.83$ & 1 & $91.12$ & 1 \\
ANP (C100) & 5\% & 4 & $74.95$ & 4 & $73.65$ & 4 \\
i-BAU (C100) & 5\% & 4 & $74.18$ & 5 & $73.19$ & 5 \\
ABL (C100) & 5\% & 4 & $85.81$ & 2 & $84.52$ & 2 \\
CLP (C100) & 5\% & 4 & $60.46$ & 9 & $59.17$ & 9 \\
DBD (C100) & 5\% & 4 & $77.61$ & 3 & $76.31$ & 3 \\
AC (C100) & 5\% & 4 & $51.47$ & 11 & $50.18$ & 11 \\
Spectral (C100) & 5\% & 4 & $50.52$ & 12 & $49.67$ & 12 \\
D-BR (C100) & 5\% & 4 & $42.93$ & 13 & $41.63$ & 13 \\
D-ST (C100) & 5\% & 4 & $55.45$ & 10 & $54.15$ & 10 \\
\midrule
FT & 10\% & 6 & $75.88$ & 4 & $75.46$ & 4 \\
FP & 10\% & 6 & $72.78$ & 5 & $71.94$ & 5 \\
NAD & 10\% & 6 & $76.83$ & 3 & $76.33$ & 3 \\
NC & 10\% & 6 & $80.95$ & 2 & $79.83$ & 2 \\
i-BAU & 10\% & 6 & $85.34$ & 1 & $84.11$ & 1 \\
\midrule
FT & 5\% & 3 & $59.88$ & 3 & $59.72$ & 3 \\
FP & 5\% & 3 & $66.11$ & 2 & $65.51$ & 2 \\
NAD & 5\% & 3 & $56.42$ & 4 & $56.22$ & 4 \\
NC & 5\% & 3 & $50.00$ & 5 & $49.28$ & 5 \\
i-BAU & 5\% & 3 & $83.24$ & 1 & $82.55$ & 1 \\
\bottomrule
\end{tabular}

%% file: supplement/sched/build/tab_sched3_body.tex
\begin{tabular}{llccccccccccc}
\toprule
attack & victim's scheduler & twin & fine-tuning & fine-pruning & NAD & clean arm (pooled) & published (pooled) & victim base & twin base & public victim & A4 & A3 (ft/fp/nad) \\
\midrule
WaNet & step $[100,\dots]$ (shipped) & 10 img & $-4.54\ \scriptstyle[-5.74,-3.34]$ & $-2.95\ \scriptstyle[-4.41,-1.50]$ & $-4.11\ \scriptstyle[-5.17,-3.04]$ & $-3.87\ \scriptstyle[-5.07,-2.66]$ & void (A0 $1/5$) & void & $65.38$ & $64.05$ & $5/5$ & $1/0/1$ \\
WaNet & cosine (swapped) & 10 img & $+0.34\ \scriptstyle[-0.15,+0.83]$ & $+4.41\ \scriptstyle[+3.68,+5.15]$ & $+0.45\ \scriptstyle[-0.10,+1.00]$ & $+1.73\ \scriptstyle[+1.43,+2.04]$ & $-0.62\ \scriptstyle[-2.02,+0.78]$$^\ddagger$ & $68.53$ & $71.39$ & --- & $5/5$ & $3/0/2$ \\
\midrule
Blended & cosine (shipped) & 1 img & $+0.46\ \scriptstyle[+0.25,+0.68]$ & $+7.55\ \scriptstyle[+6.93,+8.17]$ & $+0.28\ \scriptstyle[+0.07,+0.48]$ & $+2.76\ \scriptstyle[+2.50,+3.03]$ & $+3.54\ \scriptstyle[+3.24,+3.84]$$^\ddagger$ & $69.47$ & $70.87$ & $69.28$ & $5/5$ & $0/0/0$ \\
Blended & step $[100,\dots]$ (swapped) & 1 img & $-6.00\ \scriptstyle[-8.20,-3.80]$ & $-3.97\ \scriptstyle[-6.70,-1.25]$ & $-5.72\ \scriptstyle[-7.92,-3.51]$ & $-5.23\ \scriptstyle[-7.60,-2.86]$ & $-5.07\ \scriptstyle[-7.46,-2.68]$$^\ddagger$ & $62.30$ & $63.36$ & --- & $5/5$ & $0/0/0$ \\
\midrule
BPP & step $[100,\dots]$ (shipped) & 1 img (5\%) & $-4.89\ \scriptstyle[-6.65,-3.13]$ & $-3.28\ \scriptstyle[-5.54,-1.01]$ & $-4.62\ \scriptstyle[-6.46,-2.79]$ & $-4.26\ \scriptstyle[-6.20,-2.32]$ & $-3.96\ \scriptstyle[-4.77,-3.14]$ & $65.01$ & $65.02$ & $63.96$ & $5/5$ & $3/2/5$ \\
BPP & cosine (swapped) & 1 img (5\%) & $+0.54\ \scriptstyle[+0.12,+0.97]$ & $+6.20\ \scriptstyle[+5.23,+7.16]$ & $+0.62\ \scriptstyle[+0.21,+1.02]$ & $+2.45\ \scriptstyle[+1.95,+2.96]$ & $+2.45\ \scriptstyle[+2.22,+2.67]$ & $71.51$ & $71.77$ & --- & $5/5$ & $5/0/5$ \\
\bottomrule
\end{tabular}

%% file: supplement/sched/build/tab_sched_body.tex
\begin{tabular}{llccccccc}
\toprule
attack & victim's \texttt{lr\_scheduler} & seeds & fine-tuning & fine-pruning & NAD & pooled & twin base & sign agree \\
\midrule
WaNet & \texttt{MultiStepLR} (shipped) & $11$ & $-1.99\ \scriptstyle[-2.30,-1.68]$ & $-1.68\ \scriptstyle[-2.12,-1.24]$ & $-1.85\ \scriptstyle[-2.14,-1.56]$ & $-1.84\ \scriptstyle[-2.18,-1.50]$ & $91.50$ & $33/33$ \\
WaNet & \texttt{CosineAnnealingLR} (swapped) & $11$ & $+0.51\ \scriptstyle[+0.43,+0.59]$ & $+0.86\ \scriptstyle[+0.74,+0.97]$ & $+0.99\ \scriptstyle[+0.86,+1.12]$ & $+0.78\ \scriptstyle[+0.70,+0.87]$ & $94.18$ & $33/33$ \\
\midrule
BadNets & \texttt{CosineAnnealingLR} (shipped) & $11$ & $+0.13\ \scriptstyle[+0.06,+0.21]$ & $+1.05\ \scriptstyle[+0.84,+1.25]$ & $+0.33\ \scriptstyle[+0.24,+0.42]$ & $+0.50\ \scriptstyle[+0.42,+0.59]$ & $93.71$ & $32/33$ \\
BadNets & \texttt{MultiStepLR} (swapped) & $11$ & $-2.29\ \scriptstyle[-2.61,-1.96]$ & $-1.95\ \scriptstyle[-2.36,-1.53]$ & $-2.00\ \scriptstyle[-2.30,-1.71]$ & $-2.08\ \scriptstyle[-2.40,-1.76]$ & $91.09$ & $33/33$ \\
\bottomrule
\end{tabular}

%% file: supplement/sched/build/tab_sched2_body.tex
\begin{tabular}{llcccccccccc}
\toprule
attack & victim's scheduler & twin & fine-tuning & fine-pruning & NAD & clean arm (pooled) & published (pooled) & victim base & twin base & A4 & A3 \\
\midrule
BPP & step $[100,\dots]$ (shipped) & 1 img & $-2.18\ \scriptstyle[-2.86,-1.49]$ & $-1.87\ \scriptstyle[-2.48,-1.25]$ & $-2.00\ \scriptstyle[-2.62,-1.39]$ & $-2.01\ \scriptstyle[-2.63,-1.40]$ & $-2.01\ \scriptstyle[-2.42,-1.60]$ & $91.10$ & $91.31$ & $5/5$ & $10/15$ \\
BPP & cosine (swapped) & 1 img & $+0.55\ \scriptstyle[+0.30,+0.80]$ & $+0.97\ \scriptstyle[+0.76,+1.17]$ & $+1.11\ \scriptstyle[+0.92,+1.30]$ & $+0.87\ \scriptstyle[+0.69,+1.06]$ & $+1.28\ \scriptstyle[+1.16,+1.41]$ & $94.08$ & $94.10$ & $5/5$ & $10/15$ \\
\midrule
Input-Aware & step $[100,\dots]$ (shipped) & 10 img & $-2.14\ \scriptstyle[-2.56,-1.73]$ & $-1.93\ \scriptstyle[-2.37,-1.49]$ & $-1.93\ \scriptstyle[-2.48,-1.38]$ & $-2.00\ \scriptstyle[-2.46,-1.55]$ & $-1.84\ \scriptstyle[-2.53,-1.15]$$^\ddagger$ & $91.21$ & $90.96$ & $0/5$ & $6/15$ \\
Input-Aware & step $[30,60,90]$ (swapped) & 10 img & $+0.42\ \scriptstyle[+0.25,+0.59]$ & $+1.07\ \scriptstyle[+0.77,+1.36]$ & $+0.49\ \scriptstyle[+0.27,+0.70]$ & $+0.66\ \scriptstyle[+0.54,+0.77]$ & $+0.61\ \scriptstyle[+0.48,+0.74]$$^\ddagger$ & $93.09$ & $93.38$ & $0/5$ & $5/15$ \\
\bottomrule
\end{tabular}

%% file: supplement/sched/build/tab_sched5_body.tex
\begin{tabular}{lrrccccc}
\toprule
lr & epochs & WaNet shipped & WaNet cosine & BadNets cosine & BadNets step & range \\
\midrule
0.0001 & 5 & $91.71$ & $93.67$ & $93.71$ & $92.01$ & $2.00$ \\
0.0001 & 10 & $91.89$ & $93.68$ & $93.73$ & $92.16$ & $1.84$ \\
0.0001 & 20 & $92.10$ & $93.72$ & $93.72$ & $92.28$ & $1.62$ \\
0.0001 & 50 & $92.40$ & $93.78$ & $93.72$ & $92.49$ & $1.39$ \\
\midrule
0.001 & 5 & $92.52$ & $93.76$ & $93.70$ & $92.56$ & $1.24$ \\
0.001 & 10 & $92.81$ & $93.77$ & $93.73$ & $92.76$ & $1.01$ \\
0.001 & 20 & $93.03$ & $93.86$ & $93.71$ & $92.92$ & $0.94$ \\
0.001 & 50 & $93.19$ & $93.83$ & $93.72$ & $93.07$ & $0.76$ \\
\midrule
0.01 & 5 & $92.97$ & $93.19$ & $93.61$ & $92.90$ & $0.71$ \\
0.01 & 10 & $93.32$ & $93.34$ & $93.60$ & $93.15$ & $0.44$ \\
0.01 & 20 & $93.41$ & $93.56$ & $93.53$ & $93.29$ & $0.27$ \\
0.01 & 50 & $93.43$ & $93.64$ & $93.56$ & $93.41$ & $0.23$ \\
0.01 & 100 & $93.46$ & $93.67$ & $93.57$ & $93.45$ & $0.21$ \\
\bottomrule
\end{tabular}

%% file: supplement/sched/build/tab_sched6_body.tex
\begin{tabular}{llcccccc}
\toprule
victim's scheduler & defense & published cost & clean-arm cost & defended victim & defended twin & A3 & seeds \\
\midrule
step (shipped) & FT & $-8.44\ \scriptstyle[-20.58,+3.69]$ & $-2.30\ \scriptstyle[-2.72,-1.89]$ & $91.56$ & $91.72$ & $2/4$ & $5$ \\
 & FP & $-8.35\ \scriptstyle[-20.27,+3.57]$ & $-2.18\ \scriptstyle[-2.58,-1.78]$ & $91.46$ & $91.60$ & $2/4$ & $5$ \\
 & NAD & $-7.98\ \scriptstyle[-20.04,+4.08]$ & $-1.92\ \scriptstyle[-2.40,-1.44]$ & $91.09$ & $91.33$ & $2/4$ & $5$ \\
\midrule
cosine (swapped) & FT & $-0.76\ \scriptstyle[-5.02,+3.49]$ & $+2.09\ \scriptstyle[+1.59,+2.59]$ & $89.92$ & $90.44$ & $4/5$ & $5$ \\
 & FP & $-1.05\ \scriptstyle[-5.13,+3.04]$ & $+1.86\ \scriptstyle[+1.69,+2.03]$ & $90.20$ & $90.67$ & $3/5$ & $5$ \\
 & NAD & $-0.09\ \scriptstyle[-3.94,+3.76]$ & $+2.92\ \scriptstyle[+2.26,+3.58]$ & $89.24$ & $89.61$ & $5/5$ & $5$ \\
\bottomrule
\end{tabular}

%% file: supplement/sched/build/tab_sched9_body.tex
\begin{tabular}{lcccccc}
\toprule
defense & published cost & clean-arm cost & defended victim & defended twin & A3 & seeds \\
\midrule
FT & $-6.16\ \scriptstyle[-6.40,-5.91]$ & $+0.07\ \scriptstyle[-0.12,+0.25]$ & $90.74$ & $93.76$ & $4/5$ & $5$ \\
FP & $-5.89\ \scriptstyle[-6.25,-5.53]$ & $+1.23\ \scriptstyle[+0.79,+1.66]$ & $90.47$ & $92.61$ & $4/5$ & $5$ \\
NAD & $-5.22\ \scriptstyle[-5.38,-5.06]$ & $+0.31\ \scriptstyle[+0.04,+0.59]$ & $89.80$ & $93.52$ & $4/5$ & $5$ \\
\bottomrule
\end{tabular}

%% file: supplement/sched/build/tab_sched12_body.tex
\begin{tabular}{lrrrcccccc}
\toprule
epochs & seeds & twin & victim (A0) & FT tare & NAD tare & FP tare & FT level & NAD level & FP level \\
\midrule
$30$ & $3$ & $89.35$ & $89.54$ ($3/3$) & $-2.84\ \scriptstyle[-3.56,-2.11]$ & $-2.62\ \scriptstyle[-3.60,-1.64]$ & $-2.52\ \scriptstyle[-3.65,-1.40]$ & $92.18$ & $91.97$ & $91.87$ \\
$60$ & $3$ & $92.92$ & $91.55$ ($3/3$) & $-0.33\ \scriptstyle[-0.61,-0.06]$ & $+0.03\ \scriptstyle[-0.22,+0.28]$ & $+0.12\ \scriptstyle[-0.45,+0.68]$ & $93.25$ & $92.89$ & $92.80$ \\
$100$ & $3$ & $93.76$ & $91.62$ ($3/3$) & $+0.12\ \scriptstyle[-0.18,+0.43]$ & $+0.32\ \scriptstyle[-0.04,+0.68]$ & $+1.39\ \scriptstyle[+0.59,+2.19]$ & $93.63$ & $93.44$ & $92.37$ \\
\bottomrule
\end{tabular}

%% file: supplement/sched/build/tab_sched13_body.tex
\begin{tabular}{lrrccccccccc}
\toprule
rate & A0 & $g$ & FT tare & NAD tare & FP tare & FT published & NAD published & FP published & FT paired & NAD paired & FP paired \\
\midrule
$0.5\%$ & $0/5$ & $-0.09$ & $+0.11$ & $+0.26$ & $+1.11$ & --- & --- & --- & --- & --- & --- \\
$1\%$ & $0/5$ & $+0.24$ & $+0.11$ & $+0.26$ & $+1.11$ & --- & --- & --- & --- & --- & --- \\
$5\%$ & $4/5$ & $+1.17$ & $+0.11$ & $+0.26$ & $+1.11$ & $+1.36$ & $+2.26$ & $+0.07$ & $+2.48$ & $+3.22$ & $+0.32$ \\
$10\%$ & $5/5$ & $+2.06$ & $+0.11$ & $+0.26$ & $+1.11$ & $+0.94$ & $+2.15$ & $-0.31$ & $+2.89$ & $+3.96$ & $+0.65$ \\
$20\%$ & $5/5$ & $+3.58$ & $+0.11$ & $+0.26$ & $+1.11$ & $+0.52$ & $+1.48$ & $-0.52$ & $+3.99$ & $+4.80$ & $+1.95$ \\
\bottomrule
\end{tabular}

%% file: supplement/sched/build/tab_sched16_body.tex
\begin{tabular}{llrrccccccccc}
\toprule
attack & rate & A0 & $g$ & FT tare & NAD tare & FP tare & FT published & NAD published & FP published & FT paired & NAD paired & FP paired \\
\midrule
Blended & $1\%$ & $5/5$ & $-0.01$ & $+0.22$ & $+0.40$ & $+1.10$ & $+0.16$ & $+0.41$ & $+1.06$ & $-0.07$ & $-0.01$ & $-0.06$ \\
Blended & $5\%$ & $5/5$ & $+0.23$ & $+0.22$ & $+0.40$ & $+1.10$ & $+0.48$ & $+1.14$ & $+0.72$ & $+0.49$ & $+0.96$ & $-0.15$ \\
Blended & $10\%$ & $5/5$ & $+0.51$ & $+0.22$ & $+0.40$ & $+1.10$ & $+0.52$ & $+1.54$ & $+1.00$ & $+0.81$ & $+1.65$ & $+0.41$ \\
Blended & $20\%$ & $5/5$ & $+1.01$ & $+0.22$ & $+0.40$ & $+1.10$ & $+1.22$ & $+2.29$ & $+1.04$ & $+2.01$ & $+2.89$ & $+0.95$ \\
\midrule
LF & $1\%$ & $0/5$ & $+0.33$ & $+0.11$ & $+0.22$ & $+1.08$ & --- & --- & --- & --- & --- & --- \\
LF & $5\%$ & $5/5$ & $+0.56$ & $+0.11$ & $+0.22$ & $+1.08$ & $+0.31$ & $+1.01$ & $+0.77$ & $+0.76$ & $+1.35$ & $+0.25$ \\
LF & $10\%$ & $5/5$ & $+0.56$ & $+0.11$ & $+0.22$ & $+1.08$ & $+0.87$ & $+1.62$ & $+0.91$ & $+1.32$ & $+1.96$ & $+0.39$ \\
LF & $20\%$ & $5/5$ & $+1.06$ & $+0.11$ & $+0.22$ & $+1.08$ & $+1.26$ & $+2.02$ & $+0.95$ & $+2.21$ & $+2.86$ & $+0.93$ \\
\bottomrule
\end{tabular}

%% file: supplement/sched/build/tab_sched10_clean_body.tex
\begin{tabular}{llrrrcccccccc}
\toprule
attack & schedule & seeds & twin (\%) & A4 & FT & FP & NAD & NC$^{\dagger}$ & ANP$^{\dagger}$ & ABL$^{\dagger\natural}$ & I-BAU & SAU \\
\midrule
BadNets & cosine & $11$ & $93.71$ & $11/11$ & $+0.13$ {\tiny$\pm0.07$} & $+1.05$ {\tiny$\pm0.20$} & $+0.33$ {\tiny$\pm0.09$} & $5/11$ \textbar\ $+2.71$ & $+2.49$ {\tiny$\pm2.10$} & $+33.20$ {\tiny$\pm4.06$} & $+5.70$ {\tiny$\pm1.71$} & $+2.70$ {\tiny$\pm0.39$} \\
Blended & cosine & $8$ & $93.84$ & $8/8$ & $+0.20$ {\tiny$\pm0.11$} & $+1.12$ {\tiny$\pm0.23$} & $+0.36$ {\tiny$\pm0.15$} & $3/8$ \textbar\ $+3.06$ & $+3.86$ {\tiny$\pm2.86$} & $+27.83$ {\tiny$\pm11.98$} & $+4.95$ {\tiny$\pm1.02$} & $+3.06$ {\tiny$\pm1.76$} \\
LF & cosine & $5$ & $93.92$ & $5/5$ & $+0.11$ {\tiny$\pm0.14$} & $+1.08$ {\tiny$\pm0.46$} & $+0.22$ {\tiny$\pm0.18$} & $1/5$ \textbar\ $+3.22$ & $+3.25$ {\tiny$\pm3.08$} & $+31.28$ {\tiny$\pm15.75$} & $+5.98$ {\tiny$\pm2.26$} & $+2.83$ {\tiny$\pm0.80$} \\
FTrojanNN & cosine & $5$ & $93.84$ & $5/5$ & $+0.24$ {\tiny$\pm0.15$} & $+1.04$ {\tiny$\pm0.36$} & $+0.47$ {\tiny$\pm0.25$} & $3/5$ \textbar\ $+2.86$ & $+1.43$ {\tiny$\pm1.01$} & $+32.36$ {\tiny$\pm12.73$} & $+6.06$ {\tiny$\pm2.37$} & $+2.92$ {\tiny$\pm0.51$} \\
Refool & cosine & $5$ & $93.74$ & $5/5$ & $+0.16$ {\tiny$\pm0.15$} & $+0.91$ {\tiny$\pm0.30$} & $+0.36$ {\tiny$\pm0.16$} & $1/5$ \textbar\ $+2.79$ & $+4.72$ {\tiny$\pm3.62$} & $+35.91$ {\tiny$\pm11.28$} & $+5.66$ {\tiny$\pm1.68$} & $+3.71$ {\tiny$\pm1.95$} \\
Trojan & cosine & $5$ & $93.90$ & $0/5$ & $+0.20$ {\tiny$\pm0.14$} & $+1.01$ {\tiny$\pm0.24$} & $+0.23$ {\tiny$\pm0.15$} & $4/5$ \textbar\ $+2.58$ & $+6.53$ {\tiny$\pm1.00$} & $+37.50$ {\tiny$\pm11.55$} & $+4.82$ {\tiny$\pm1.51$} & $+2.56$ {\tiny$\pm0.40$} \\
SIG & cosine (damaged) & $5$ & $93.83$ & $1/5$ & $+0.07$ {\tiny$\pm0.18$} & $+1.23$ {\tiny$\pm0.44$} & $+0.31$ {\tiny$\pm0.28$} & $1/5$ \textbar\ $+3.98$ & $+7.27$ {\tiny$\pm1.75$} & $+34.59$ {\tiny$\pm10.97$} & $+6.64$ {\tiny$\pm1.42$} & $+3.91$ {\tiny$\pm4.61^{\flat}$} \\
WaNet & never & $11$ & $91.50$ & $11/11$ & $-1.99$ {\tiny$\pm0.31$} & $-1.68$ {\tiny$\pm0.44$} & $-1.85$ {\tiny$\pm0.29$} & $3/11$ \textbar\ $-1.30$ & $+1.78$ {\tiny$\pm1.26$} & $+27.77$ {\tiny$\pm6.49$} & $+0.23$ {\tiny$\pm0.53$} & $-0.15$ {\tiny$\pm0.62$} \\
BPP & never & $5$ & $91.48$ & $5/5$ & $-2.07$ {\tiny$\pm0.75$} & $-1.82$ {\tiny$\pm1.07$} & $-1.84$ {\tiny$\pm0.64$} & $0/5$ \textbar\ --- & $+4.28$ {\tiny$\pm2.48$} & $+29.83$ {\tiny$\pm12.18$} & $+0.71$ {\tiny$\pm1.67$} & $-0.60$ {\tiny$\pm0.96$} \\
\bottomrule
\end{tabular}

%% file: supplement/sched/build/tab_sched10_pub_body.tex
\begin{tabular}{llrrrcccccccc}
\toprule
attack & schedule & seeds & A0 & victim (\%) & FT & FP & NAD & NC$^{\dagger}$ & ANP$^{\dagger}$ & ABL$^{\dagger\natural}$ & I-BAU & SAU \\
\midrule
BadNets & cosine & $11$ & $11/11$ & $91.64$ & $+1.01$ & $-0.22$ & $+2.10$ & $+1.11$ & $+6.50$ & $+9.39$ & $+3.85$ & $+1.39$ \\
Blended & cosine & $8$ & $8/8$ & $93.41$ & $+0.62^{\ddagger}$ & $+0.97^{\ddagger}$ & $+1.58^{\ddagger}$ & $+1.28^{\ddagger}$ & $+7.03$ & $+20.31$ & $+4.10$ & $+4.12$ \\
LF & cosine & $5$ & $5/5$ & $93.36$ & $+0.87^{\ddagger}$ & $+0.91^{\ddagger}$ & $+1.62^{\ddagger}$ & $+2.43$ & $+6.78$ & $+24.40^{\ddagger}$ & $+4.81^{\ddagger}$ & $+2.52$ \\
FTrojanNN & cosine & $5$ & $5/5$ & $93.46$ & $+0.70^{\ddagger}$ & $+0.81^{\ddagger}$ & $+1.36^{\ddagger}$ & $+1.90$ & $+6.84$ & $+16.51$ & $+4.36^{\ddagger}$ & $+2.19$ \\
Refool & cosine & $5$ & $5/5$ & $92.06$ & $+0.18^{\ddagger}$ & $+0.08^{\ddagger}$ & $+0.68^{\ddagger}$ & $+0.00^{\ddagger}$ & $+6.33$ & $+39.14^{\ddagger}$ & $+4.25^{\ddagger}$ & $+1.76^{\ddagger}$ \\
Trojan & cosine & $5$ & $5/5$ & $93.30$ & $+0.86^{\ddagger}$ & $+0.86^{\ddagger}$ & $+1.79^{\ddagger}$ & $+1.46$ & $+7.29$ & $+20.85$ & $+4.01$ & $+8.18$ \\
SIG & cosine (damaged) & $5$ & $5/5$ & $84.58$ & $-6.16$ & $-5.89$ & $-5.22$ & $-2.21^{\ddagger}$ & $+5.30$ & $+32.41^{\ddagger}$ & $-1.60$ & $+5.68$ \\
WaNet & never & $11$ & $8/11$ & $90.05$ & $-3.14$ & $-3.15$ & $-2.92$ & $-1.45$ & $+5.15$ & $+20.72^{\ddagger}$ & $-0.96$ & $-1.68$ \\
BPP & never & $5$ & $5/5$ & $91.10$ & $-2.22$ & $-1.93^{\ddagger}$ & $-1.88$ & $-1.59$ & $+6.42$ & $+27.40^{\ddagger}$ & $+0.40$ & $-0.41$ \\
\bottomrule
\end{tabular}

%% file: supplement/sched/build/tab_sched11_clean_body.tex
\begin{tabular}{llrrrccccc}
\toprule
attack & schedule & seeds & twin (\%) & A4 & CLP & NPD & MCR & RNP$^{\dagger}$ & D-BR$^{\natural}$ \\
\midrule
BadNets & cosine & $11$ & $93.71$ & $11/11$ & $+0.74$ {\tiny$\pm0.42$} & $+2.64$ {\tiny$\pm0.30$} & $-0.01$ {\tiny$\pm0.07$} & $+1.76$ {\tiny$\pm3.00^{\flat}$} ($n{=}7$) & --- \\
Blended & cosine & $8$ & $93.84$ & $8/8$ & $+1.32$ {\tiny$\pm0.83$} & $+2.49$ {\tiny$\pm0.29$} & $-0.04$ {\tiny$\pm0.08$} & $+3.98$ {\tiny$\pm5.58^{\flat}$} ($n{=}7$) & --- \\
LF & cosine & $5$ & $93.92$ & $5/5$ & $+0.96$ {\tiny$\pm0.39$} & $+2.75$ {\tiny$\pm0.41$} & $-0.01$ {\tiny$\pm0.06$} & $+0.10$ {\tiny$\pm0.31^{\flat}$} ($n{=}4$) & --- \\
FTrojanNN & cosine & $5$ & $93.84$ & $5/5$ & $+0.73$ {\tiny$\pm0.37$} & $+2.71$ {\tiny$\pm0.53$} & $+0.05$ {\tiny$\pm0.09$} & $+4.01$ {\tiny$\pm7.44$} & --- \\
Refool & cosine & $5$ & $93.74$ & $5/5$ & $+1.15$ {\tiny$\pm0.82$} & $+2.51$ {\tiny$\pm0.51$} & $+0.01$ {\tiny$\pm0.09$} & $+0.52$ {\tiny$\pm1.13^{\flat}$} ($n{=}4$) & --- \\
Trojan & cosine & $5$ & $93.90$ & $0/5$ & $+1.83$ {\tiny$\pm1.14$} & $+2.95$ {\tiny$\pm0.69$} & $+0.01$ {\tiny$\pm0.14$} & $+13.01$ {\tiny$\pm13.50^{\flat}$} ($n{=}4$) & --- \\
SIG & cosine (damaged) & $5$ & $93.83$ & $1/5$ & $+0.47$ {\tiny$\pm0.22$} & $+2.51$ {\tiny$\pm0.68$} & $+0.04$ {\tiny$\pm0.09$} & $+1.83$ {\tiny$\pm5.54^{\flat}$} ($n{=}4$) & --- \\
WaNet & never & $11$ & $91.50$ & $11/11$ & $+0.96$ {\tiny$\pm0.34$} & $+8.06$ {\tiny$\pm3.88$} & $-1.78$ {\tiny$\pm0.28$} & $+12.38$ {\tiny$\pm2.89^{\flat}$} ($n{=}10$) & $+10.07$ {\tiny$\pm2.64$} ($n{=}10$) \\
BPP & never & $5$ & $91.48$ & $5/5$ & $+1.54$ {\tiny$\pm1.77$} & $+5.45$ {\tiny$\pm5.40$} & $-1.83$ {\tiny$\pm0.91$} & $+11.45$ {\tiny$\pm7.63^{\flat}$} ($n{=}4$) & $+8.96$ {\tiny$\pm3.28$} \\
\bottomrule
\end{tabular}

%% file: supplement/sched/build/tab_sched11_pub_body.tex
\begin{tabular}{llrrrccccc}
\toprule
attack & schedule & seeds & A0 & victim (\%) & CLP & NPD & MCR & RNP$^{\dagger}$ & D-BR$^{\natural}$ \\
\midrule
BadNets & cosine & $11$ & $11/11$ & $91.64$ & $+1.82^{\ddagger}$ & $+3.46^{\ddagger}$ & $+0.12$ & $+57.70^{\ddagger}$ & $+75.45^{\ddagger}$ \\
Blended & cosine & $8$ & $8/8$ & $93.41$ & $+1.19^{\ddagger}$ & $+2.39^{\ddagger}$ & $+0.40^{\ddagger}$ & $+61.51^{\ddagger}$ & $+80.43^{\ddagger}$ \\
LF & cosine & $5$ & $5/5$ & $93.36$ & $+1.85^{\ddagger}$ & $+2.54$ & $+0.25^{\ddagger}$ & $+57.17^{\ddagger}$ & $+78.96^{\ddagger}$ \\
FTrojanNN & cosine & $5$ & $5/5$ & $93.46$ & $+0.96$ & $+2.45$ & $+0.39^{\ddagger}$ & $+34.55^{\ddagger}$ & $+83.01^{\ddagger}$ \\
Refool & cosine & $5$ & $5/5$ & $92.06$ & $+0.78^{\ddagger}$ & $+2.34^{\ddagger}$ & $-0.35^{\ddagger}$ & $+72.80^{\ddagger}$ & $+72.00^{\ddagger}$ \\
Trojan & cosine & $5$ & $5/5$ & $93.30$ & $+0.81^{\ddagger}$ & $+2.23^{\ddagger}$ & $+0.40$ & $+79.21^{\ddagger}$ & $+81.73^{\ddagger}$ \\
SIG & cosine (damaged) & $5$ & $5/5$ & $84.58$ & $+0.87^{\ddagger}$ & $-4.16^{\ddagger}$ & $-5.96$ & $+46.21^{\ddagger}$ & $+67.75^{\ddagger}$ \\
WaNet & never & $11$ & $8/11$ & $90.05$ & $+2.34$ & $+0.20^{\ddagger}$ & $-2.97$ & $+40.38^{\ddagger}$ & $+15.52^{\ddagger}$ \\
BPP & never & $5$ & $5/5$ & $91.10$ & $+1.41$ & $+17.79^{\ddagger}$ & $-1.77$ & $+35.61^{\ddagger}$ & $+5.70^{\ddagger}$ \\
\bottomrule
\end{tabular}

%% file: supplement/sched/build/tab_sched10_der_body.tex
\begin{tabular}{lcccccccc}
\toprule
DER (mean over the 9 attacks) & FT & FP & NAD & NC & ANP & ABL & I-BAU & SAU \\
\midrule
published (clipped, victim-referenced) & $82.1$ & $84.3$ & $86.3$ & $79.6$ & $95.4$ & $75.4$ & $93.9$ & $95.8$ \\
prototype-referenced & $81.6$ & $83.8$ & $85.7$ & $78.8$ & $94.3$ & $74.3$ & $93.0$ & $95.0$ \\
\bottomrule
\end{tabular}

%% file: supplement/sched/build/tab_sched7_body.tex
\begin{tabular}{llcccccc}
\toprule
method & victim's scheduler & published cost & clean-arm cost & defended victim & defended twin & A3 & seeds \\
\midrule
TSBD & step (shipped) & $-2.90\ \scriptstyle[-3.49,-2.32]$ & $-1.11\ \scriptstyle[-1.86,-0.35]$ & $92.94$ & $92.56$ & $4/4$ & $5$ \\
TSBD & cosine (swapped) & $+0.43\ \scriptstyle[-0.61,+1.47]$ & $+2.59\ \scriptstyle[+1.37,+3.82]$ & $92.06$ & $91.64$ & $5/5$ & $5$ \\
\midrule
CGD & step (shipped) & $-3.69\ \scriptstyle[-4.22,-3.16]$ & $-2.43\ \scriptstyle[-3.00,-1.86]$ & $93.73$ & $93.88$ & $4/4$ & $5$ \\
CGD & cosine (swapped) & $-1.00\ \scriptstyle[-2.10,+0.10]$ & $+0.41\ \scriptstyle[+0.26,+0.57]$ & $93.49$ & $93.82$ & $5/5$ & $5$ \\
\bottomrule
\end{tabular}

%% file: supplement/sched/build/tab_sched7_sweep_body.tex
\begin{tabular}{lcccc}
\toprule
reinit.\ ratio $n$ & shipped victim & shipped twin & cosine victim & cosine twin \\
\midrule
0.01 & $-3.09$ & $-1.93$ & $+0.19$ & $+1.43$ \\
0.05 & $-3.16$ & $-1.78$ & $+0.08$ & $+1.68$ \\
0.1 & $-3.02$ & $-1.52$ & $+0.34$ & $+2.08$ \\
0.15 (default) & $-2.90$ & $-1.11$ & $+0.43$ & $+2.59$ \\
0.2 & $-2.77$ & $-0.74$ & $+0.48$ & $+3.06$ \\
0.25 & $-2.60$ & $-0.24$ & $+0.88$ & $+3.57$ \\
0.3 & $-2.49$ & $+0.15$ & $+0.88$ & $+4.31$ \\
0.4 & $-1.53$ & $+1.24$ & $+1.66$ & $+5.88$ \\
0.5 & $-0.60$ & $+2.91$ & $+2.92$ & $+10.10$ \\
0.7 & $+2.69$ & $+10.40$ & $+6.26$ & $+20.23$ \\
0.9 & $+18.30$ & $+31.57$ & $+27.33$ & $+37.46$ \\
\bottomrule
\end{tabular}